\documentclass[11pt,aps,prd,preprint,groupedaddress,tightenlines,nofootinbib,epsf,floatfix]{revtex4}
\usepackage{amsmath,amssymb,amsbsy}
\usepackage{psfrag}
\usepackage[dvipdfmx]{graphicx}

\newcommand{\nn}{\nonumber}

\newcommand{\ovl}[1]{\overline{#1}}
\newcommand{\wt}[1]{\widetilde{#1}}
\newcommand{\wh}[1]{\widehat{#1}}
\newcommand{\eqn}[1]{(\ref{#1})}
\newcommand{\p}{\partial}
\newcommand{\bpsi}{{\overline{\psi}}}

\newcommand{\bq}{{\overline{q}}}
\newcommand{\oP}{{\overline{\Pi}}}
\newcommand{\vev}[1]{\left\langle #1 \right\rangle}

\newcommand{\pslash}{p\kern-1ex /}
\newcommand{\Dslash}{{\cal D}\kern-1.5ex /}
\newcommand{\tr}{{\rm tr}}

\newcommand{\be}{\begin{equation}}
\newcommand{\ee}{\end{equation}}
\newcommand{\bea}{\begin{eqnarray}}
\newcommand{\eea}{\end{eqnarray}}
\newcommand{\beal}{\begin{align}}
\newcommand{\enal}{\end{align}}
\newcommand{\bi}{\begin{itemize}}
\newcommand{\ei}{\end{itemize}}

\newcommand{\la}{\langle}
\newcommand{\ra}{\rangle}
\newcommand{\bs}{\overline{s}}
\newcommand{\kzbar}{\bar{K}^0}
\newcommand{\msbar}{\overline{\rm MS}}

\begin{document}

\begin{flushright}
{\normalsize UTHEP-557}\\
{\normalsize UTCCS-P-40}\\
\end{flushright}

\title{
Precise determination of $B_K$ and light quark masses
in quenched domain-wall QCD
}
\author{Yousuke Nakamura$^{1}$, Sinya Aoki$^{1,2}$, Yusuke
Taniguchi$^{1,3}$ and Tomoteru Yoshi\'e$^{1,3}$\\
for the CP-PACS collaboration
}
\affiliation{
$^1$Graduate School of Pure and Applied Sciences,
University of Tsukuba,
Tsukuba 305-8571, Japan\\
$^2$Riken BNL Research Center, BNL, Upton, NY 11973, USA \\
$^3$Center for Computational Physics,
University of Tsukuba,
Tsukuba 305-8577, Japan
}

\date{\today}

%
\begin{abstract}
%
We calculate  non-perturbative renormalization factors at hadronic scale
for $\Delta S=2$
four-quark operators in quenched domain-wall QCD using the
Schr\"{o}dinger functional method.
Combining them with the non-perturbative renormalization group running
by the Alpha collaboration, our result yields the fully non-perturbative 
renormalization factor, which converts the lattice bare $B_K$ to the 
renormalization group  invariant (RGI) $\wh{B}_K$.
Applying this to the bare $B_K$ previously obtained by the CP-PACS 
collaboration  at $a^{-1}\simeq 2, 3, 4$ GeV,
we obtain  $\wh{B}_K=0.782(5)(7)$ (equivalent to
 $B_K^{\overline{\rm MS}}({\rm
 NDR}, 2 {\rm GeV}) = 0.565(4)(5)$ by 2-loop running) in the continuum
 limit, 
where the first error is statistical and the second is systematic due to
the continuum extrapolation.
Except the quenching error, the total error we have achieved is less than 2\%,
which is much smaller than the previous ones.
Taking the same procedure, 
we obtain
 $m_{u,d}^{\rm RGI}=5.613(66)$ MeV and $m_s^{\rm RGI}=147.1(17)$ MeV
(equivalent to 
 $m_{u,d}^{\overline{\rm MS}}(2 {\rm GeV})=4.026(48)$ MeV and
 $m_{s}^{\overline{\rm MS}}(2 {\rm GeV})=105.6(12)$ MeV by 4-loop
 running) in the continuum limit.
\end{abstract}

\maketitle

\section{Introduction}

In the standard model, the dimension-six four-quark operator, 
\begin{eqnarray}
\mathcal{O}_{LL}=\bs\gamma_\mu(1-\gamma_5)d\cdot\bs\gamma_\mu(1-\gamma_5)d,
\end{eqnarray}
of the low-energy effective Hamiltonian induces the $K^0-\kzbar$ mixing,
and the estimation of its hadronic matrix element $\la\kzbar|\mathcal{O}_{LL}|K^0\ra$
is required to extract Cabibbo-Kobayashi-Maskawa matrix elements from 
the experimental value of the indirect $CP$ violation parameter $\epsilon_K$.
The hadronic matrix element is parametrized by the kaon $B$
parameter $B_K$, defined by
\begin{eqnarray}
B_K = \frac{\la\kzbar|\bs\gamma_{\mu}(1-\gamma_5)d\cdot
\bs\gamma_{\mu}(1-\gamma_5)d|K^0\ra}
{(8/3)\la\kzbar|\bs\gamma_\mu\gamma_5d|0\ra\la0|\bs\gamma_\mu\gamma_5d|K^0\ra},
\label{eqn:BK}
\end{eqnarray}
and lattice QCD can provide the first principle calculation of it. 
In the past decades much effort have been devoted to the estimation of
$B_K$ by employing various quark and gauge actions
\cite{Blum:1996jf,Blum:1997mz,Aoki:1997nr,Aoki:1997ts,Aoki:1999gw,AliKhan:2001wr,Garron:2003cb,DeGrand:2003in,Becirevic:2004aj,Lee:2004qv,Aoki:2005ga,bk_tm,Dimopoulos:2007cn}.
Recently it is recognized that
an essential step toward the precise determination of $B_K$ is to
control the systematic error associated with the renormalization,
and for the precision now required, 
the non-perturbative renormalization seems necessary
\cite{Aoki:1997nr,Aoki:2005ga,bk_tm,Dimopoulos:2007cn}.
Among several non-perturbative schemes on the lattice
the Schr\"odinger functional (SF) scheme 
\cite{Luscher:1992an,Luscher:1996sc,Luscher:1996vw,Sint:2000vc}
has an advantage that systematic errors can be unambiguously controlled:
A unique renormalization scale is introduced through the box size to reduce the lattice artifact and
a large range of the renormalization scale can be covered by the step scaling function
(SSF) technique.

A few years ago the CP-PACS collaboration has calculated $B_K$ using
the quenched domain-wall QCD (DWQCD) with the Iwasaki gauge action
\cite{AliKhan:2001wr}, and
a good scaling behavior with small statistical errors has been observed.
Systematic errors associated with the perturbative renormalization
factor at one loop, however, can not be precisely estimated.
A main purpose of this paper is to  remove this uncertainty of the
renormalization factor,
by evaluating it non-perturbatively.

We adopt the SF scheme to control systematic uncertainties due to the
finite lattice spacing.
In the SF schemes,
the renormalization factor $\mathcal{Z}_{B_K}(g_0)$,
which convert the bare $B_K$ to the renormalization group invariant
(RGI) $\wh{B}_K$, 
is decomposed into three steps as
\begin{equation}
\mathcal{Z}_{B_K}(g_0)=
Z_{VA+AV}^{\rm PT}(\infty,\mu_{\rm max})
Z_{VA+AV}^{\rm NP}(\mu_{\rm max},\mu_{\rm min})
Z_{B_K}^{\rm NP}(g_0,a\mu_{\rm min})
\end{equation}
at a given bare coupling. 

The first one is the renormalization factor at the hadronic scale 
$\mu_{\rm min}$, which is given by
\begin{equation}
Z_{B_K}^{\rm NP}(g_0,a\mu_{\rm min}) =
\frac{Z_{VV+AA}(g_0,a\mu_{\rm min})}{Z_A^2(g_0)},
\end{equation}
where $Z_{VV+AA}$ and $Z_A$ are the renormalization factors for the
parity even part of ${\mathcal O}_{LL}$ and for the axial vector
current, respectively.
Here $a\mu_{\rm min}\ll1$ should always be satisfied to keep
the lattice artifact small enough for the reliable continuum extrapolation.
This factor depends on both renormalization scheme and lattice
regularization.
Multiplying it by the lattice bare operator, the regularization
dependence is removed and only the scheme dependence remains.

$Z_{VA+AV}^{\rm NP}(\mu_{\rm max},\mu_{\rm mix})$ represents
non-perturbative RG running for the parity odd part of ${\mathcal O}_{LL}$,
from the low energy scale $\mu_{\rm min}$ to the high
energy scale $\mu_{\rm max}=2^7\mu_{\rm min}$ where perturbation theory
can be safely applied.
Among three steps this part requires the most extensive calculation.
Since this factor does not depend on a specific lattice regularization after the continuum extrapolation,
we can employ $Z_{VA+AV}^{\rm NP}(\mu_{\rm max},\mu_{\rm min})$
evaluated previously by the Alpha collaboration with the improved Wilson
fermion action\cite{Guagnelli:2005zc}, instead of calculating it by ourselves.
Note that  the renormalization factors for the parity even and the parity odd
parts agree after the continuum extrapolation, thanks to the chiral symmetry.

The last factor $Z_{VA+AV}^{\rm PT}(\infty,\mu_{\rm max})$ is the RG
evolution from the high energy scale $\mu_{\rm max}$ to infinity, which
absorbs the scale dependence to give the RGI operator.
Since we are already deep in the perturbative region at $\mu_{\rm max}$, we
can evaluate this factor perturbatively, using
the two loop calculation in Ref.~\cite{Palombi:2005zd}.
Note that the scheme dependence is also removed at this stage and the RGI
operator becomes scheme independent.

Our target in this paper is the calculation of the first factor
$Z_{B_K}^{\rm NP}(g_0,a\mu_{\rm min})$.
In order to further reduce the computational cost, we use a relation that
$Z_{V}=Z_A$ implied by the chiral symmetry of the DWQCD in SF scheme
\cite{CPPACSZvZa03}, together with another one that  $Z_{VV+AA} =
Z_{VA+AV}$, which will be checked numerically in this paper. 
Therefore, throughout this paper,
we adopt the following definition, 
\begin{equation}
Z_{B_K}(g_0,\mu) = \frac{Z_{VA+AV}(g_0,a\mu)}{Z_V^2(g_0)}.
\label{eqn:ZBK}
\end{equation}

This paper is organized as follows.
In section \ref{sec:sf-scheme} we introduce the SF renormalization
scheme and RGI operator for ${B}_K$ following the Alpha collaboration.
Numerical simulation details are described in section \ref{sec:action}.
In section \ref{sec:NPRBK} we present our main results for the
non-perturbatively renormalized RGI $\wh{B}_K$, 
and we discuss its continuum extrapolation. 
We have also made several numerical checks of our formulation.
Section \ref{sec:NPRmass} is devoted to the non-perturbative
renormalization of light quark masses.
Our conclusion and discussion are given in section \ref{sec:conclusion}.

\section{Schr\"odinger functional scheme and RGI operator}
\label{sec:sf-scheme}

\subsection{Renormalization group invariant operator}

A bare $n$-point correlation function on the lattice,
\begin{eqnarray}
G_0(x_1,\cdots,x_n;a;g_0,m_0)=\vev{O_1(x_1)\cdots O_n(x_n)},
\end{eqnarray}
is multiplicatively renormalized in the mass
independent scheme as
\begin{eqnarray}
G_R(x_1,\cdots,x_n;\mu;g_R(\mu),m_R(\mu))
=\left(\prod_{i=1}^nZ_{O_i}(g_0,a\mu)\right)G_0(x_1,\cdots,x_n;a;g_0,m_0),
\end{eqnarray}
where $g_R$ and $m_R$ are the gauge coupling and the quark mass
respectively, 
while corresponding bare quantities have the subscript $0$.

The RG equation for the $n$-point function 
reads
\begin{eqnarray}
&&
\left(\mu\frac{\p}{\p\mu}+\beta(g_R)\frac{\p}{\p g_R}
+\tau(g_R)m_R\frac{\p}{\p m_R}
-\sum_{i=1}^{n}\gamma_{O_i}(g_R)\right)
G_R(x_1,\cdots,x_n;\mu;g_R(\mu),m_R(\mu))=0,
\nn\\
\end{eqnarray}
where
\begin{eqnarray}
&&
\beta(g_R)=\mu\frac{\p g_R(\mu)}{\p\mu},
\qquad
\tau(g_R)=\frac{\mu}{m_R(\mu)}\frac{\p m_R(\mu)}{\p\mu},
\\&&
\gamma_{O_i}(g_R)=
\lim_{a\to0}\frac{1}{Z_{O_i}(g_0,a\mu)}\mu\frac{\p Z_{O_i}(g_0,a\mu)}{\p\mu}.
\end{eqnarray}
From the RG equation,
the finite scale evolution of $G_R$ from $\mu$ to $\mu'$ is calculated as
\begin{eqnarray}
G_R(x_1,\cdots,x_n;\mu';g_R(\mu'),m_R(\mu'))
=\left(\prod_{i=1}^nU_i(\mu',\mu)\right)
G_R(x_1,\cdots,x_n;\mu;g_R(\mu),m_R(\mu)),
\end{eqnarray}
where
\begin{eqnarray}
U_i(\mu',\mu)=\exp\left(
\int_{\ovl{g}(\mu)}^{\ovl{g}(\mu')}dg\frac{\gamma_{O_i}(g)}{\beta(g)}\right)
=\lim_{a\to0}\frac{Z_{O_i}(g_0,a\mu')}{Z_{O_i}(g_0,a\mu)}
\end{eqnarray}
is the scale evolution for each operator in the continuum limit.
Using this factor we can define
the RGI operator $\wh{O}(x)$ as
\begin{eqnarray}
\wh{O}(x)=
\left(\frac{g_R^2(\mu)}{4\pi}\right)^{-\frac{\gamma_{O}^{(0)}}{2b_0}}
\exp\left(-\int_{0}^{g_R(\mu)}dg\left(\frac{\gamma_{O}(g)}{\beta(g)}
-\frac{\gamma_{O}^{(0)}}{b_0g}\right)\right)
O_R(x;\mu),
\end{eqnarray}
where 
$b_0$ and $\gamma_{O}^{(0)}$ are given by
\begin{eqnarray}
&&
\beta(g)=-b_0g^3-b_1g^5-b_2g^7+\cdots
\\&&
\gamma_{O}(g)=-\gamma_{O}^{(0)}g^2-\gamma_{O}^{(1)}g^4-\gamma_{O}^{(2)}g^6
+\cdots,
\end{eqnarray}
and $O_R(x;\mu)$ is the renormalized operator at some scale $\mu$.

As mentioned in the introduction, the evaluation of the RGI operator in  the
SF scheme is decomposed into three steps.
The lattice bare operator is renormalized at scale $\mu_{\rm min}$
non-perturbatively with the first factor $Z_{O}^{\rm NP}(g_0,a\mu_{\rm min})$.
The scale evolution from $\mu_{\rm min}$ to $\mu_{\rm max}$ is given 
by the second one,
\begin{eqnarray}
Z_{O}^{\rm NP}(\mu_{\rm max},\mu_{\rm min})=U(\mu_{\rm max},\mu_{\rm min}),
\end{eqnarray}
which can be evaluated non-perturbatively using the step scaling function.
The last factor is the running from $\mu_{\rm max}$ to infinity, which can be  calculated safely by the 
perturbative expansion as 
\begin{equation}
Z_{O}^{\rm PT}(\infty,\mu_{\rm max})=
\left(\frac{g_R^2(\mu_{\rm max})}{4\pi}\right)^{-\frac{\gamma_{O}^{(0)}}{2b_0}}
\exp\left(-\int_{0}^{g_R(\mu_{\rm max})}dg\left(\frac{\gamma_{O}(g)}{\beta(g)}
-\frac{\gamma_{O}^{(0)}}{b_0g}\right)\right).
\end{equation}

\subsection{Schr\"{o}dinger functional scheme}
\label{subsec:sfscheme}

In the SF method, the renormalization scheme is specified by the choice of the
correlation function in the finite box.
Since we rely on the result by the Alpha collaboration~\cite{Guagnelli:2005zc}
for the RG running from $\mu_{\rm min}$ to $\mu_{\rm max}$,
the same correlation function must be taken as the renormalization scheme  for our definition of
$Z_{B_K}^{\rm NP}(g_0,\mu_{\rm min})$.

We here consider The following form of the correlation function,
\begin{equation}
\mathcal{F}_{\Gamma_A\Gamma_B\Gamma_C}^\pm(x_0) =
\frac{1}{L^3}\langle\mathcal{O}_{21}[\Gamma_A]\mathcal{O}_{45}[\Gamma_B]
O_{VA+AV}^\pm(x)\mathcal{O}'_{53}[\Gamma_C]\rangle,
\label{eqn:correlation}
\end{equation}
where subscripts $1\sim 5$ represent quark flavours, and
\begin{eqnarray}
O_{VA+AV}^\pm&=&
\frac{1}{2}\Bigl(
\left(\bpsi_1\gamma_\mu\psi_2\right)
\left(\bpsi_3\gamma_\mu\gamma_5\psi_4\right)
+\left(\bpsi_1\gamma_\mu\gamma_5\psi_2\right)
\left(\bpsi_3\gamma_\mu\psi_4\right)
\nn\\&&
\pm\left(
\left(\bpsi_1\gamma_\mu\psi_4\right)
\left(\bpsi_3\gamma_\mu\gamma_5\psi_2\right)
+\left(\bpsi_1\gamma_\mu\gamma_5\psi_4\right)
\left(\bpsi_3\gamma_\mu\psi_2\right)
\right)\Bigr)
\end{eqnarray}
is the parity odd four quark operator made of four different flavors.
Boundary operators $\mathcal{O}_{ij}$ and $\mathcal{O}'_{ij}$ are given
in terms of boundary fields $\zeta$ and $\zeta'$ \cite{Sint:2000vc} as
\begin{equation}
\mathcal{O}_{ij}[\Gamma]=
a^6\sum_{\vec{x}\vec{y}}\bar{\zeta}_i(\vec{x})\Gamma\zeta_j(\vec{y}),
\hspace{0.5cm}
\mathcal{O}'_{ij}[\Gamma]=
a^6\sum_{\vec{x}\vec{y}}\bar{\zeta}'_i(\vec{x})\Gamma\zeta'_j(\vec{y}).
\label{eqn:boundary-operator}
\end{equation}
Due to the SF boundary condition for fermion fields the boundary
operator should be parity odd and we then have two independent choices,
$\Gamma=\gamma_5$ and $\Gamma=\gamma_k$ $(k=1,2,3)$.
For the correlation function to be totally parity-even we need at least
three boundary operators as in (\ref{eqn:correlation}).

The Alpha collaboration has adopted five independent choices
for the correlation function
\begin{eqnarray}
F_1^\pm(x_0) &=& \mathcal{F}^\pm_{\gamma_5,\gamma_5,\gamma_5}(x_0),
\notag \\
F_2^\pm(x_0) &=& \frac{1}{6}\sum_{j,k,l=1}^3
\epsilon_{jkl}\mathcal{F}^\pm_{\gamma_j,\gamma_k,\gamma_l}(x_0),
\notag \\
F_3^\pm(x_0) &=& \frac{1}{3}\sum_{k=1}^3
\mathcal{F}^\pm_{\gamma_5,\gamma_k,\gamma_k}(x_0),
\label{eq:crrelfunc.f} \\
F_4^\pm(x_0) &=& \frac{1}{3}\sum_{k=1}^3
\mathcal{F}^\pm_{\gamma_k,\gamma_5,\gamma_k}(x_0),
\notag \\
F_5^\pm(x_0) &=& \frac{1}{3}\sum_{k=1}^3
\mathcal{F}^\pm_{\gamma_k,\gamma_k,\gamma_5}(x_0).
\notag
\end{eqnarray}
To remove logarithmic divergences of boundary fields $\zeta$'s from
these correlation functions,
one can consider 
the following 9 ratio of the correlation functions, 
\begin{eqnarray}
h_i^\pm(x_0) &=& \frac{F_i^\pm(x_0)}{f_1^{3/2}},\hspace{0.5cm} i=1,\dots,5,
\notag \\
h_6^\pm(x_0) &=& \frac{F_2^\pm(x_0)}{k_1^{3/2}},
\label{eq:crrelfunc.h}\\
h_{i+4}^\pm(x_0) &=& \frac{F_i^\pm(x_0)}{f_1^{1/2}k_1},\hspace{0.5cm}
 i=3,4,5 ,
\notag 
\end{eqnarray}
where
\begin{eqnarray}
&&
f_1=-\frac{1}{2L^6}
\langle\mathcal{O}'_{12}[\gamma_5]\mathcal{O}_{21}[\gamma_5]\rangle,
\qquad
k_1=-\frac{1}{6L^6}\sum_{k=1}^{3}
\langle\mathcal{O}'_{12}[\gamma_k]\mathcal{O}_{21}[\gamma_k]\rangle,
\label{eqn:k1}
\end{eqnarray}
are the boundary-boundary correlation functions.
Each ratio, distinguished by the label $s=1,\cdots,9$,  gives a different renormalization scheme.
Among these 9 choices, $h_{s=1,3,7}^+(x_0)$ and
$h_{s=2,4,5,6,8,9}^-(x_0)$ define good schemes
\cite{Palombi:2005zd}, whose scaling violations are perturbatively shown to be small. 

The renormalization factor we need in this study is defined by
\begin{equation}
Z_{VA+AV;s}^\pm(g_0,\mu)h_s^{\pm}(x_0=L/2;g_0) =
 h_s^{\pm ({\rm tree})}(x_0=L/2),
\label{eqn:renormalization-condition}
\end{equation}
where $s$ labels the scheme, and $h_s^{\pm ({\rm tree})}$ is the correlation function at the tree level in the continuum theory.

According to Ref.~\cite{Luscher:1996jn},
the renormalization factor for the local vector current is defined
through the Ward-Takahashi identity  as
\begin{eqnarray}
&&
Z_V=\frac{f_1}{f_V(x_0=L/2)},
\qquad
f_V(x_0)=-\frac{a^3}{2L^6}\sum_{\vec{x}}
\vev{{\cal O'}_{12}[\gamma_5]V_0(\vec{x},x_0){\cal O}_{31}[\gamma_5]},
\label{eqn:zv}
\end{eqnarray}
where 
$
V_\mu(x)=\bpsi_2(x)\gamma_\mu\psi_3(x)
$ .

\section{Numerical simulation details}
\label{sec:action}

\subsection{Gauge action}

The theory is defined on an $L^3\times T$ lattice of $L=T=N_La=N_Ta$ \cite{Guagnelli:2005zc},
with the periodic boundary condition in the spatial directions
and the Dirichlet boundary condition in the temporal direction.
The dynamical gauge variables are spatial links $U_k(x)$ at
$x_0=1,\dots,N_T-1$ and temporal ones $U_0(x)$ at $x_0=0,\dots,N_T-1$.
The Dirichlet boundary condition is imposed on the spatial link at
$x_0=0$ and $N_T$ as
\begin{eqnarray}
U_k(\vec{x},x_0=0) = \exp[aC_k],\quad
U_k(\vec{x},x_0=N_T) = \exp[aC'_k],
\label{eq:sfbc.gauge}
\end{eqnarray}
where $C_k$ and $C'_k$ are anti-Hermitian diagonal matrices
\cite{Luscher:1992an,Luscher:1996sc}, which we set to zero in our
simulation.

We employ the renormalization group improved gauge action ,
\begin{eqnarray}
S_{\rm gluon}[U] = \frac{2}{g_0}\left(
W_P(g_0^2)\sum_{\rm plaquette}{\rm Re}\, \tr(I-U_{\rm pl})
+W_R(g_0^2)\sum_{\rm rectangle}{\rm Re}\, \tr(I-U_{\rm rtg})\right)
\end{eqnarray}
where $U_{\rm pl}$ represents the standard plaquette and $U_{\rm rtg}$
an $1\times2$ six-link rectangle.
The $O(a)$ lattice artifact due to temporal boundary is removed
by setting weight factors $W_P$ and $W_R$ as
\begin{eqnarray}
W_P(g_0^2) &=&
 \begin{cases}
  c_0c_t^P(g_0^2) \hspace{0.5cm}
  \text{Set of temporal plaquettes that just touch one of the boundaries}  \\
  c_0\phantom{c_t^P(g_0^2)}  \hspace{0.5cm} \text{Otherwise}
 \end{cases}
\nn\\
W_R(g_0^2) &=& 
 \begin{cases}
  c_1c_t^R(g_0^2) \hspace{0.5cm}
  \text{Set of temporal rectangles that have exactly two links on a boundary}\\
  c_1\phantom{c_t^R(g_0^2)} \hspace{0.5cm} \text{Otherwise}
 \end{cases} .
\nn
\end{eqnarray}
The coefficients $c_0$ and $c_1$ are normalized such that $c_0+8c_1=1$.
In this paper we take $c_1=-0.331$ (the Iwasaki gauge action) \cite{Iwasaki:1983ck}.
The boundary coefficients are expanded perturbatively as
\begin{eqnarray}
&&
c_0c^P_t(g_0^2) = c_0\left(1+c_t^{P(1)}g_0^2+O(g_0^4)\right),
\\&&
c_1c^R_t(g_0^2) = c_1\left(\frac{3}{2}+c_t^{R(1)}g_0^2+O(g_0^4)\right).
\end{eqnarray}
Since only a single improvement condition
\begin{eqnarray}
c_0c_t^{P(1)}+4c_1c_t^{R(1)} = 0.1518
\end{eqnarray}
is available\cite{Takeda:2003he},  there exists no unique choice.
Therefore, in this study,  we adopt the condition A~\cite{Takeda:2004xh} that
$ c_t^{R(1)} = 2c_t^{P(1)}$ .

\subsection{Fermion action}

In this paper we adopt the orbifolding construction of the
SF formalism for the domain-wall fermion
\cite{Taniguchi:2004gf,Taniguchi:2005gh,Taniguchi:2006qw}.
Instead of folding the temporal direction as was discussed in
Ref.~\cite{Taniguchi:2006qw}, we keep both positive and negative regions in 
the temporal direction,
in order to implement the even-odd preconditioning in five dimensions.
The gauge link in the negative region is defined  to satisfy the time reflection symmetry as
\begin{eqnarray}
U_k(\vec{x},x_0) = U_k(\vec{x},-x_0),\quad
U_0(\vec{x},x_0) = U_0^\dagger(\vec{x},-x_0-1).
\label{eq:orbfld.gauge}
\end{eqnarray}

We implement the Shamir's domain-wall fermion
action~\cite{Shamir:1993zy,Furman:1994ky} on
$2N_T\times N_L^3\times N_5$ lattice,
\begin{eqnarray}
S_{\rm dwf} = \sum_{\vec{x},\vec{y}}\sum_{x_0,y_0=-N_T+1}^{N_T}
\sum_{s,t=1}^{N_5}\bpsi(x,s)D_{\rm dwf}(x,s;y,t)\psi(y,t),
\end{eqnarray}
where the temporal coordinates $x_0$ and $y_0$  run from $-N_T+1$ to $N_T$, while 
the fifth dimensional coordinates $s$ and $t$ from $1$ to $N_5$.
For the orbifolding we set the anti-periodic boundary condition in the temporal
direction,
\begin{eqnarray}
\psi(\vec{x},x_0+2N_T,s) = -\psi(\vec{x},x_0,s),\quad
\bpsi(\vec{x},x_0+2N_T,s) = -\bpsi(\vec{x},x_0,s).
\end{eqnarray}
On the other hand,
the periodic boundary condition with the phase $\theta=1/2$ 
in spatial directions\cite{Luscher:1996sc,Guagnelli:2005zc}, 
\begin{eqnarray}
\psi({x}_k+N_L,x_0,s) = e^{i\theta}\psi({x}_k,x_0,s),\quad
\bpsi({x}_k+N_L,x_0,s) = e^{-i\theta}\bpsi({x}_k,x_0,s),
\end{eqnarray}
is imposed by replacing the spatial gauge link as $
U_k(x)\rightarrow e^{i\theta/N_L}U_k(x)
$.
We set the physical quark mass to be zero for the mass independent
(massless) scheme.

The physical quark field is defined in the standard manner as
\begin{eqnarray}
&&
q(x) = (P_L\delta_{s,1} + P_R\delta_{s,N_5})\psi(x,s),\quad
\ovl{q}(x) = \bpsi(x,s)(\delta_{s,N_5}P_L + \delta_{s,1}P_R),
\label{eq:physquark}
\end{eqnarray}
with $P_{L/R} = (1\pm\gamma_5)/2$, and 
its propagator on the $2N_T\times N_L^3\times N_5$ lattice
is given by 
\begin{eqnarray}
&&
G_{\rm quark}(x,y)=
\left(P_L\delta_{s,1}+P_R\delta_{s,N_5}\right)
\left(\frac{1}{D_{\rm dwf}}\right)_{x,y;s,t}
\left(\delta_{t,N_5}P_L+\delta_{t,1}P_R\right).
\end{eqnarray}
Imposing the orbifolding projection we get the physical quark propagator 
in the SF formalism as
\begin{eqnarray}
&&
G_{\rm quark}^{\rm SF}(x,y)=2(\Pi_-G_{\rm quark}\Pi_+)_{x,y},
\qquad
\Pi_\pm =\frac{1\pm\gamma_0R}{2},
\label{eq:quark-prop}
\end{eqnarray}
where $R$ is the time reflection operator:
$R_{x_0,y_0}q(\vec{x},y_0)=q(\vec{x},-x_0)$.
Due to the projection, the physical quark fields satisfy the proper
homogeneous SF Dirichlet boundary condition at $x_0=0,N_T$ such that
\begin{eqnarray}
&&
P_+q(x)|_{x_0=0}=0,\quad
P_-q(x)|_{x_0=N_T}=0,
\\&&
\bq(x)P_-|_{x_0=0}=0,\quad
\bq(x)P_+|_{x_0=N_T}=0.
\label{eqn:DirichletBC}
\end{eqnarray}
As usual,
the boundary-bulk and boundary-boundary propagator are constructed in terms of the SF
quark propagator \eqn{eq:quark-prop} \cite{Luscher:1996vw}.

\subsection{Parameters}
\label{sec:parameter}

The CP-PACS collaboration has calculated the lattice bare $B_K$
in quenched DWQCD with the Iwasaki gauge action at the domain wall height
$M=1.8$ and the fifth dimensional length $N_5=16$ \cite{AliKhan:2001wr}.
In order to renormalize this $B_K$ we have to take the same lattice
formulation.
The bare value of $B_K$, calculated at 
three lattice spacings $\beta=2.6$, $2.9$
and $3.2$\footnote{The data at $\beta=3.2$ is new and not published in
\cite{AliKhan:2001wr}.
Its numerical analysis is briefly given in appendix \ref{sec:appendixB}.}
($a^{-1}\sim 2$, $3$ and $4$ GeV)
in the previous simulation\cite{AliKhan:2001wr},
is listed in table \ref{tab:bare_bk}.

The  renormalization scale at the low energy (hadronic scale) is
introduced as $1/\mu_{\rm min}=2L_{\rm max}$,
where $L_{\rm max}$ is defined through the renormalized coupling
$\ovl{g}^2(1/L_{\rm max})=3.480$ in the SF scheme
\cite{Guagnelli:2005zc}, and
$L_{\rm max}/r_0=0.749(18)$~\cite{Takeda:2004xh} 
in the continuum limit. ($\mu_{\rm min}=1/2L_{\rm max}\sim263$ MeV
for $r_0=0.5$ fm). 
At $\beta=2.6$, $2.9$ and $3.2$,
$N_L$ which satisfies $aN_L=2L_{\rm max}=1.498r_0$ can be estimated, using
the interpolation formula \cite{Takeda:2004xh},
\begin{equation}
\ln\left(\frac{a}{r_0}\right) = -2.193 -1.344(\beta-3) + 0.191(\beta-3)^2, 
\label{eq:interpolate}
\end{equation}
valid at $2.456\leq\beta\leq3.53$.
To cover the resulting lattice sizes, $N_L = 7.60625, 11.7144, 17.4317$,  
we take
7 lattice sizes, $N_L=6,\cdots,18$, and  using the formula (\ref{eq:interpolate}) again,
 we tune $\beta$ so that the physical box size satisfies $aN_L=2L_{\rm max}=1.498r_0$ at each $N_L$.

Quenched gauge configurations are generated by the HMC algorithm.
First $2000$ trajectories are discarded for thermalization, and
the correlation functions are calculated every $200$ trajectories. 
By the jackknife analysis we found that each configuration separated by 200 
trajectories is almost independent.
A value of $\beta$  and a number of configurations at each lattice size
are listed in table \ref{tab:beta-latsize}.

\section{Non-perturbative renormalization of $B_K$}
\label{sec:NPRBK}

\subsection{Extraction of renormalization factors}

\begin{figure}
    \begin{center}
      \scalebox{0.32}{\includegraphics{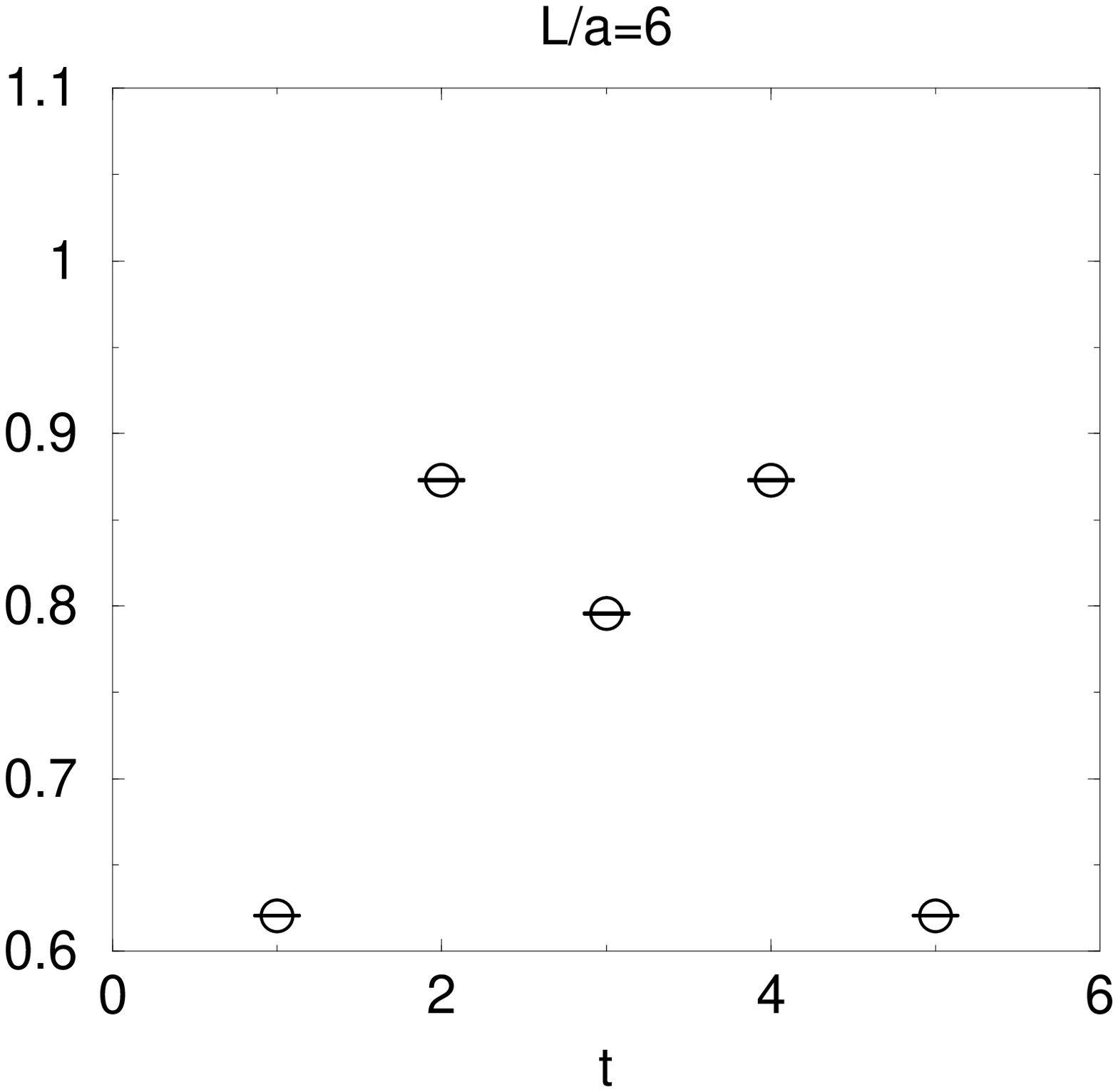}}
\qquad
      \scalebox{0.32}{\includegraphics{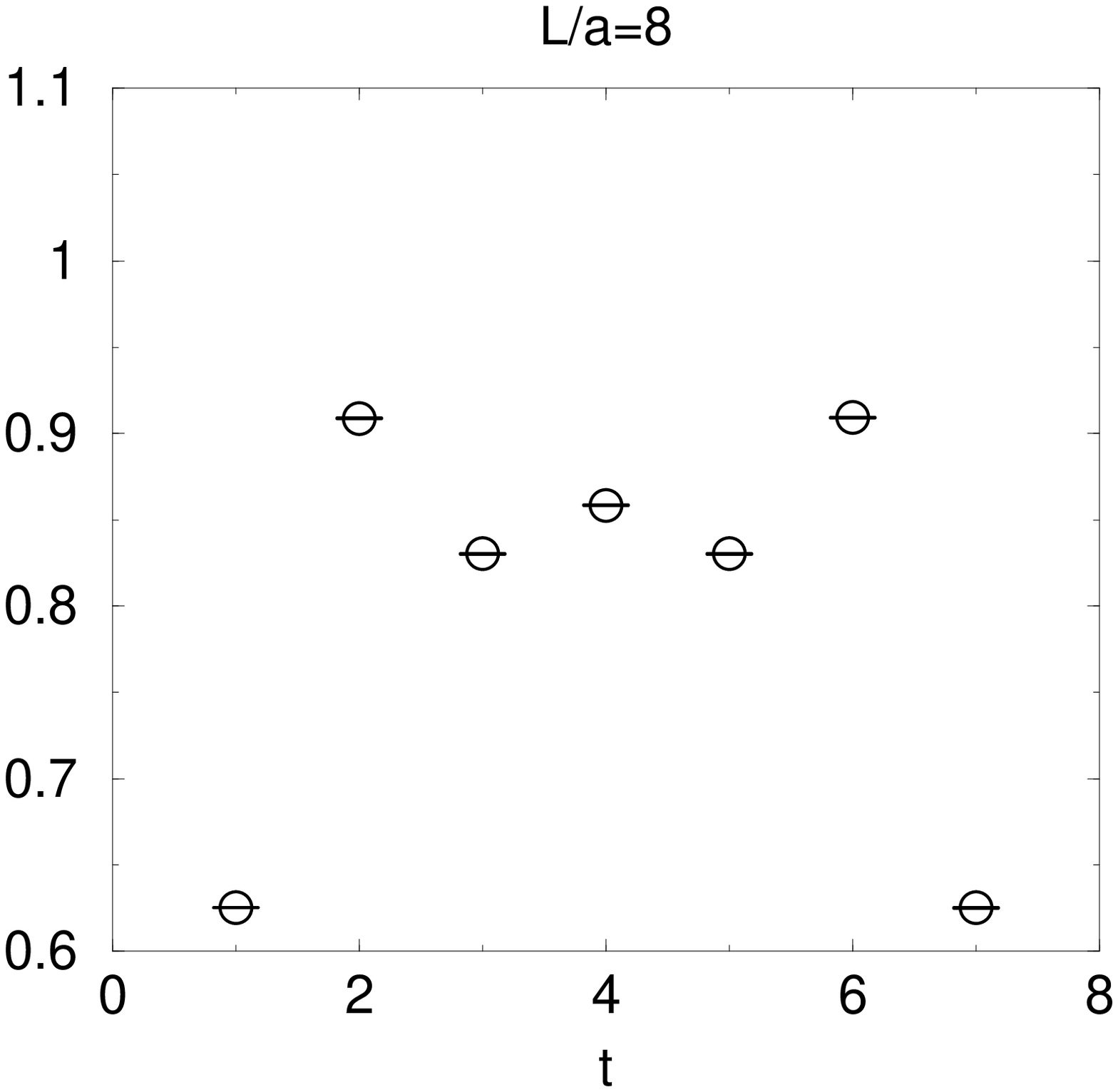}}
    \end{center}
    \begin{center}
      \scalebox{0.32}{\includegraphics{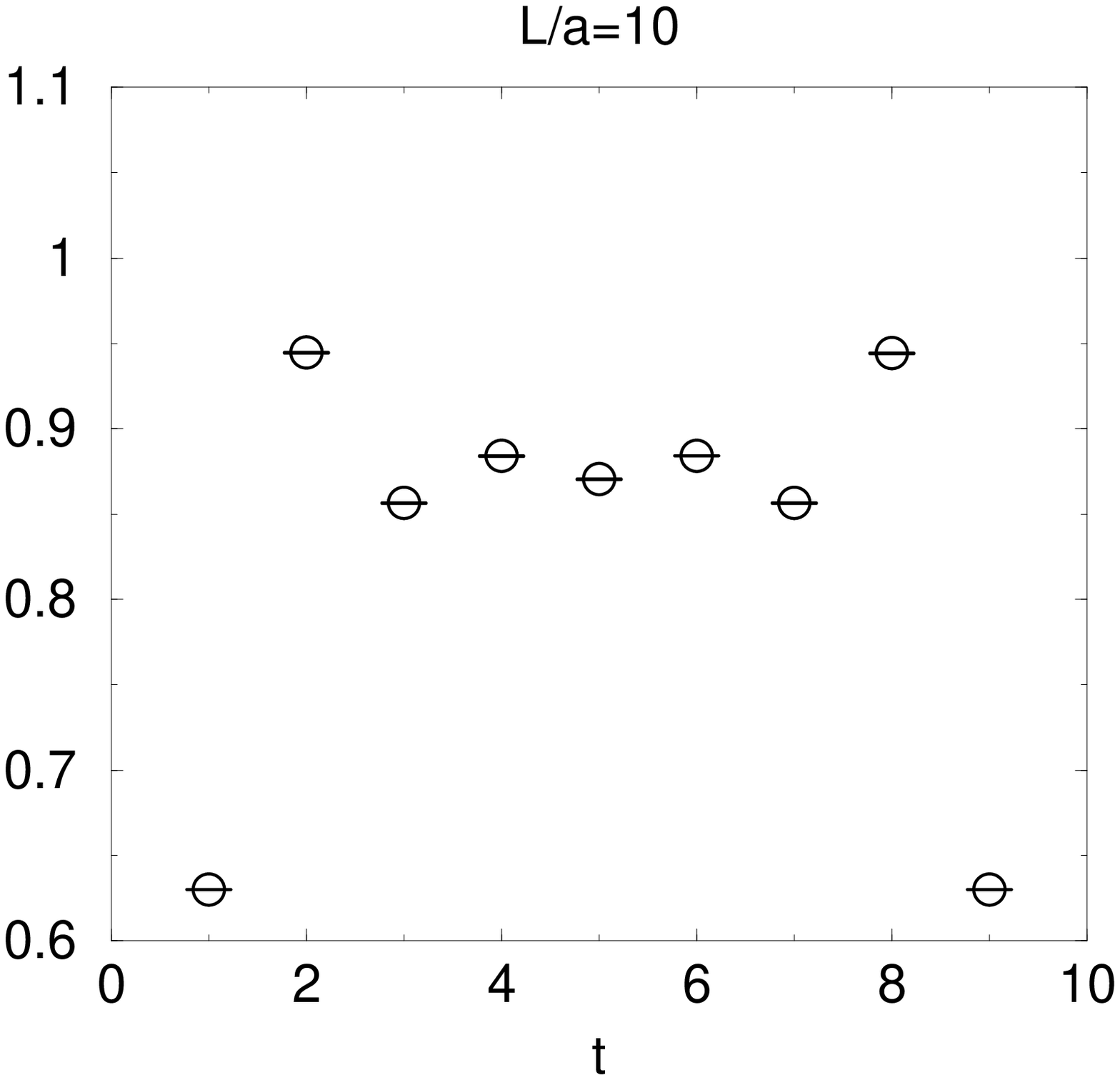}}
\qquad
      \scalebox{0.32}{\includegraphics{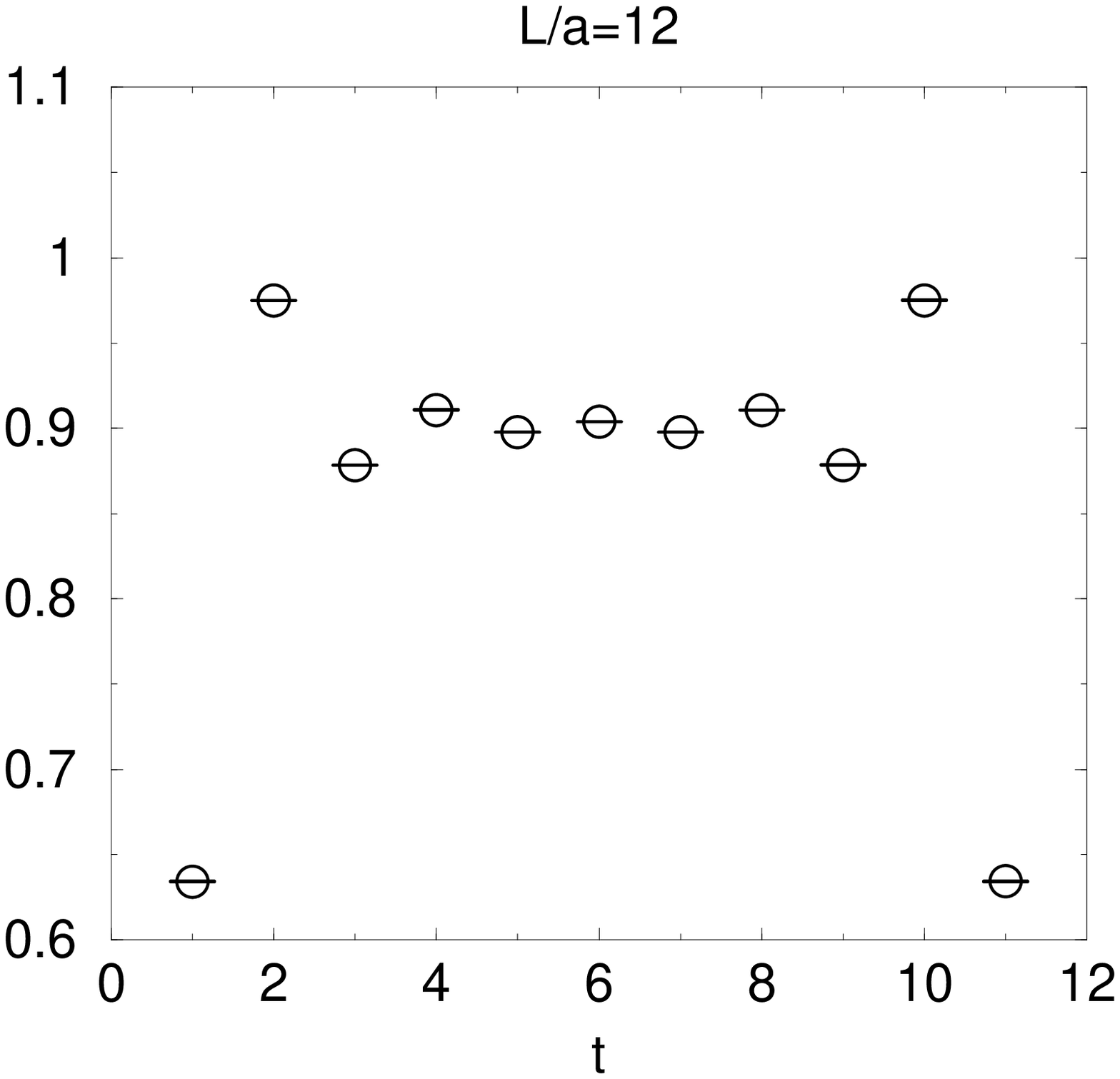}}
    \end{center}
    \begin{center}
      \scalebox{0.32}{\includegraphics{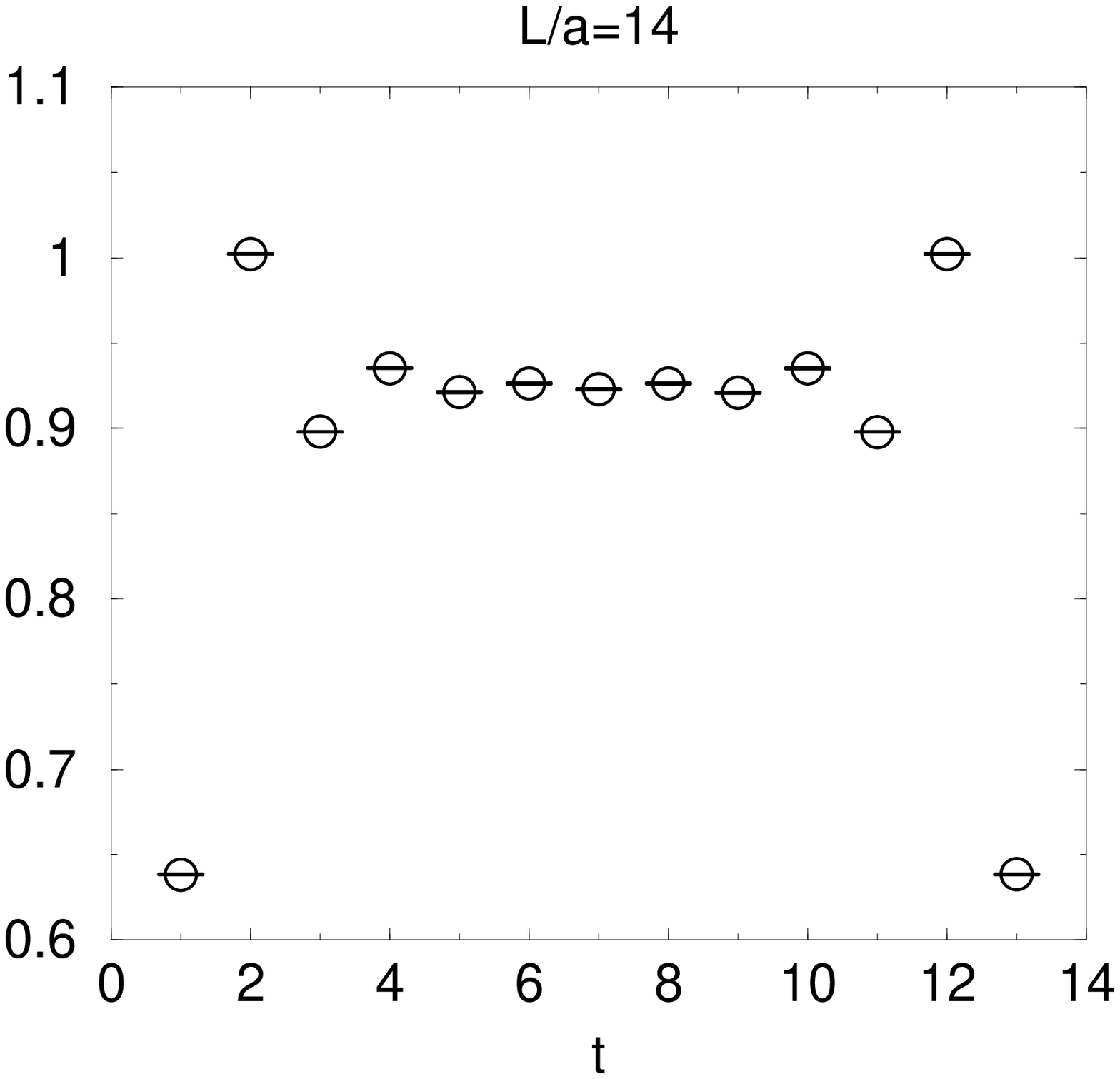}}
\qquad
      \scalebox{0.32}{\includegraphics{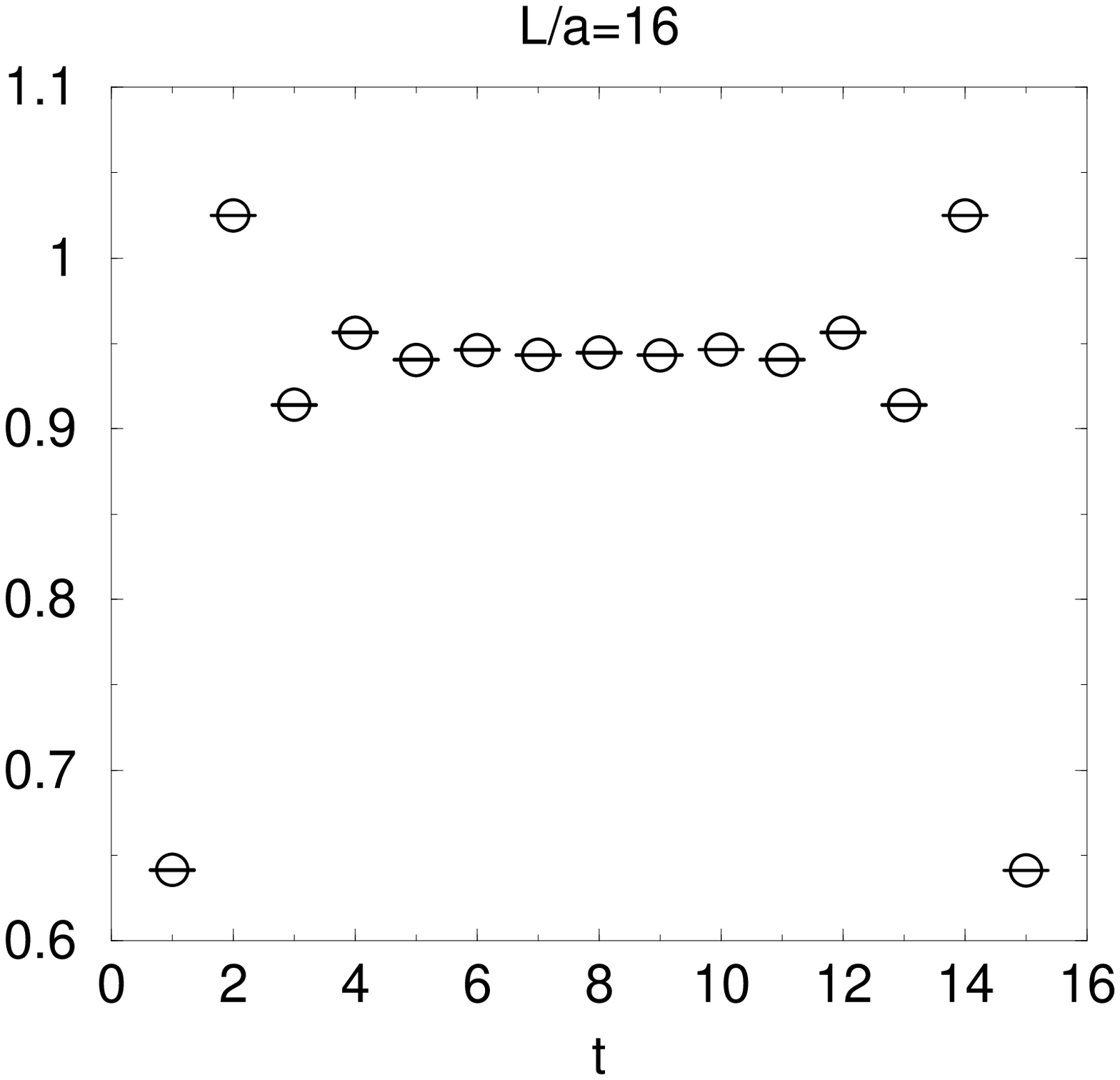}}
    \end{center}
    \begin{center}
      \scalebox{0.32}{\includegraphics{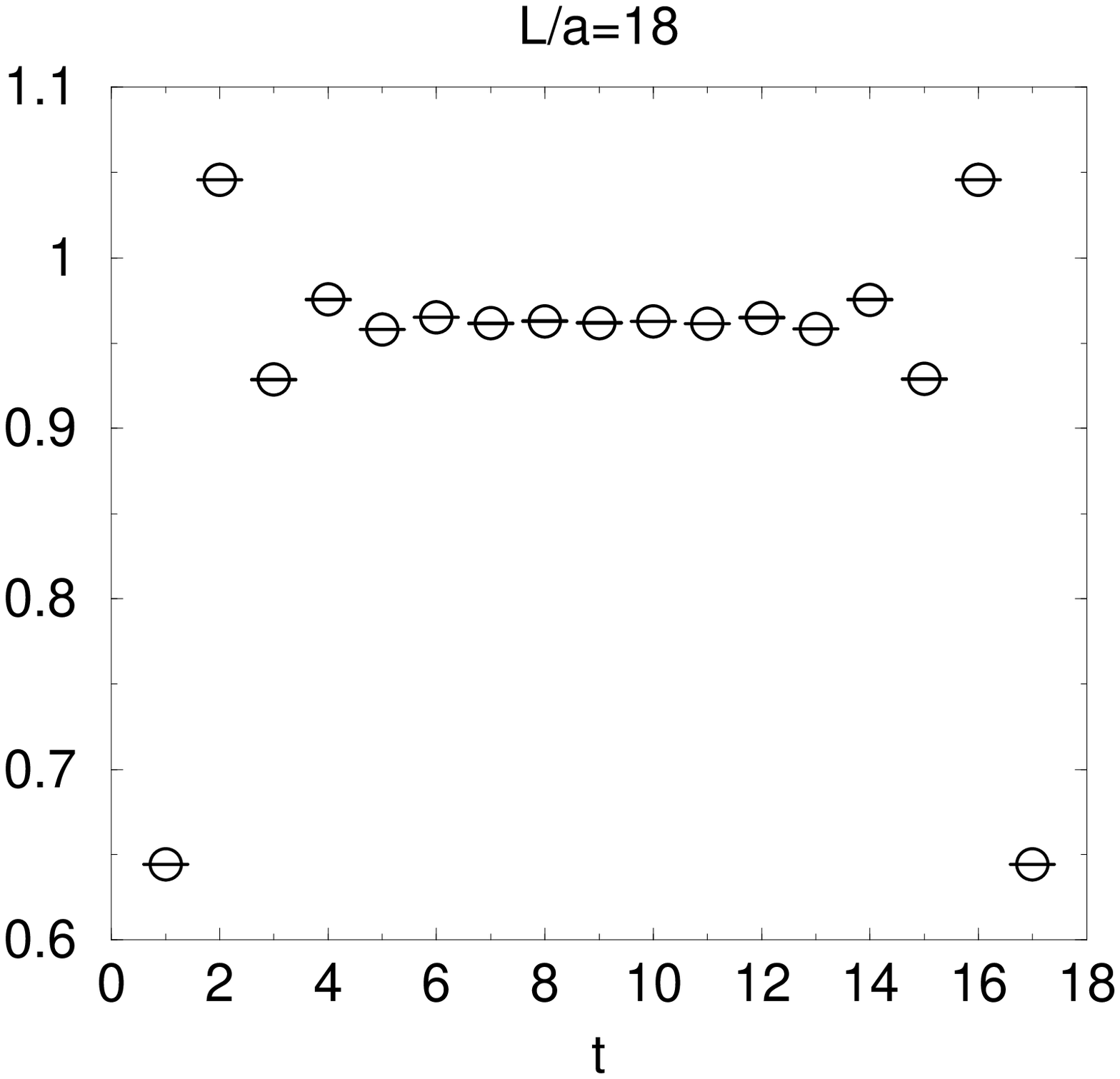}}
\qquad
      \scalebox{0.32}{\includegraphics{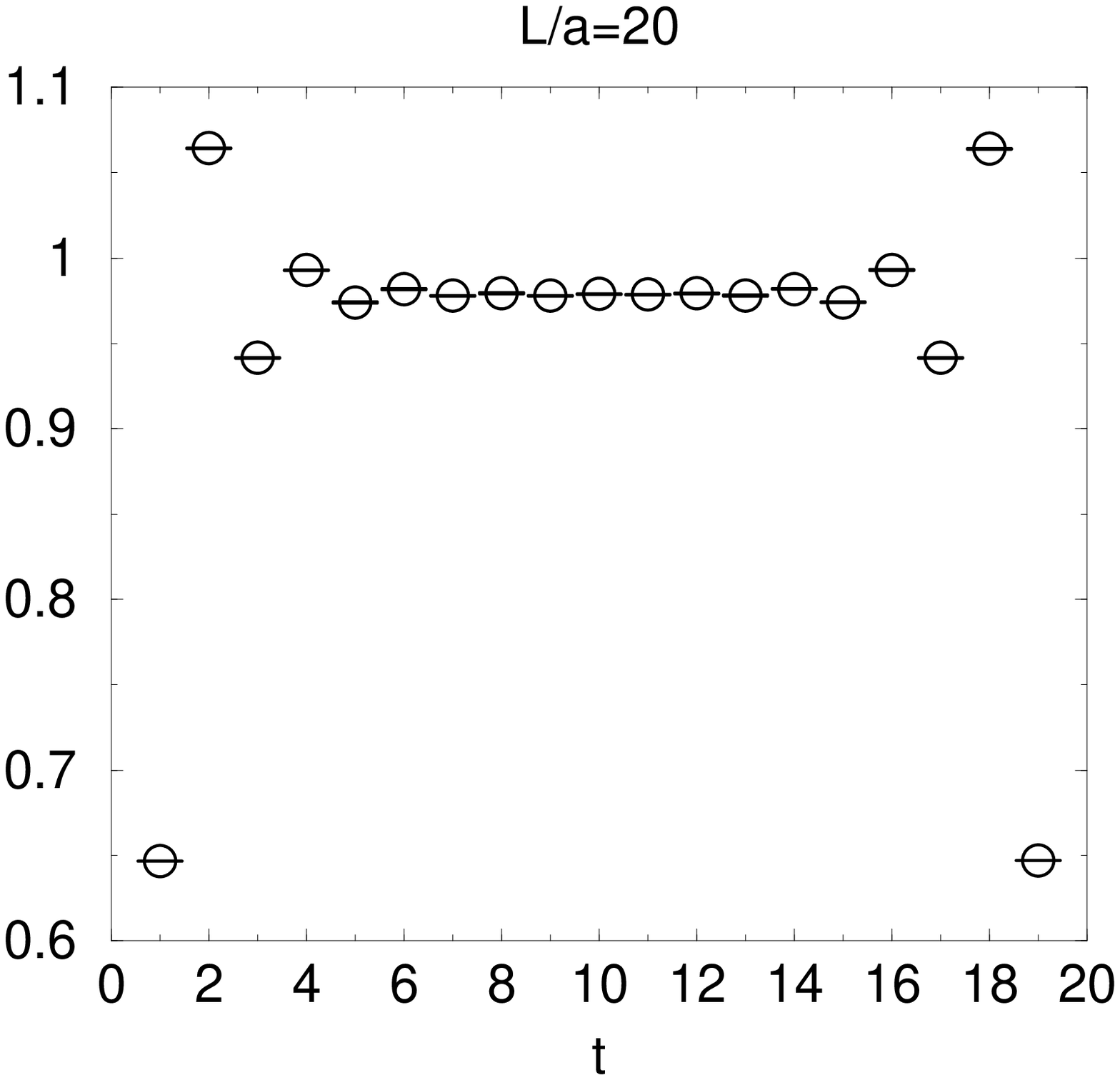}}
    \end{center}
 \caption{$x_0$ dependence of $Z_V(g_0)$ at various lattice sizes.}
 \label{fig:zv.x0dep}
\end{figure}

\begin{figure}
 \begin{center}
  \scalebox{0.32}{\includegraphics{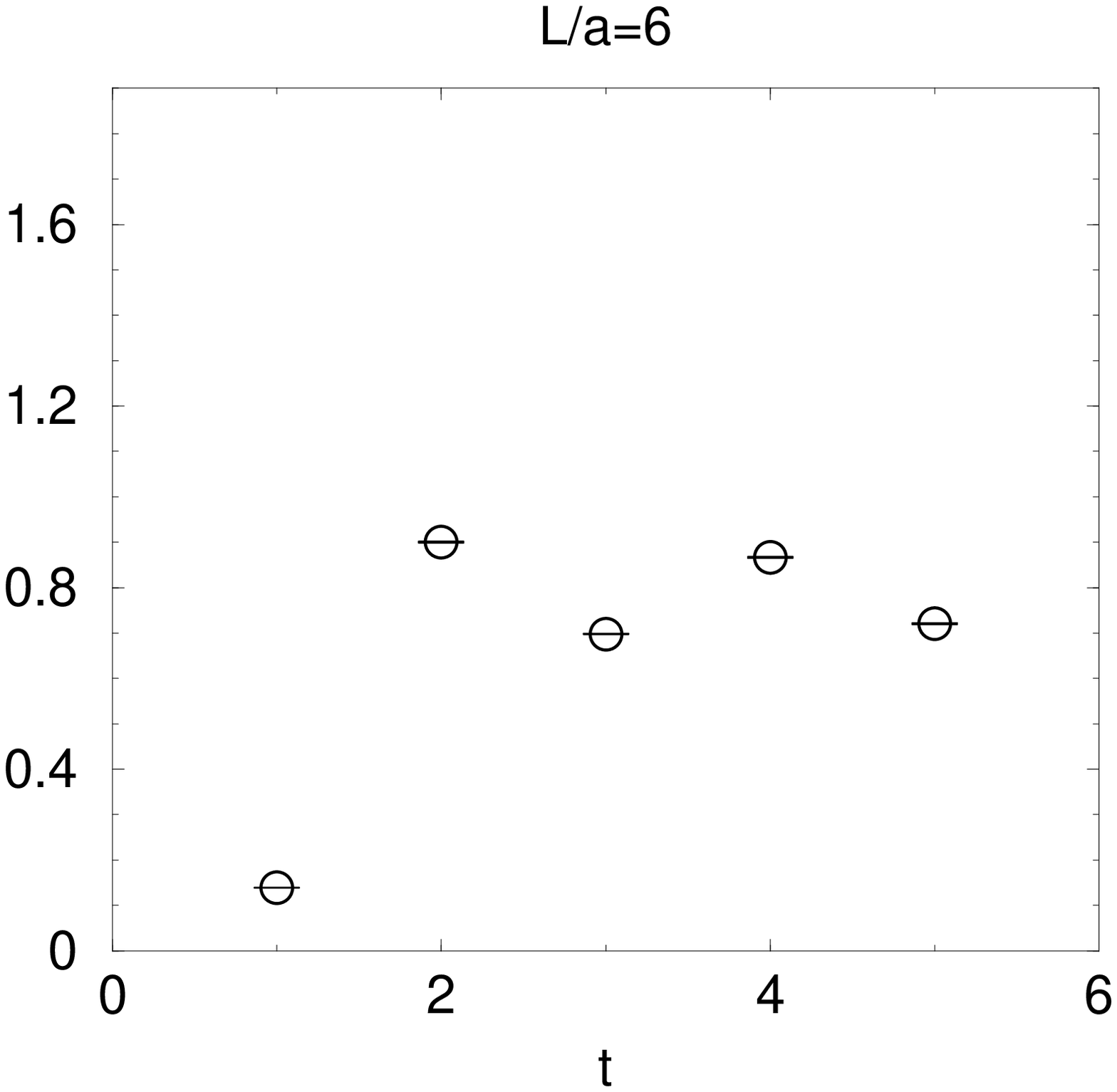}}
\qquad
  \scalebox{0.32}{\includegraphics{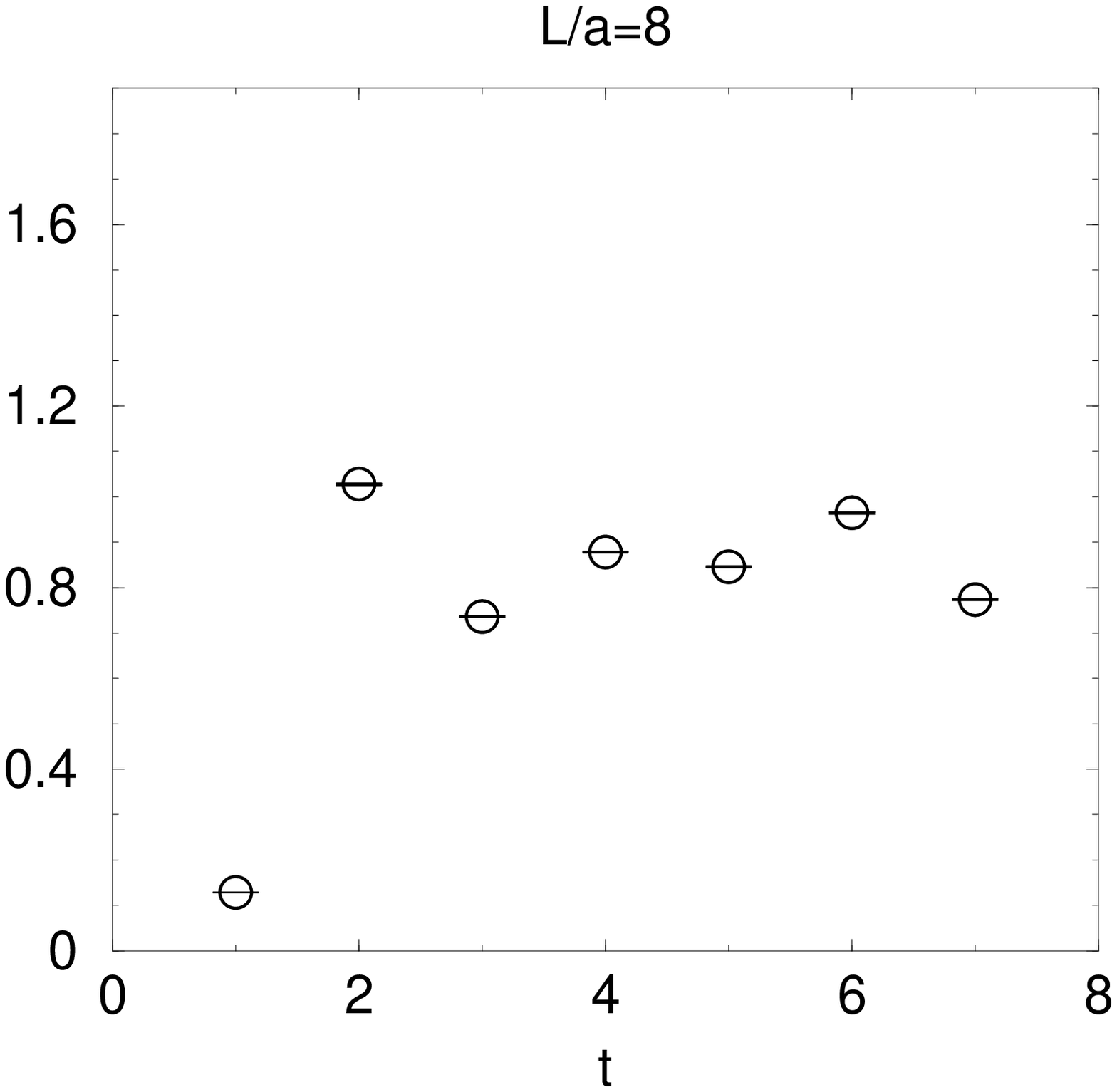}}
 \end{center}
 \begin{center}
  \scalebox{0.32}{\includegraphics{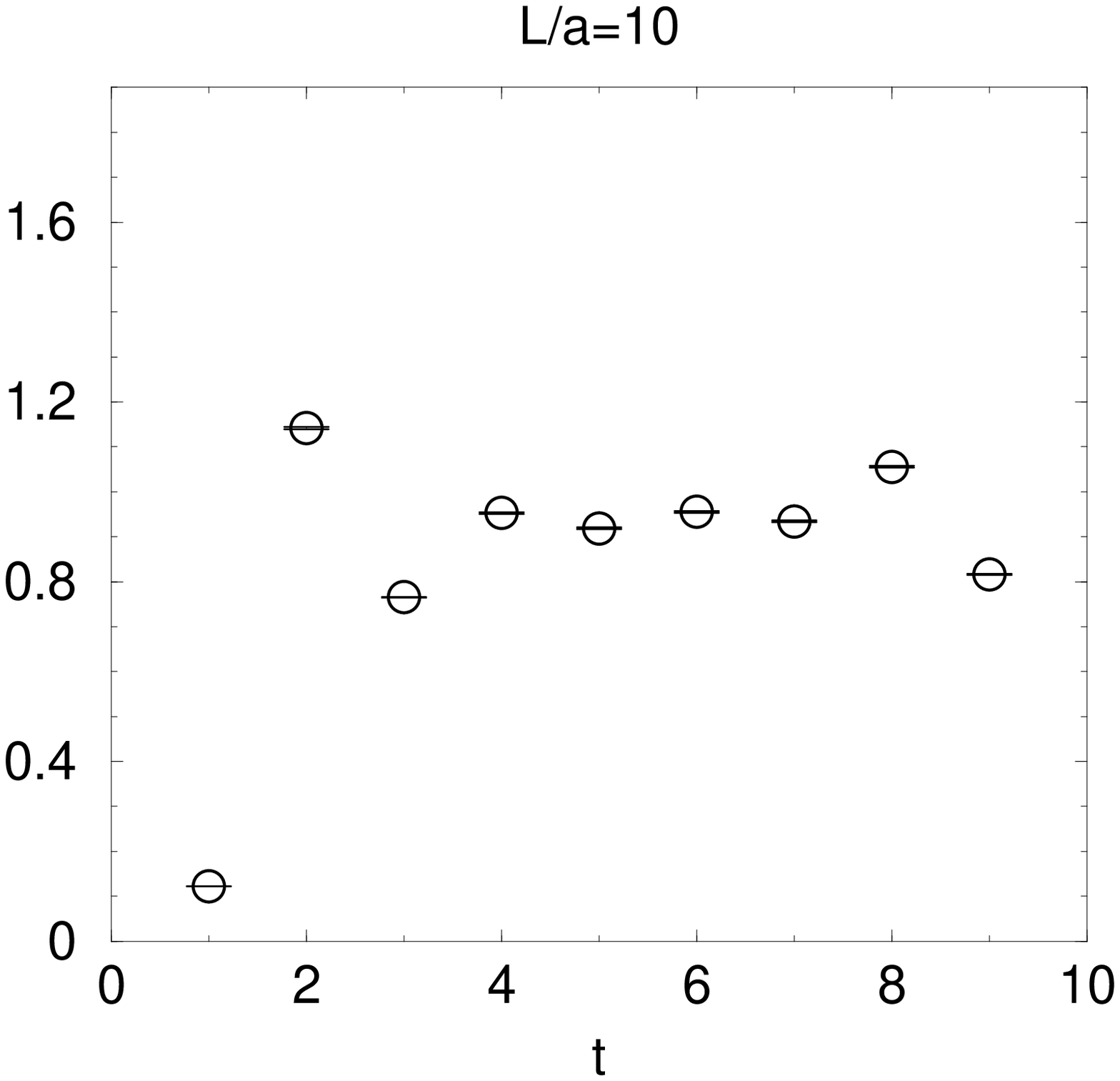}}
\qquad
  \scalebox{0.32}{\includegraphics{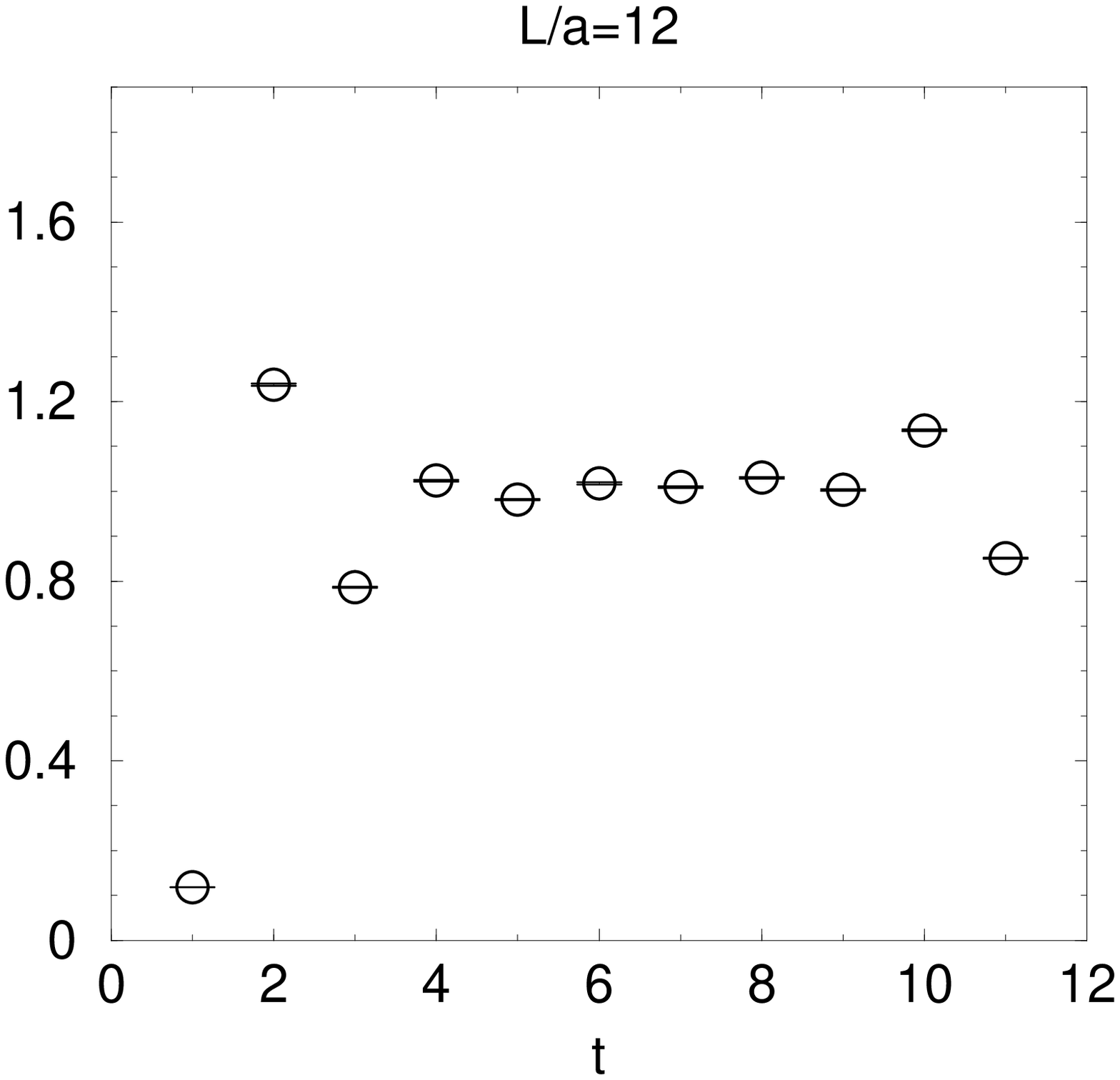}}
 \end{center}
 \begin{center}
  \scalebox{0.32}{\includegraphics{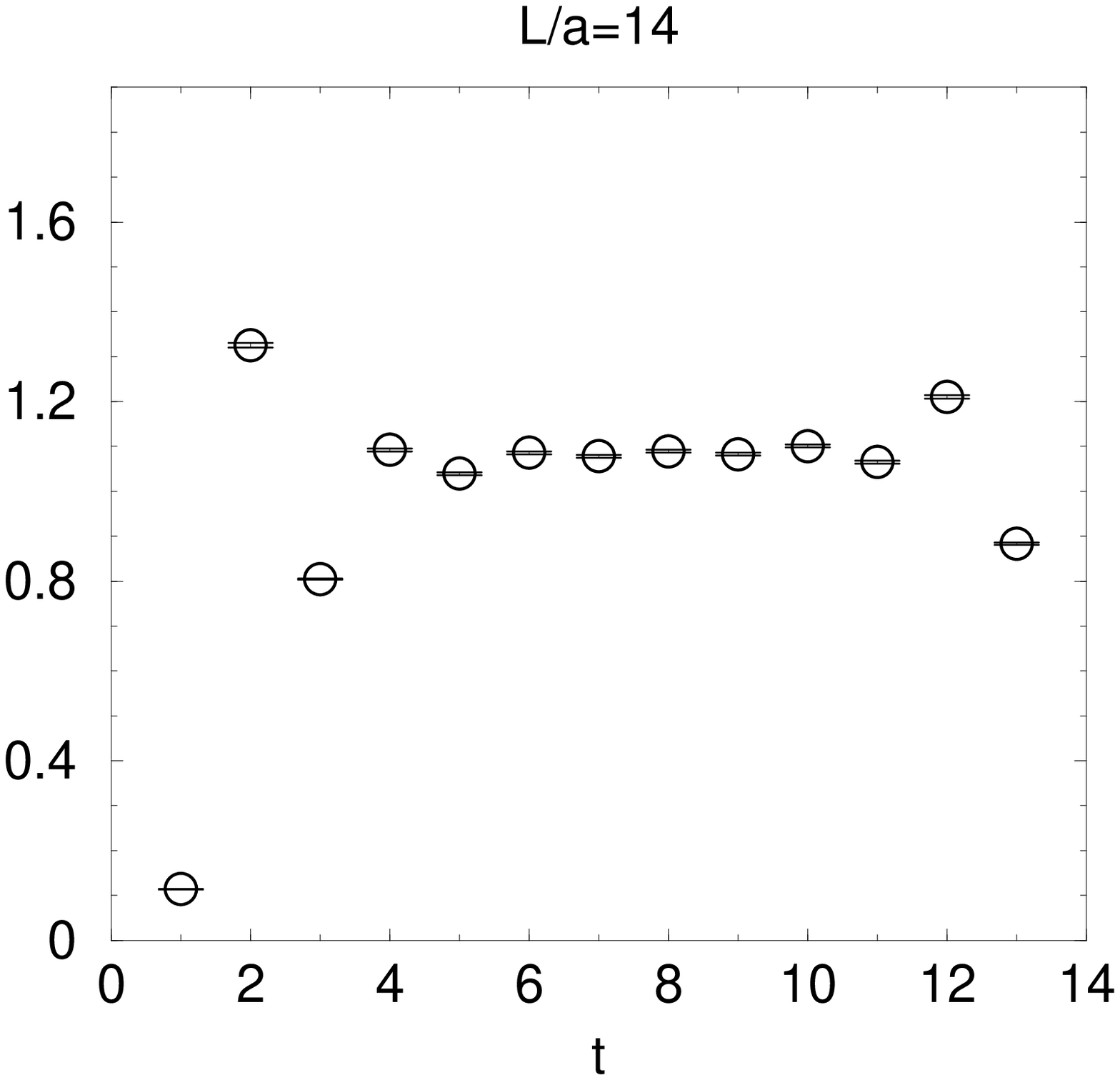}}
\qquad
  \scalebox{0.32}{\includegraphics{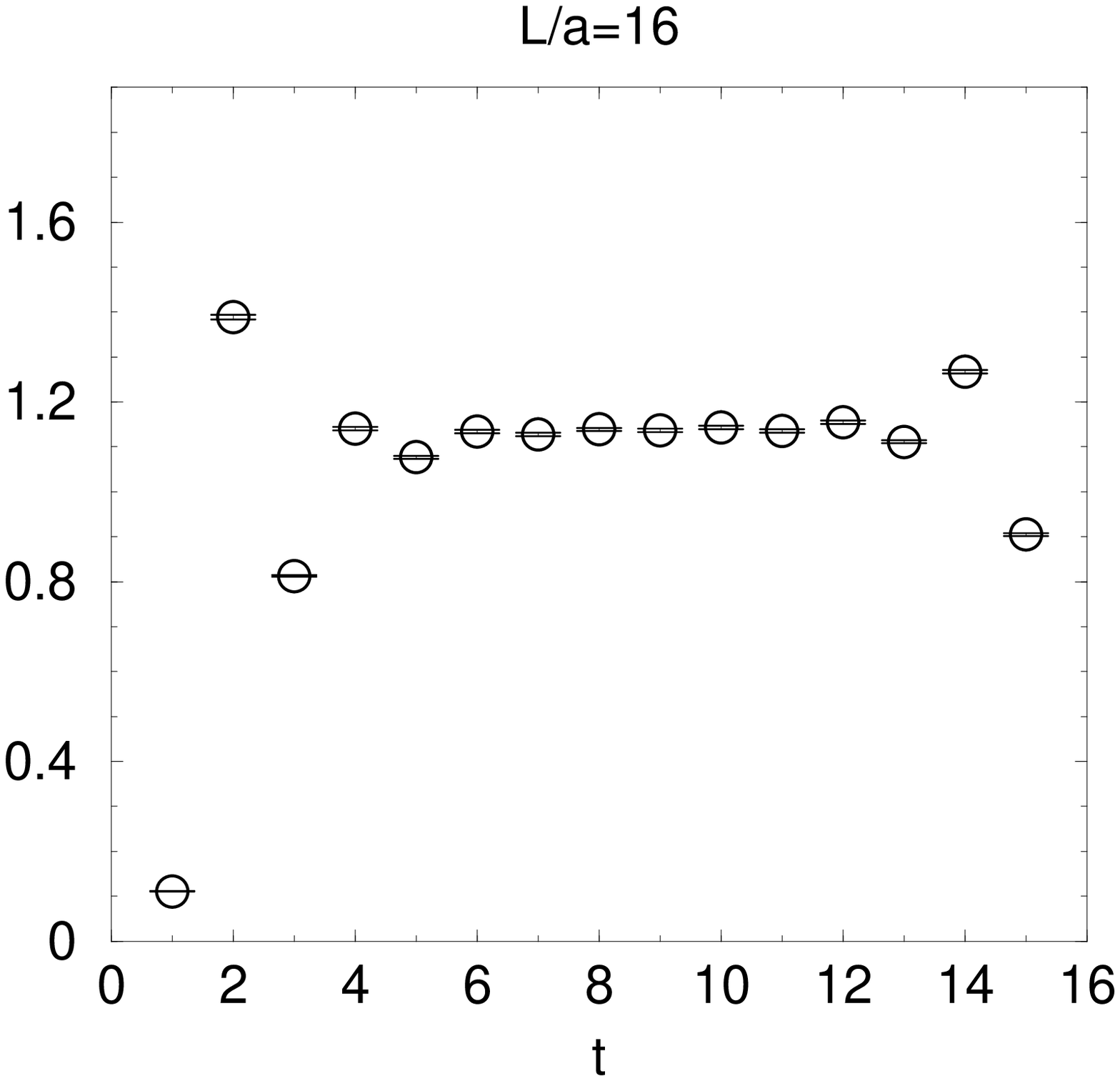}}
 \end{center}
 \begin{center}
  \scalebox{0.32}{\includegraphics{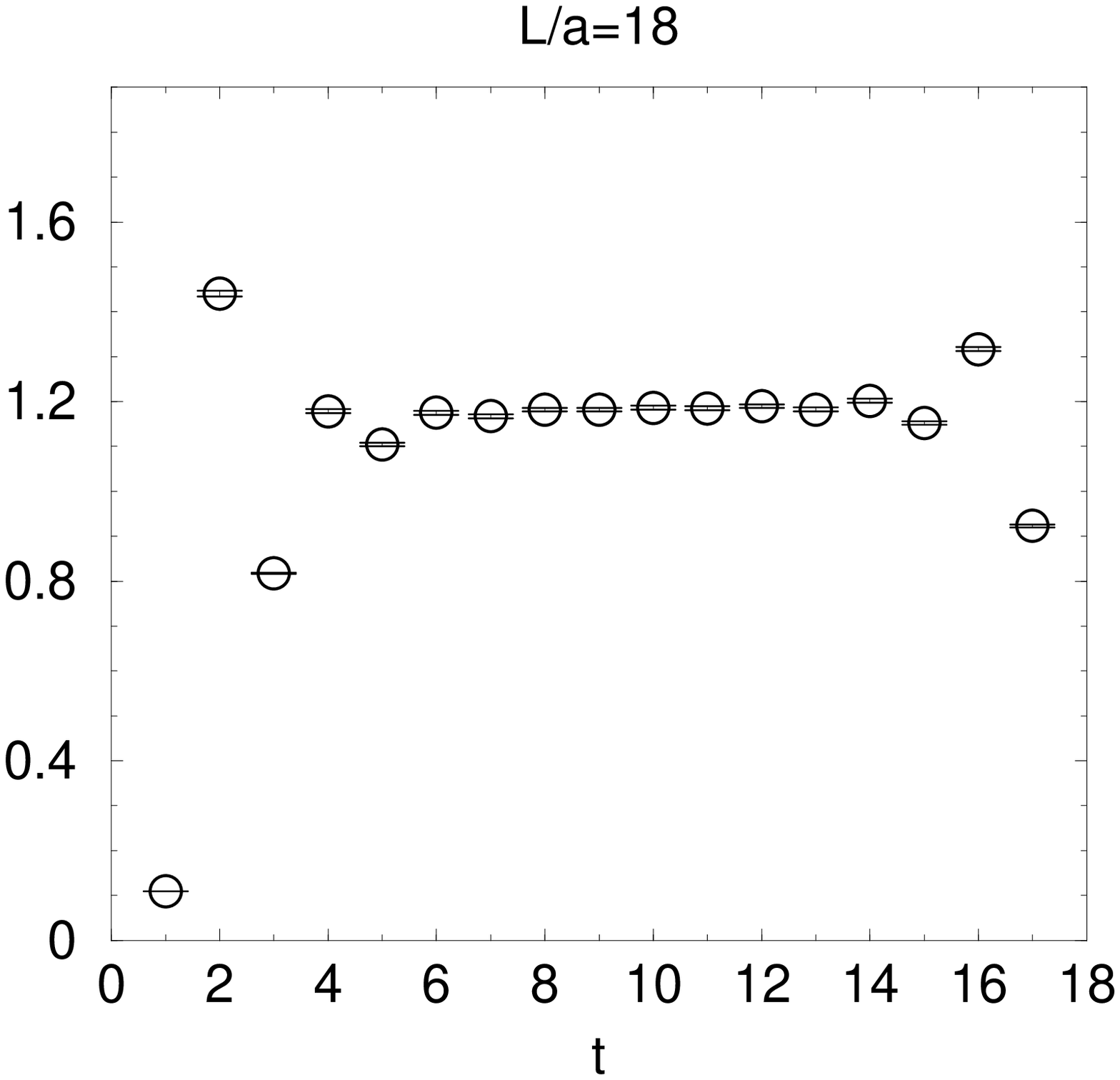}}
\qquad
  \scalebox{0.32}{\includegraphics{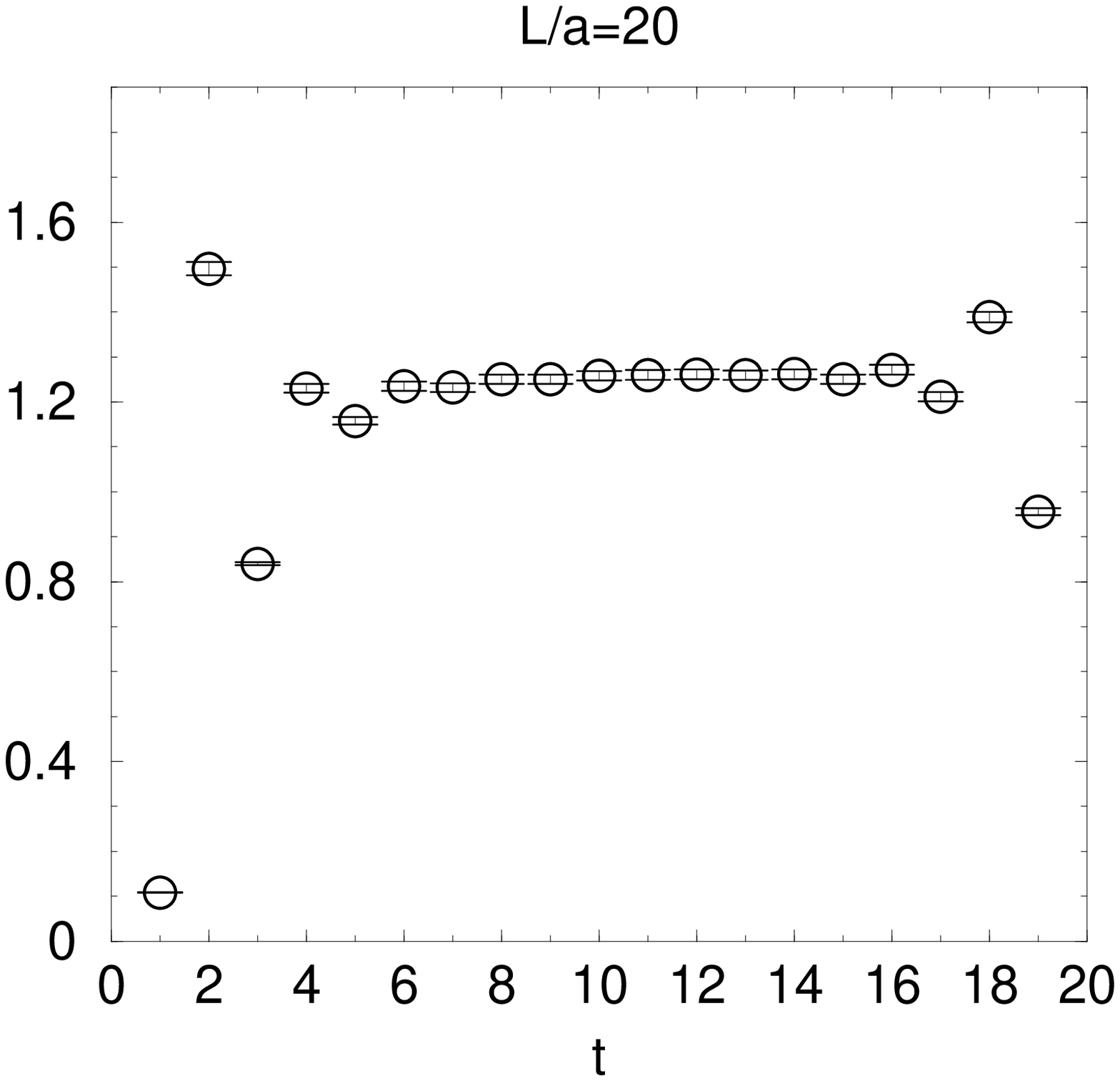}}
 \end{center}
 \caption{$x_0$ dependence of $Z_{VA+AV;1}^+(g_0,a\mu_{\rm min})$ for the
scheme $1$.}
 \label{fig:zvap1.x0dep}
\end{figure}

\begin{figure}
 \begin{center}
  \scalebox{0.32}{\includegraphics{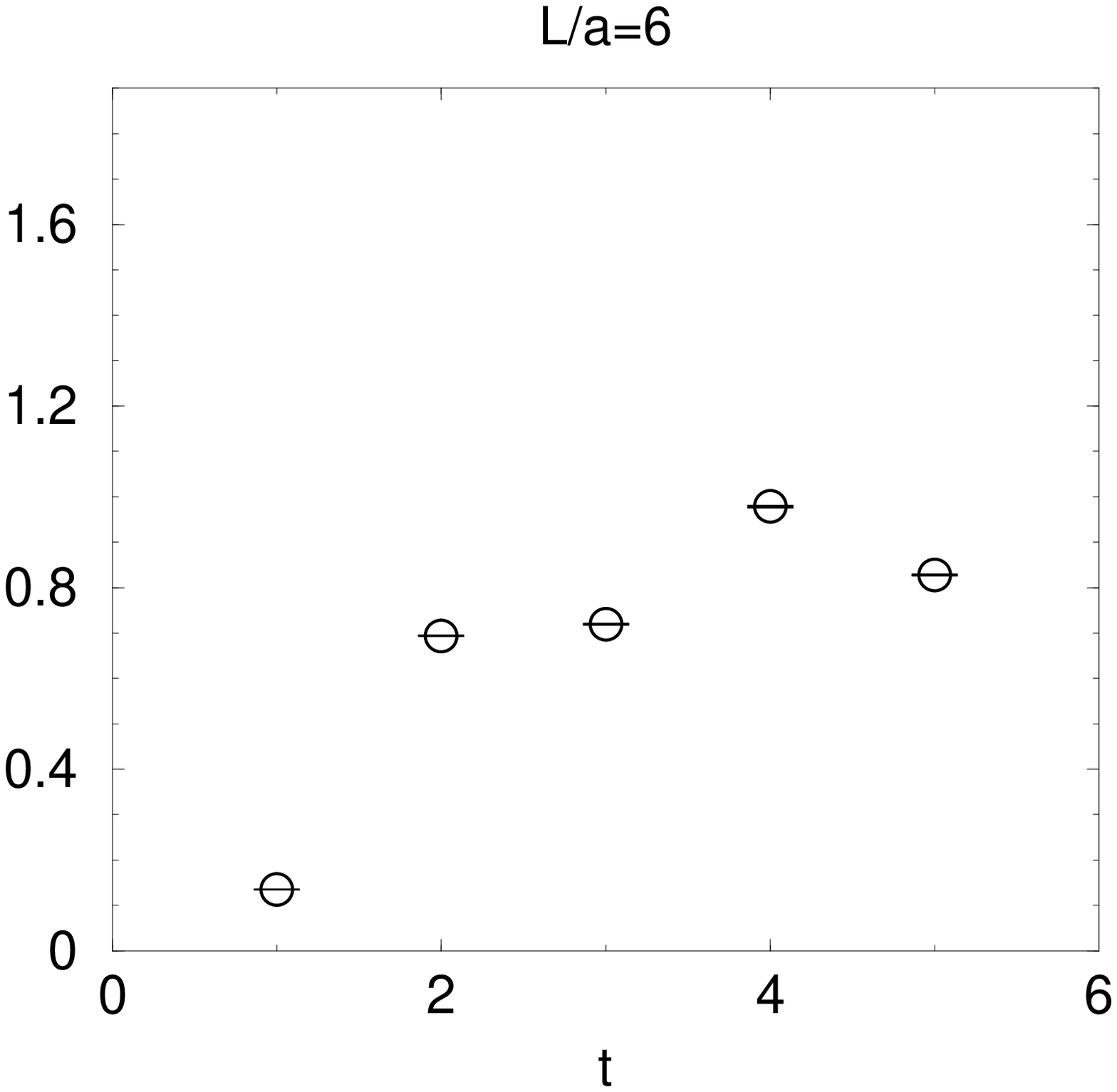}}
\qquad
  \scalebox{0.32}{\includegraphics{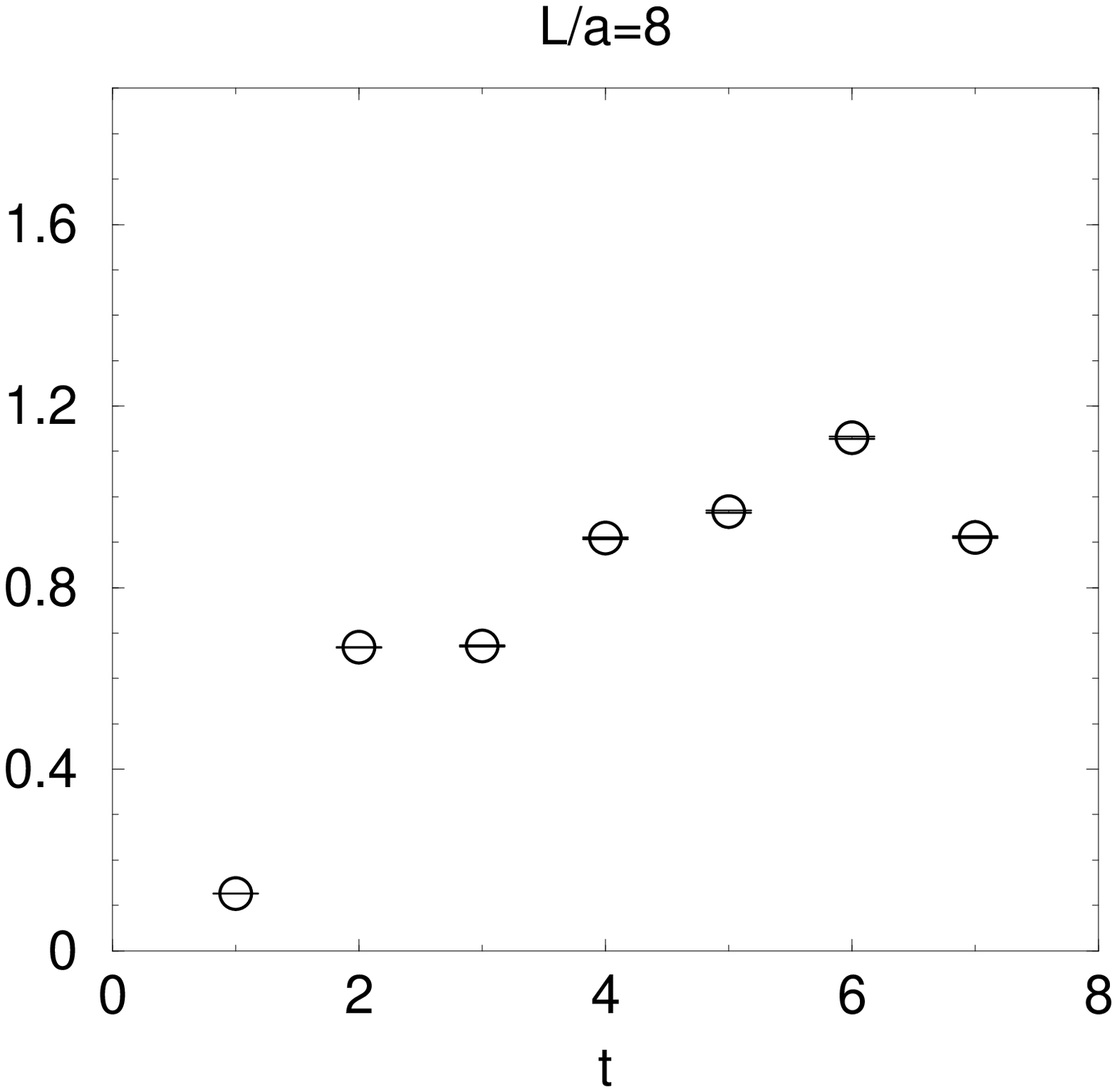}}
 \end{center}
 \begin{center}
  \scalebox{0.32}{\includegraphics{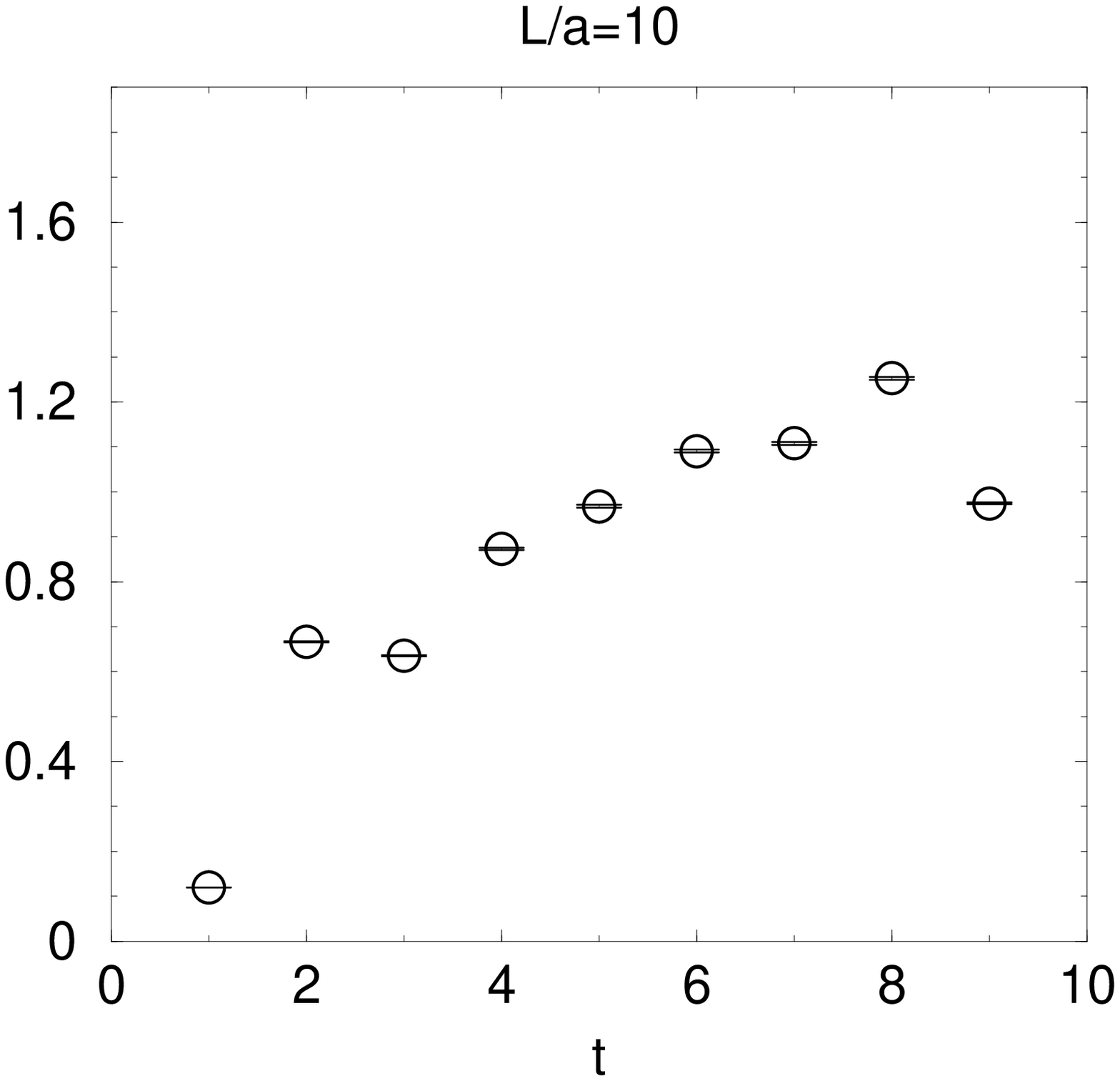}}
\qquad
  \scalebox{0.32}{\includegraphics{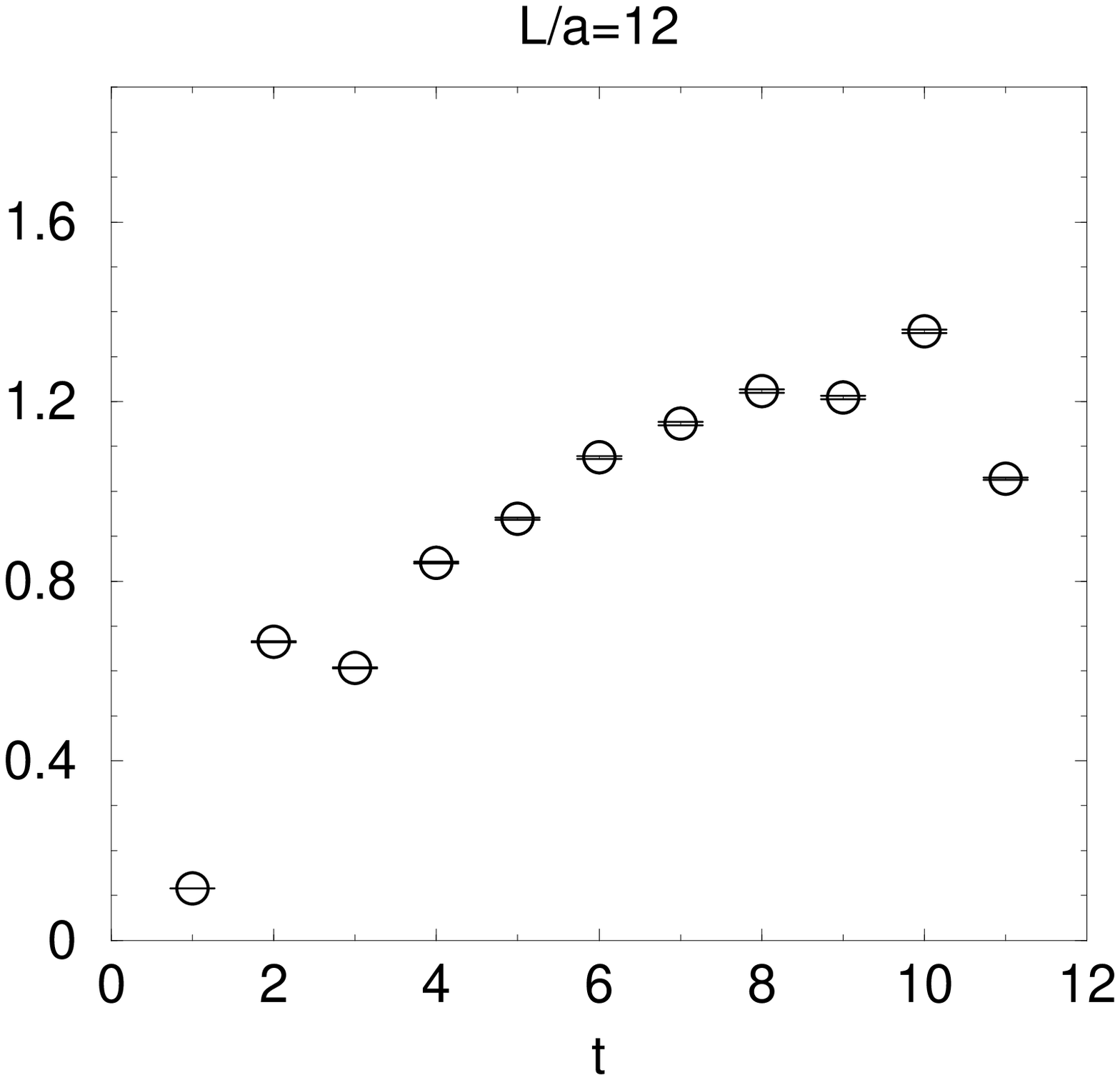}}
 \end{center}
 \begin{center}
  \scalebox{0.32}{\includegraphics{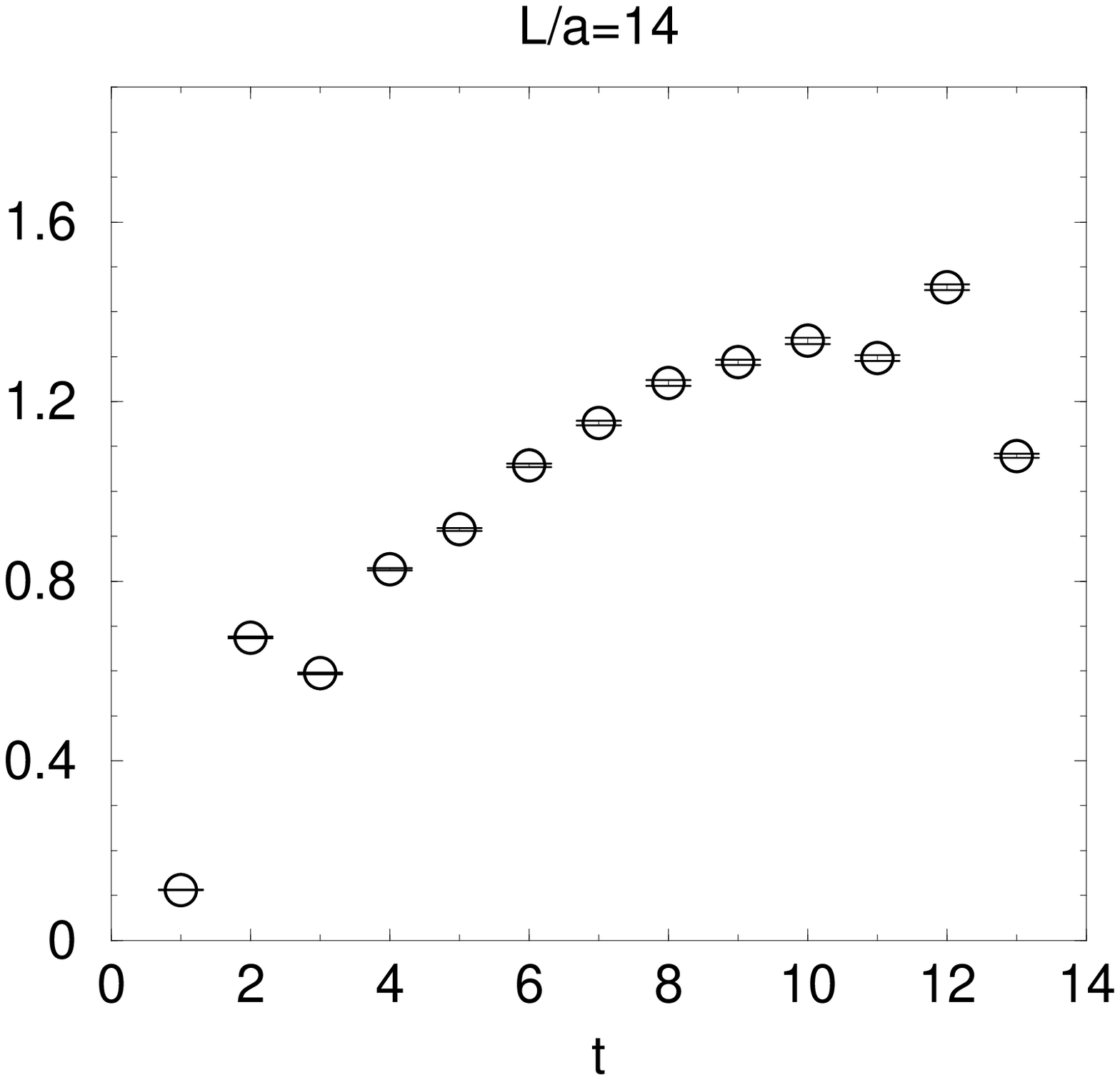}}
\qquad
  \scalebox{0.32}{\includegraphics{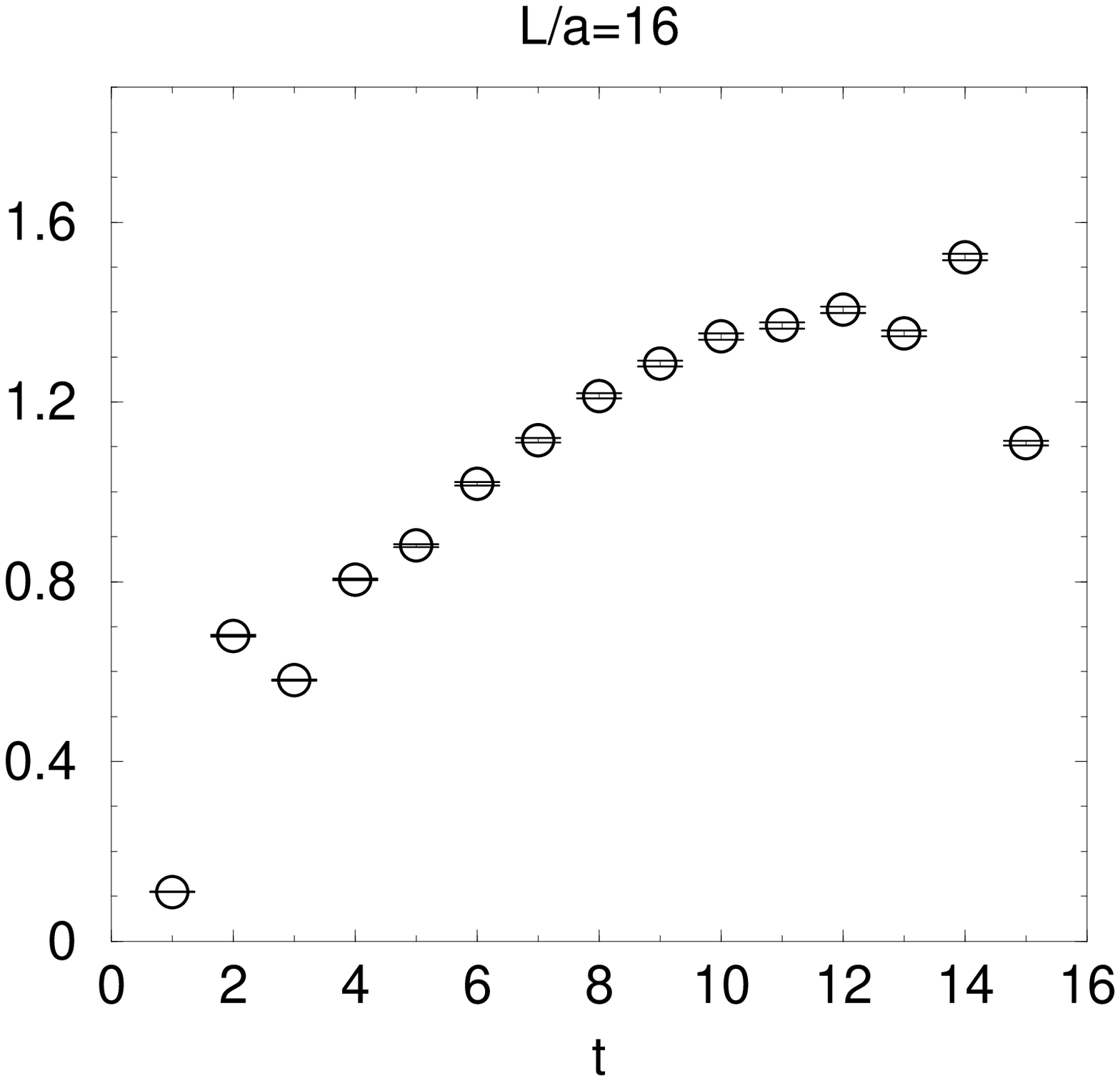}}
 \end{center}
 \begin{center}
  \scalebox{0.32}{\includegraphics{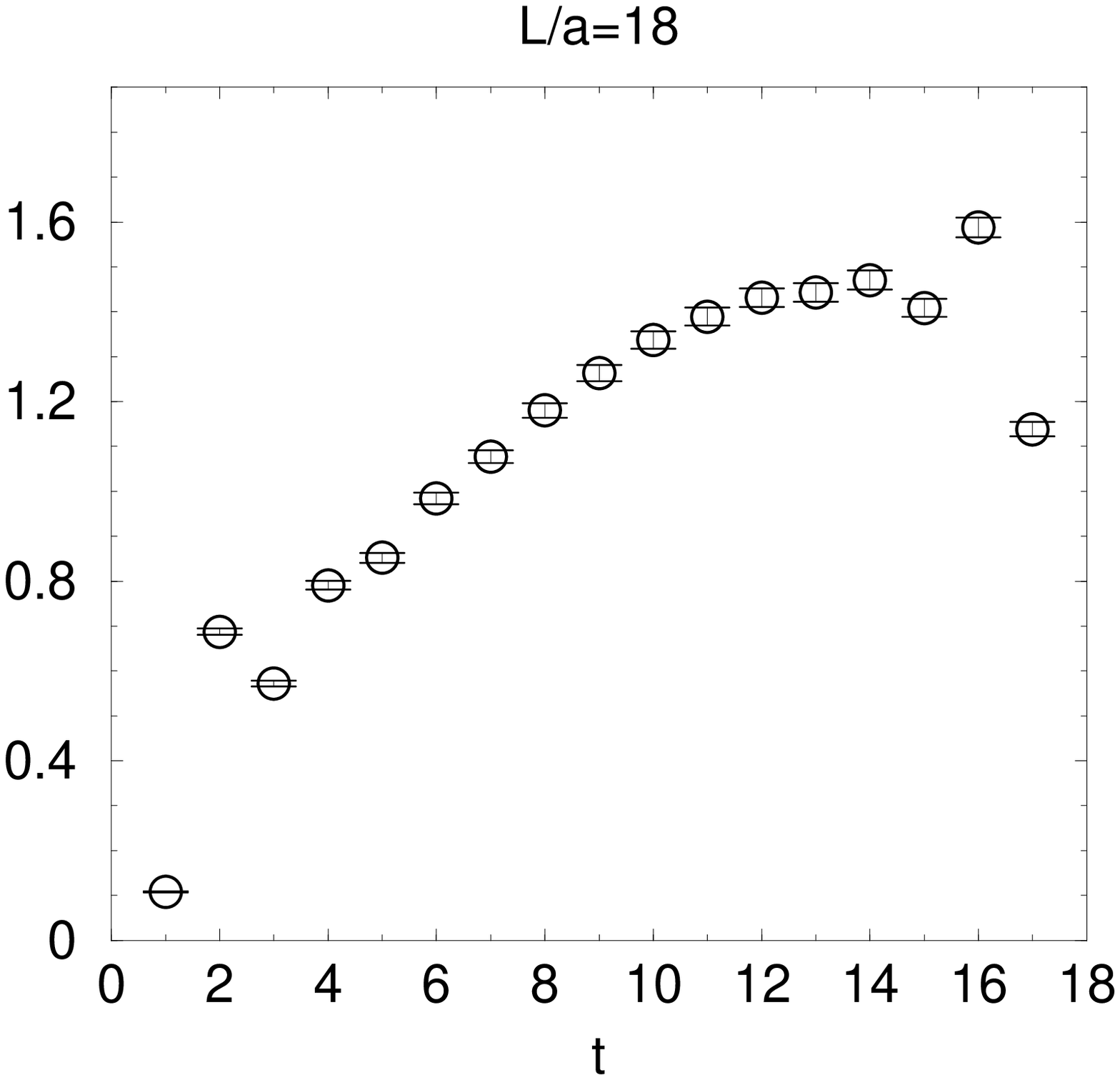}}
\qquad
  \scalebox{0.32}{\includegraphics{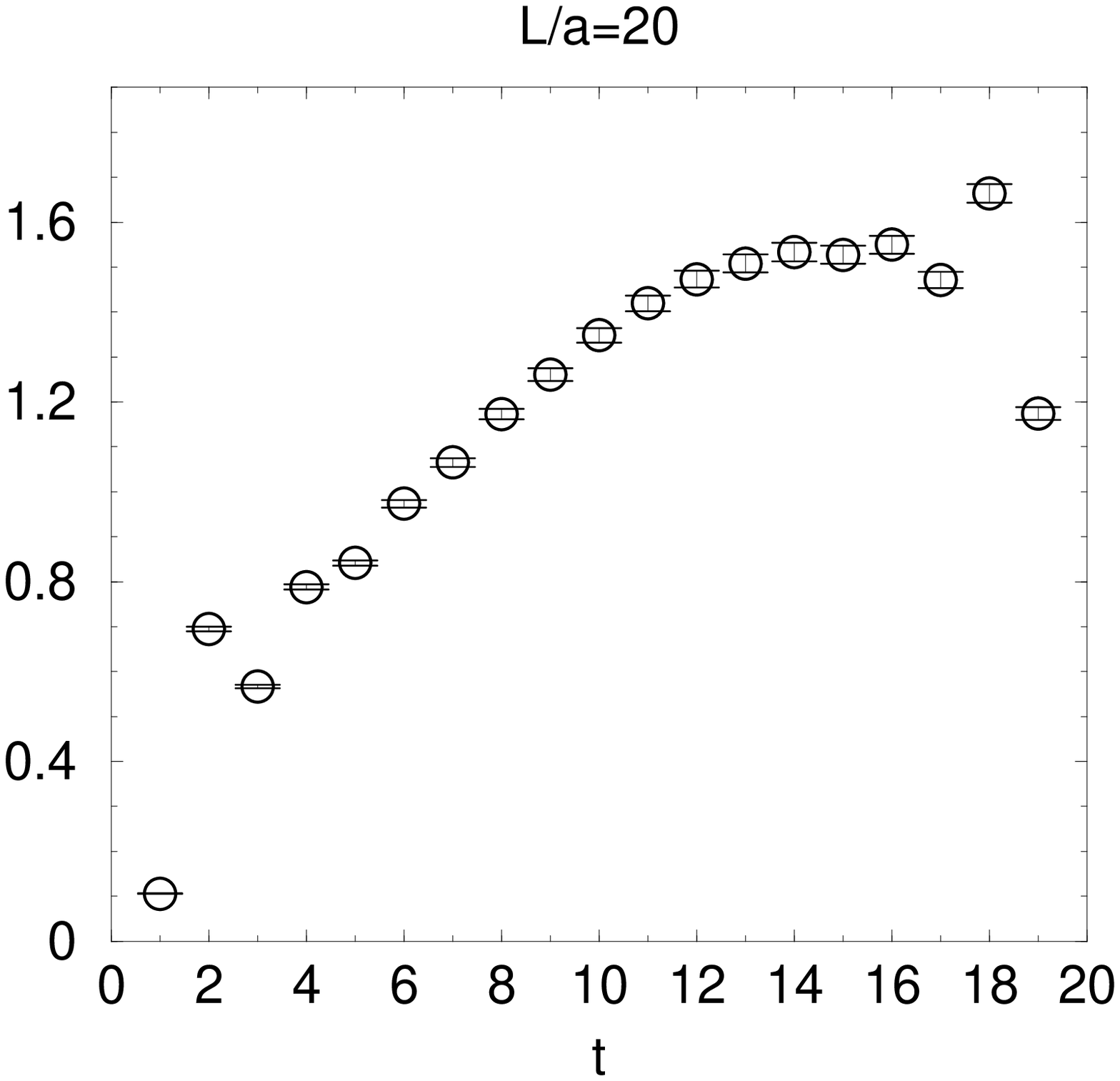}}
 \end{center}
 \caption{$x_0$ dependence of $Z_{VA+AV;2}^+(g_0,a\mu_{\rm min})$ for the 
scheme $2$.}
 \label{fig:zvap2.x0dep}
\end{figure}

\begin{figure}
 \begin{center}
  \scalebox{0.32}{\includegraphics{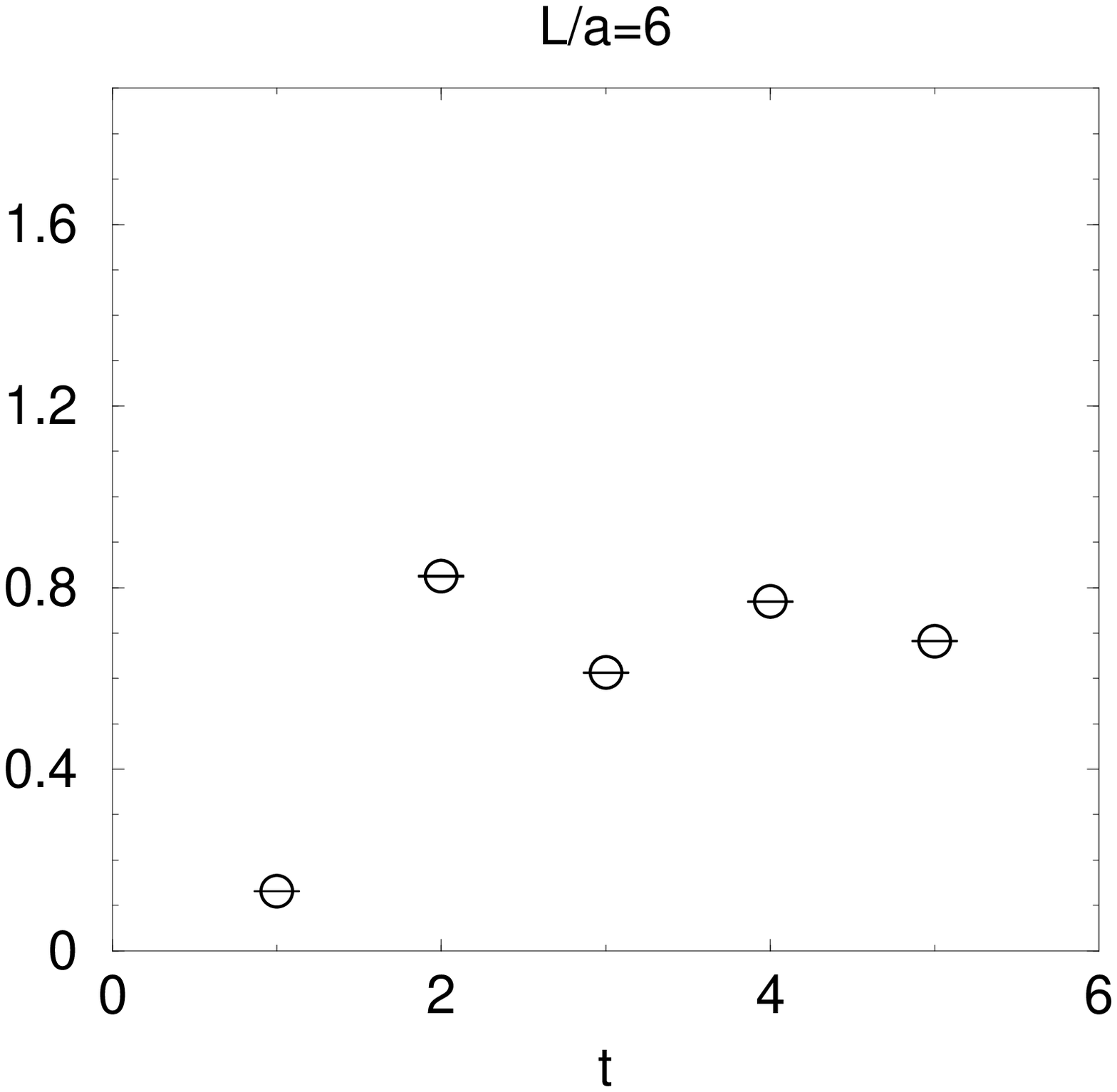}}
\qquad
  \scalebox{0.32}{\includegraphics{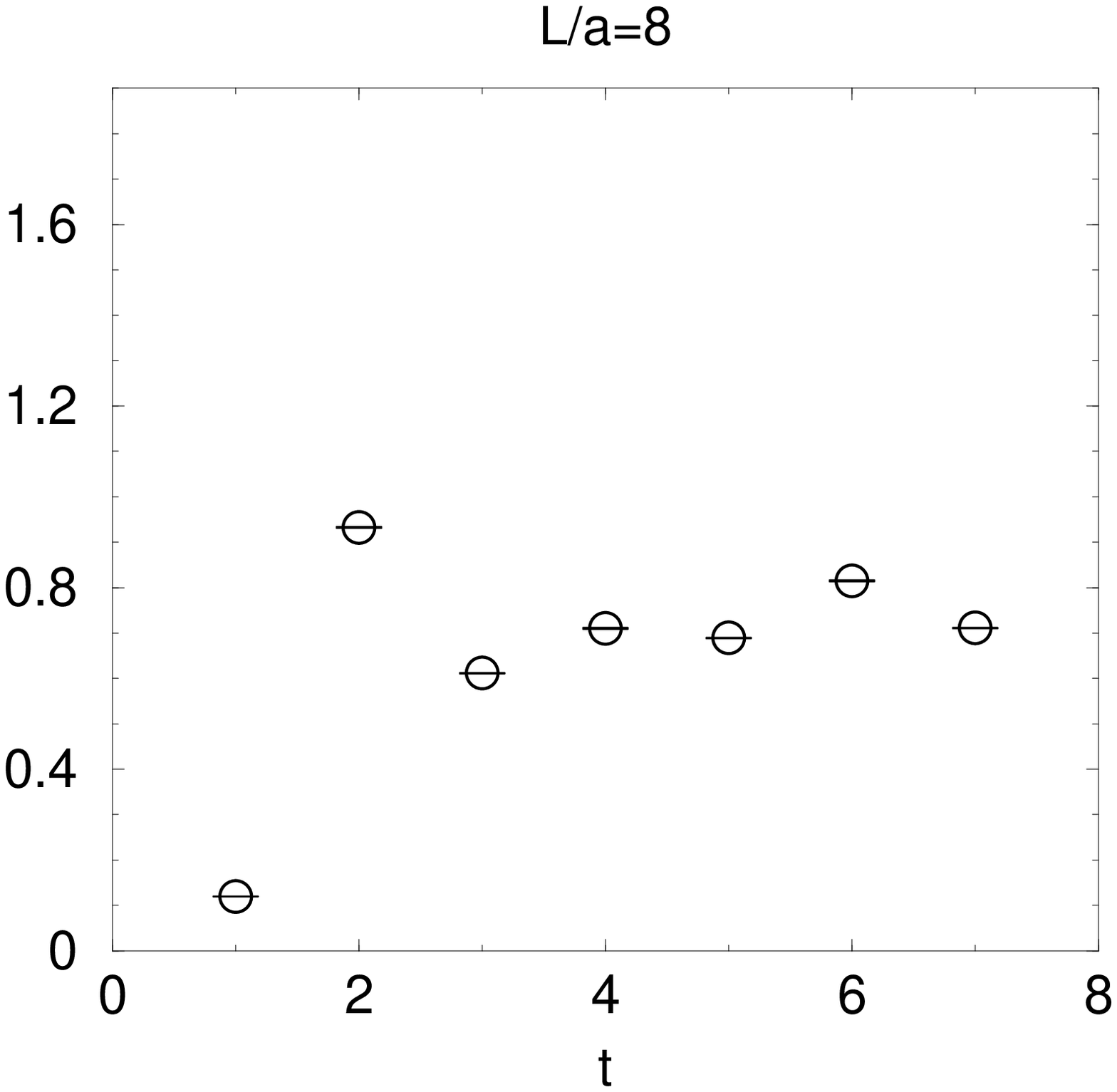}}
 \end{center}
 \begin{center}
  \scalebox{0.32}{\includegraphics{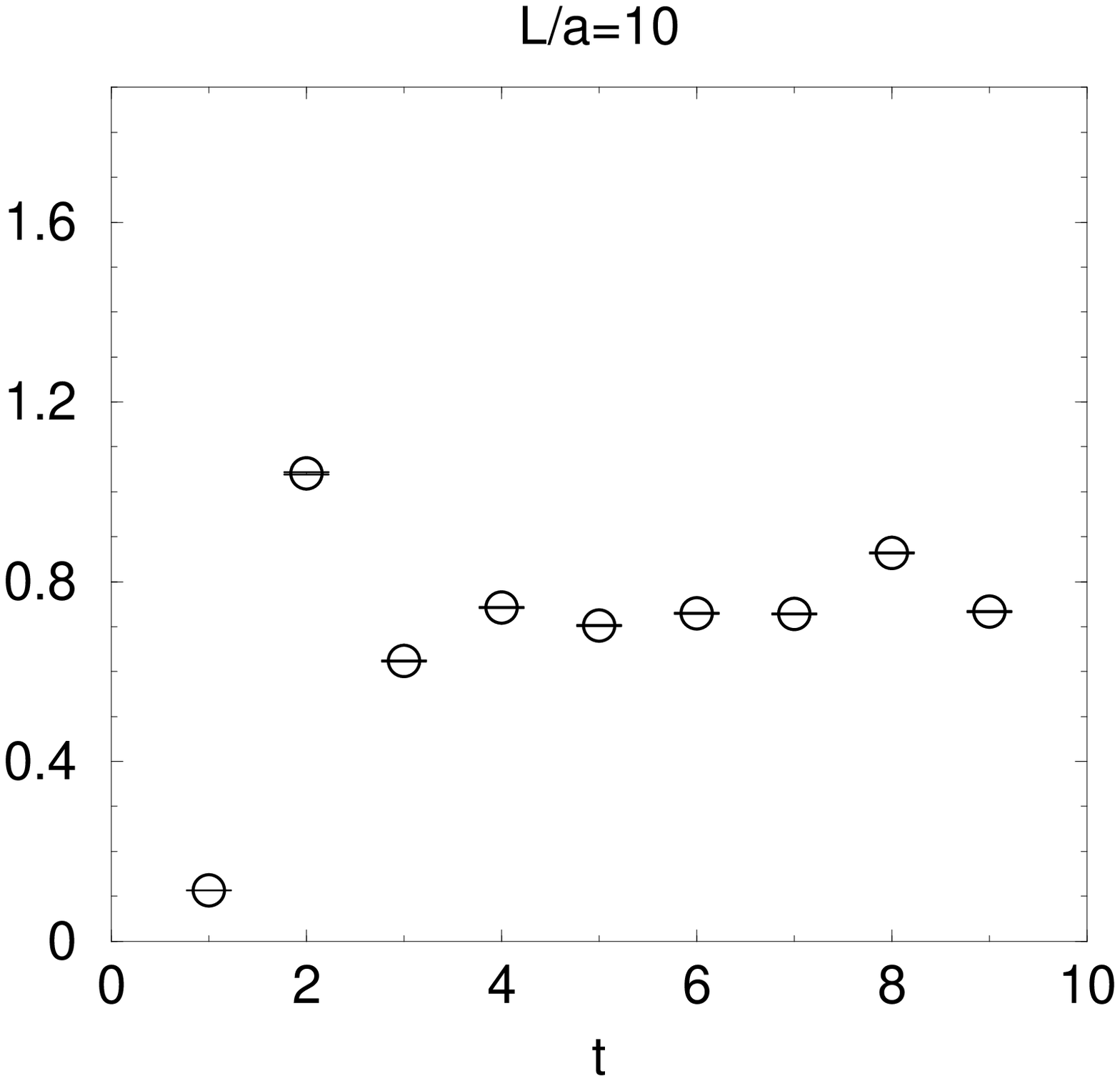}}
\qquad
  \scalebox{0.32}{\includegraphics{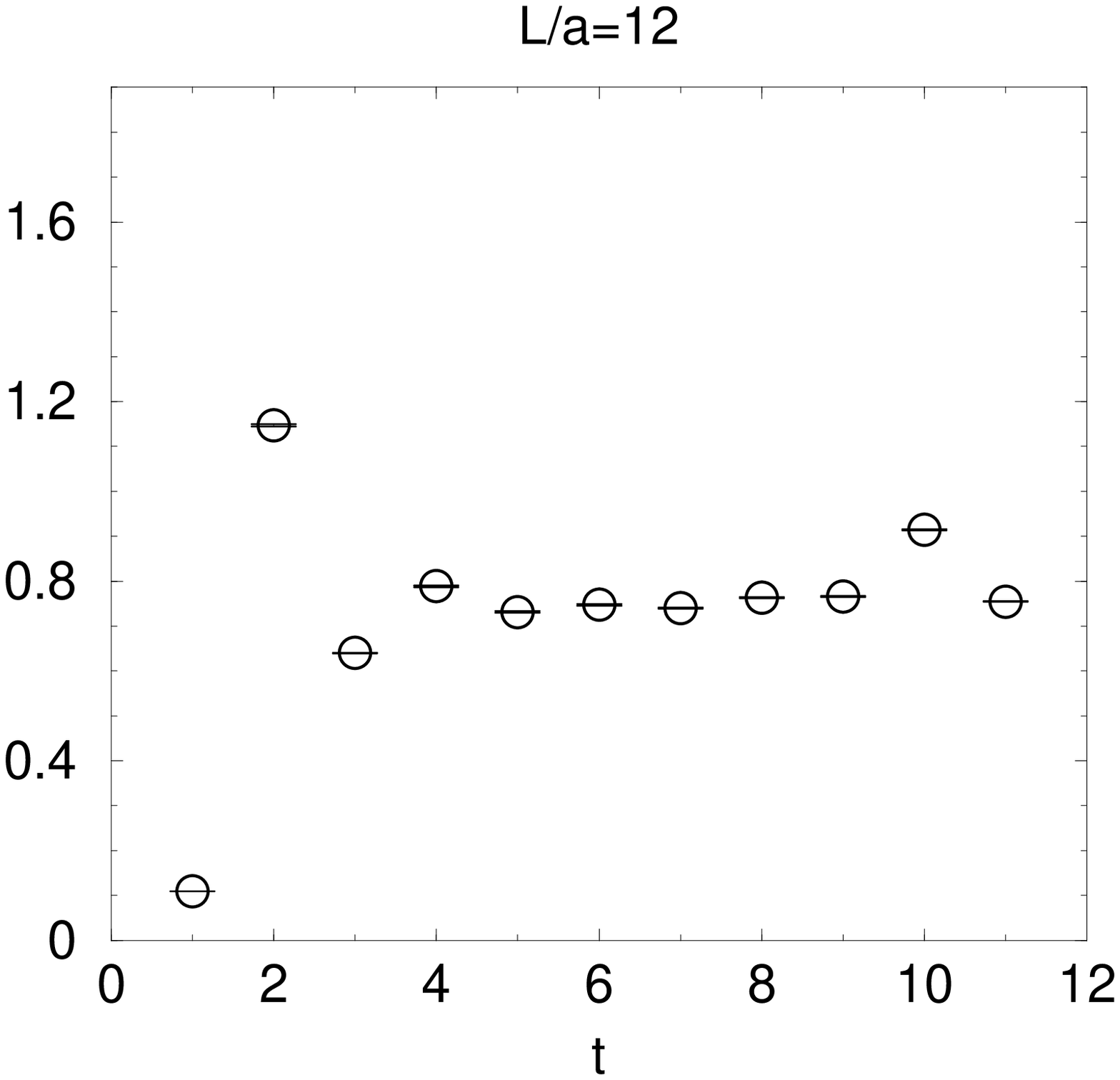}}
 \end{center}
 \begin{center}
  \scalebox{0.32}{\includegraphics{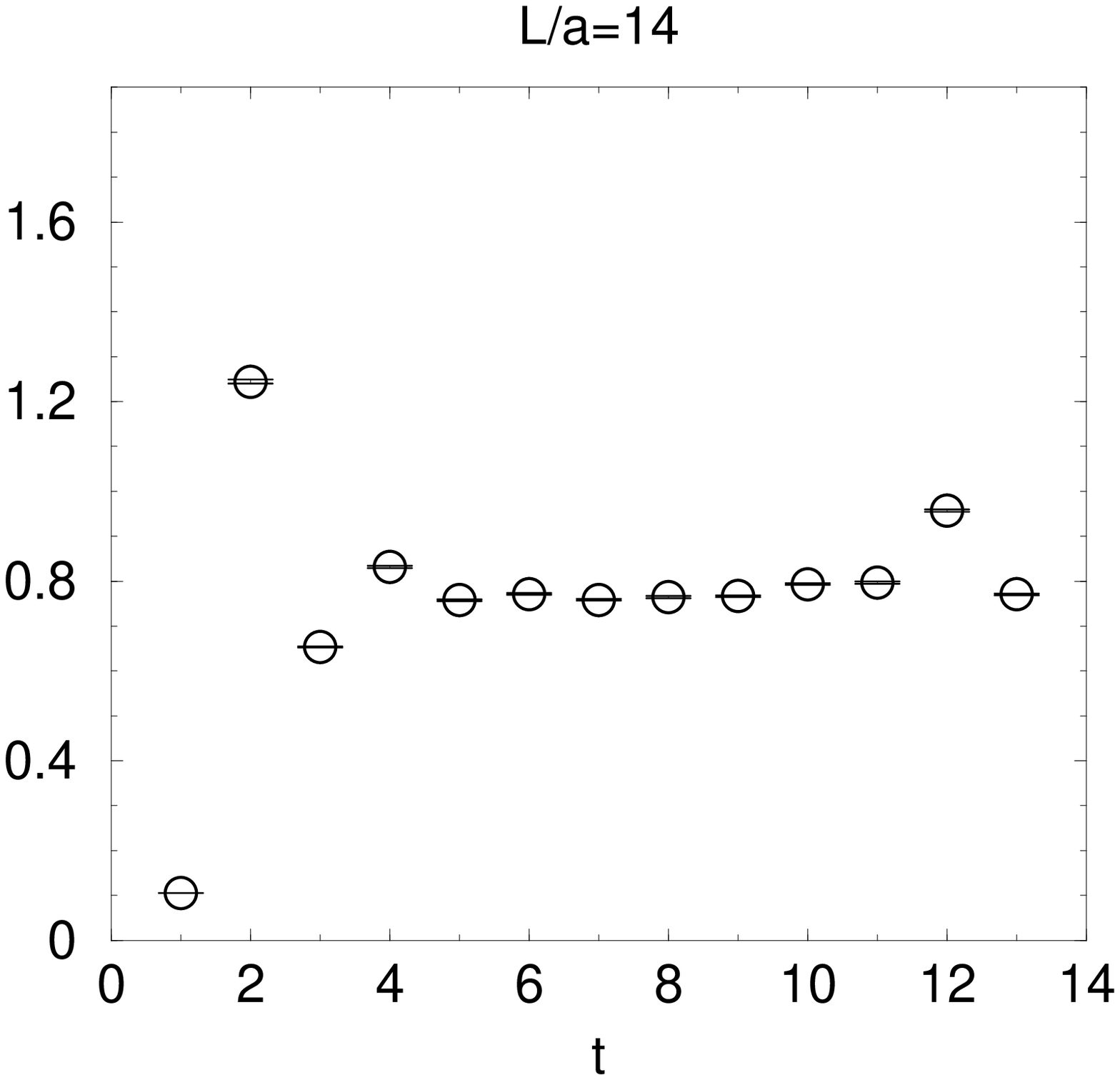}}
\qquad
  \scalebox{0.32}{\includegraphics{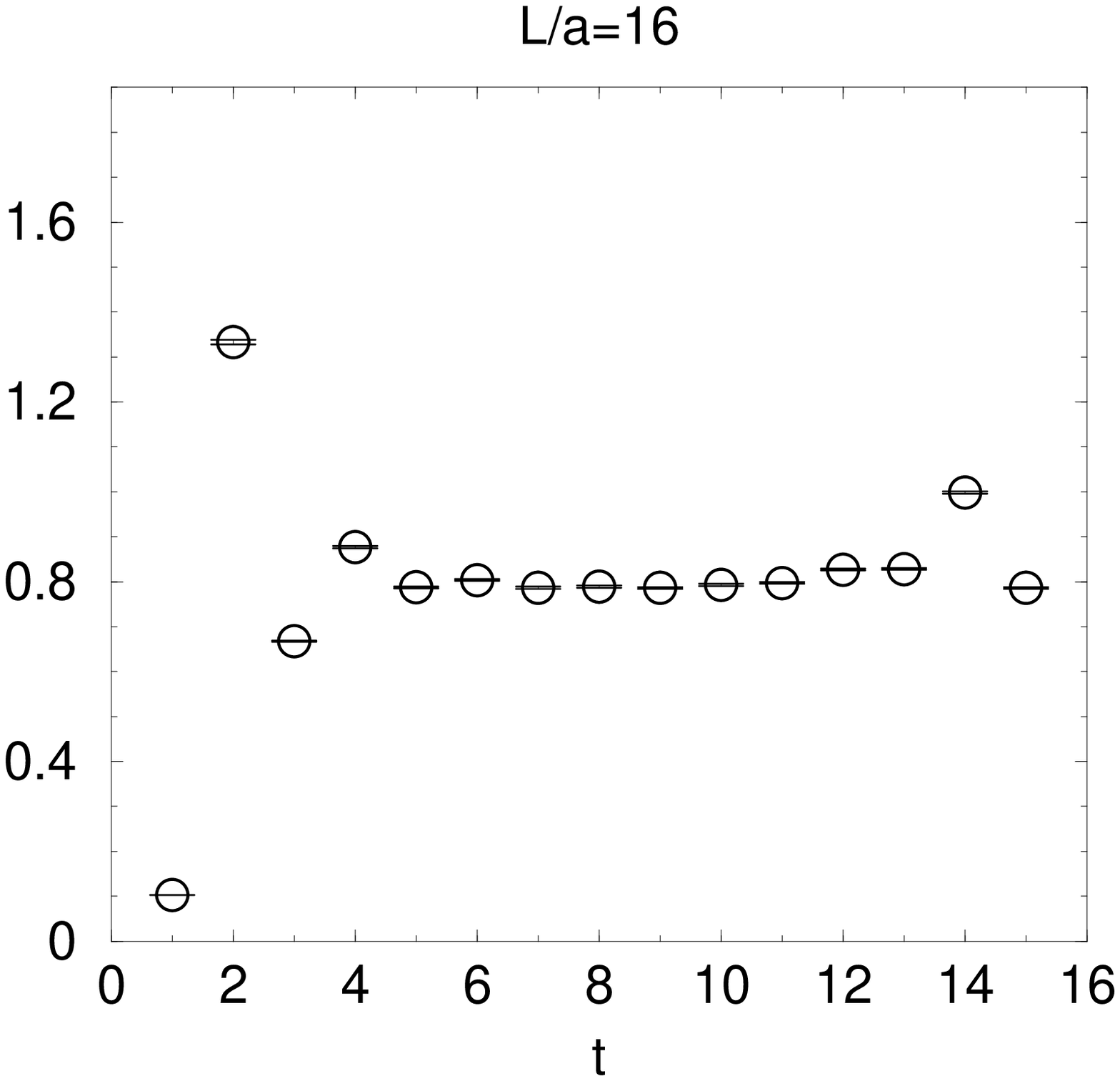}}
 \end{center}
 \begin{center}
  \scalebox{0.32}{\includegraphics{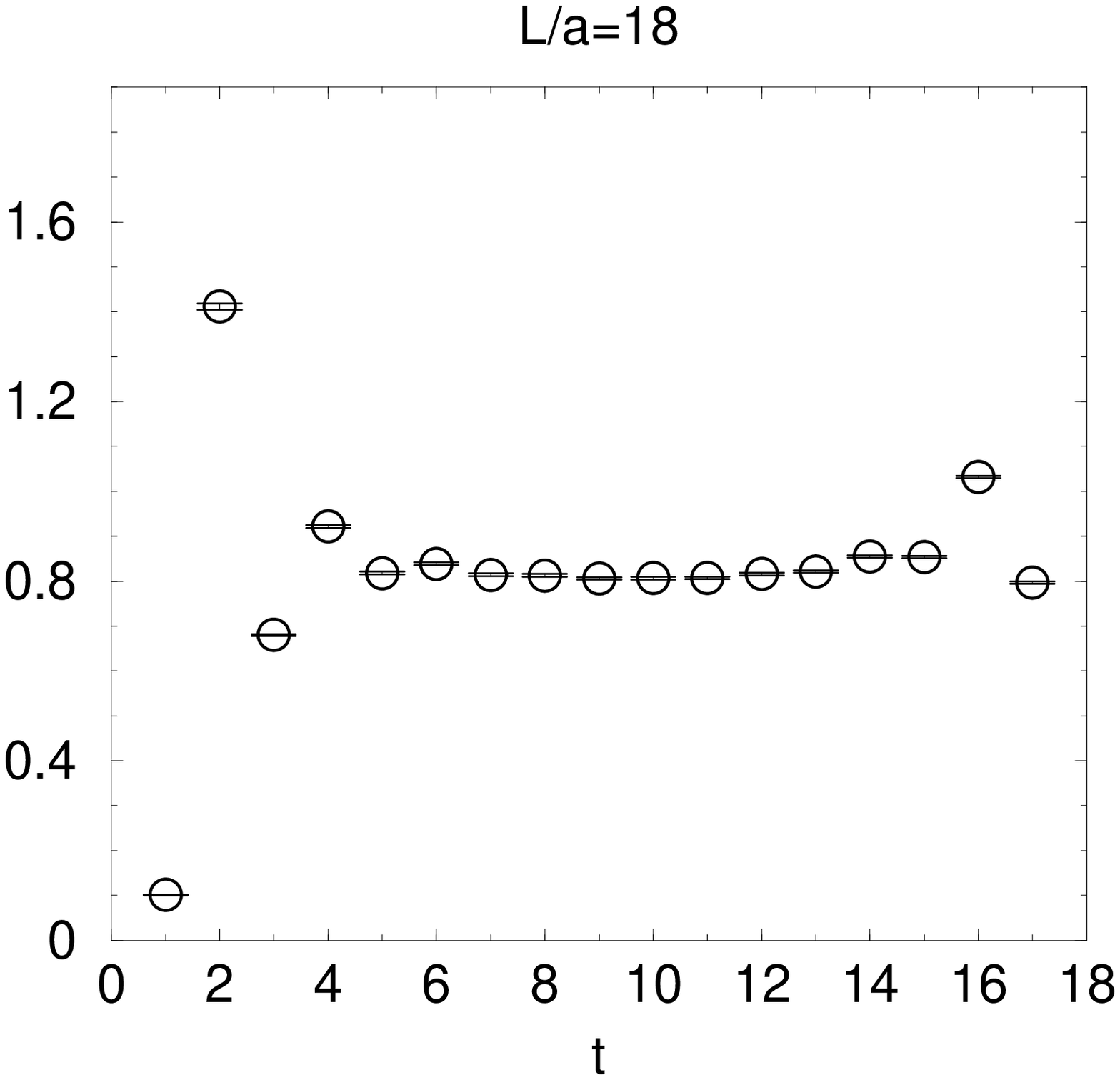}}
\qquad
  \scalebox{0.32}{\includegraphics{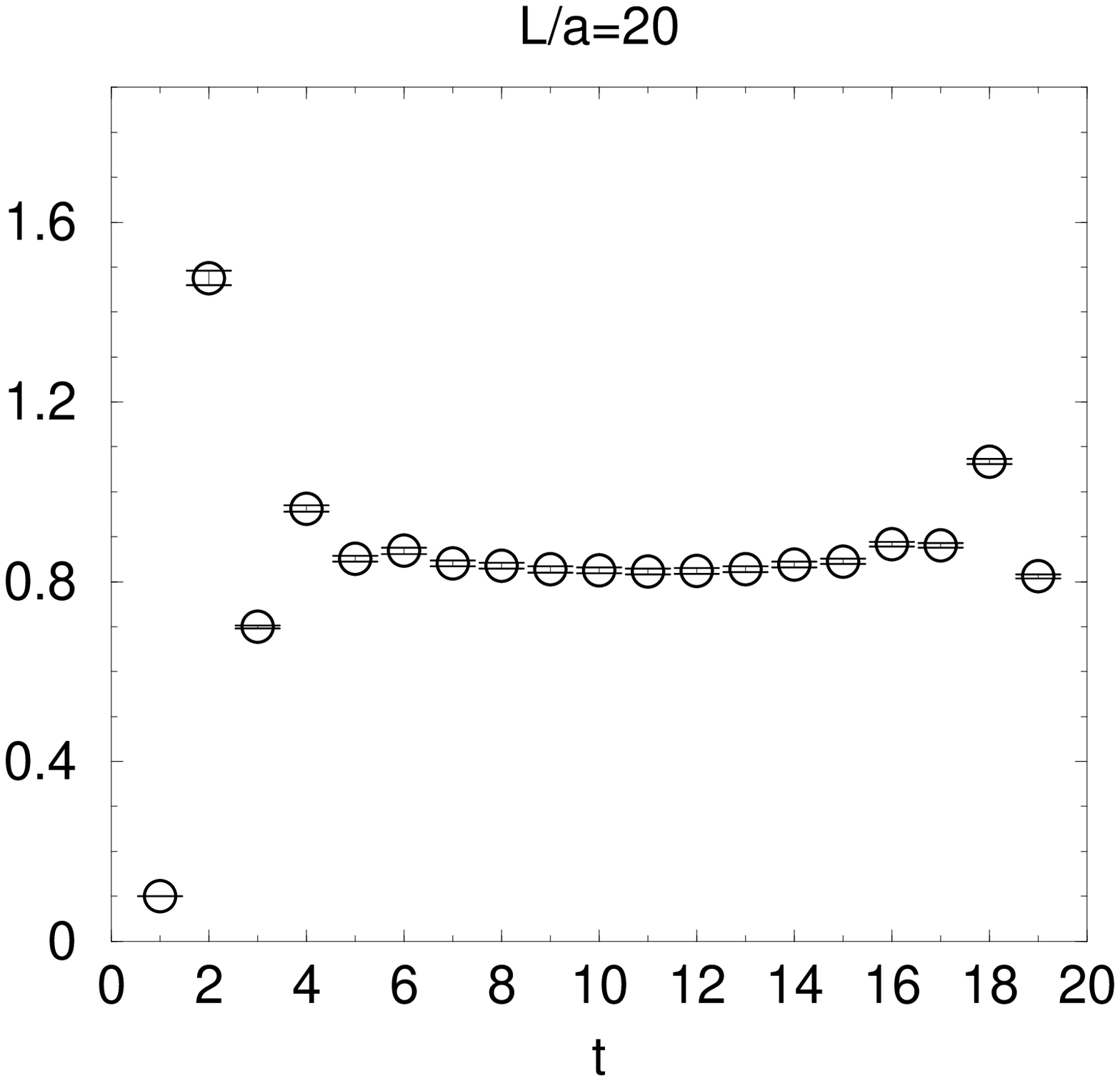}}
 \end{center}
 \caption{$x_0$ dependence of $Z_{VA+AV;1}^-(g_0,a\mu_{\rm min})$ for the 
scheme $1$.}
 \label{fig:zvam1.x0dep}
\end{figure}

\begin{figure}
 \begin{center}
  \scalebox{0.32}{\includegraphics{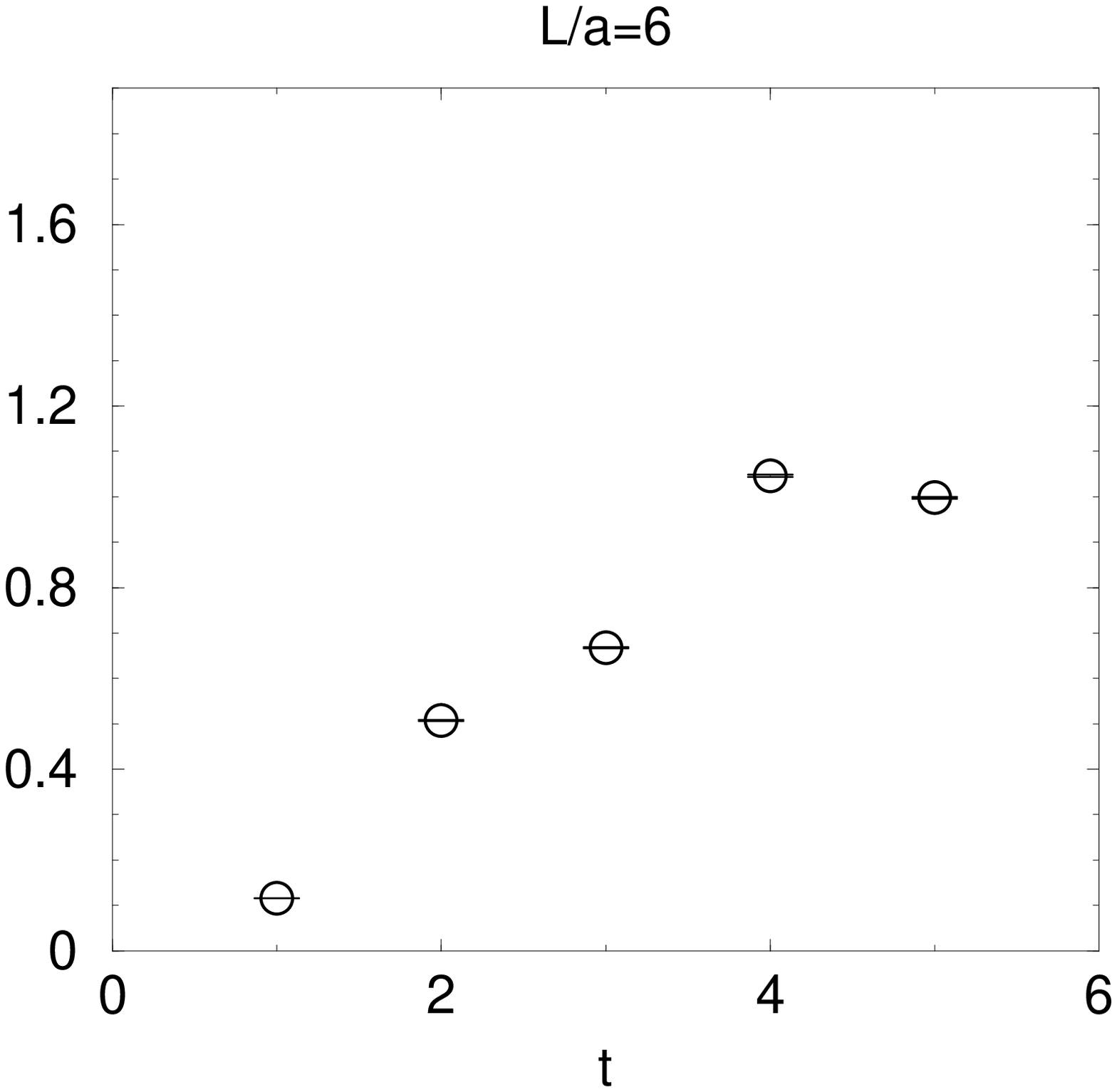}}
\qquad
  \scalebox{0.32}{\includegraphics{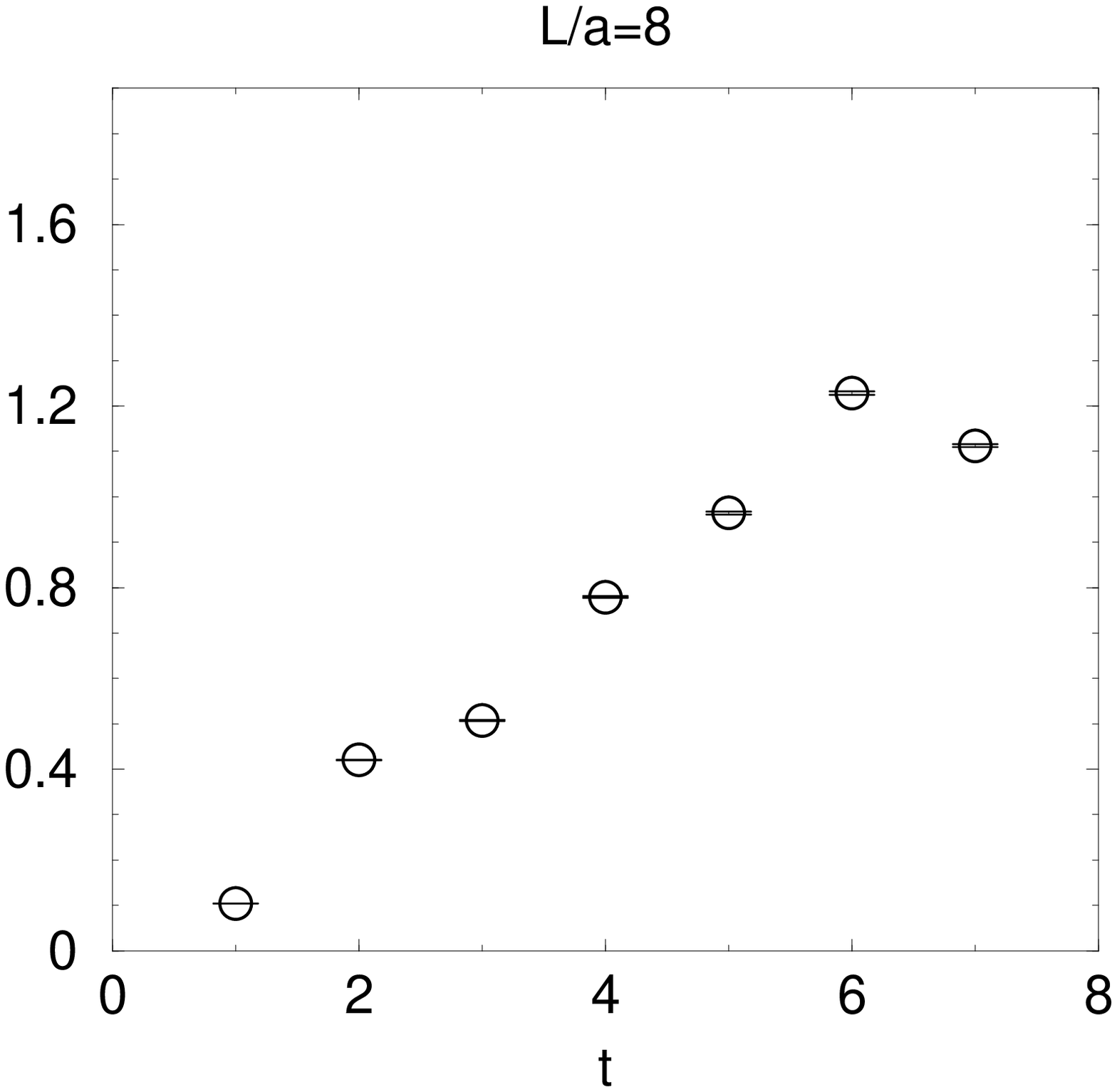}}
 \end{center}
 \begin{center}
  \scalebox{0.32}{\includegraphics{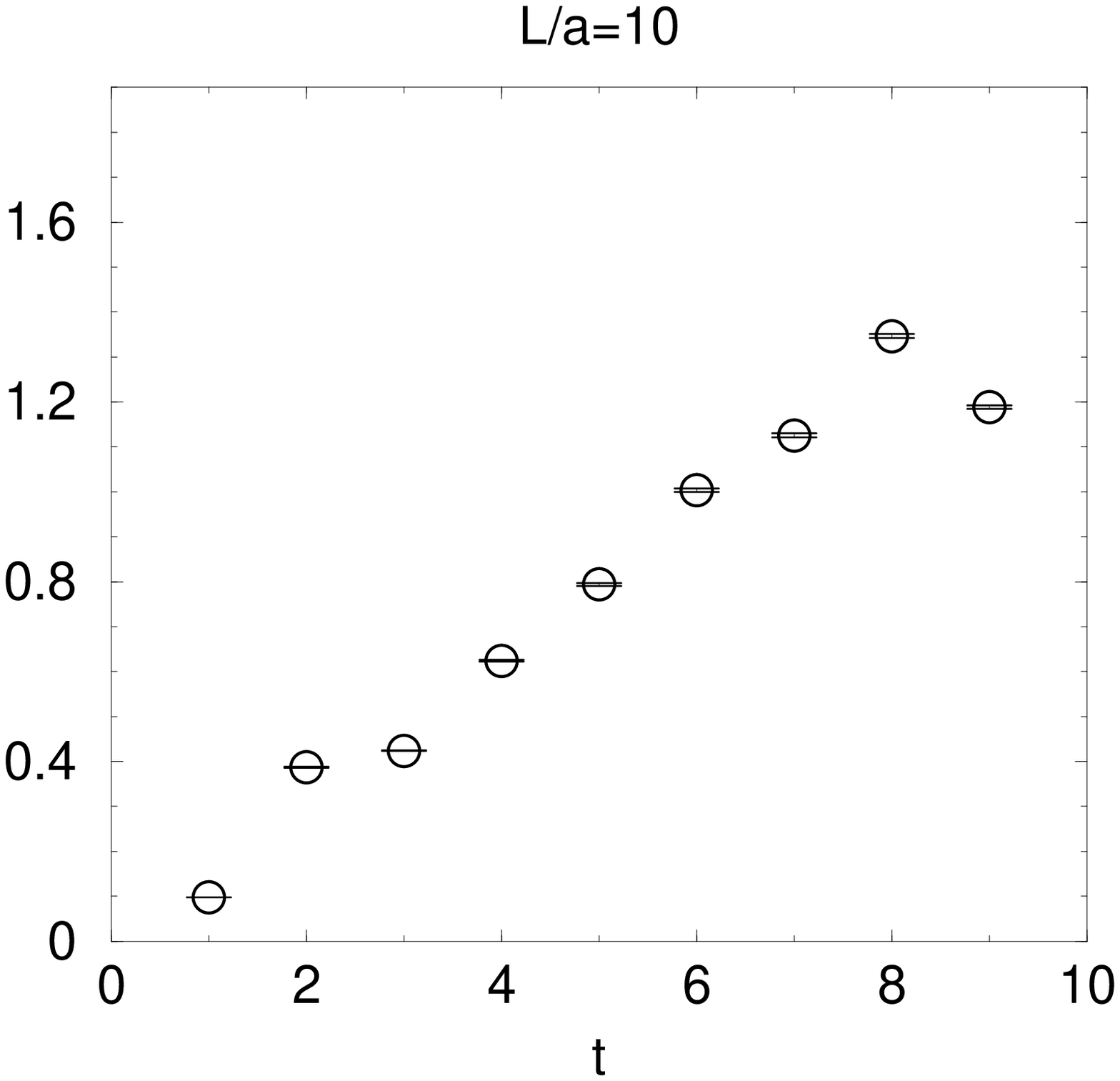}}
\qquad
  \scalebox{0.32}{\includegraphics{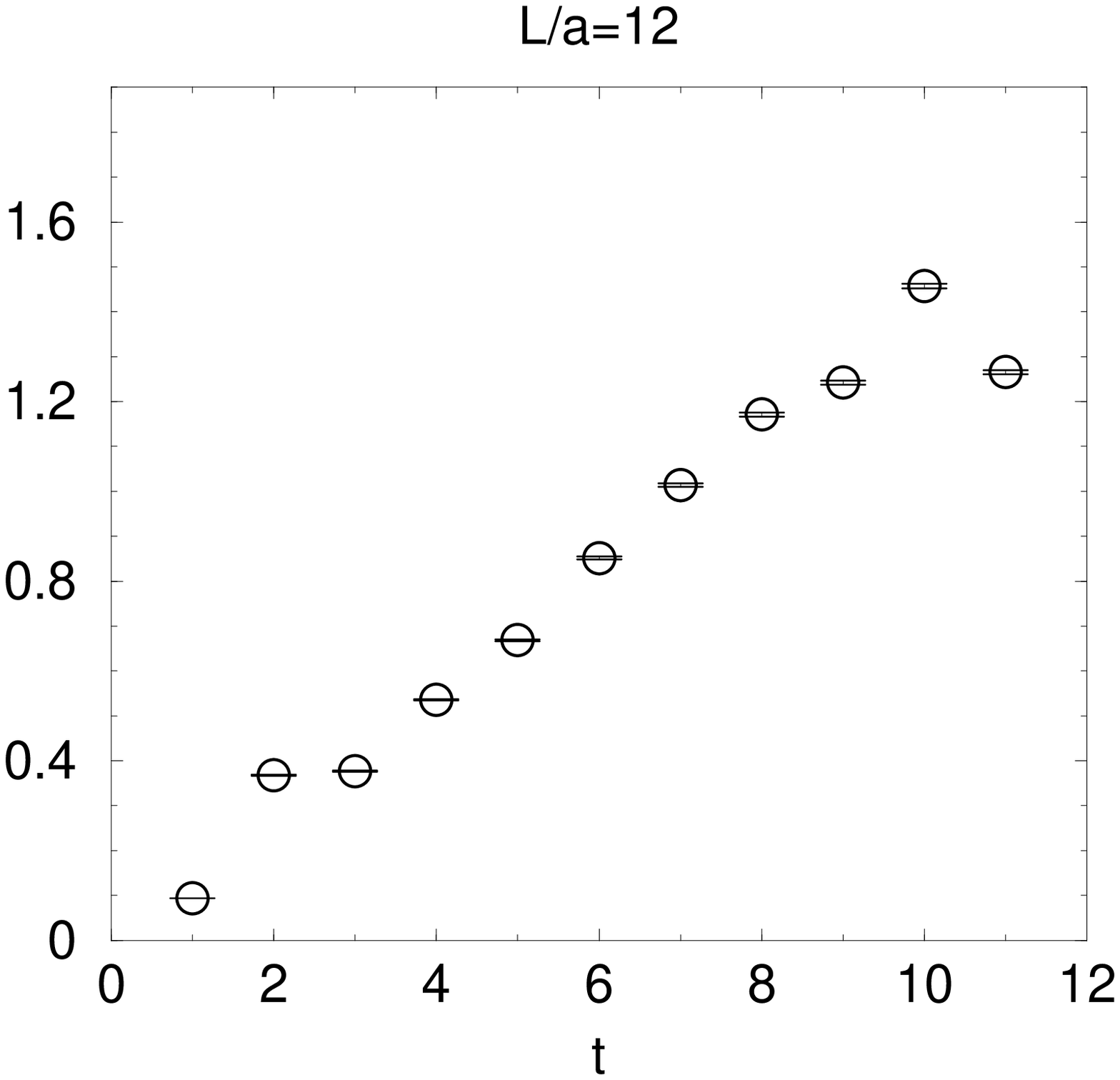}}
 \end{center}
 \begin{center}
  \scalebox{0.32}{\includegraphics{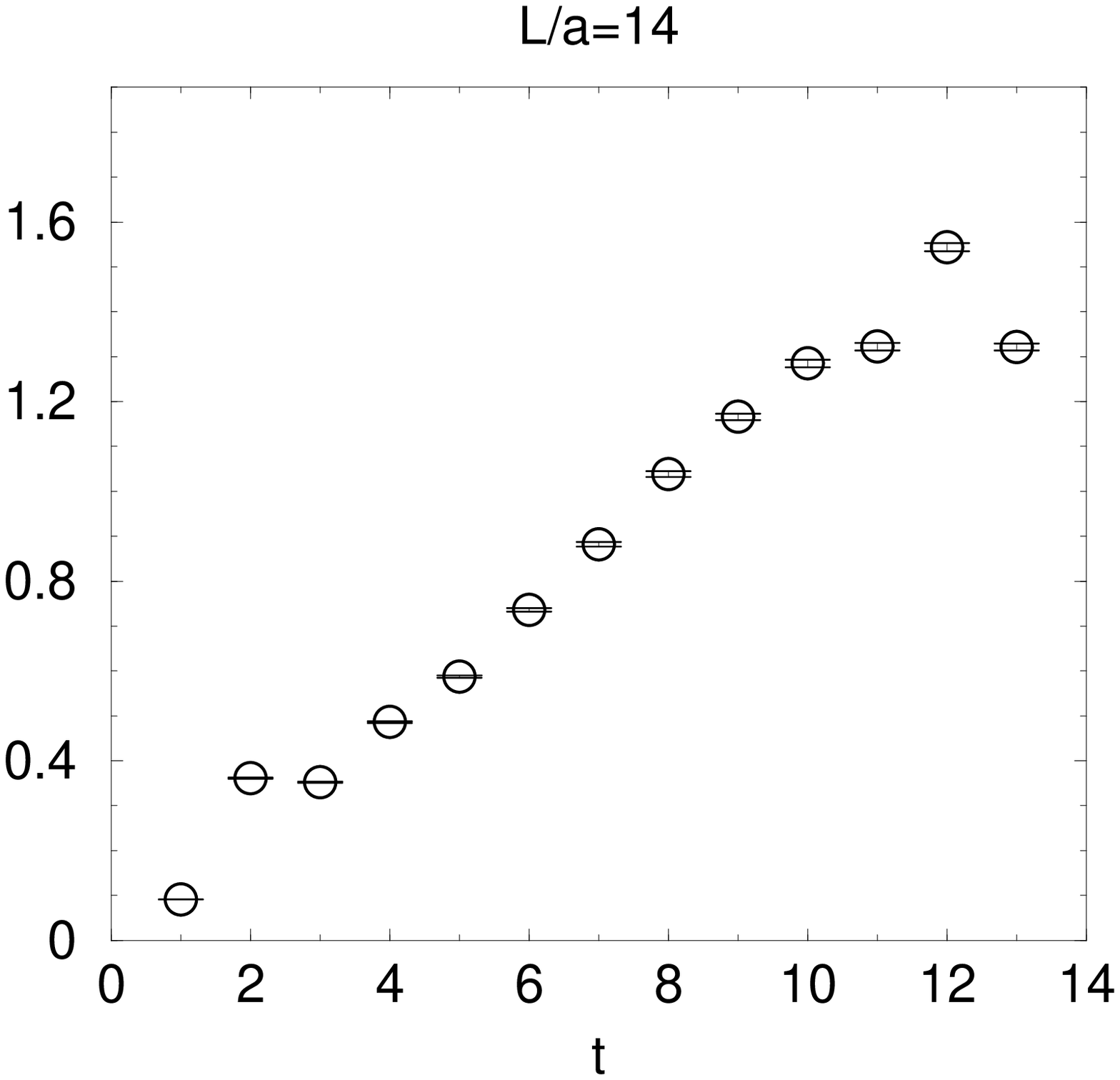}}
\qquad
  \scalebox{0.32}{\includegraphics{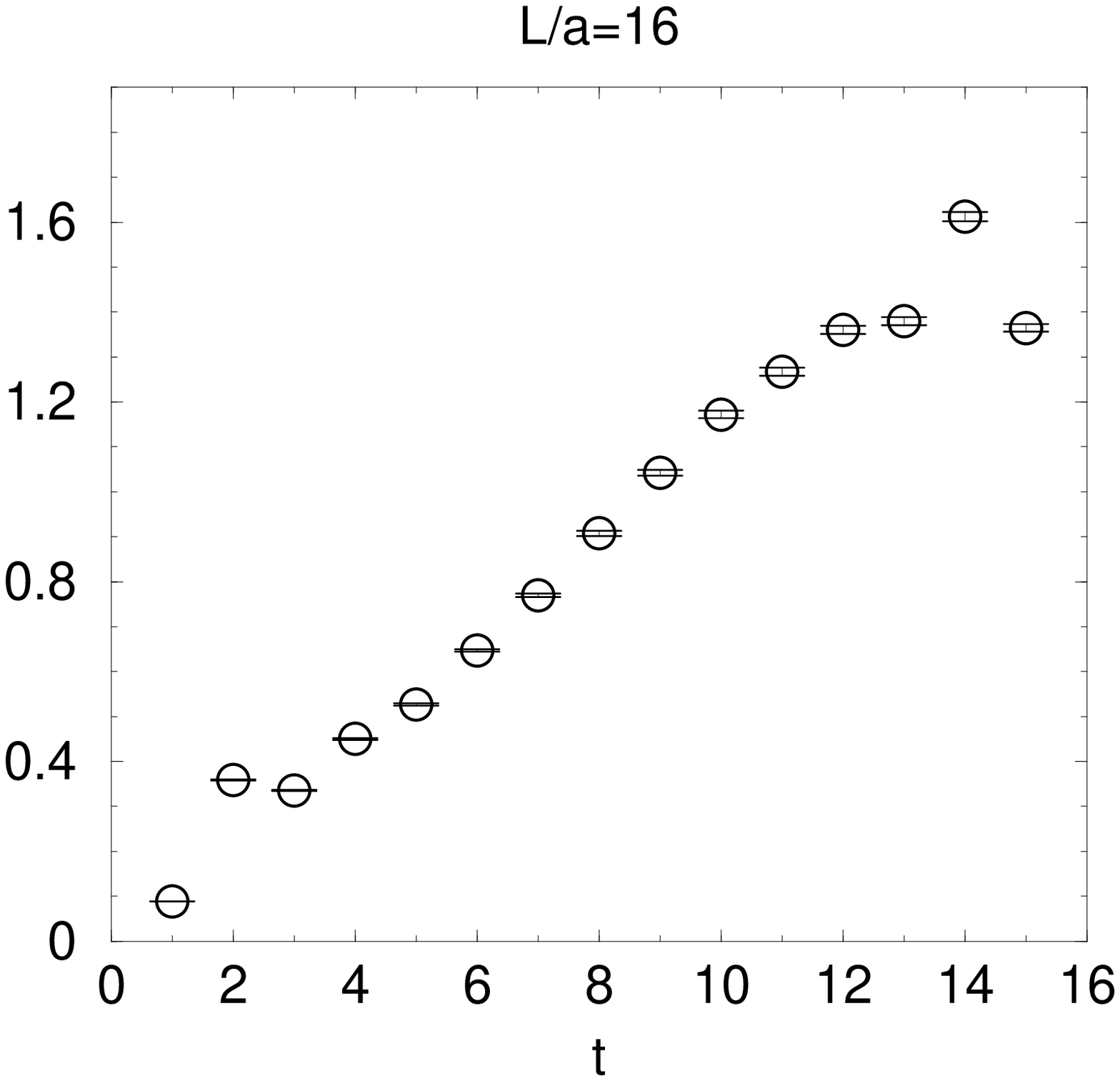}}
 \end{center}
 \begin{center}
  \scalebox{0.32}{\includegraphics{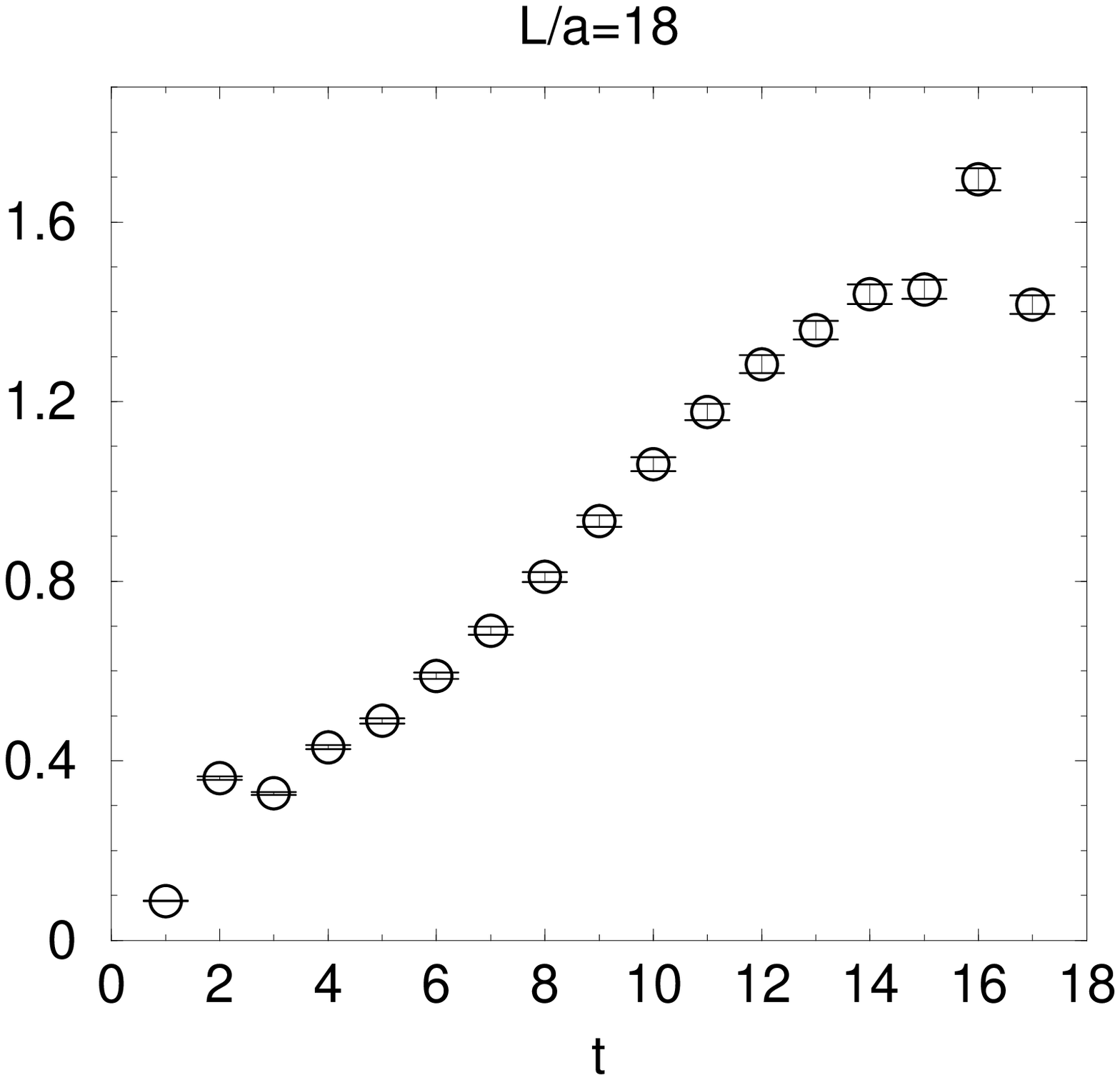}}
\qquad
  \scalebox{0.32}{\includegraphics{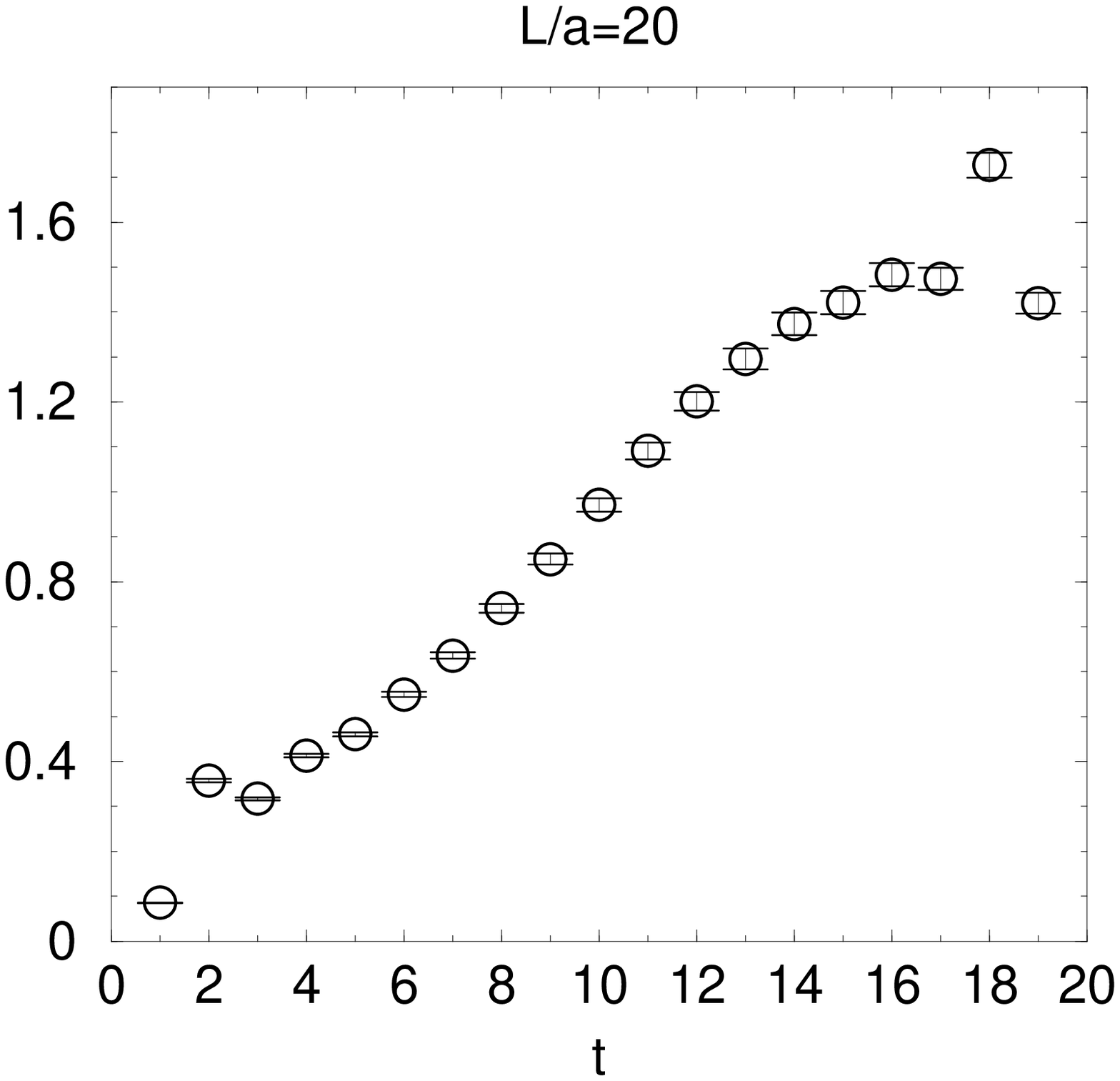}}
 \end{center}
 \caption{$x_0$ dependence of $Z_{VA+AV;2}^-(g_0,a\mu_{\rm min})$ for the 
scheme $2$.}
 \label{fig:zvam2.x0dep}
\end{figure}

The behavior of $Z_{V}$ given in  \eqn{eqn:zv} is plotted as a function
of time $x_0$ in Fig.~\ref{fig:zv.x0dep} at  $N_L=6\sim20$. As $a$
decreases($N_L$ increases), $Z_V$ becomes flatter in $x_0$. 
Typical behaviors of $Z_{VA+AV;s}^+$ \eqn{eqn:renormalization-condition}
is given for the schemes $s=1$ in Fig.~\ref{fig:zvap1.x0dep} and the
schemes $s=2$  in Fig.~\ref{fig:zvap2.x0dep}, and  $Z_{VA+AV;s}^-$ for 
$s=1$ in Fig.~\ref{fig:zvam1.x0dep} and $s=2$ in Fig.~\ref{fig:zvam2.x0dep}.
Both renormalization factors are almost $x_0$ independent for $s=1$, 
while they strongly depend on $x_0$ for $s=2$.

Taking the value at $x_0=L/2$\cite{Guagnelli:2005zc}, we get
renormalization factors, whose
numerical value are listed in table \ref{tab:zvaav.mumin}.
Combining $Z_{VA+AV;s}^+$ and $Z_V$,
we get the renormalization factors for $B_K$ in \eqn{eqn:ZBK},
which is also listed in the table \ref{tab:zvaav.mumin}.
All errors in the table are evaluated by a single elimination jackknife
procedure.

\subsection{Scaling behavior of the step scaling function at $L_{\rm max}$}
In this subsection,
we discuss universality of the scale evolution function $Z_{VA+AV}^{\rm
NP}(\mu_{\rm max}, \mu_{\rm min})$.
More explicitly
we calculate the SSF at the largest coupling  $u=\bar{g}^2(1/L_{\rm max})=3.480$ for four values of  $\beta$'s (lattice spacings) to make the continuum extrapolation,
and  compare the result with that by the Alpha collaboration~\cite{Guagnelli:2005zc}.
The SSF of the four-fermion operator is defined as a
ratio of the renormalization factors at two different box sizes:
\begin{eqnarray}
\Sigma_{VA+AV;s}^\pm(u,a/L) =
\frac{Z_{VA+AV;s}^\pm(g_0,a/(2L) )}
{Z_{VA+AV;s}^\pm(g_0,a/L)}\biggl|_{m=0,\ovl{g}^2(1/L)=u}.
\end{eqnarray}
Since we have already calculated $Z^\pm(g_0,a/(2L_{\rm max}) )$ in the previous subsection,
we need to calculate $Z^\pm(g_0,a/L_{\rm max})$ except $N_L=20 = 2L_{\rm max}/a$.
Number of configuration is fixed to $100$ for all $N_L$.
Values of $Z^\pm(g_0,a/L_{\rm max})$ at $N_L =4,6,8,10$ and
$Z^\pm(g_0,a/(2L_{\rm max}) )$ at $N_L=20$ are given in table
\ref{tab:zvaav.ssf}.

\begin{figure}
 \begin{center}
  \scalebox{0.4}{\includegraphics{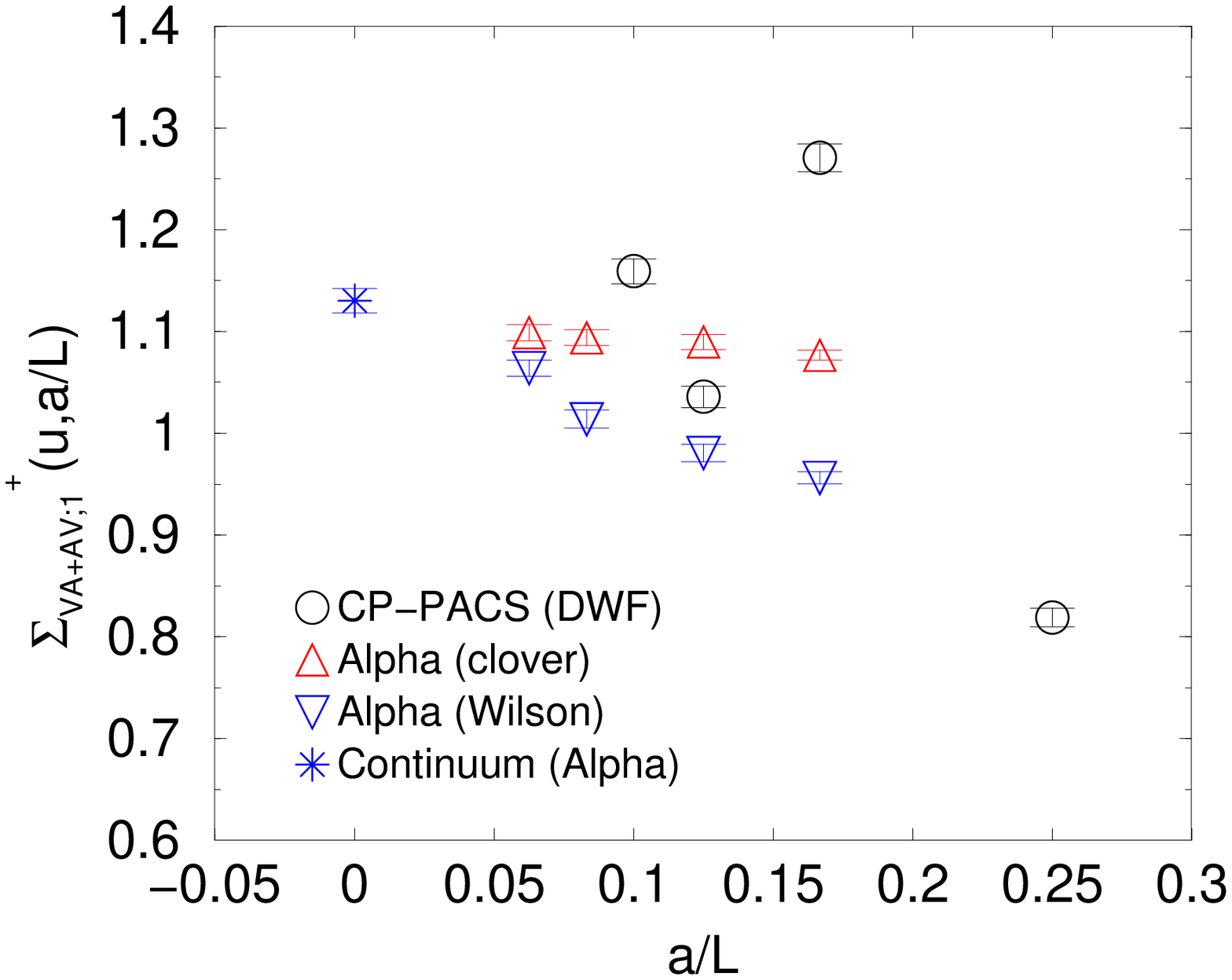}}
  \scalebox{0.4}{\includegraphics{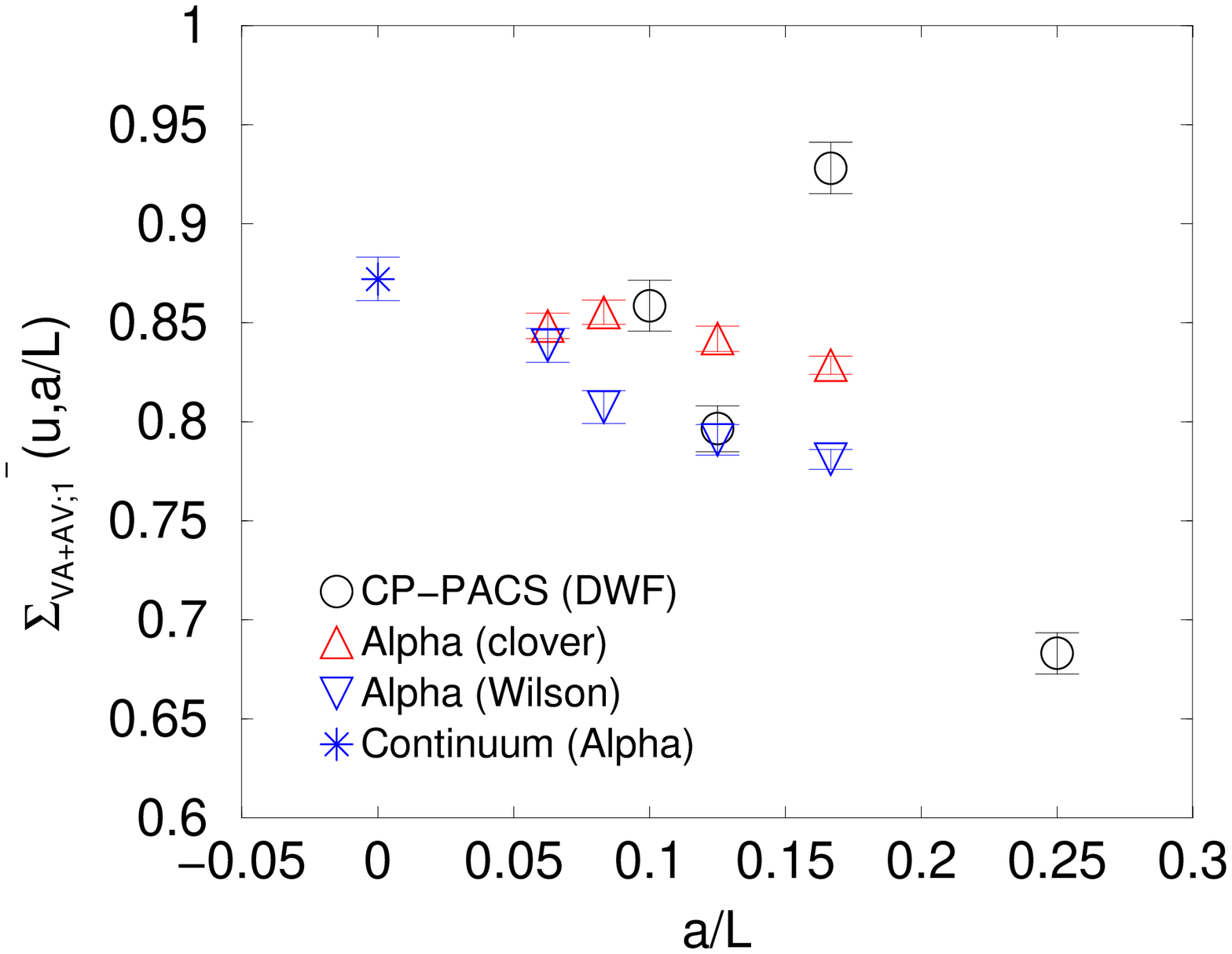}}
  \caption{Scaling behaviors of the SSF $\Sigma_{VA+AV;1}^+(u,a/L)$ (left)
  and $\Sigma_{VA+AV;8}^-(u,a/L)$ (right).
  Filled circle shows our result with the domain-wall fermion at $M=1.8$.
  Open up and down triangles show results by the Alpha collaboration
  with improved and ordinary Wilson fermion actions,
  together with the combined continuum limit(star)\cite{Guagnelli:2005zc}.}
  \label{fig:SSF.scheme1}
 \end{center}
\end{figure}

\begin{figure}
 \begin{center}
  \scalebox{0.4}{\includegraphics{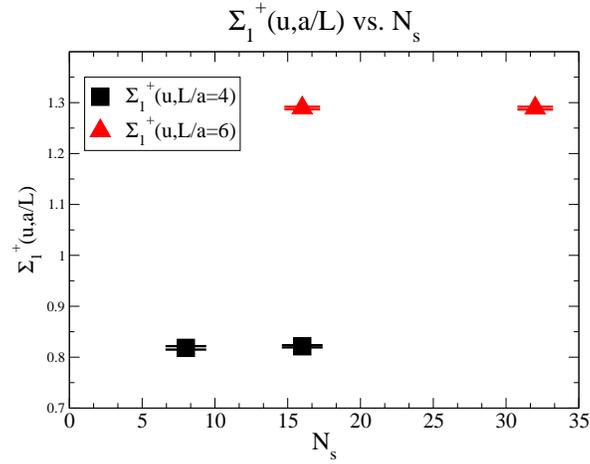}}
  \caption{The SSF $\Sigma_{VA+AV;1}^+(u,a/L)$ as a function of the fifth
  dimensional length $N_5$.
  Triangles(squares) are results at $L/a=4(6)$.} 
  \label{fig:SSF.ns}
 \end{center}
\end{figure}

In Fig.~\ref{fig:SSF.scheme1}, we compare our SSF (filled circles) for
$\Sigma_{VA+AV;1}^+(u,a/L)$ (Left) and $\Sigma_{VA+AV;8}^-(u,a/L)$
(Right) with those by the Alpha collaboration calculated with the clover
fermion (triangle up) and the Wilson fermion (triangle down) as a
function of $a/L$,
together with their combined continuum limit by star~\cite{Guagnelli:2005zc}.
We have surprisingly found that the scaling violation of our SSF is large and 
they seem to approach their continuum limits with oscillation.
To check whether this oscillating behavior is caused by the 
$\mathcal{O}(a)$ bulk chiral symmetry breaking effect of the DWQCD at finite $N_5$
or not, we investigate the $N_5$ dependence of the SSF at $u=3.480$.
As is shown in Fig.~\ref{fig:SSF.ns}, comparisons between $N_5=8$ and
$N_5=16$ for $\Sigma_{VA+AV;1}(u,L/a=4)$ and between $N_5=32$ and
$N_5=16$ for $\Sigma_{VA+AV;1}(u,L/a=6)$ indicate no $N_5$ dependence
within statistical errors. The $\mathcal{O}(a)$ bulk chiral symmetry breaking effect
has nothing to do with the oscillating behavior.

\begin{figure}
 \begin{center}
  \scalebox{0.4}{\includegraphics{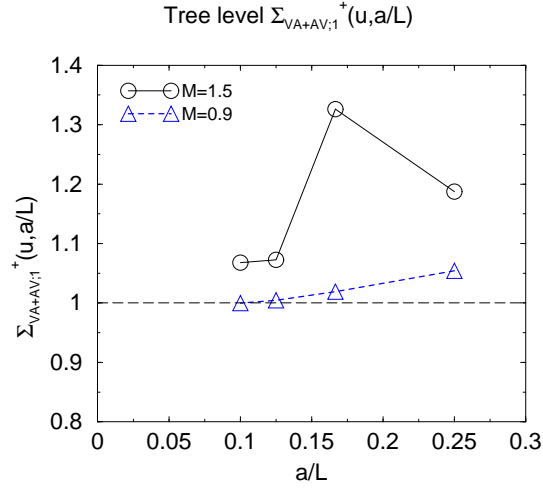}}
  \caption{Scaling behaviors of the tree level SSF.
  Open circles are results at $M=1.5$ while open triangle at $M=0.9$.
  The continuum value is represented by the dotted line.}
  \label{fig:SSF.tree}
 \end{center}
\end{figure}

We then suspect that the bad scaling behavior is caused by the
$\mathcal{O}(a)$ boundary effect in the SF scheme of DWQCD.
To confirm this,  we calculate the tree level SSF on the lattice,
\begin{eqnarray}
\Sigma_{VA+AV;1}^{+}(a/L)
=\frac{Z_{VA+AV;1}^+(a/2L)^{\rm lattice}_{\rm tree}}
{Z_{VA+AV;1}^+(a/L)^{\rm lattice}_{\rm tree}},
\end{eqnarray}
where $N_5\to\infty$ limit is already taken.
At tree level, we have $\Sigma_{VA+AV;1}^+(a/L)=\Sigma_{VA+AV;8}^-(a/L)$, which of course
approach to 1 in the continuum limit. 
In this calculation,
we take the tadpole improved value $M_{\rm tad} = 1.5$ at $\beta\sim 2.9$ for the value of
$M$ instead of the tree level value $M=1.8$,
in order to take into account an additive shift of $M$ caused by the quantum correction.
We plot the scaling behavior of  $\Sigma_{VA+AV;1}^{+}(a/L)$
by open circles in Fig.~\ref{fig:SSF.tree}, 
which shows an oscillation similar to one in Fig.~\ref{fig:SSF.scheme1}.
On the other hand, if we take $M$ is close to but smaller than unity,
the scaling behavior is much improved without oscillation, as is shown by open 
triangles at $M=0.9$ in Fig~\ref{fig:SSF.tree}.

\begin{figure}
 \begin{center}
  \scalebox{0.4}{\includegraphics{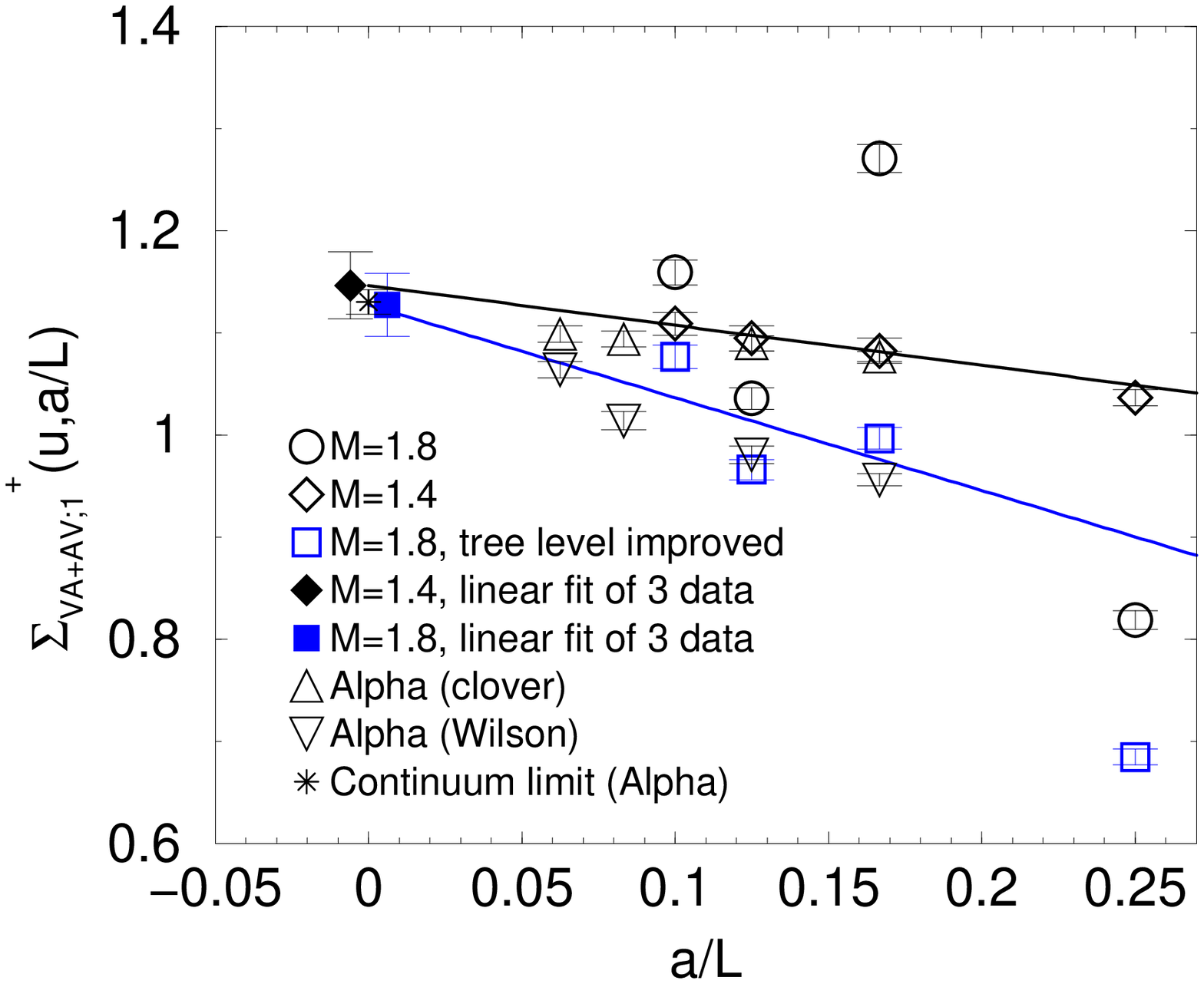}}
  \scalebox{0.4}{\includegraphics{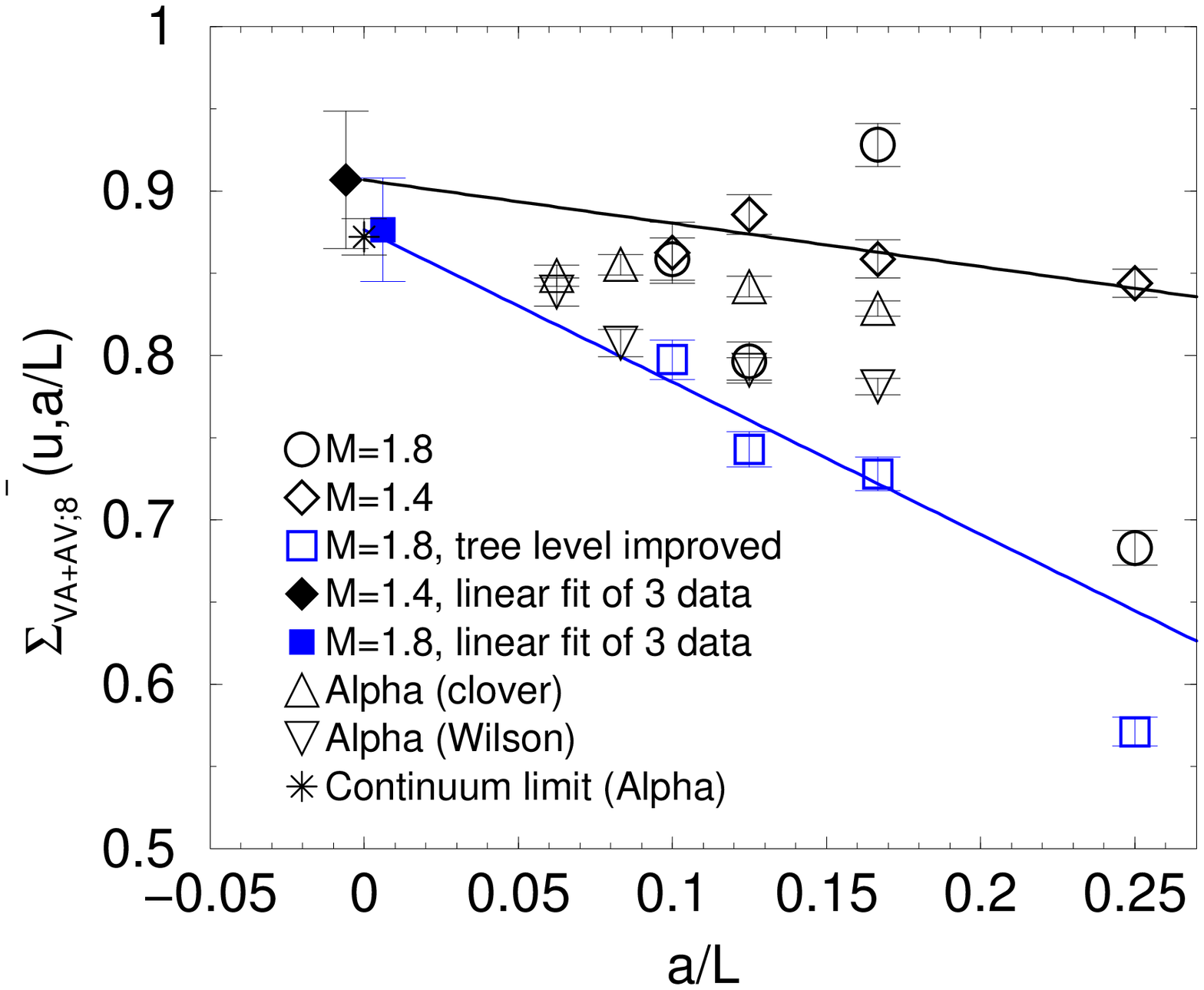}}
  \caption{Scaling behaviors of the SSF $\Sigma_{VA+AV;1}^+(u,a/L)$ (left)
  and $\Sigma_{VA+AV;8}^-(u,a/L)$ (right).
  Open diamonds represent our result with the domain-wall fermion at $M=1.4$.
  Open squares are result at $M=1.8$ with a tree level improved
  definition Eq.~\eqref{eq:treeimp}.
  Corresponding filled symbols denote linear continuum extrapolations with
  last three data.
  Open up and down triangles show results by the Alpha collaboration
  with improved and ordinary Wilson fermion actions, together
  with the combined continuum limit(star)\cite{Guagnelli:2005zc}.}
  \label{fig:SSF.imp}
 \end{center}
\end{figure}

The tree level analysis indicates that the scaling behavior can be improved by changing the domain-wall height $M$ so that the tadpole improved value becomes close to unity.
Motivated by this, we have recalculated the non-perturbative SSF at $M=1.4$, 
which corresponds to $M_{\rm tad} \simeq 1.0$ at the range of our $\beta$. 
Results are given in table \ref{tab:zvaav.ssf.M14}, and are plotted by open
diamonds in Fig~\ref{fig:SSF.imp}.
It is clearly seen that the scaling behavior at $M=1.4$ is much improved, so that the linear 
continuum extrapolation can be made using last three points.
The value in the continuum limit (filled symbol) 
is consistent with the previous one by the Alpha collaboration (star).

We explore a different method to improve the scaling behavior of the SSF, without performing 
new simulations at different value of $M$. 
A main idea is to cancel the oscillating behavior of the SSF by that at tree level, 
changing the renormalization condition from \eqn{eqn:renormalization-condition} to
\begin{eqnarray}
Z_{VA+AV;s}^\pm(g_0,\mu)h_s^{\pm}(x_0=L/2;g_0) =
 h_{s\ {\rm (lattice)}}^{\pm ({\rm tree})}(x_0=L/2),
\label{eq:treeimp}
\end{eqnarray}
where the tree level correlation function, evaluated at
$M=M_{\rm tad}$ for corresponding $\beta$, is used in the right-hand side.
We call this method the tree level improvement.
Results are given in table \ref{tab:zvaav.imp} and
\ref{tab:zvaav.ssf.imp}, and are plotted by open squares in the
Fig~\ref{fig:SSF.imp}.
We find that the magnitude of oscillation is reduced, so that 
a linear continuum extrapolation using last three data becomes possible.
The value in the continuum limit is consistent with both one by the Alpha collaboration and one at $M=1.4$.

\begin{figure}
 \begin{center}
  \scalebox{0.45}{\includegraphics{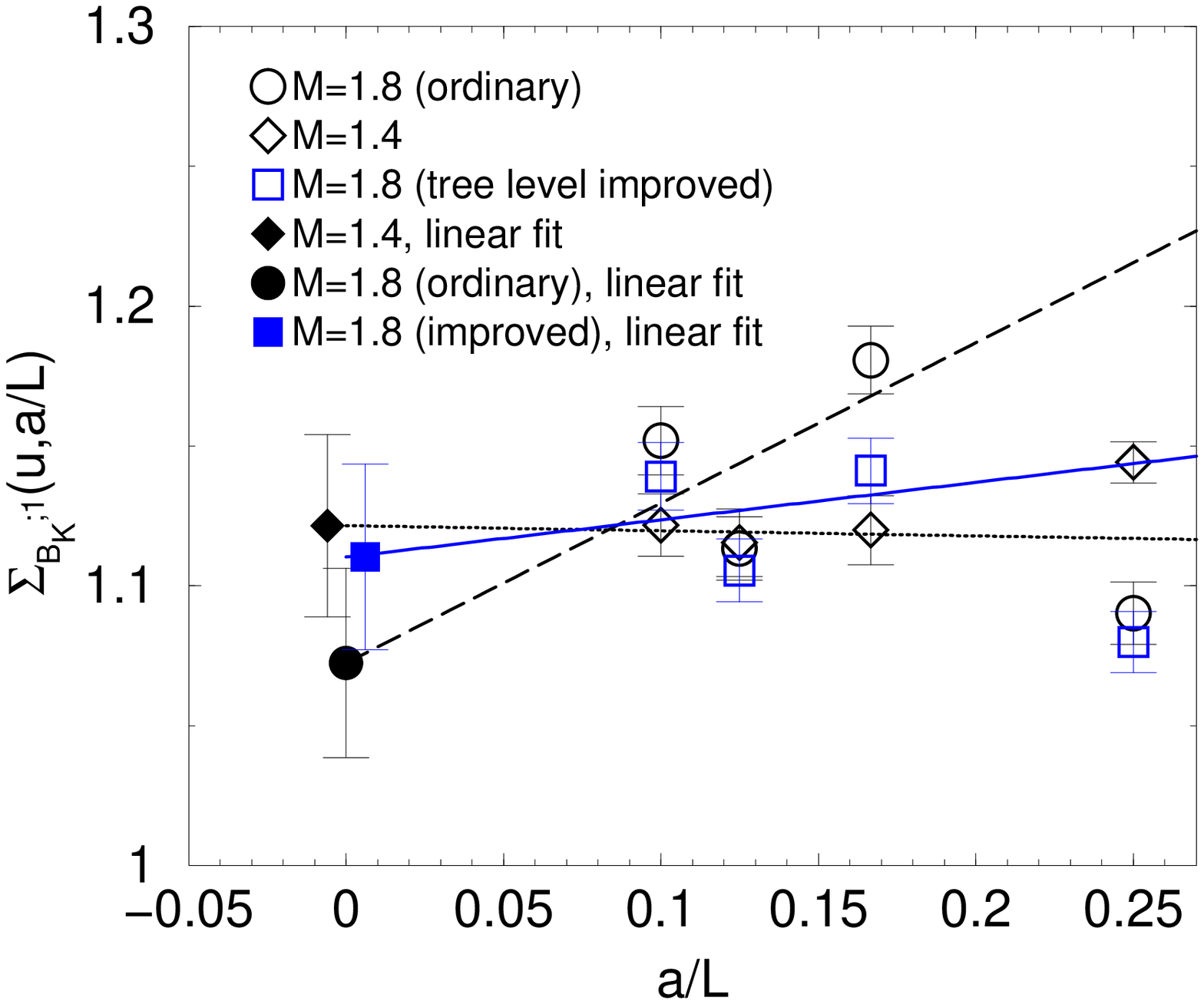}}
  \scalebox{0.45}{\includegraphics{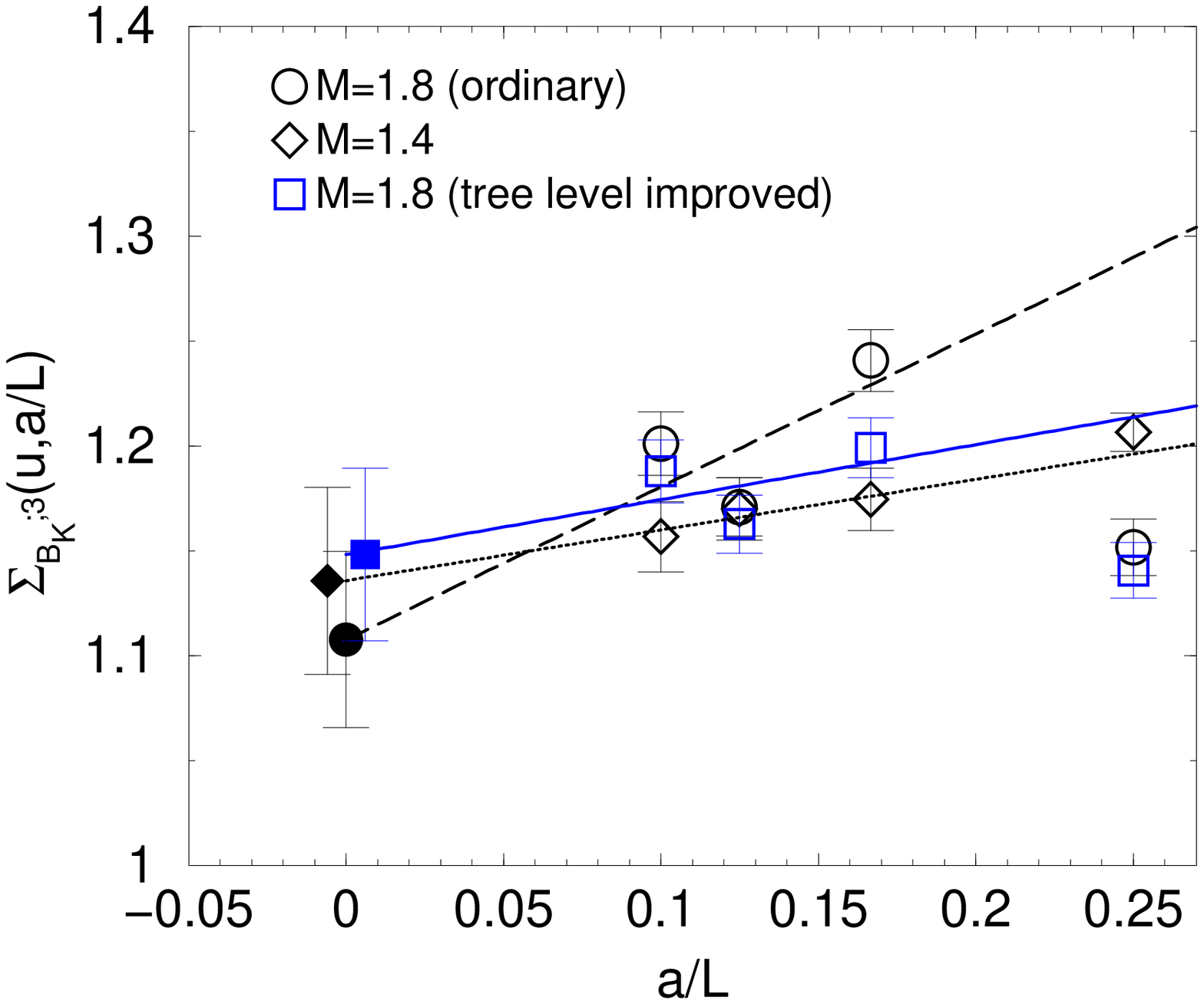}}
  \scalebox{0.45}{\includegraphics{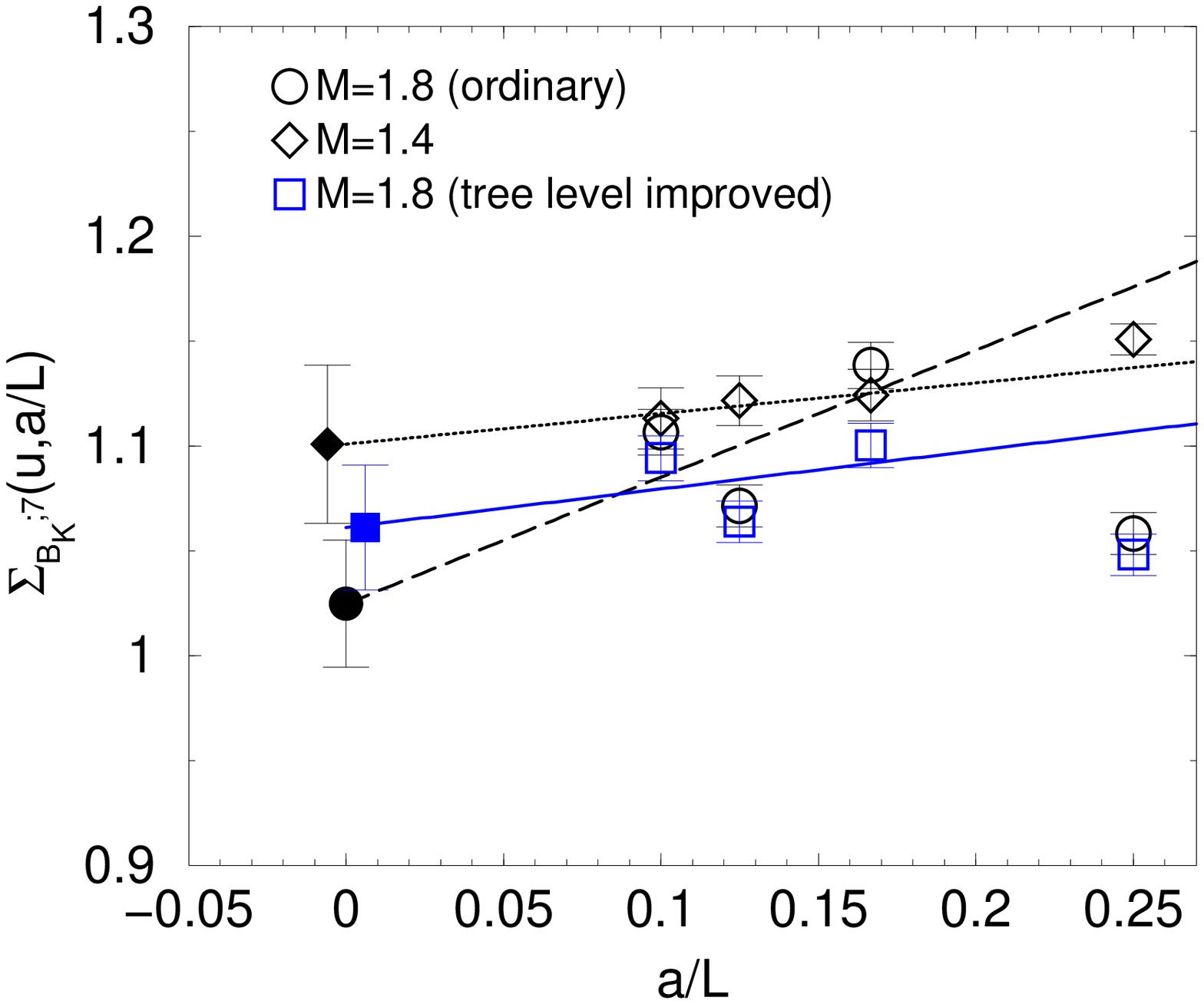}}
  \caption{Scaling behaviors of the SSF $\Sigma_{B_K}(u,a/L_{\rm max})$ of
  $B_K$ for schemes $s=1, 3, 7$.
  Open circles represent results at $M=1.8$ with the ordinary renormalization
  condition and open squares are results at $M=1.8$ with the improved
  condition.
  Results at $M=1.4$ are given by open diamonds.
  Linear continuum extrapolations are made using data at finest three lattice
  spacings.}
  \label{fig:SSFzBK}
 \end{center}
\end{figure}

In addition to the SSF of $VA+AV$, we have also considered 
the SSF of $B_K$ defined by 
\begin{eqnarray}
\Sigma_{B_K}(u,a/L)=\frac{Z_{B_K}(g_0,a/2L)}{Z_{B_K}(g_0,a/L)}
\end{eqnarray}
at $L=L_{\rm max}$.
Results are plotted as a function of $a/L$ in Fig.~\ref{fig:SSFzBK} for three ``good
schemes''.
In each figure,
results at $M=1.8$ with and without the tree level improvement
are represented by open squares and open circles, respectively, while
the result at $M=1.4$ by open diamonds.
The scaling behaviors are reasonably  well with the tree level improvement or at $M=1.4$.
Even without improvement, the oscillation is not so large.
Linear extrapolations with three data at finest lattice spacings give
consistent results among all three cases.
The large oscillating behavior seems to be partly canceled between
$Z_{VA+AV}^+$ and $Z_V$ in $Z_{B_K}$.

We finally perform combined linear fits of the $M=1.4$ data and the tree level improved data
using the finest three lattice spacings.
Values in the continuum limit of all SSF are given in table \ref{tab:ssf.cont}.

\subsection{Renormalization of $B_K$}

In this subsection we evaluate the renormalization factor
$\mathcal{Z}_{B_K;s}(g_0)$ which convert the lattice bare $B_K(g_0)$ of
DWQCD to the RGI $\wh{B}_K$. 
As suggested in the previous subsection,
we here employ the renormalization factor obtained with the tree level improved condition, 
hoping that this also improves the scaling behavior of $B_K$.
Combining our renormalization factor
$Z_{B_K;s}^{\rm NP}(g_0,\mu_{\rm min})$ in table \ref{tab:zvaav.imp}
with the RG running factor
$Z_{VA+AV}^{\rm PT}(\infty,\mu_{\rm max})
Z_{VA+AV}^{\rm NP}(\mu_{\rm max},\mu_{\rm min})$
given by the Alpha collaboration \cite{Guagnelli:2005zc}, we obtain the
renormalization factor $\mathcal{Z}_{B_K;s}(g_0)$ at each $\beta$ in table \ref{tab:RGIZBK.mumin}.

\begin{figure}
 \begin{center}
  \scalebox{0.45}{\includegraphics{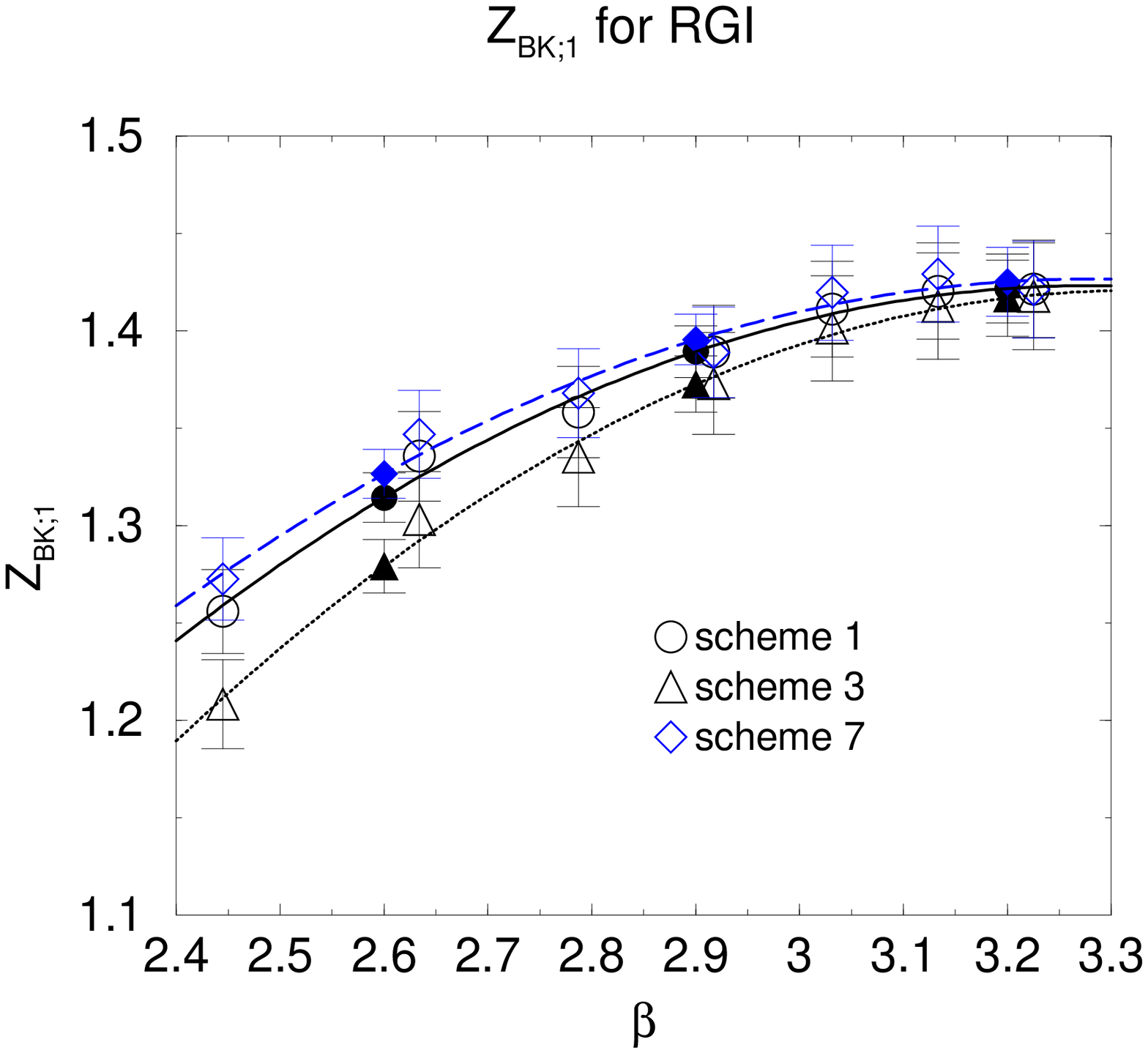}}
  \scalebox{0.45}{\includegraphics{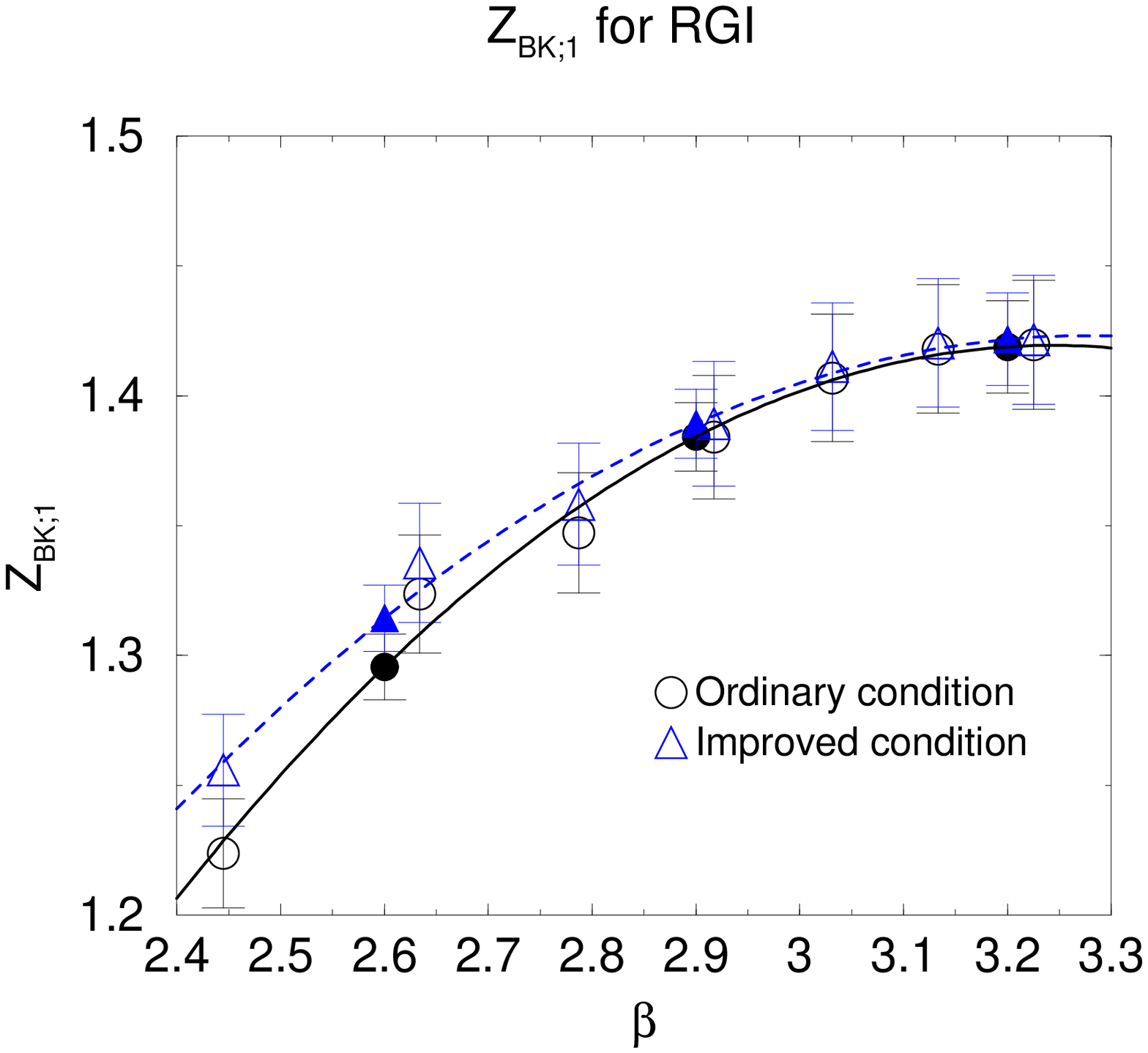}}
  \caption{$\beta$ dependence of the renormalization factor
  $\mathcal{Z}_{B_K}$ for the RGI operator with the polynomial fit.
  Filled symbols represent fitted values at $\beta=2.6$, $2.9$ and $3.2$.
  The left panel shows a comparison between schemes $s=1,3,7$ with the improved
  renormalization condition.
  The right panel show a comparison between improved and ordinary conditions
  for the scheme $1$.}
     \label{fig:RGIZBK.betadep}
 \end{center}
\end{figure}

In order to obtain the renormalization factors at $\beta=2.6$, $2.9$ and
$3.2$, we interpolate the result at each scheme $s$ by the polynomial,  
\begin{eqnarray}
\mathcal{Z}_{B_K;s}(\beta) = a_s + b_s(\beta-3) +c_s(\beta-3)^2,
\label{eqn:polynomial-form}
\end{eqnarray}
which is shown in the left panel of Fig.~\ref{fig:RGIZBK.betadep} for schemes 1, 3, 7, 
together with  interpolated values at three $\beta$'s by solid symbols.
 As is shown in the figure the polynomial interpolation works very well with
small $\chi^2/{\rm dof}\sim0.1$.
 Since the renormalization factor $\mathcal{Z}_{B_K;s}$ should not depend on schemes,
 the discrepancy between three schemes is considered to be
the lattice artifact and therefore it should disappear at high $\beta$, as seen in the figure.
The renormalization factors with and without the tree level improvement  are compared  
in the right panel of Fig.~\ref{fig:RGIZBK.betadep}.
Two renormalization factors are consistent within
statistical errors, and agree completely at high $\beta$ as expected.

\begin{figure}
  \begin{center}
   \scalebox{0.45}{\includegraphics{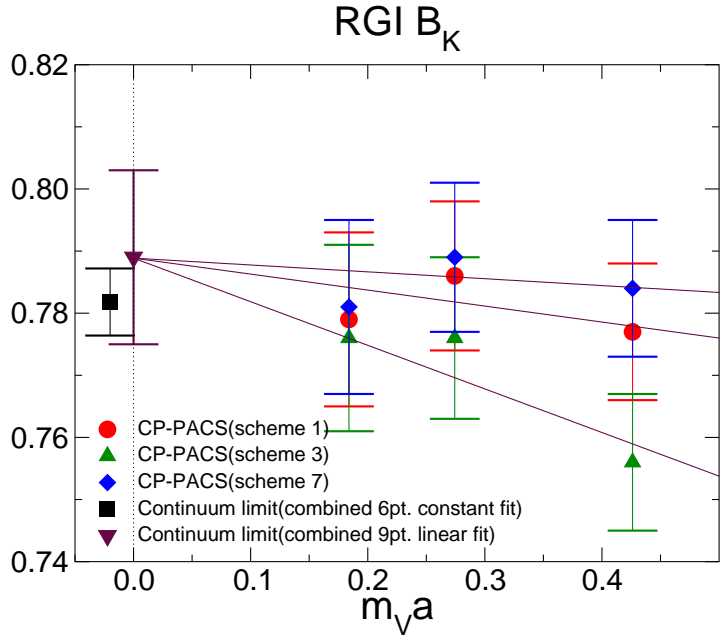}}
   \caption{Scaling behaviors of the RGI $\wh{B}_K$.
   Our results are given by filled circles, up triangle and diamond
   symbols for schemes $s=1, 3, 7$, respectively.
   A filled square is the continuum limit by the combined fit using 
   6 data of three schemes at finest two lattice spacings, while
   a filled down triangle shows the continuum limit by the combined 
   fit using all 9 data.
   }
   \label{fig:RGIBK}
  \end{center}
\end{figure}

Multiplying the renormalization factor to the bare $B_K(g_0)$, we obtain
the RGI $\wh{B}_K$ in table \ref{tab:RGIbk}. 
The scaling behavior of $\wh{B}_K$ is shown in the Fig.~\ref{fig:RGIBK} for
$s=1,3,7$, as a function of $m_V a$.
Note that the scaling behavior of $\wh{B}_K$ with other  (bad)
schemes is indeed bad,  therefore we do not use them in our analysis.
Since the scaling violations are small, we have made the constant continuum extrapolation using the last two data points.
We arrive at 
\begin{eqnarray}
\wh{B}_K &=& 0.783(9)\quad\quad\text{for Scheme 1},
\\
\wh{B}_K &=& 0.776(10)\quad\quad\text{for Scheme 3},
\\
\wh{B}_K &=& 0.786(9)\quad\quad\text{for Scheme 7}.
\end{eqnarray}
Since the values in the three schemes agree within errors in the continuum limit,
we have made the combined constant fit for all three schemes, which gives
\begin{eqnarray}
\wh{B}_K = 0.782(5)\quad(\chi^2/{\rm dof}=0.87).
\end{eqnarray}
To estimate the ambiguity of the continuum extrapolation, we have also made the
combined linear extrapolation using all 9 data points, and we obtain
 \begin{eqnarray}
\wh{B}_K = 0.789(14)\quad(\chi^2/{\rm dof}=0.16).
\end{eqnarray}
Now the final result we obtain leads
\begin{eqnarray}
\wh{B}_K = 0.782(5)(7)
\end{eqnarray} 
where the central value and the first error are taken from the combined constant fit,
while the systematic error, given in the second, is estimated by the difference between the constant and the linear fits.
Our result is consistent with previous results nonperturbatively renormalized 
by DWQCD~\cite{Aoki:2005ga} with the RI/MOM scheme
($\wh{B}_K=0.786(31)$) and by tmQCD~\cite{bk_tm,Dimopoulos:2007cn} with the SF
scheme ($\wh{B}_K=0.735(71)$).

\begin{figure}
 \begin{center}
  \scalebox{0.45}{\includegraphics{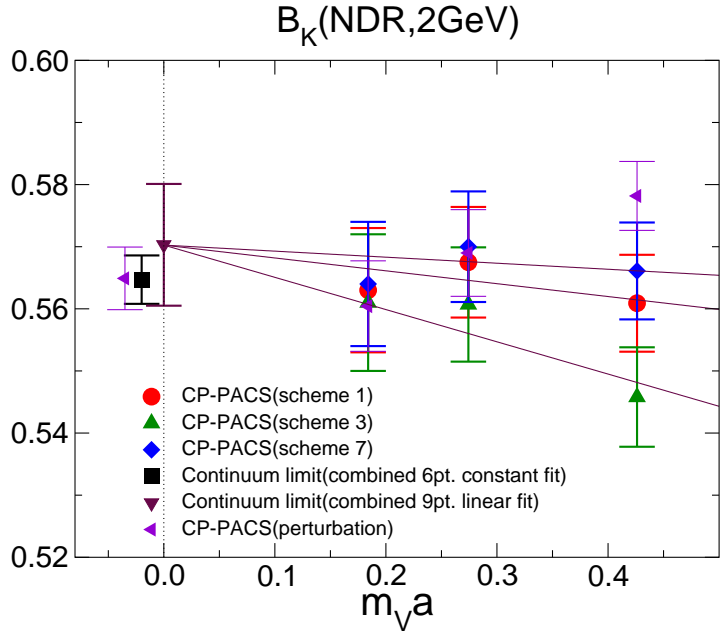}}
 \caption{Scaling behaviors of the renormalized $B_K(\msbar,\mu=2{\rm GeV})$. 
   Our results are given by filled circle, up triangle and diamond
   symbols for schemes $s=1, 3, 7$, respectively.
   A filled square symbol is the continuum limit by the combined fit using 
   6 data of three schemes, while a filled down triangle shows 
   the continuum limit by the combined fit with all 9 data.
   For a comparison previous results with the perturbative renormalization 
   factor~\cite{AliKhan:2001wr} are given by left triangle. }
 \label{fig:BKNDR}
 \end{center}
\end{figure}

For the latter convenience, we convert $\wh{B}_K$ to
the renormalized $B_K$ in $\overline{\rm MS}$ scheme
with the naive dimensional regularization (NDR) at a scale $\mu=2$ GeV.
The renormalized operator in $\msbar$ scheme is obtained by inverting the
definition of the RGI operator as
\begin{eqnarray}
O_{\msbar}(x;\mu)=
\left(\frac{g_{\msbar}^2(\mu)}{4\pi}\right)^{\frac{\gamma_{O}^{(0)}}{2b_0}}
\exp\left(\int_{0}^{g_{\msbar}(\mu)}dg
\left(\frac{\gamma_{O}^{\msbar}(g)}{\beta_{\msbar}(g)}
-\frac{\gamma_{O}^{(0)}}{b_0g}\right)\right)
\wh{O}(x),
\label{eqn:MSBK}
\end{eqnarray}
where $\beta_{\msbar}(g)$ and $\gamma_{O}^{\msbar}(g)$ are
renormalization group functions in $\msbar$ scheme, which are estimated
at four loops for $\beta_{\msbar}(g)$ \cite{vanRitbergen:1997va} and at
two loops for $\gamma_{O}^{\msbar}(g)$ \cite{Buras:1989xd}.
The gauge coupling $g_{\msbar}(\mu)$ in $\msbar$ scheme is given in terms of  $\Lambda_{\msbar}$ as
\begin{eqnarray}
&&
\Lambda_{\msbar} = \mu(b_0g^2_{\msbar}(\mu))^{-\frac{b_1}{2b_0^2}}
\exp\left[-\frac{1}{2b_0g^2_{\msbar}(\mu)}\right]
\exp\left[-\int_0^{g_{\msbar}(\mu)}dg\left(\frac{1}{\beta_{\msbar}(g)}
+\frac{1}{b_0g^3}-\frac{b_1}{b_0^2g}\right)\right].
\nn\\
\end{eqnarray}
We adopt a value $\Lambda_{\msbar}=0.586(48)/r_0$ in
Ref.~\cite{NeccoSommer}, and take $r_0 = 0.5$ fm to set a scale.
Multiplying $\mathcal{Z}_{B_K;s}(g_0)$ with the factor in
\eqn{eqn:MSBK}, we obtain the renormalization factor ${Z}_{B_K}^{\msbar}(g_0,
 {\rm 2 GeV})$,
which is listed in table \ref{tab:MSbk}.
The scaling behavior of $B_K^{\msbar}$ is given in the Fig.~\ref{fig:BKNDR}.
Note that
our non-perturbative result differs from at $\beta=2.9$ but agrees with at $\beta=3.2$ 
with the previous result \cite{AliKhan:2001wr} of the DWQCD with
perturbative renormalization \cite{Aoki:1998vv,Aoki:1999ky,Aoki:2002iq}.
The continuum extrapolation has been made as before, and we obtain the final result,
\begin{eqnarray}
B_K^{\msbar}({\rm NDR},2{\rm GeV}) &=& 0.565(4)(5).
\end{eqnarray}

\subsection{Chiral symmetry breaking effect}

As a further check,
we investigate whether the assumption that $Z_{VV+AA}=Z_{VA+AV}$ holds
or not in our DWQCD.
We give only a result here and a detail of 
the SF formalism relevant in the analysis can be found in appendix A.

If the chiral symmetry were exact, 
under chiral rotation of the first flavor
\begin{eqnarray}
q_1\to\wt{q}_1 = i\gamma_5{q}_1,\quad
\zeta_1\to\wt{\zeta}_1=-i\gamma_5{\zeta}_1,\quad
\zeta'_1\to\wt{\zeta}'_1=-i\gamma_5{\zeta}'_1,
\label{eq:chiral}
\end{eqnarray}
we could have the chiral Ward-Takahashi identity,
\begin{eqnarray}
\langle{O}_{VA+AV}\mathcal{O}[\zeta]\rangle_{S} =
\langle{O}_{VV+AA}\wt{\mathcal{O}}[\zeta]\rangle_{S},
\label{eqn:WTid}
\end{eqnarray}
where $\wt{\mathcal{O}}[\zeta]$ is the chirally rotated boundary
operator of (\ref{eqn:boundary-operator}).
A subscript $S$ represents an action under which the expectation value is
evaluated.
Note that the boundary fields are also rotated, which satisfies
opposite SF boundary condition to \eqn{eqn:DirichletBC}.
Therefore
both actions are identical in the bulk but have opposite temporal boundary
conditions. From this WT identity we obtain 
$
Z_{VV+AA} =Z_{VA+AV}
$.

Unfortunately the domain-wall fermion action has a non-invariant part under
the chiral rotation as
\begin{eqnarray}
S_{\rm dwf}\rightarrow S_{\rm dwf}+Y,
\end{eqnarray}
where $Y=\bar{\psi}X\psi$ is the bulk chiral symmetry violating term at the
middle of the fifth dimension.
Therefore the WT identity becomes
\begin{eqnarray}
\langle{O}_{VA+AV}\mathcal{O}[\zeta]\rangle_{S} =
\langle{O}_{VV+AA}\tilde{\mathcal{O}}[\zeta]\rangle_{S+Y} \neq
\langle{O}_{VV+AA}\tilde{\mathcal{O}}[\zeta]\rangle_{S}.
\label{eqn:WTid-dwf}
\end{eqnarray}
A possible chiral symmetry violation comes from the contribution of $Y$,
which is expected to be suppressed exponentially  in $N_5$. 
We estimate the violating effect by comparing
$\langle{O}_{VA+AV}\mathcal{O}[\zeta]\rangle_{S}$ with
$\langle{O}_{VV+AA}\tilde{\mathcal{O}}[\zeta]\rangle_{S}$ directly.

\begin{figure}
  \begin{center}
  \begin{tabular}{ll}
  \begin{minipage}{65mm}
    \begin{center}
      \scalebox{0.45}{\includegraphics{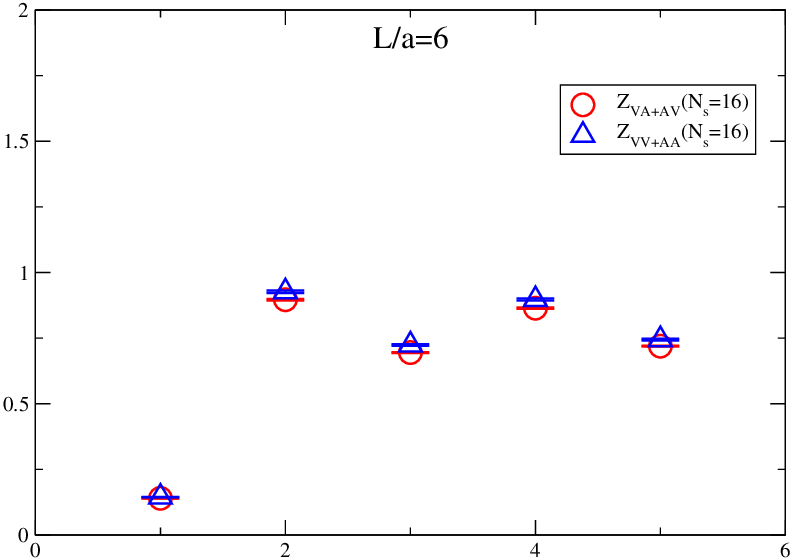}}
    \end{center}
  \end{minipage}&
\hspace{0.35cm}
  \begin{minipage}{65mm}
    \begin{center}
      \scalebox{0.45}{\includegraphics{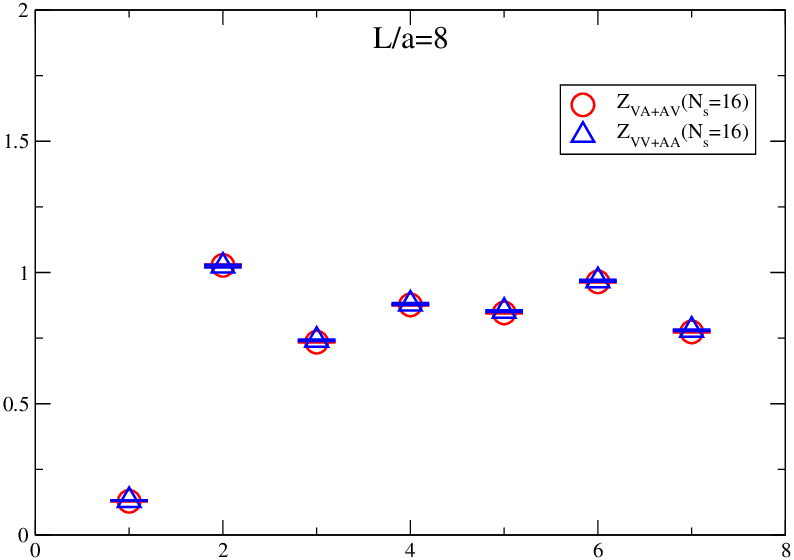}}
    \end{center}
  \end{minipage}\\[3.cm]
  \begin{minipage}{65mm}
    \begin{center}
      \scalebox{0.45}{\includegraphics{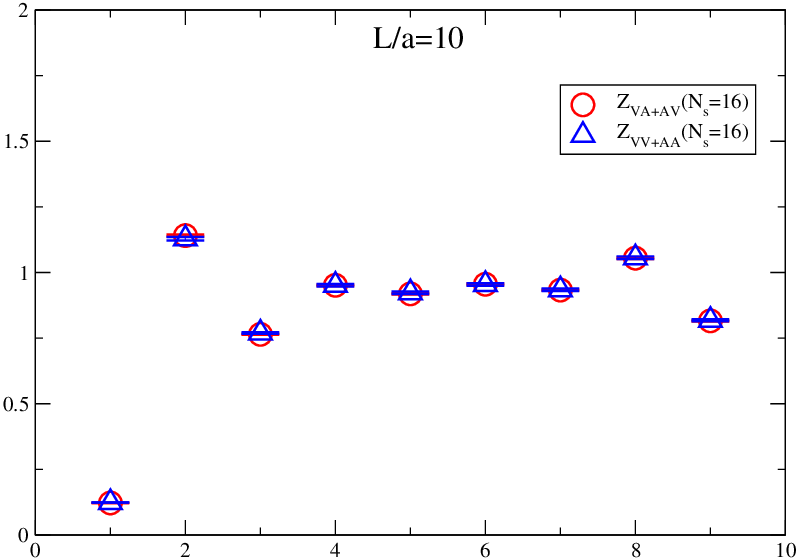}}
    \end{center}
  \end{minipage}&
\hspace{0.35cm}
  \begin{minipage}{65mm}
    \begin{center}
      \scalebox{0.45}{\includegraphics{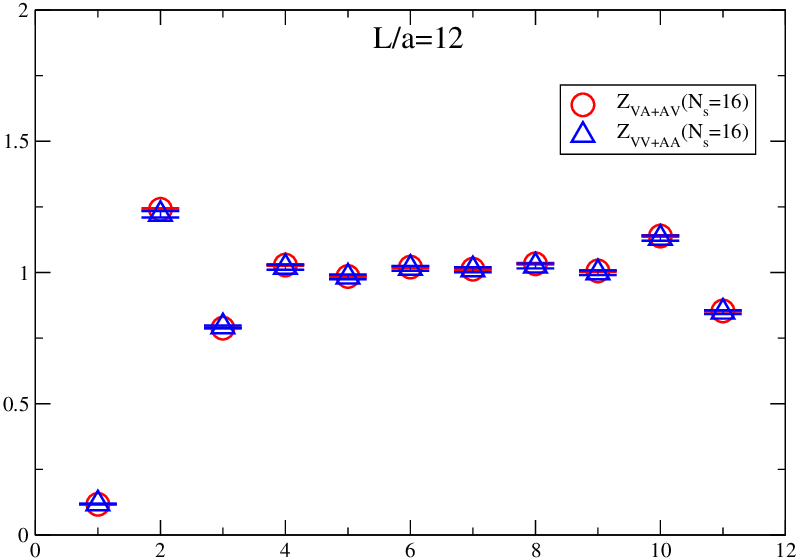}}
    \end{center}
  \end{minipage}\\[3.cm]
  \begin{minipage}{65mm}
    \begin{center}
      \scalebox{0.45}{\includegraphics{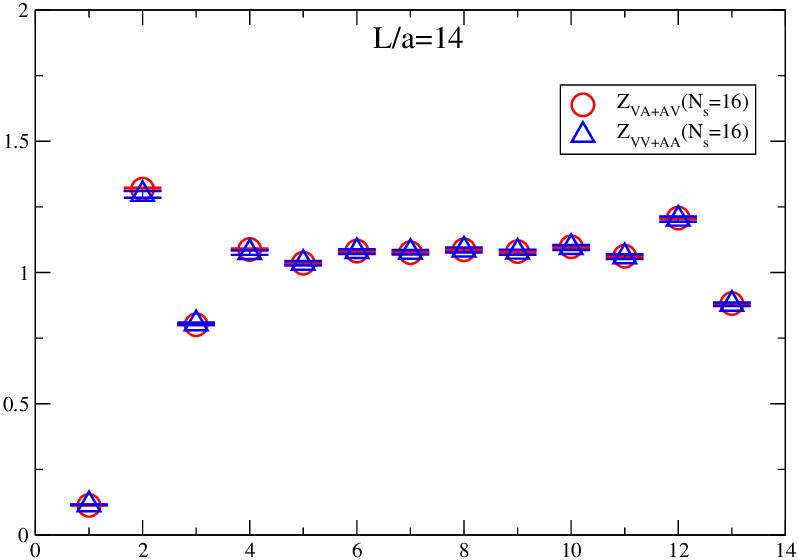}}
    \end{center}
  \end{minipage}&
\hspace{0.35cm}
  \begin{minipage}{65mm}
    \begin{center}
      \scalebox{0.45}{\includegraphics{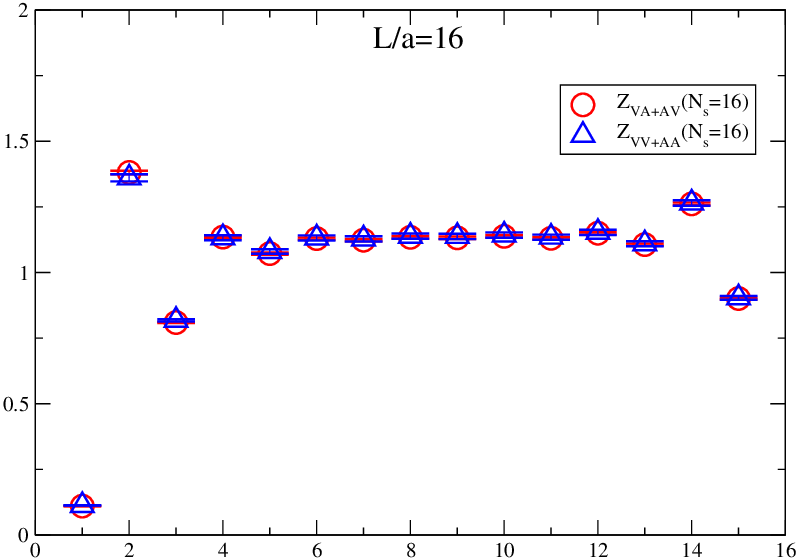}}
    \end{center}
  \end{minipage}\\[3.cm]
  \begin{minipage}{65mm}
    \begin{center}
      \scalebox{0.45}{\includegraphics{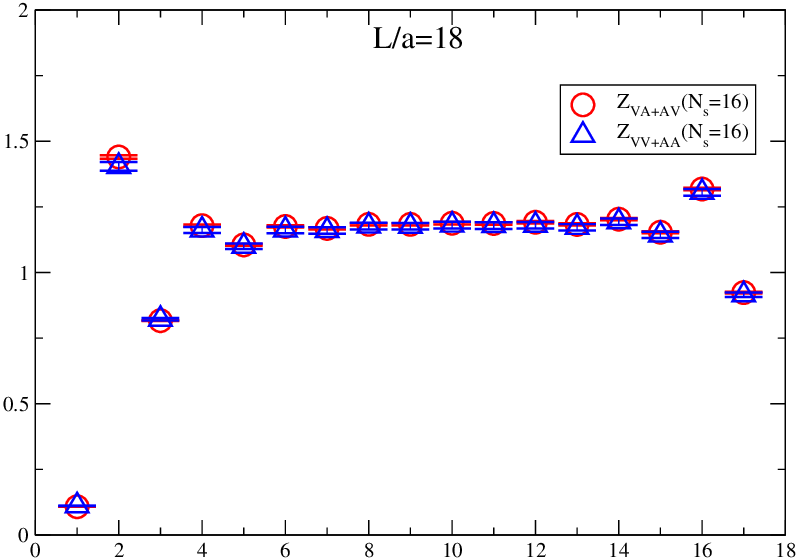}}
    \end{center}
  \end{minipage}&
   \end{tabular}
\end{center}
 \caption{A comparison between two renormalization factors
 $Z_{VA+AV;1}^+(g_0,a\mu_{\rm min})$(open circle) and
 $Z_{VV+AA;1}^+(g_0,a\mu_{\rm min})$ (open up triangle)
 as  a function of $x_0$ for the scheme $1$ at various lattice sizes.
}
 \label{fig:z1pzvvaa}
\end{figure}

We evaluate the renormalization factor $Z_{VV+AA}(g_0,\mu_{\rm min})$
using the chirally rotated boundary condition and correlation functions of 
\eqn{eq:crrelfunc.f} and \eqn{eqn:k1} with the statistics of 100 configurations.
The results are listed in table \ref{tab:zvvaa1.mumin} for the unimproved
renormalization condition, and the time dependences of $Z^\pm_{VA+AV}$ and
$Z^\pm_{VV+AA}$  are shown in 
Fig.~\ref{fig:z1pzvvaa} and~\ref{fig:z8mzvvaa} for schemes 1 and 8, 
respectively.
We observe good agreements between them at all $L/a$,
and a similar results are obtained at other schemes.
This investigation concludes that the relation
$Z_{VV+AA}=Z_{VA+AV}$ holds within statistical errors in our simulations.

\section{Non-perturbative renormalization of quark masses}
\label{sec:NPRmass}

From the relation derived from the axial-vector Ward-Takahashi identity in
DWQCD\cite{AliKhan:2001wr} that
\begin{eqnarray}
Z_m=\frac{1}{Z_P},
\end{eqnarray}
we obtain the renormalization factor of quark masses from $Z_P$, 
which can easily be extracted as a by-product of the calculation in 
the previous sections.
In this section, we report our results for the nonperturbative 
renormalization of quark masses. 

\subsection{Renormalization group invariant quark mass}

The renormalization group invariant quark mass is defined by
\begin{eqnarray}
M&=&\ovl{m}(\mu)\left(2b_0\ovl{g}^2(\mu)\right)^{-\frac{d_0}{2b_0}}
\exp\left(-\int_{0}^{\ovl{g}(\mu)}dg\left(\frac{\tau(g)}{\beta(g)}
-\frac{d_0}{b_0g}\right)\right) ,
\end{eqnarray}
where $\ovl{m}(\mu)$ is a renormalized mass in some scheme at
scale $\mu$.
We evaluate the renormalization factor ${\cal Z}_M$, which converts
the bare quark mass on the lattice in DWQCD to the RGI quark mass.
A strategy to derive the renormalization factor is the same as that for
$B_K$, and we write
\begin{equation}
\mathcal{Z}_{M}(g_0)=
Z_{m}^{\rm PT}(\infty,\mu_{\rm max})
Z_{m}^{\rm NP}(\mu_{\rm max},\mu_{\rm min})
Z_{m}^{\rm NP}(g_0,a\mu_{\rm min}).
\end{equation}
The first two factors  have already been calculated by the Alpha collaboration as
\cite{Sint:1998iq,Capitani:1998mq}
\begin{eqnarray}
Z_{m}^{\rm PT}(\infty,\mu_{\rm max})Z_{m}^{\rm NP}(\mu_{\rm max},\mu_{\rm min})
=\frac{M}{\ovl{m}(\mu_{\rm min})}=1.157(12)
\label{eqn:RGmass}
\end{eqnarray}
at the same scale $\mu_{\rm min}=1/(2L_{\rm max})$ as $B_K$.
As in the case for $B_K$ what we need to calculate is the third factor,
\begin{eqnarray}
Z_m(g_0,a\mu_{\rm min})=\frac{1}{Z_P(g_0,\mu_{\rm min})}.
\end{eqnarray}

The Alpha collaboration adopted the definition for the 
renormalization factor of the pseudoscalar density such that 
\begin{eqnarray}
Z_P(1/L)=\frac{\sqrt{3f_1}}{f_P(x_0=L/2)},\quad
f_P(x_0)=-\frac{1}{3}\vev{P_0^a(x_0,\vec{x}){\cal O}_0^a},\quad
f_1=-\frac{1}{3L^6}\vev{{\cal O'}_0^a{\cal O}_0^a},
\label{eqn:zp}
\end{eqnarray}
where
\begin{eqnarray}
P^a(x)=\bq(x)\gamma_5\tau^aq(x),\quad
\mathcal{O}^a=
a^6\sum_{\vec{x}\vec{y}}\bar{\zeta}(\vec{x})\gamma_5\tau^a\zeta(\vec{y})
\end{eqnarray}
are bulk and boundary pseudoscalar densities.

\begin{figure}
    \begin{center}
      \scalebox{0.32}{\includegraphics{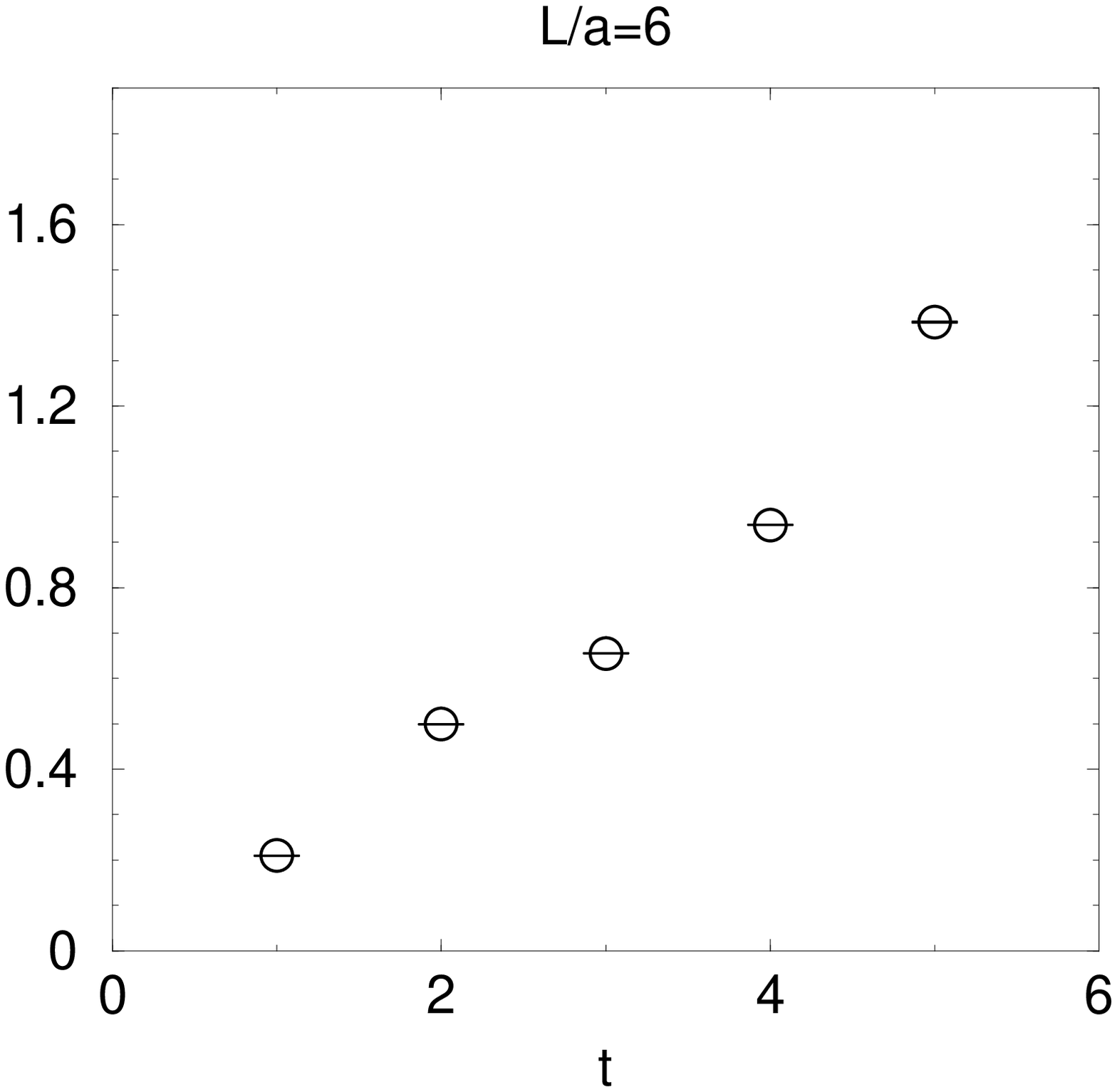}}
\qquad
      \scalebox{0.32}{\includegraphics{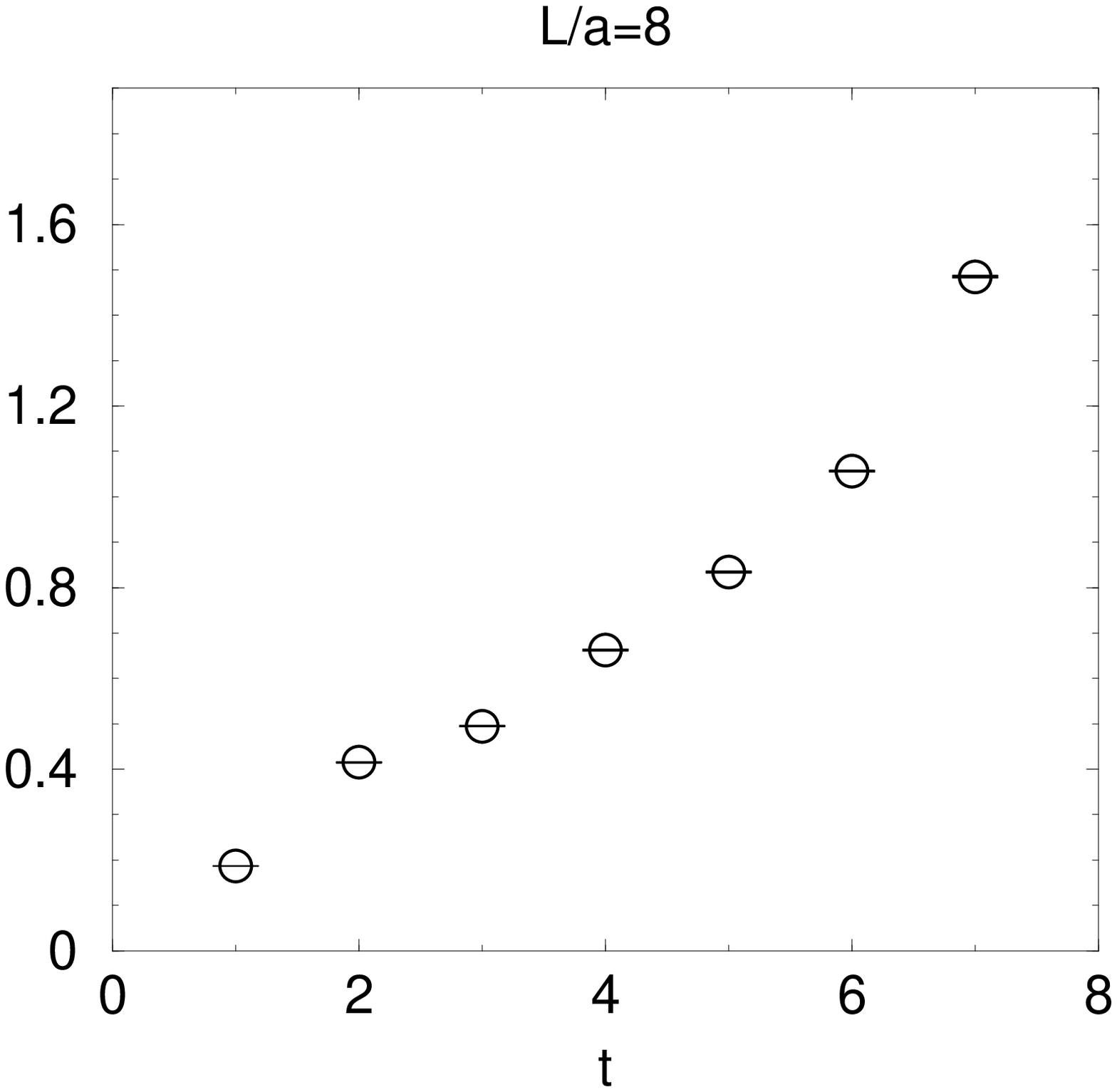}}
    \end{center}
    \begin{center}
      \scalebox{0.32}{\includegraphics{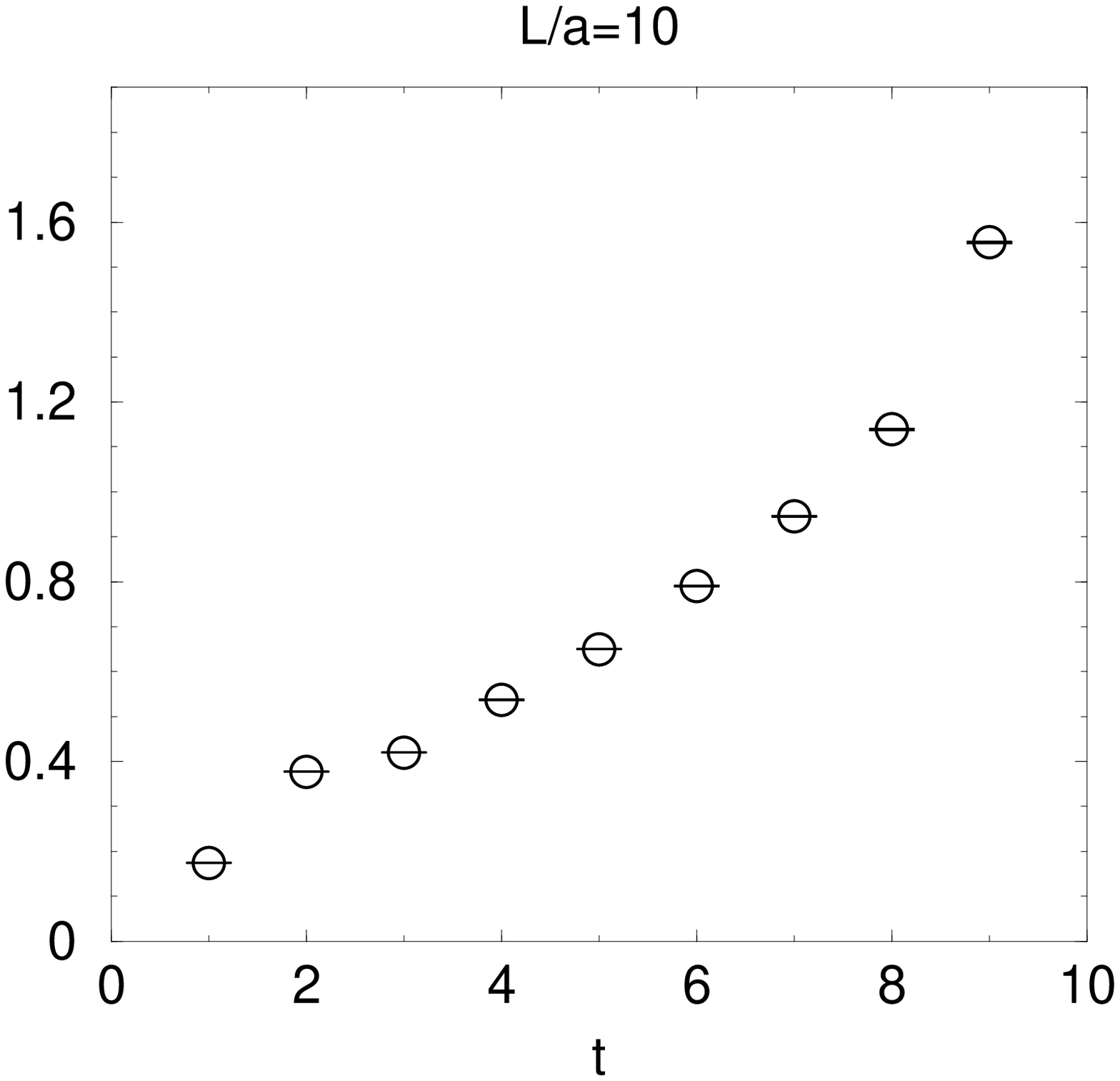}}
\qquad
      \scalebox{0.32}{\includegraphics{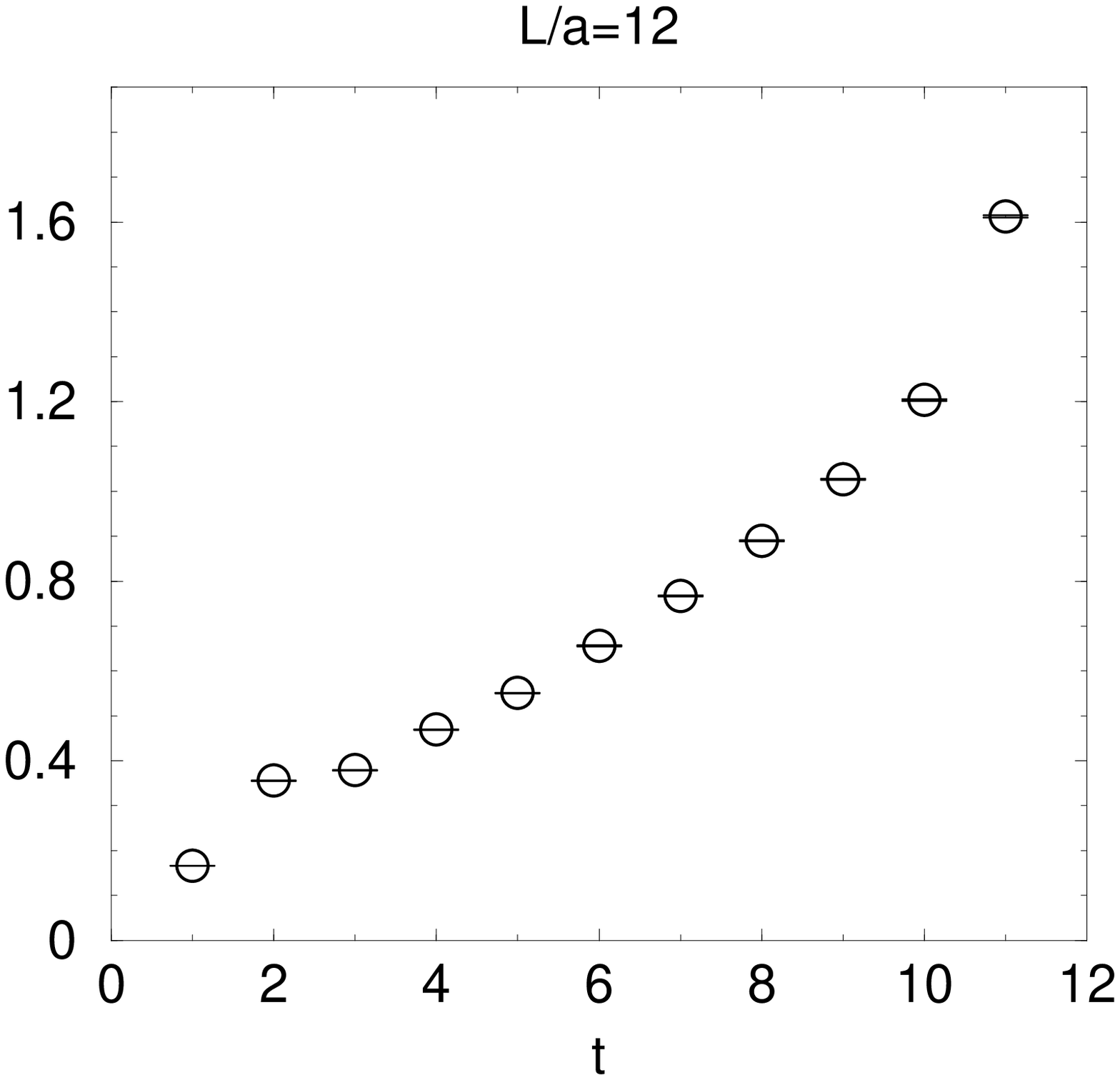}}
    \end{center}
    \begin{center}
      \scalebox{0.32}{\includegraphics{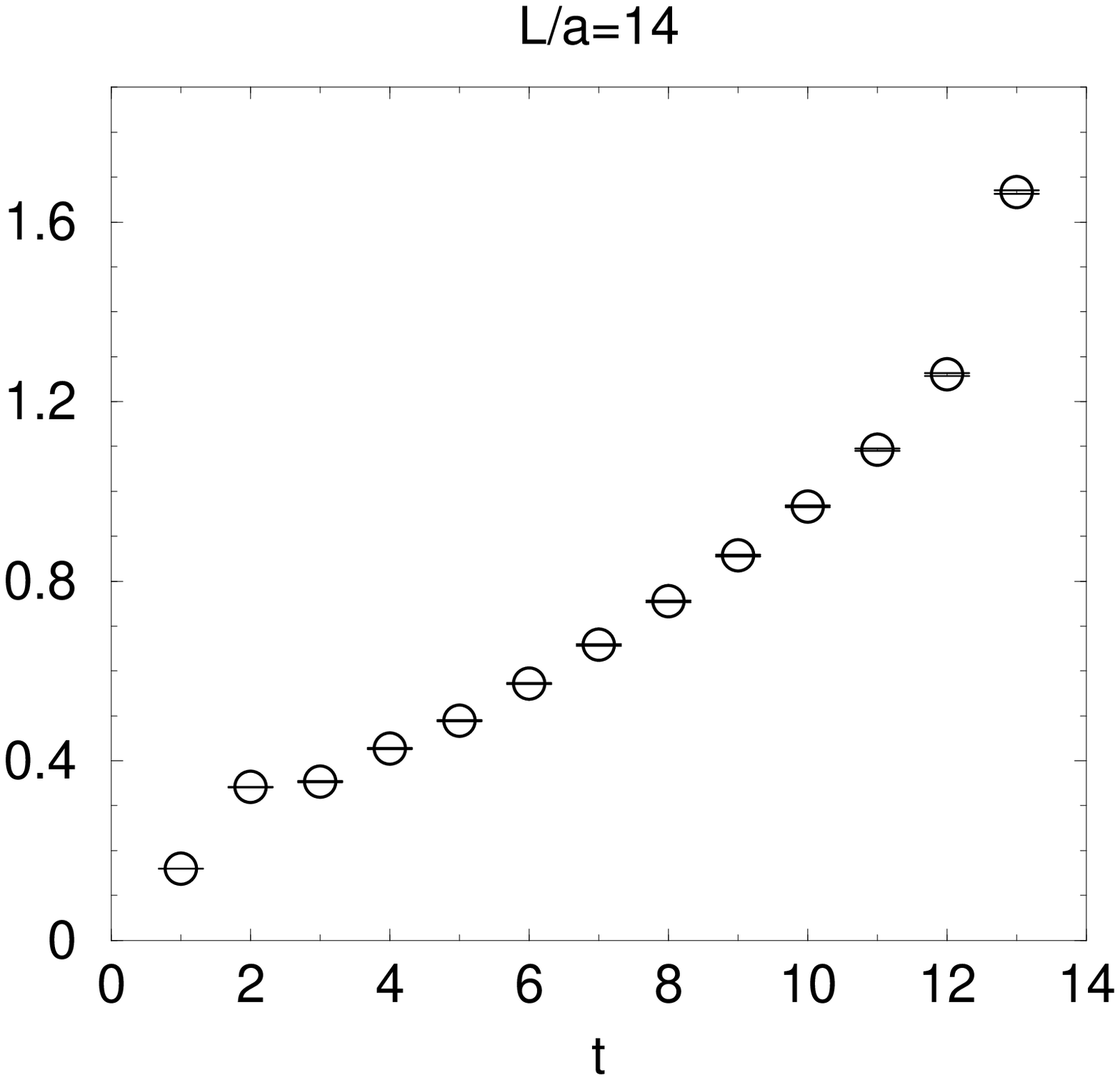}}
\qquad
      \scalebox{0.32}{\includegraphics{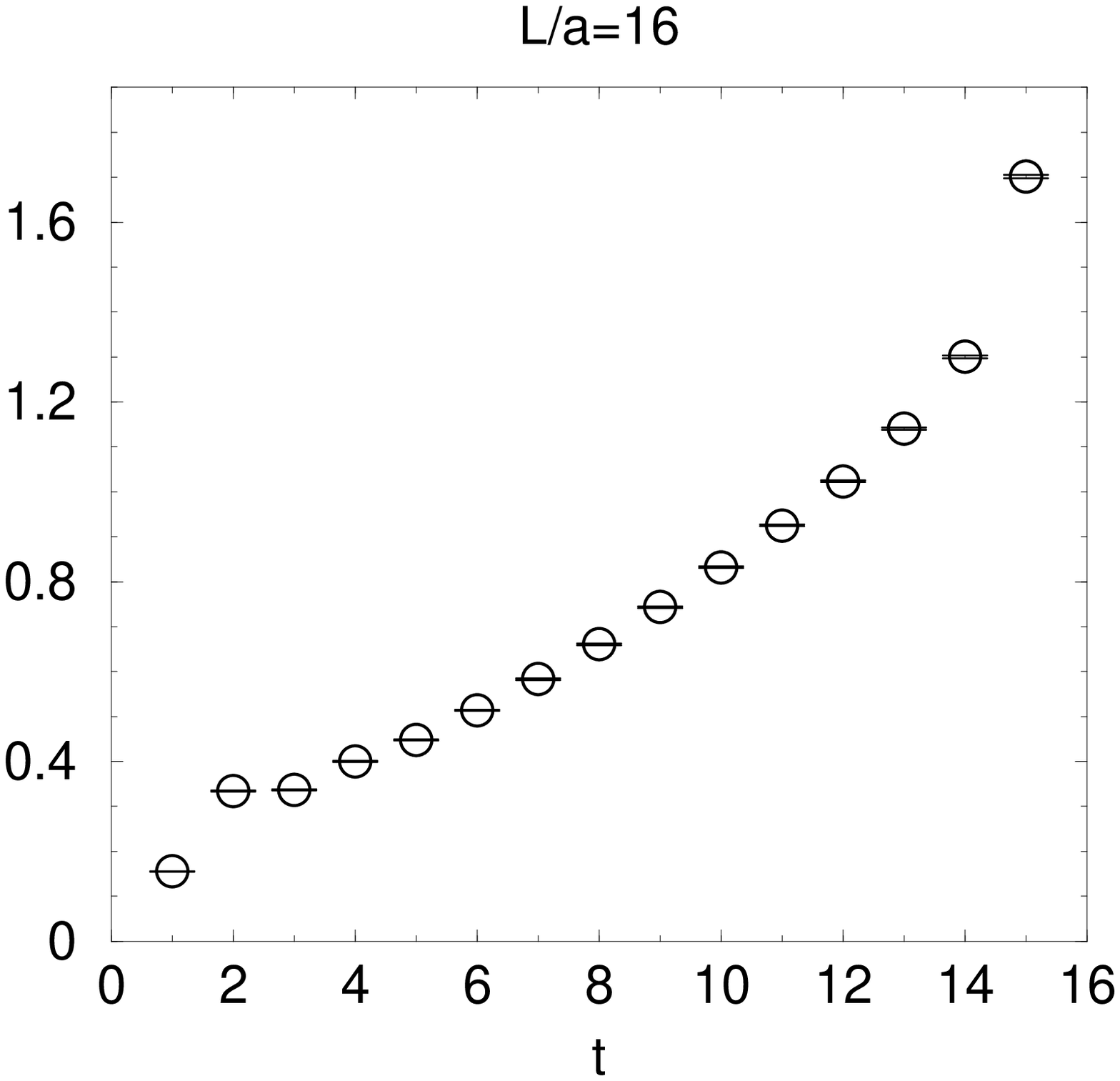}}
    \end{center}
    \begin{center}
      \scalebox{0.32}{\includegraphics{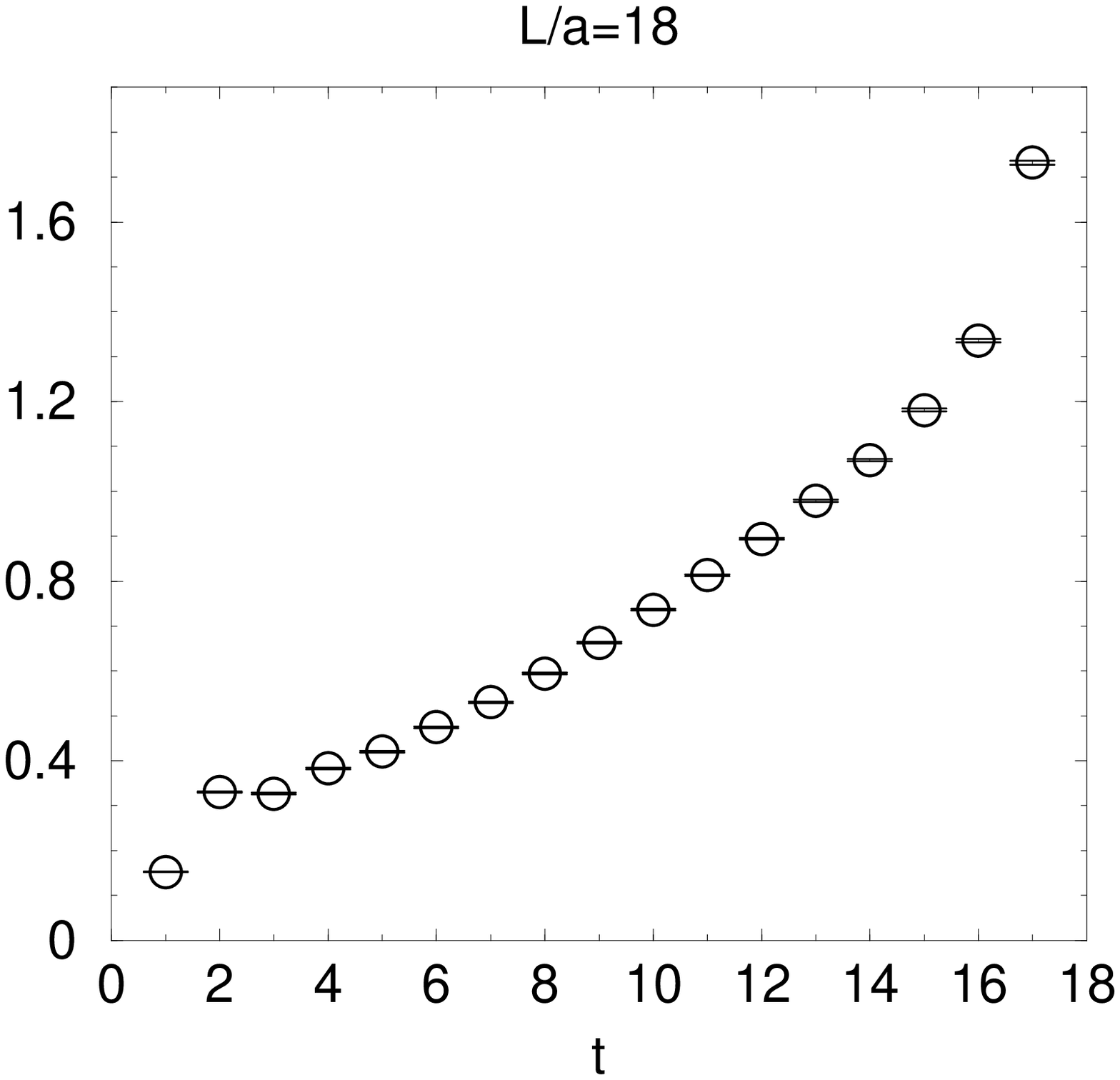}}
\qquad
      \scalebox{0.32}{\includegraphics{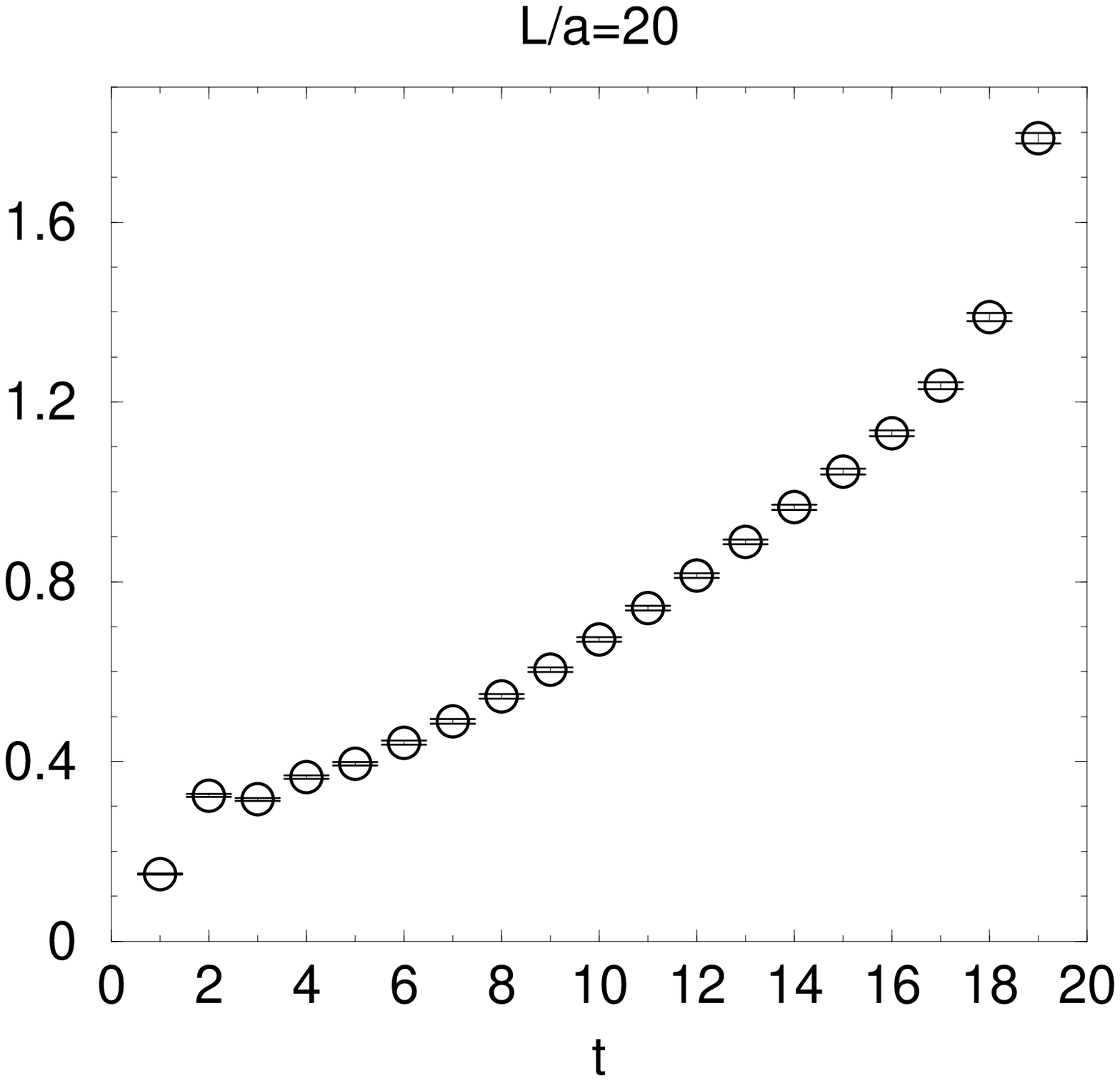}}
    \end{center}
 \caption{$x_0$ dependence of $Z_P(g_0,1/2L_{\rm max})$ at various 
lattice sizes.}
 \label{fig:zp.x0dep}
\end{figure}

Parameters for numerical simulations are same as those in section
\ref{sec:parameter}.
A typical behavior of $Z_{P}$ as a function of time is shown in
Fig.~\ref{fig:zp.x0dep}.
Values of $Z_P$ and $Z_m$ at $x_0=L/2$ are listed
in table \ref{tab:zvaav.mumin}, with
errors evaluated by a single elimination jackknife procedure.

\subsection{Scaling behavior of the step scaling function at $L_{\rm max}$}

\begin{figure}
 \begin{center}
  \scalebox{0.5}{\includegraphics{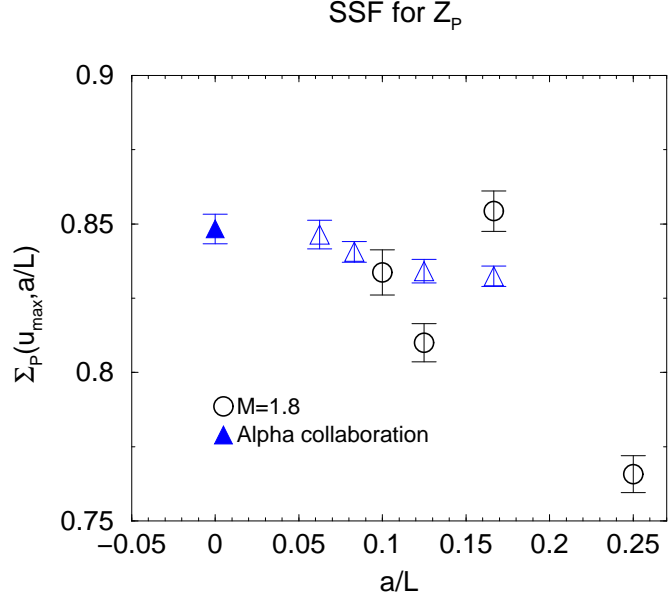}}
  \caption{The scaling behavior of the SSF $\Sigma_P(u,a/L)$ with the ordinary
  renormalization condition.
  Open circles are our results with the domain-wall fermion at $M=1.8$, while
  triangles show results by the Alpha collaboration with the improved Wilson
  fermion action. Corresponding continuum limits are represented by 
  filled symbols.}
  \label{fig:SSF.zp}
 \end{center}
\end{figure}

We again study the scaling behavior of the SSF, 
\begin{eqnarray}
\Sigma_{P}(u,a/L) =
\frac{Z_P(g_0,a/2L)}{Z_P(g_0,a/L)}\biggl|_{m=0,\ovl{g}^2(1/L)=u}
\end{eqnarray}
at $L=L_{\rm max}$, which can be calculated from data in
table \ref{tab:zvaav.mumin} and \ref{tab:zvaav.ssf}.
As seen in  Fig.\ref{fig:SSF.zp},  the scaling violation in this case is also large and it  
seems to approach the continuum limit with oscillation.

\begin{figure}
 \begin{center}
  \scalebox{0.5}{\includegraphics{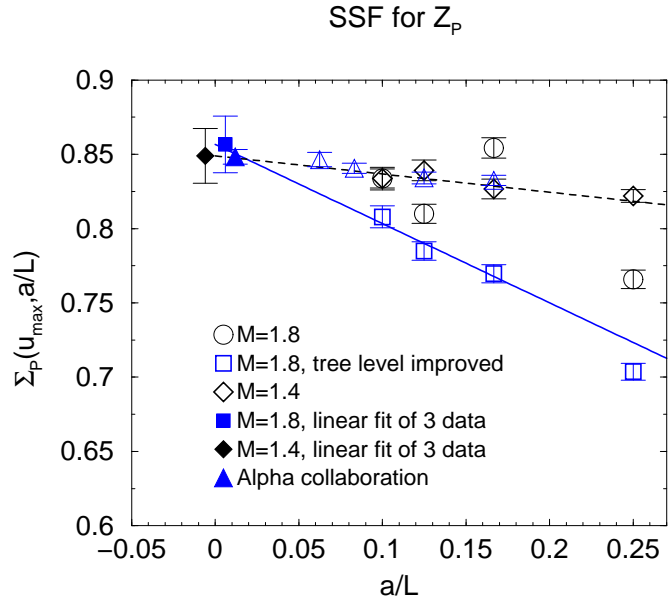}}
  \caption{Scaling behaviors of the SSF $\Sigma_P(u,a/L)$ for the 
  domain-wall fermion at $M=1.8$ with the ordinary renormalization condition
  (circles) and the improved condition (squares), and
  at $M=1.4$ (diamonds), together with results by the Alpha collaboration
  (triangles). Filled symbols are continuum limits.}
  \label{fig:SSF.zp.imp}
 \end{center}
\end{figure}

As before, we try to improve the scaling behavior by either taking $M=1.4$ or using the tree level improvement with the renormalization condition that
\begin{eqnarray}
Z_P(1/L) \frac{f_P(x_0=L/2)}{\sqrt{f_1}} &=& \left(Z_P\right)_{({\rm lattice})}^{({\rm tree})}
\frac{f_P(x_0=L/2)_{({\rm lattice})}^{({\rm tree})}}
{\sqrt{\left(f_1\right)_{({\rm lattice})}^{({\rm tree})}}}, 
\label{eqn:impM}
\end{eqnarray}
where
$
\left(Z_P\right)_{({\rm lattice})}^{({\rm tree})}
=\displaystyle\frac{1}{M_{\rm tad}\left(2-M_{\rm tad}\right)}
$.
The SSF obtained from the ratio of $Z_P$  in  tables
\ref{tab:zvaav.ssf.M14}, \ref{tab:zvaav.imp} and \ref{tab:zvaav.ssf.imp} 
is plotted in  Fig.~\ref{fig:SSF.zp.imp} for data at $M=1.4$(open diamonds)
and for the tree level improvement (open squares). 
In both cases the scaling behaviors are improved  and the linear continuum 
extrapolation using the finest three data in each case are consistent with the value 
by the Alpha collaboration.
A combined linear fit to both data
gives
$\sigma_m(u)=0.853(13)$
with $\chi^2/{\rm dof}=0.71$.

\subsection{Renormalization of quark masses}

\begin{figure}
 \begin{center}
  \scalebox{0.5}{\includegraphics{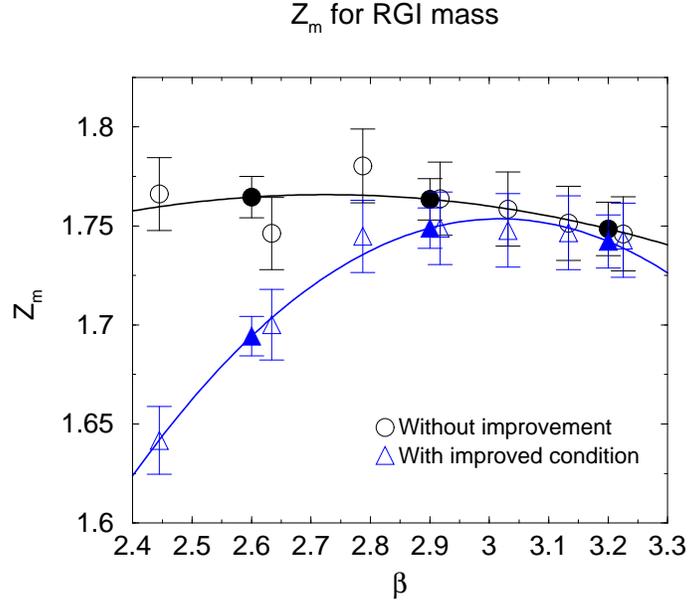}}
  \caption{$\beta$ dependences of the renormalization factor
  $\mathcal{Z}_{m}$ for the RGI operator with the polynomial fit.
  Filled symbols represent interpolated values at $\beta=2.6$, $2.9$ and
  $3.2$.
  A comparison is made between the ordinary renormalization condition
  (circle) and the improved condition (triangle).}
  \label{fig:RGIZm.betadep}
 \end{center}
\end{figure}

We employ the tree level improved condition \eqn{eqn:impM} for the renormalization of quark masses.
Multiplying $Z_{m}^{\rm NP}(g_0,a\mu_{\rm min})$ in table
\ref{tab:zvaav.imp} with the RG running factor \eqn{eqn:RGmass}, we obtain
the renormalization factor ${\cal Z}_M$  in table \ref{tab:RGIZm},
which is plotted in Fig.~\ref{fig:RGIZm.betadep} by open triangles, together
with data in table \ref{tab:RGImass} at $\beta=2.6$, $2.9$ and $3.2$ (filled symbols) by the quadratic interpolation.
Data without the improvement are also shown in the figure.
A discrepancy between the two are clearly observed at low $\beta$.

We convert the bare masses in table \ref{tab:bare_bk} to the RGI light quark masses, which is listed
in table \ref{tab:RGImass}. Here 
$m_{ud}$ is the up and down averaged quark mass determined  by $\pi$, while $m_s(K)$
or $m_s(\phi)$ is the strange quark mass by $K$ or $\phi$, respectively, and
$-m_{\rm res}$ is the  residual mass of the DWQCD at which the pion mass vanishes.
Following the previous paper \cite{AliKhan:2001wr}, we adopt
$m_q+m_{\rm res}$ for our definition of quark masses.
Lattice spacing is given with $\rho$ meson input.

\begin{figure}
 \begin{center}
  \scalebox{0.45}{\includegraphics{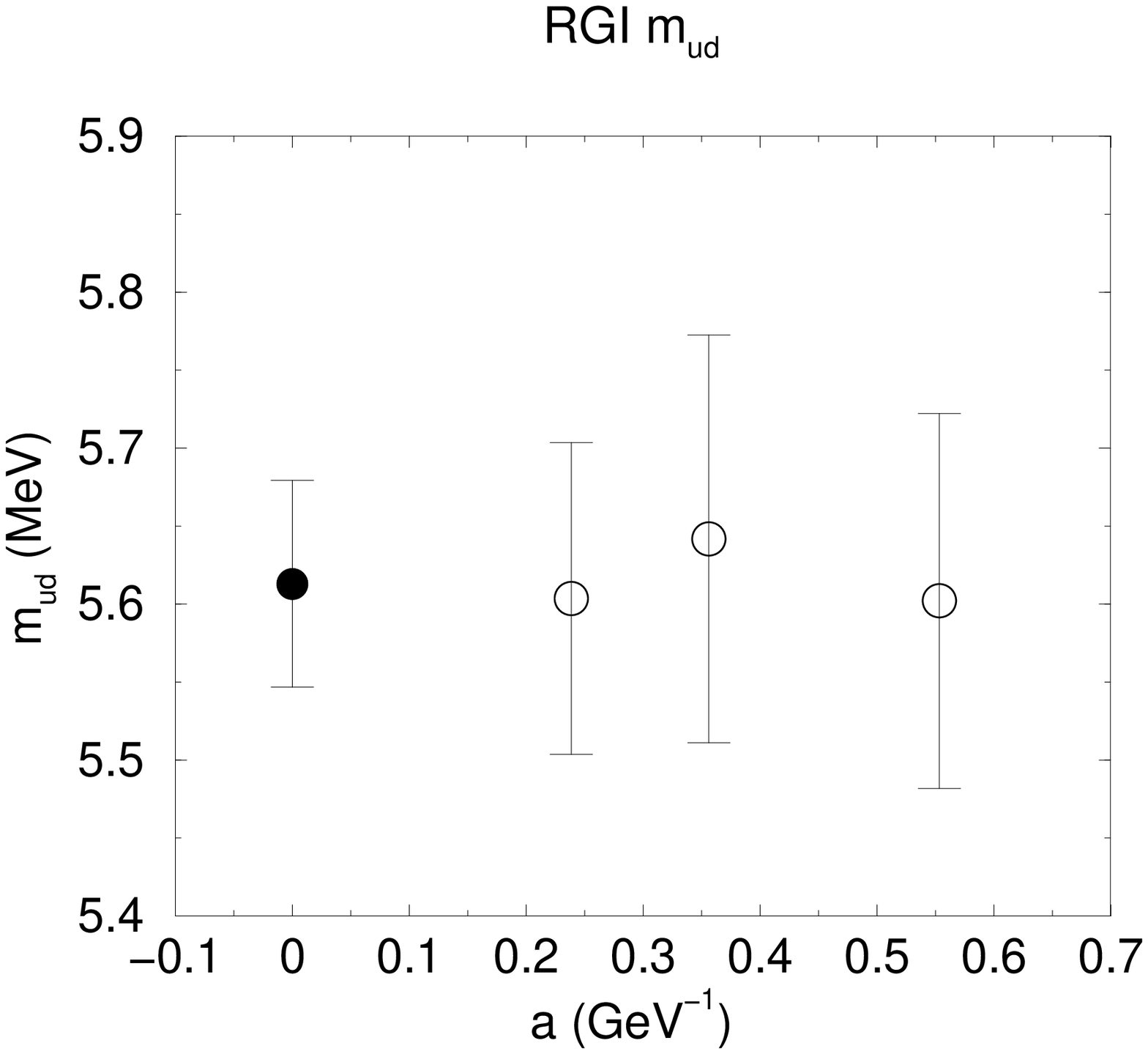}}\\
  \scalebox{0.45}{\includegraphics{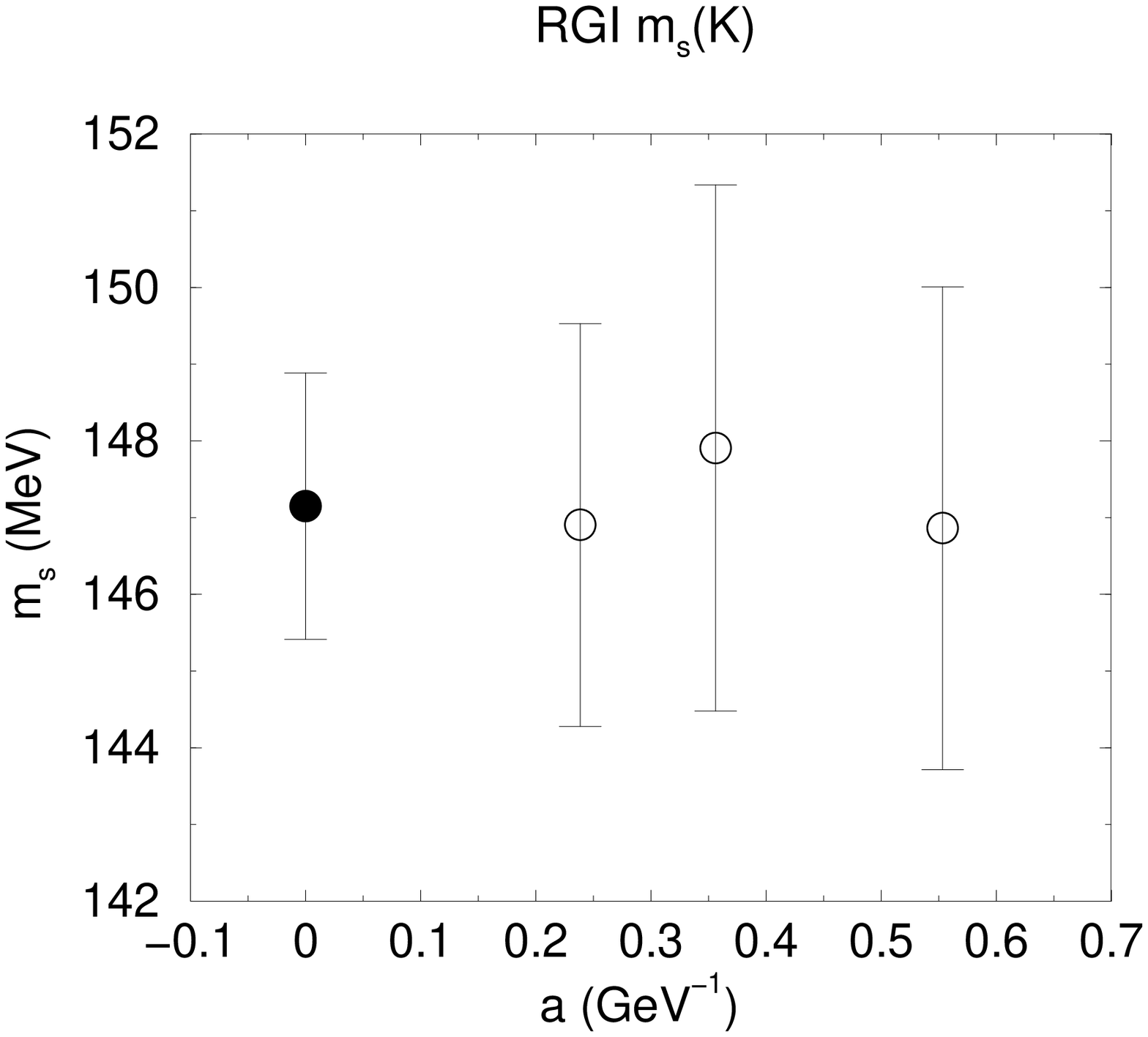}}
  \scalebox{0.45}{\includegraphics{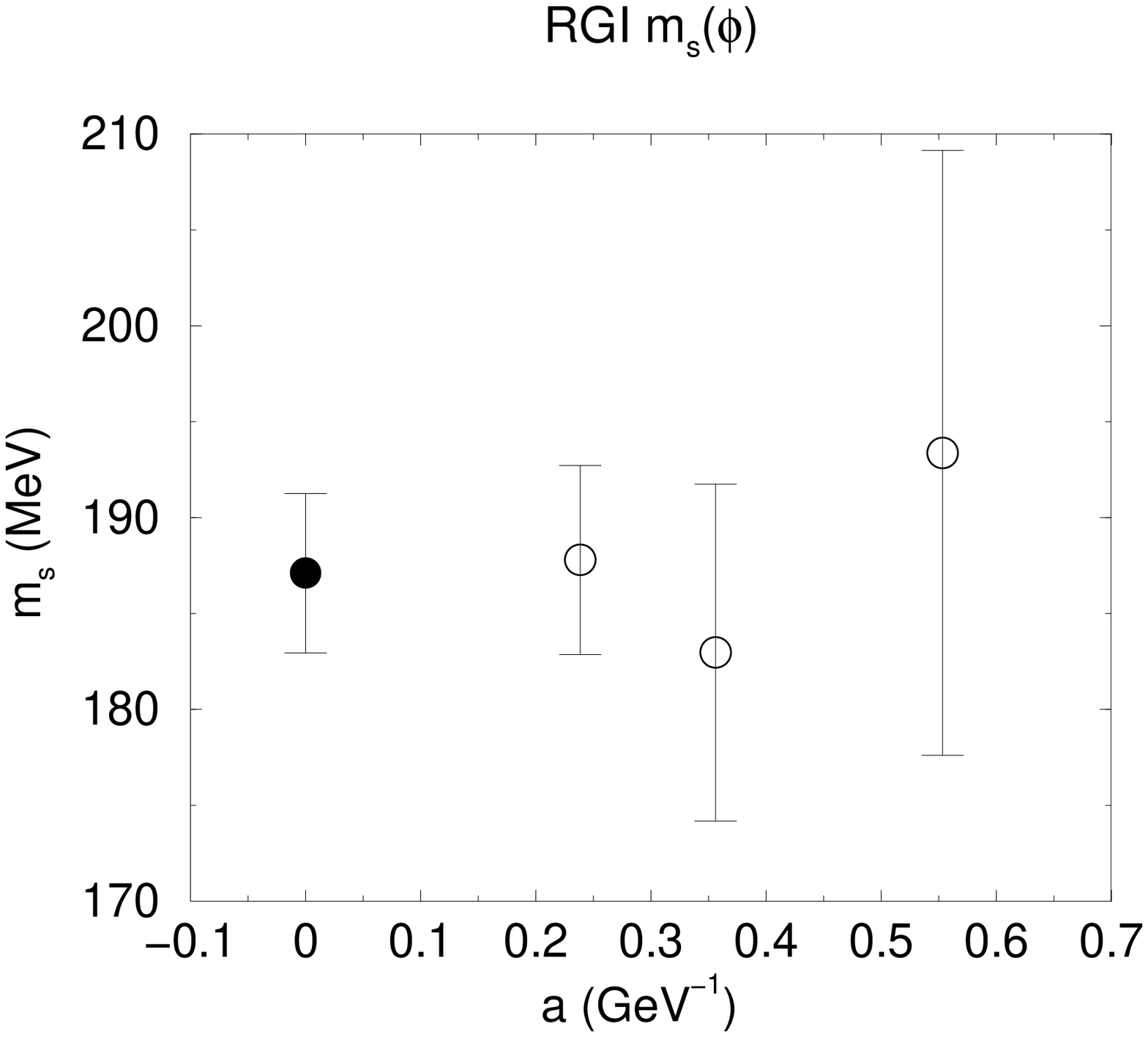}}
  \caption{Scaling behaviors of RGI quark masses and continuum
  extrapolations.
  $u,d$ quark mass and strange quark mass from the $K$ input are defined by
  $m_q+m_{\rm res}$.}
  \label{fig:RGImass}
 \end{center}
\end{figure}

Since the scaling behavior of the RGI quark masses is reasonably good as shown in 
Fig.~\ref{fig:RGImass},
we take the constant continuum extrapolations, which give
\begin{eqnarray}
&&
\hat{m}_{ud}^{\rm RGI}=5.613(66)\;({\rm MeV}),
\\&&
\hat{m}_s^{\rm RGI}(K)=147.1(17)\;({\rm MeV}),
\\&&
\hat{m}_s^{\rm RGI}(\phi)=187.1(41)\;({\rm MeV}).
\end{eqnarray}

\begin{figure}
 \begin{center}
  \scalebox{0.45}{\includegraphics{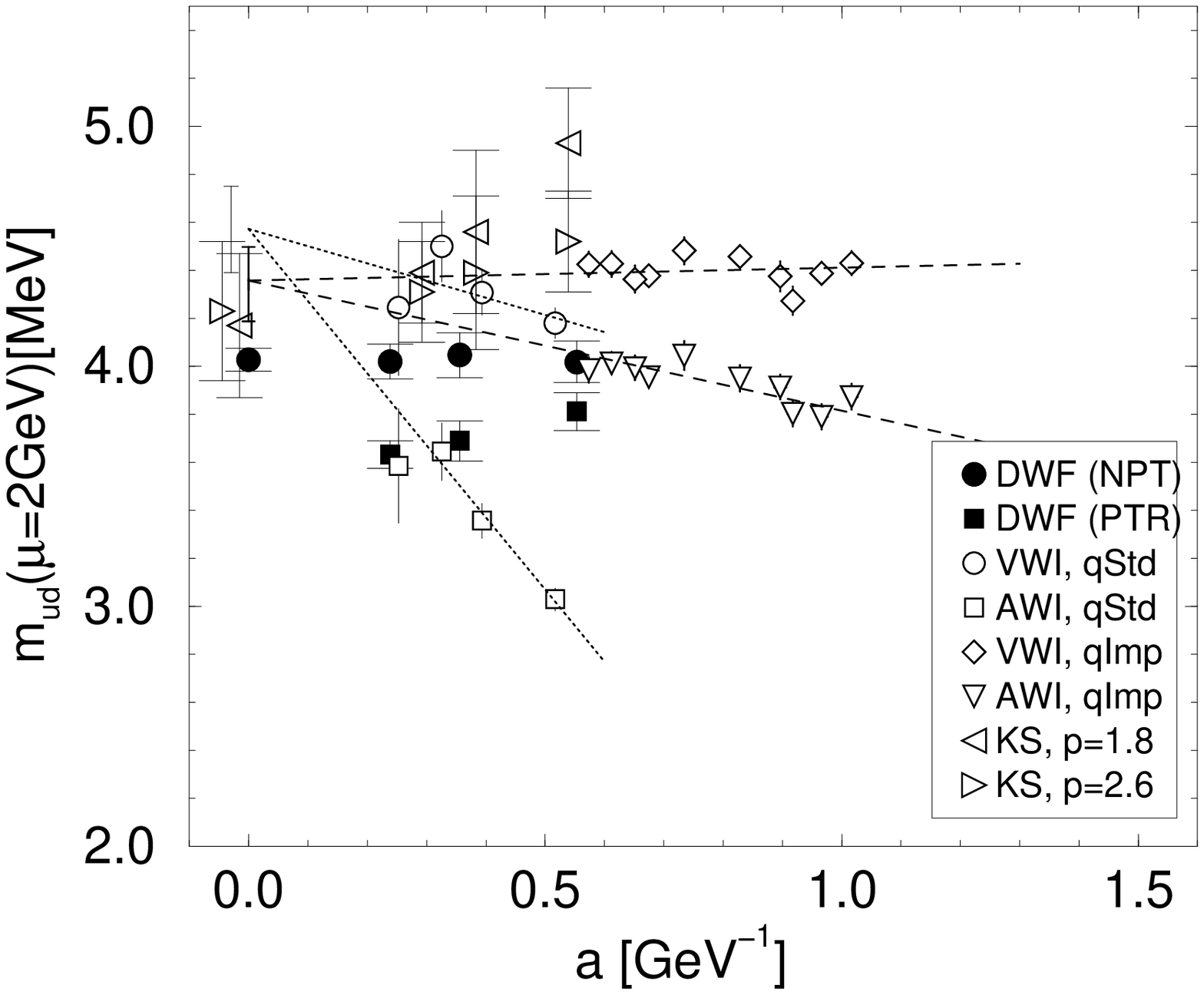}}\\
  \scalebox{0.45}{\includegraphics{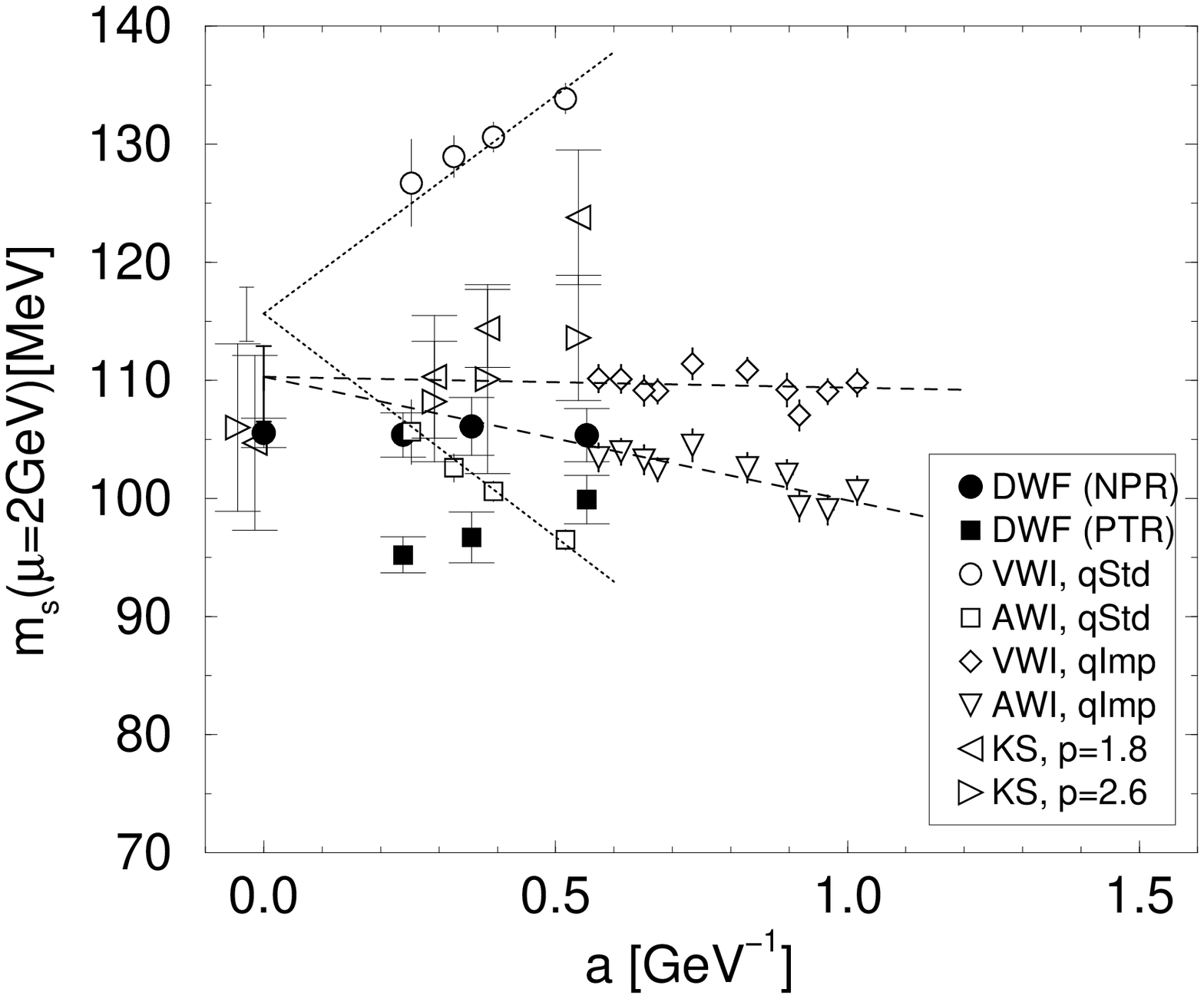}}
  \scalebox{0.45}{\includegraphics{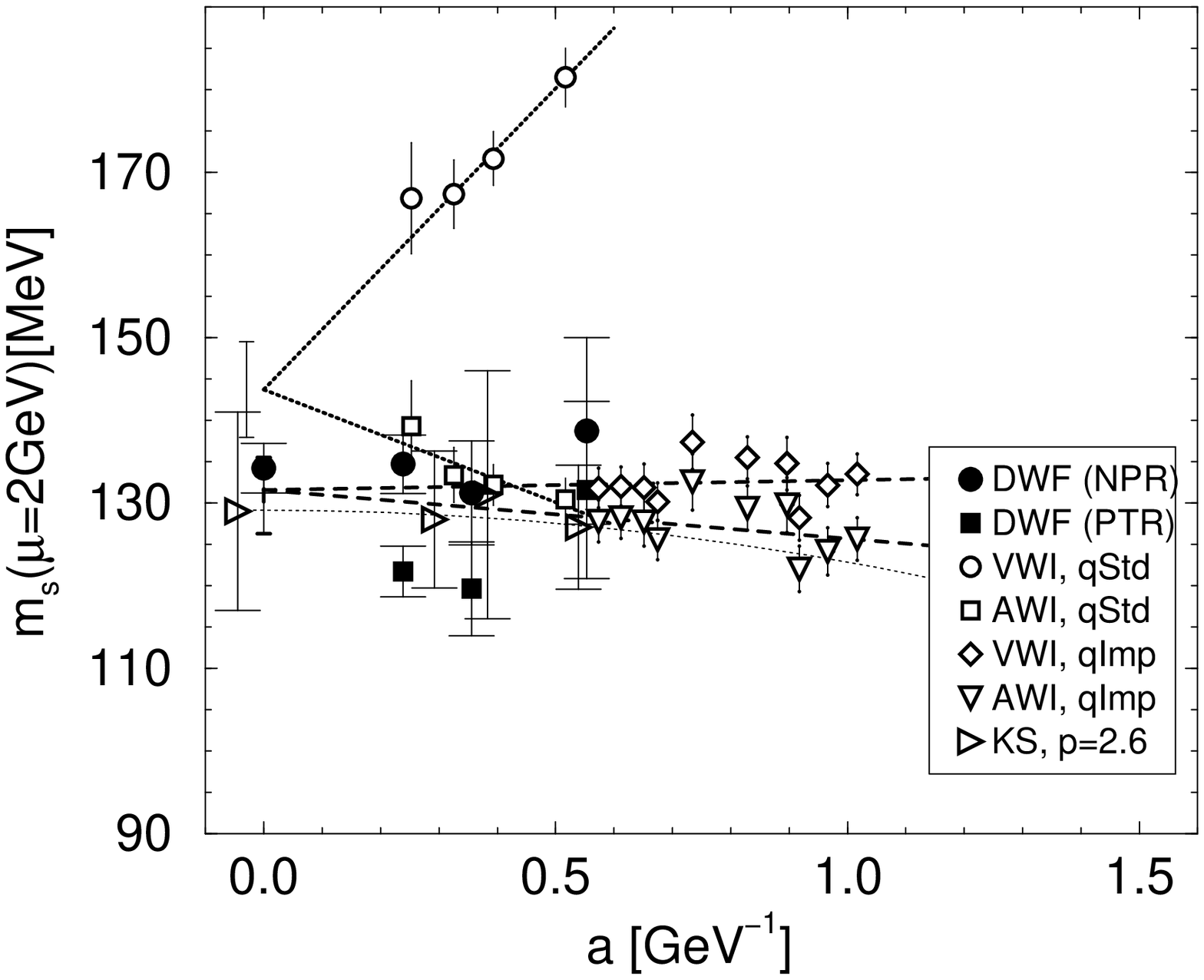}}
  \caption{Scaling behaviors of renormalized quark masses in $\msbar$
  scheme and continuum extrapolations.
  $u,d$ quark mass (upper panel) and strange quark masses from the $K$ input
  (lower left panel) and the $\phi$ input (lower right panel).
  Results of this paper are represented  by filled circles, together with
  the CP-PACS result with the perturbative 
  renormalization\cite{AliKhan:2001wr} (solid squares).
  The VWI(AWI) quark mass with the standard (qStd)\cite{Aoki:1999yr} and the
  improved (qImp)\cite{AliKhan:2000mw} Wilson fermions are represented by
  open circles(squares) and diamonds(down triangles), respectively, while
  results from the staggered fermion (KS)\cite{Aoki:1999mr} by open triangles. 
}
  \label{fig:MSmass}
 \end{center}
\end{figure}

To compare the previous results,
these values are converted to the $\ovl{\rm MS}$ scheme by
\begin{eqnarray}
m^{\ovl{\rm MS}}(\mu)=
M\left(2b_0{\left(g^{\ovl{\rm MS}}(\mu)\right)}^2\right)^{\frac{d_0}{2b_0}}
\exp\left(\int_{0}^{g^{\ovl{\rm MS}}(\mu)}dg\left(
\frac{\tau^{\ovl{\rm MS}}(g)}{\beta^{\ovl{\rm MS}}(g)}
-\frac{d_0}{b_0g}\right)\right),
\end{eqnarray}
where four loop expression is used for renormalization group functions
$\beta$ and $\tau$ in the $\msbar$ scheme\cite{vanRitbergen:1997va,Chetyrkin:1999pq}.
The results in Fig.~\ref{fig:MSmass} show good scalings, and the constant 
continuum extrapolations give
\begin{eqnarray}
&&
{m}_{ud}^{\msbar}({\rm NDR},2{\rm GeV})=4.026(48)\;({\rm MeV}),
\\&&
{m}_s^{\msbar}(K)({\rm NDR},2{\rm GeV})=105.6(12)\;({\rm MeV}),
\\&&
{m}_s^{\msbar}(\phi)({\rm NDR},2{\rm GeV})=134.2(30)\;({\rm MeV}).
\end{eqnarray}
Contrary to the case of $B_K$, perturbatively renormalized quark masses 
of the previous CP-PACS result (filled squares) are  underestimated,
as seen in the figure. 
This clearly shows that the necessity of the non-perturbative renormalization 
for precision calculations in lattice QCD.  We think that the large effects of 
the renormalizations are mainly canceled in the ratio of the  $B_K$ definition,
and therefore, such cancellations can not be expected for general operators.

\section{Conclusion and Discussions}
\label{sec:conclusion}

In this paper we have performed the non-perturbative renormalization
of $B_K$ and quark masses in the quenched domain-wall QCD using the
Schr\"{o}dinger functional method.
Combined with the non-perturbative running obtained by Alpha
collaboration, we have obtained the renormalization factors,  which convert the
lattice bare $B_K$ and quark masses previously obtained by the
CP-PACS collaboration to the RGI values. We obtain 
\begin{eqnarray}
&&
\wh{B}_K = 0.782(5)(7),
\\&&
\hat{m}_{ud}^{\rm RGI}=5.613(66)\;({\rm MeV}),
\\&&
\hat{m}_s^{\rm RGI}(K)=147.1(17)\;({\rm MeV}),
\\&&
\hat{m}_s^{\rm RGI}(\phi)=187.1(41)\;({\rm MeV}).
\end{eqnarray}
in the continuum limit.
These values correspond to
renormalized values in $\msbar$ scheme with the naive dimensional
regularization are given as
\begin{eqnarray}
&&
B_K^{\overline{\rm MS}}({\rm NDR},2{\rm GeV}) =0.565(4)(5),
\\&&
{m}_{ud}^{\msbar}({\rm NDR},2{\rm GeV})=4.026(48)\;({\rm MeV}),
\\&&
{m}_s^{\msbar}(K)({\rm NDR},2{\rm GeV})=105.6(12)\;({\rm MeV}),
\\&&
{m}_s^{\msbar}(\phi)({\rm NDR},2{\rm GeV})=134.2(30)\;({\rm MeV}).
\end{eqnarray}

\begin{figure}
 \begin{center}
  \includegraphics[width=8cm]{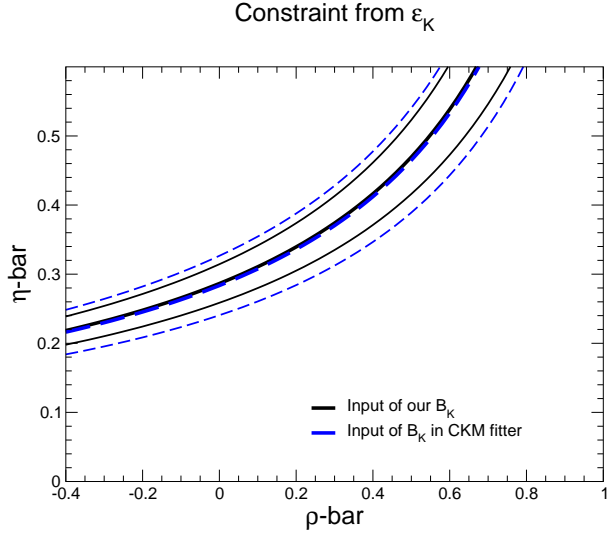}
  \caption{Constraint bands in the CKM triangle from $\epsilon_K$.
  The solid lines are central value and one standard deviation of the
  constraint with our $B_K$.
  The dashed lines are results with $B_K$ adopted by the CKM fitter
  group \cite{Charles:2004jd}.
  For other inputs we use those given by the CKM fitter.}
  \label{fig:CKM}
 \end{center}
\end{figure}

With the non-perturbative renormalization in the DWQCD and data at
$a^{-1}\simeq 2,3,4$ GeV, we can extract $B_K$ within 2\% errors, except
the quenching errors. 
The error in $B_K$ is directly reflected to that in the CKM triangle
constraint from $\epsilon_K$.
If we adopt our $B_K$ for an input, the error of the constraint is
improved as is shown in Fig.~\ref{fig:CKM}.
The solid lines are central value and one standard deviation of the
constraint from $\epsilon_K$ with our $B_K$.
The dashed lines are results with $B_K$ adopted by the CKM fitter
group \cite{Charles:2004jd}.
For other inputs we used those given by the CKM fitter.
\footnote{
We use the standard formula for $\epsilon_K$ as is given in
Ref.~\cite{Charles:2004jd}.
Note however that our figure is not a global fit.}

Although the perturbative renormalization can also achieve the same level of
accuracy for $B_K$, it is clearly shown that the nonperturbative 
renormalization
is indeed necessary for the precise determination of quark masses in
lattice QCD.

%
\section*{Acknowledgments}
%
This work is supported in part by Grants-in-Aid of the Ministry
of Education (NO.15540251, 13135204, 18740130).
%


%
\newcommand{\J}[4]{{#1} {\bf #2} (#3) #4}
\newcommand{\AP}{Ann.~Phys.}
\newcommand{\CMP}{Commun.~Math.~Phys.}
\newcommand{\EUR}{Eur.~Phys.J}
\newcommand{\IJMP}{Int.~J.~Mod.~Phys.}
\newcommand{\MPL}{Mod.~Phys.~Lett.}
\newcommand{\NP}{Nucl.~Phys.}
\newcommand{\NPSup}{Nucl.~Phys.~B (Proc.~Suppl.)}
\newcommand{\PL}{Phys.~Lett.}
\newcommand{\PR}{Phys.~Rev.}
\newcommand{\PRL}{Phys.~Rev.~Lett.}
\newcommand{\PTP}{Prog. Theor. Phys.}
\newcommand{\Suppl}{Prog. Theor. Phys. Suppl.}

\newpage

\appendix
\section{Chiral Ward-Takahashi identity for SF formalism with DWF}

In this appendix we derive the Ward-Takahashi identity
\eqn{eqn:WTid-dwf} for the Schr\"odinger functional formalism with
domain-wall fermion.
An explicit form of the chirally rotated correlation function used for
evaluation of $Z_{VV+AA}$ is presented.
We use the same notation of Ref.~\cite{Taniguchi:2006qw} for the SF 
formalism  in this appendix.

In this paper we adopt following massless DWF action with an orbifolding
projection
\begin{eqnarray}
&&
S_{\rm dwf}^{\rm SF}=a^4\sum\frac{1}{2}\bpsi\oP_+D_{\rm dwf}\psi
\label{eqn:massless}
\\&&
\oP_\pm=\frac{1\pm\gamma_0\gamma_5PQR}{2} .
\end{eqnarray}
Here $P$ is a parity transformation in fifth direction
$P_{s,t}\psi(\vec{x},x_0,t)=\psi(\vec{x},x_0,N_5-s+1)$, 
and $R$ is a time reflection operator acting on the temporal direction
$R_{x_0,y_0}\psi(\vec{x},y_0,s)=\psi(\vec{x},-x_0,s)$.
$Q$ is the vector charge matrix for the chiral transformation
\cite{Furman:1994ky}
\begin{eqnarray}
Q_{s,t}=\begin{pmatrix}
1 & 0 & 0 & 0 & 0 & 0\cr
0 & 1 & 0 & 0 & 0 & 0\cr
0 & 0 & 1 & 0 & 0 & 0\cr
0 & 0 & 0 &-1 & 0 & 0\cr
0 & 0 & 0 & 0 &-1 & 0\cr
0 & 0 & 0 & 0 & 0 &-1\cr
\end{pmatrix},\quad
({\rm for}\; N_5=6).
\label{eqn:Q}
\end{eqnarray}
For this action the physical quark propagator is given by
\begin{eqnarray}
&&
G_{\rm quark}^{\rm SF}(x,y)=2(G_{\rm quark}\Pi_+)_{x,y},
\quad
\Pi_\pm =\frac{1\pm\gamma_0R}{2}.
\end{eqnarray}
This propagator is shown to agree with \eqn{eq:quark-prop}
numerically and we employ the latter in our numerical simulations.

We consider the chiral rotation of the first flavour
\begin{eqnarray}
\psi_1(x,s)\to\wt{\psi}_1(x,s)
=\left(e^{-i\alpha Q}\right)_{st}{\psi}_1(x,t),\quad
\bpsi_1(x,s)\to\wt{\bpsi}_1(x,s)
={\bpsi}_1(x,t)\left(e^{i\alpha Q}\right)_{ts},
\end{eqnarray}
under which the physical quark and boundary quark fields are rotated as
\begin{eqnarray}
&&
\wt{q}_1(x)=e^{i\alpha\gamma_5}{q}_1(x),\quad
\wt{\bq}_1(x)={\bq}_1(x)e^{i\alpha\gamma_5},
\\&&
\wt{\zeta}_1(x)=e^{-i\alpha\gamma_5}{\zeta}_1(x),\quad
\wt{\ovl{\zeta}}_1(x)={\ovl{\zeta}}_1(x)e^{-i\alpha\gamma_5},
\\&&
\wt{\zeta}'_1(x)=e^{-i\alpha\gamma_5}{\zeta}'_1(x),\quad
\wt{\ovl{\zeta}}'_1(x)={\ovl{\zeta}}'_1(x)e^{-i\alpha\gamma_5}.
\end{eqnarray}
For $\alpha=\pi/2$ the action for the first flavour is transformed as
\begin{eqnarray}
&&
S_{\rm dwf}^{\rm SF}\to\wt{S}_{\rm dwf}^{\rm SF}+Y,
\\&&
\wt{S}_{\rm dwf}^{\rm SF}
=a^4\sum\frac{1}{2}\wt{\bpsi}{\oP}_-{D}_{\rm dwf}\wt{\psi},
\\&&
Y=a^4\sum\wt{\bpsi}{\oP}_-X\wt{\psi},
\\&&
aX=\left(P_L\delta_{s,\frac{N_5}{2}}\delta_{t,\frac{N_5}{2}+1}
+P_R\delta_{s,\frac{N_5}{2}+1}\delta_{t,\frac{N_5}{2}}\right)\delta_{x,y}.
\end{eqnarray}
We notice that the orbifolding projection in the rotated action
$\wt{S}_{\rm dwf}^{\rm SF}$ has an opposite sign and the rotated quark
fields satisfy the opposite Dirichlet boundary condition to
\eqn{eqn:DirichletBC} 
\begin{eqnarray}
&&
P_-\wt{q}_1(x)|_{x_0=0}=0,\quad
P_+\wt{q}_1(x)|_{x_0=N_T}=0,
\label{eqn:DBC1}
\\&&
\wt{\bq}_1(x)P_+|_{x_0=0}=0,\quad
\wt{\bq}_1(x)P_-|_{x_0=N_T}=0.
\label{eqn:DBC2}
\end{eqnarray}

Now we derive the Ward-Takahashi identity for the four point function
\eqn{eqn:correlation} and two point functions \eqn{eqn:k1}
as
\begin{eqnarray}
\mathcal{F}_{\Gamma_A\Gamma_B\Gamma_C}^\pm&=&
\frac{1}{L^3}\langle\mathcal{O}_{21}[\Gamma_A]\mathcal{O}_{45}[\Gamma_B]
 O_{VA+AV}^\pm(x)\mathcal{O}'_{53}[\Gamma_C]\rangle_{S}
\nn\\&=&
\frac{1}{L^3}\langle\wt{\mathcal{O}}_{21}[\Gamma_Ai\gamma_5]
\mathcal{O}_{45}[\Gamma_B]i\wt{O}_{VV+AA}^\pm(x)
\mathcal{O}'_{53}[\Gamma_C]\rangle_{\tilde{S}+Y},
\\
f_1&=&
-\frac{1}{2L^6}\langle\mathcal{O}_{12}'[\gamma_5]
\mathcal{O}_{21}[\gamma_5]\rangle_S      
=-\frac{1}{2L^6}\langle i\wt{\mathcal{O}}_{12}'[1]
i\wt{\mathcal{O}}_{21}[1]\rangle_{\tilde{S}+Y},
\\
k_1&=&-\frac{1}{6L^6}\sum_{k=1}^{3}
\langle\mathcal{O}'_{12}[\gamma_k]\mathcal{O}_{21}[\gamma_k]\rangle_S
=-\frac{1}{6L^6}\sum_{k=1}^{3}
\langle\wt{\mathcal{O}}'_{12}[i\gamma_5\gamma_k]
\wt{\mathcal{O}}_{21}[\gamma_ki\gamma_5]\rangle_{\tilde{S}+Y},
\end{eqnarray}
where
\begin{eqnarray}
\wt{O}^\pm_{VV+AA}&=&
\left(\wt{\bq}_1\gamma_\mu q_2\right)\left(\bq_3\gamma_\mu q_4\right)
+\left(\wt{\bq}_1\gamma_\mu\gamma_5q_2\right)
 \left(\bq_3\gamma_\mu\gamma_5q_4\right)
\nn\\&&
\pm\left\{
\left(\wt{\bq}_1\gamma_\mu q_4\right)\left(\bq_3\gamma_\mu q_2\right)
+\left(\wt{\bq}_1\gamma_\mu\gamma_5q_4\right)
 \left(\bq_3\gamma_\mu\gamma_5q_2\right)
\right\}.
\end{eqnarray}
Operators with tilde consist of the chirally rotated field for the first
flavour, which satisfies the opposite SF Dirichlet boundary condition.
Subscript $S$ and $\tilde{S}+Y$ mean the action under which the VEV is
taken.

The chiral symmetry breaking effect comes from a contribution of $Y$ on the
right hand side of the WT identity, which generates an operator mixing with
the parity even operator.
The contribution should be suppressed exponentially in the physical quark
propagator and this is the case at tree level.
We evaluate the effect by comparing two correlation functions directly
\begin{eqnarray}
\la\mathcal{O}_{21}[\Gamma_A]\mathcal{O}_{45}[\Gamma_B]O_{VA+AV}^\pm(x)
\mathcal{O}'_{53}[\Gamma_C]\ra_S
\leftrightarrow
\la\wt{\mathcal{O}}_{21}[\Gamma_Ai\gamma_5]\mathcal{O}_{45}[\Gamma_B]
i\wt{O}_{VV+AA}^\pm(x)\mathcal{O}'_{53}[\Gamma_C]\ra_{\tilde{S}}.
\end{eqnarray}
For this purpose we define the renormalization factor for the parity
even four fermi operator as
\begin{eqnarray}
&&
Z_{VV+AA;s}^\pm(g_0,a\mu)
=\frac{\wt{h}_s^{\pm({\rm tree})}(x_0=L/2)}{\wt{h}_s^\pm(x_0=L/2;g_0)},
\\&&
\wt{h}_1^\pm(x_0)
=\frac{\wt{\mathcal{F}}_{\gamma_5,\gamma_5,\gamma_5}^\pm(x_0)}{\wt{f}_1^{3/2}},
\\&&
\wt{\mathcal{F}}_{\Gamma_A\Gamma_B\Gamma_C}^\pm(x_0)
=\frac{1}{L^3}\langle \wt{\mathcal{O}}_{21}[\Gamma_Ai\gamma_5]
\mathcal{O}_{45}[\Gamma_B]i\wt{O}_{VV+AA}^\pm(x)
\mathcal{O}'_{53}[\Gamma_C]\rangle_{\tilde{S}},
\\&&
\wt{f}_1=-\frac{1}{2L^6}\langle i\wt{\mathcal{O}}_{12}'[1]
i\wt{\mathcal{O}}_{21}[1]\rangle_{\tilde{S}}
\end{eqnarray}
and compare it with that for the parity odd operator.

\section{Numerical analysis of $B_K$ for $\beta=3.2$}
\label{sec:appendixB}

Since the numerical data of CP-PACS collaboration at $\beta=3.2$ is new
and not published,
we give a short summary of its numerical analysis in this appendix.

\subsection{Run parameters and measurements}

We carry out run at $\beta=3.2$, corresponding to a lattice spacing
$a^{-1}=4.19(6)$ GeV determined from the the $\rho$ meson mass
$m_\rho=770$ MeV.
We use the lattice size 
$N_\sigma^3 \times N_t \times N_5 = 48^3\times 80\times 16$.
This lattice has a reasonably large spatial size of
$aN_\sigma\approx 2.3$~fm and fifth dimensional length $N_5=16$, which
has been confirmed to be enough for $B_K$ \cite{AliKhan:2001wr}.
In this numerical simulation the domain wall height is taken to be
$M=1.8$.

We take degenerate quarks in our calculations.  
The value of bare quark mass is chosen to be 
$m_fa=0.009, 0.018, 0.027, 0.036$,
which covers the range that $m_{PS}/m_V\approx 0.5334 - 0.8224$.

Quenched gauge configurations are generated on four-dimensional
lattices.
A sweep of gauge update contains one pseudo-heatbath and four 
overrelaxation steps.
After a thermalization of 2000 sweeps hadron propagators and 3-point 
functions necessary to evaluate $B_K$ are calculated
at every 200th sweep.
The gauge configuration on each fifth dimensional coordinate $s$
is identical and is fixed to the Coulomb gauge.

The domain-wall quark propagator needed to extract the $B_K$
is calculated by the conjugate gradient algorithm with an even-odd 
pre-conditioning.  
Two quark propagators are evaluated for each configuration corresponding 
to the wall sources placed at either $t=4$ or $77$
in the time direction with the Dirichlet boundary condition, while 
the periodic boundary condition is imposed in the spatial directions.  
The two quark propagators are combined to form the kaon Green function 
with an insertion of the four-quark operator at time slices 
$1\leq t\leq N_t$ in a standard manner.
We employ the same quark propagators to evaluate pseudoscalar and
vector meson propagators, and extract their masses.

\subsection{Pseudo scalar and vector meson masses}

\begin{figure}
 \begin{center}
\includegraphics[width=6.5cm]{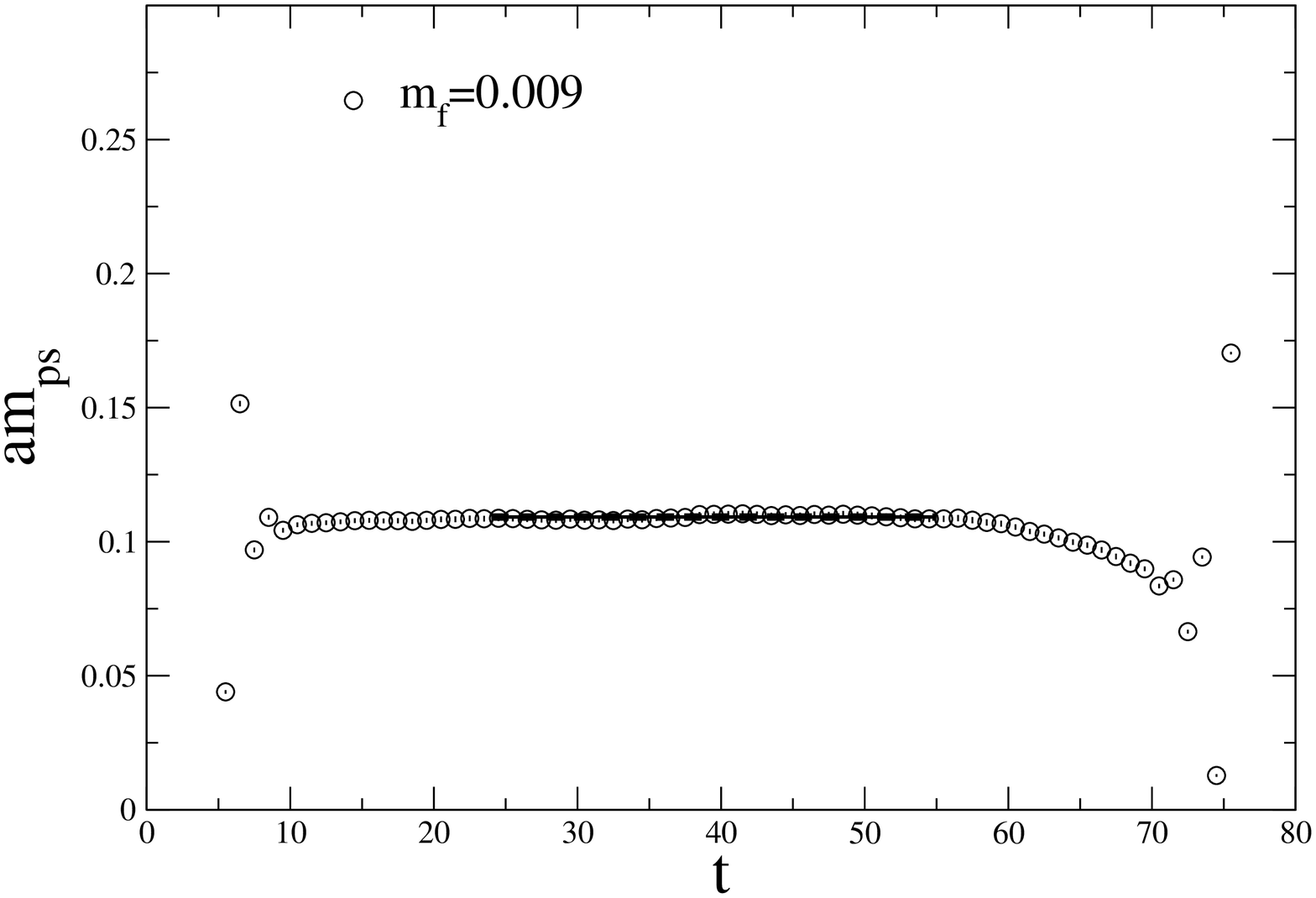}
\includegraphics[width=6.5cm]{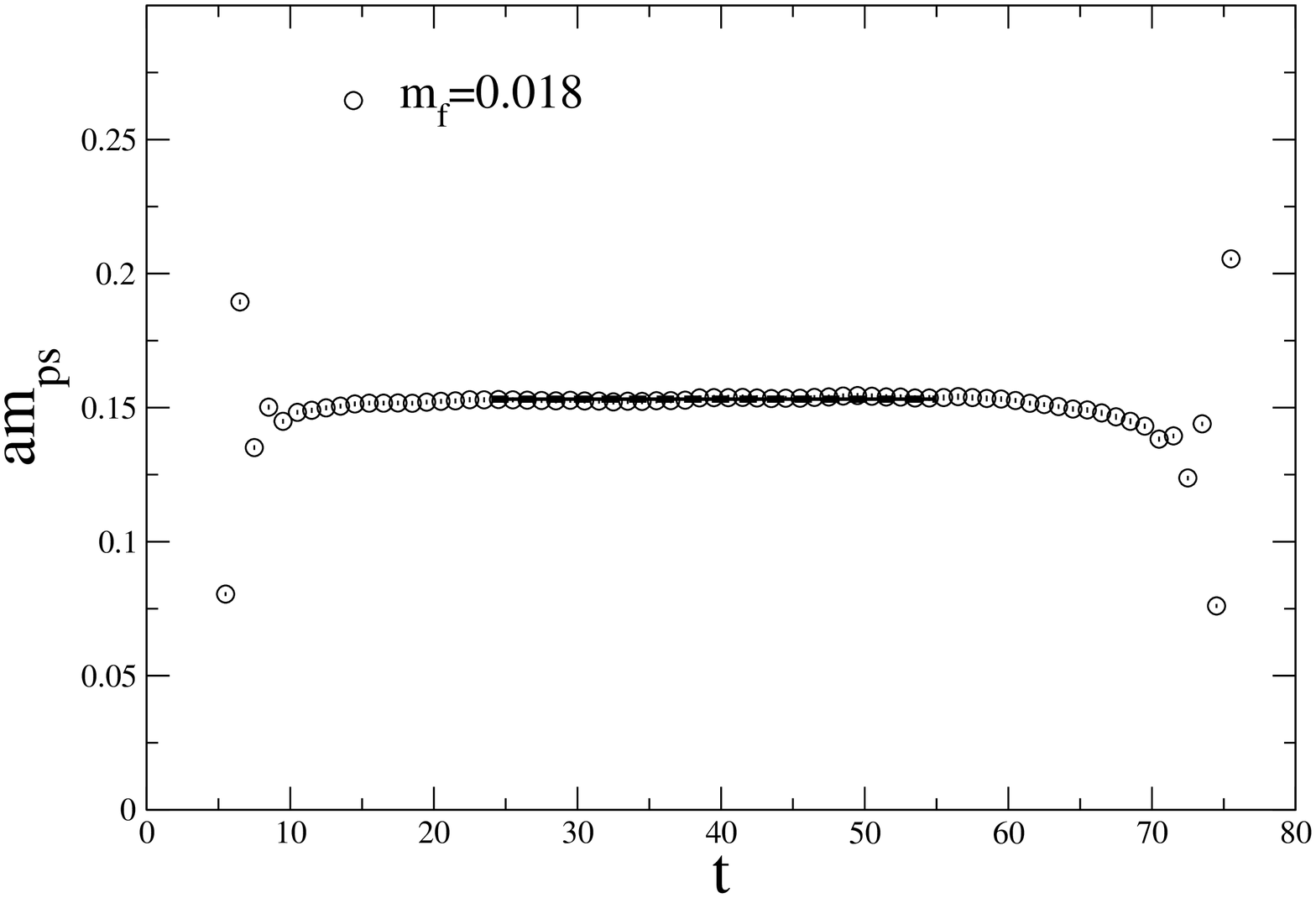}
\includegraphics[width=6.5cm]{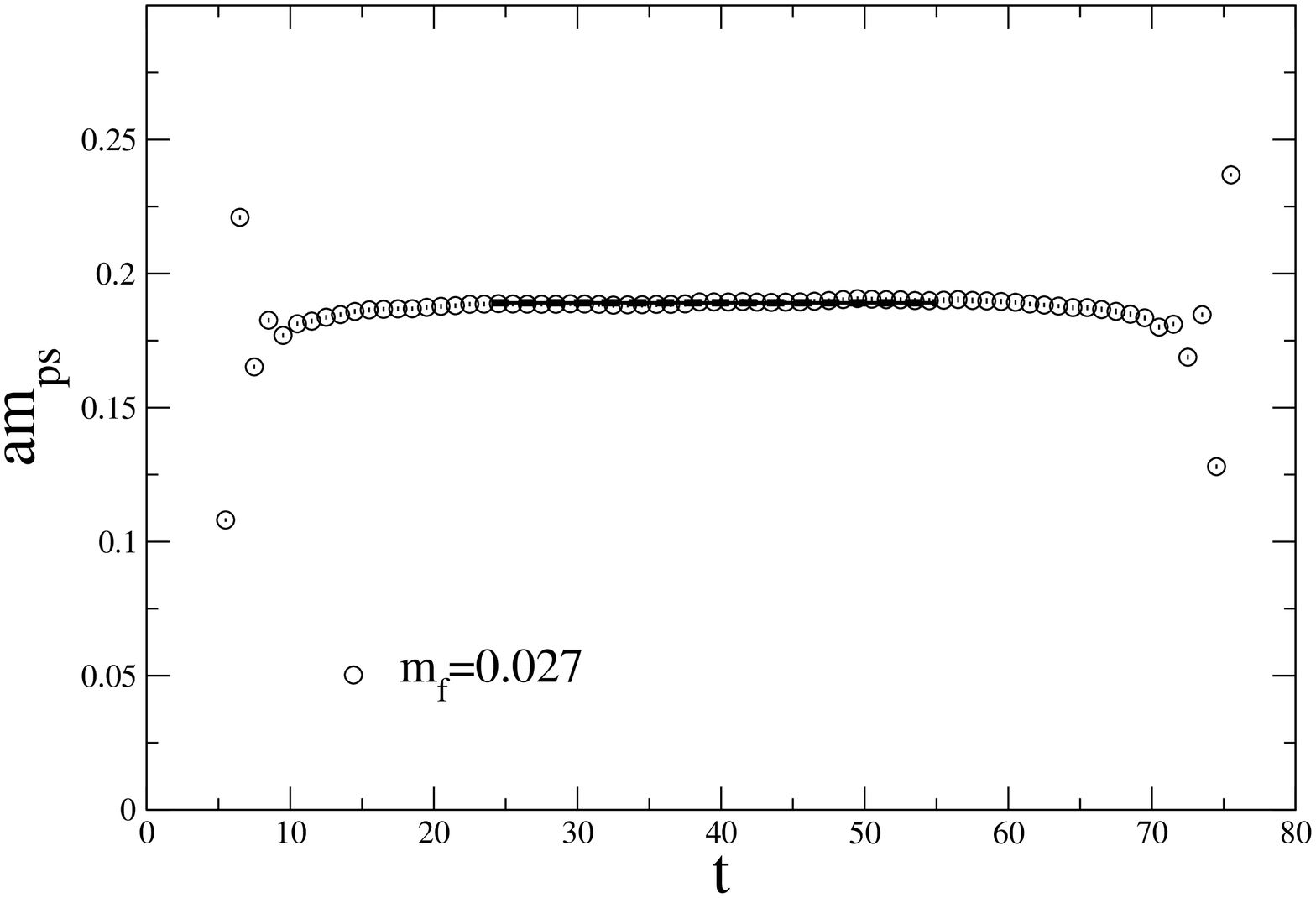}
\includegraphics[width=6.5cm]{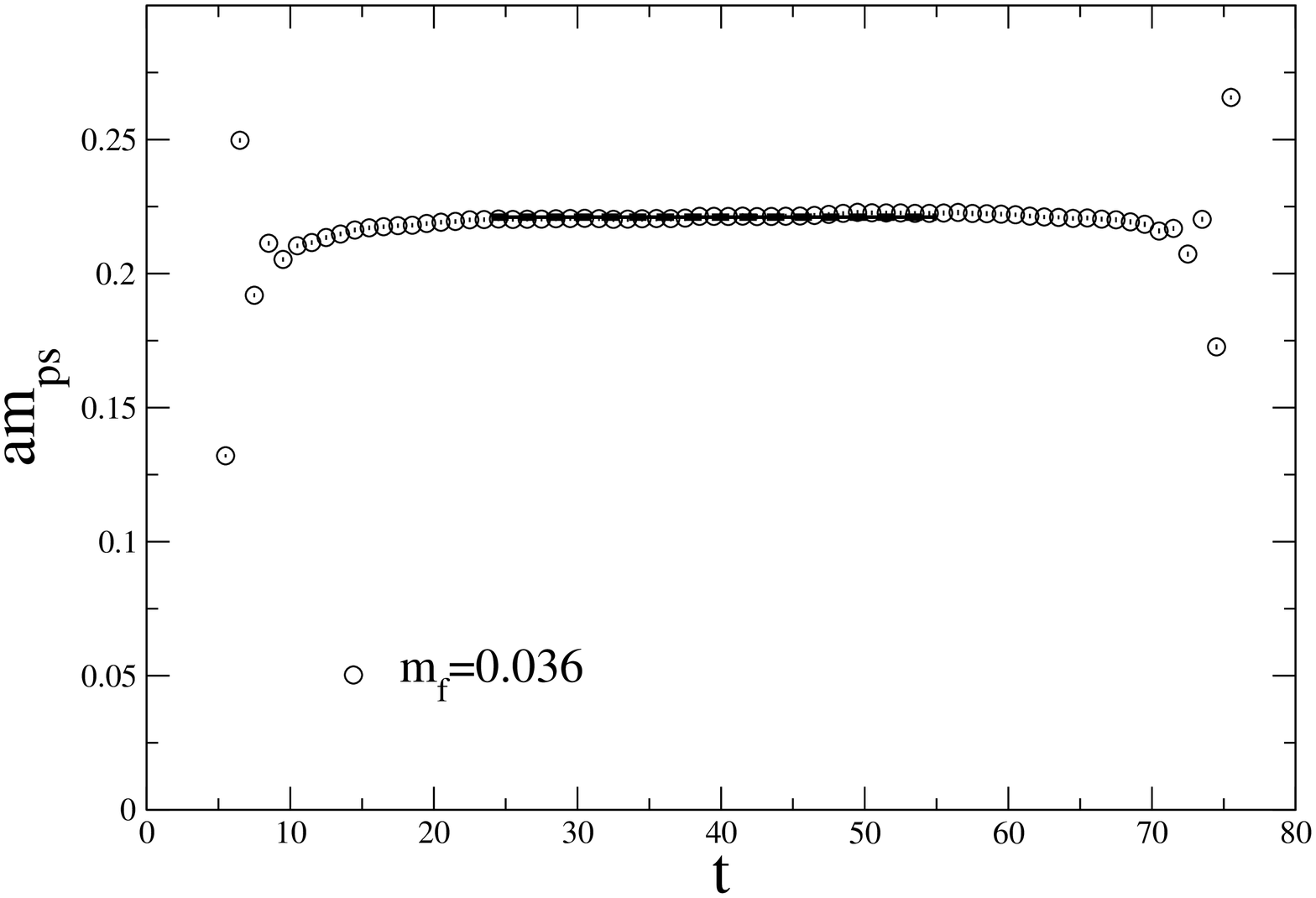}
  \caption{Effective masses of the pseudoscalar meson as a function of the 
  temporal distance $t$ at each quark mass.
  Lines represent the central value of the exponential fit of the propagator and its fitting range.}
  \label{fig:mpi-t}
 \end{center}
\end{figure}

\begin{figure}
 \begin{center}
\includegraphics[width=6.5cm]{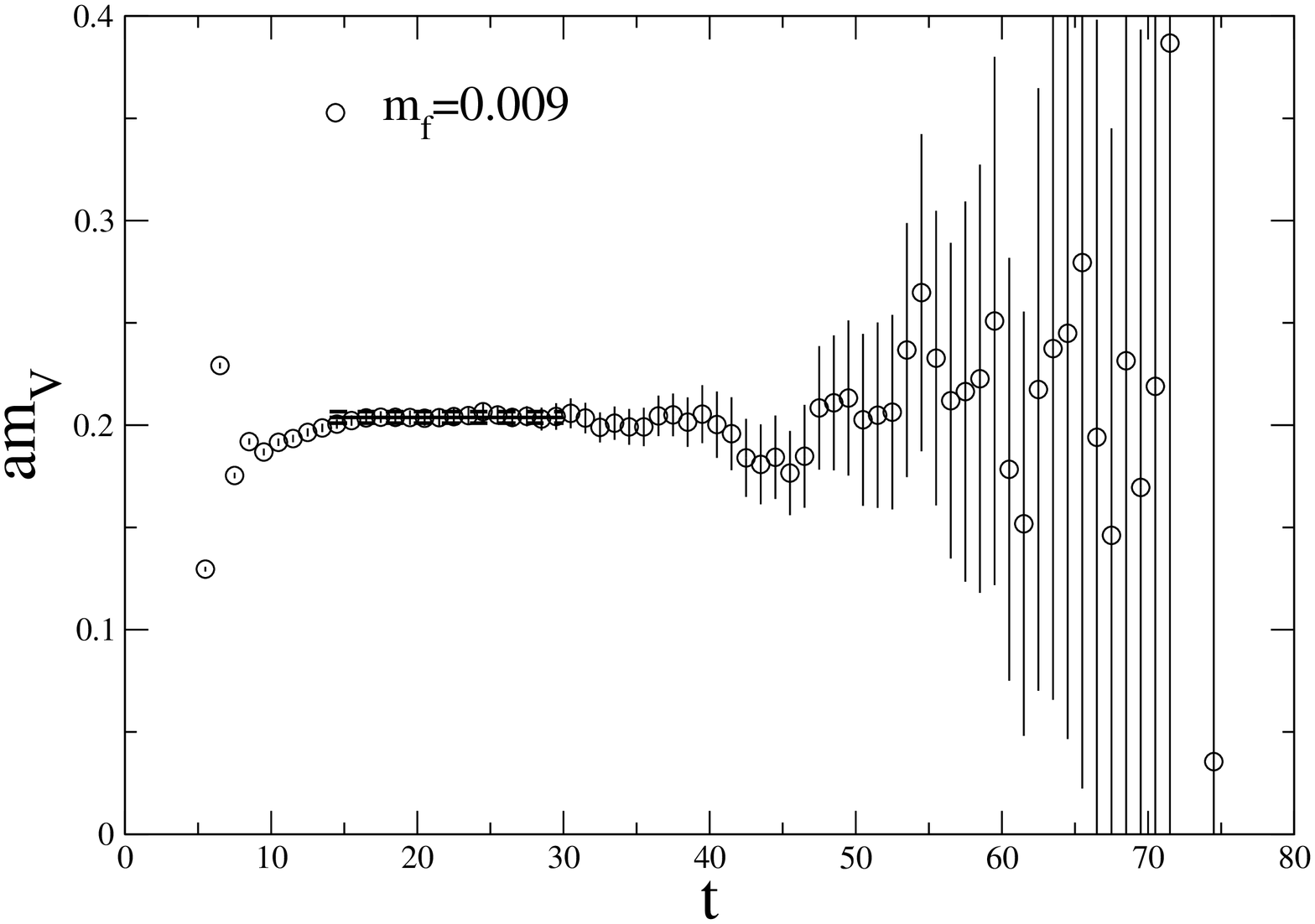}
\includegraphics[width=6.5cm]{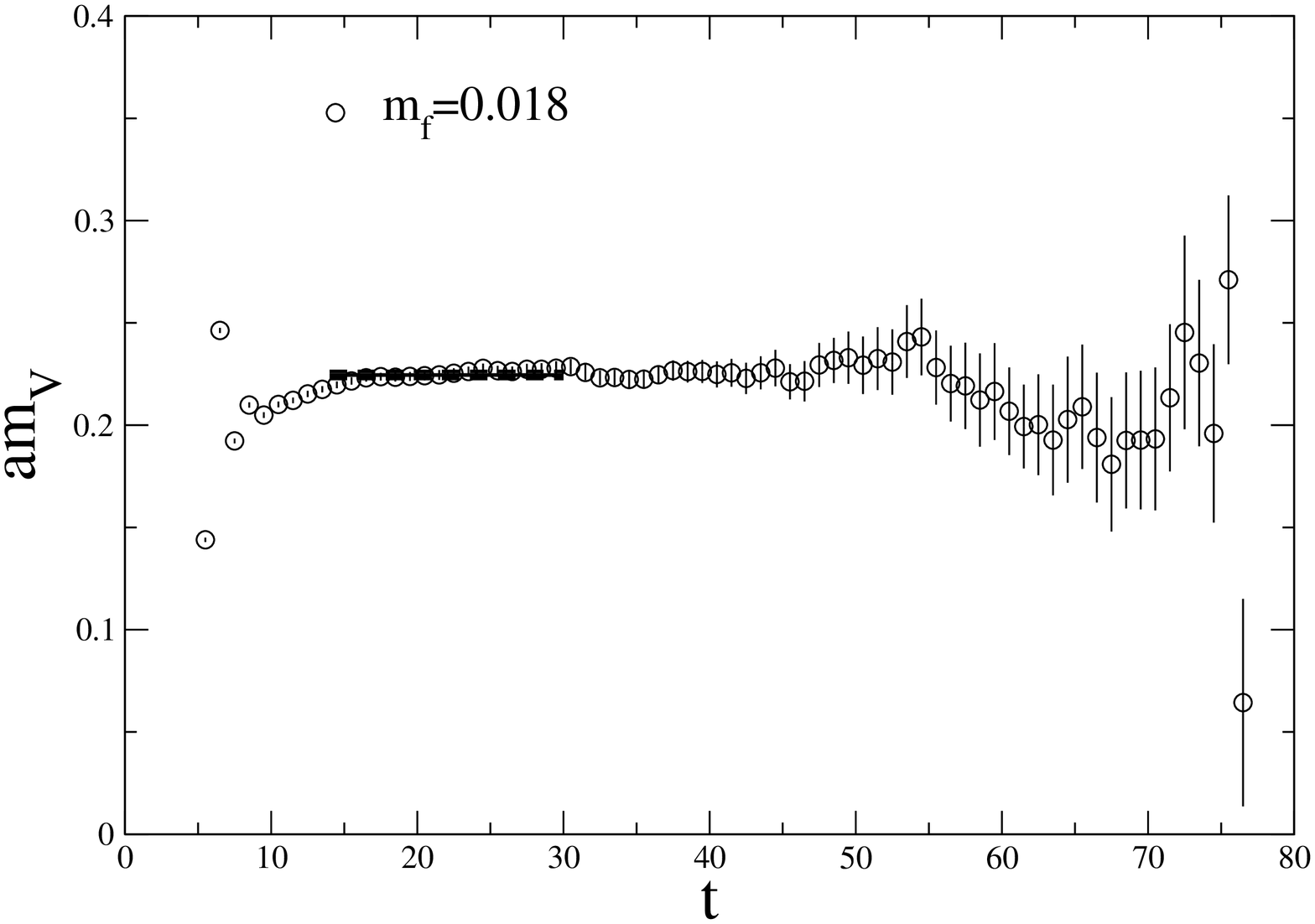}
\includegraphics[width=6.5cm]{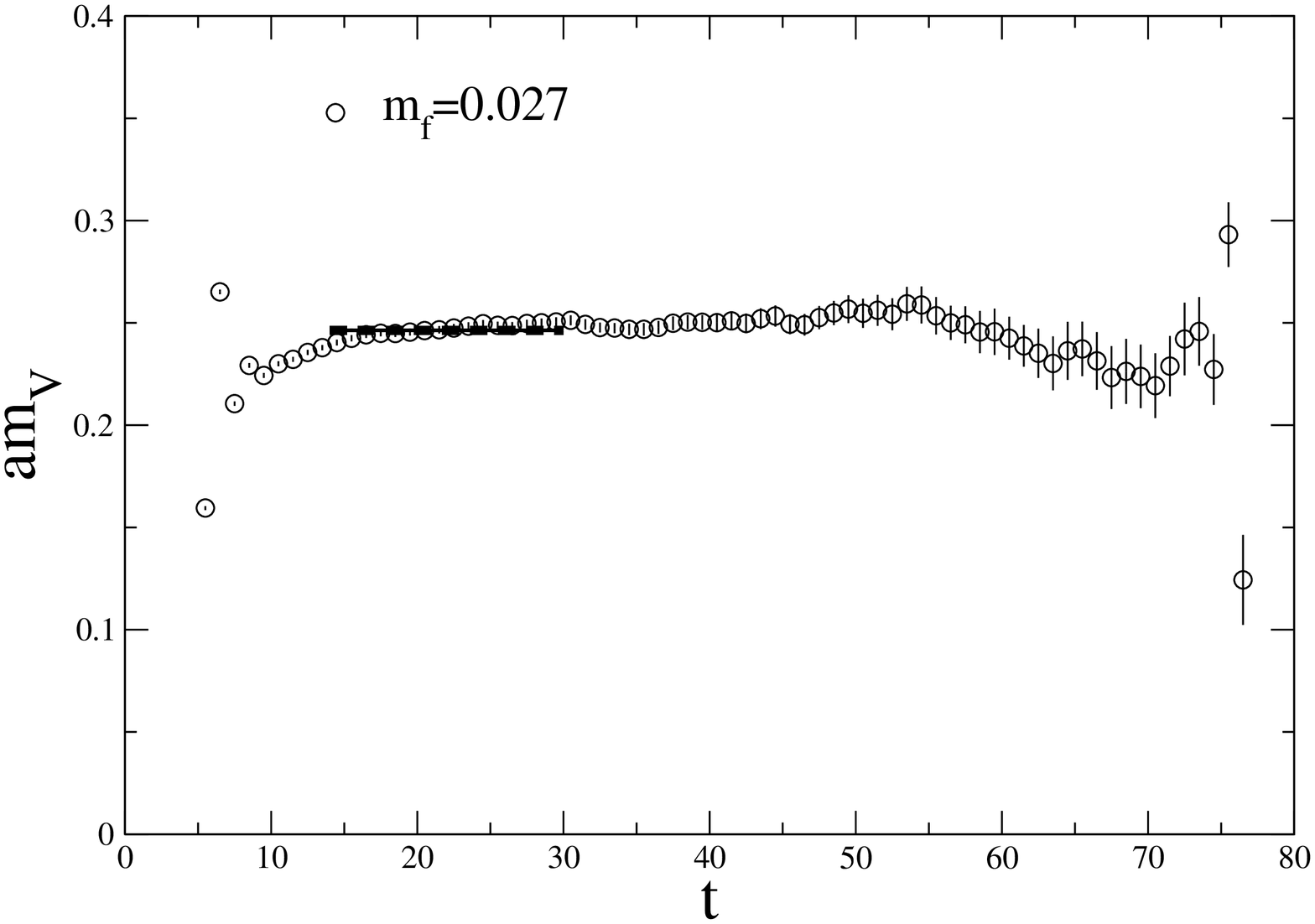}
\includegraphics[width=6.5cm]{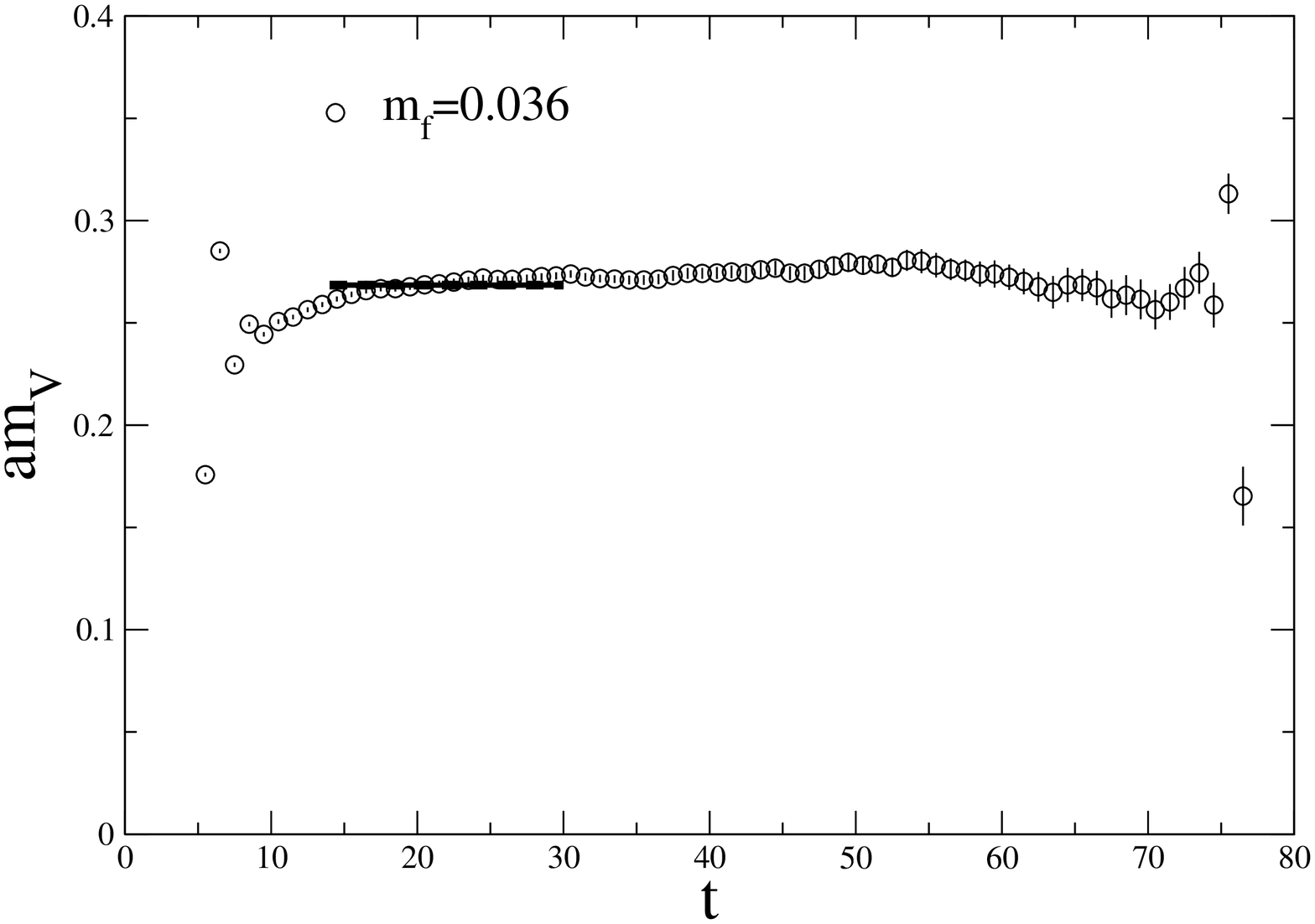}
  \caption{Effective masses of the vector meson as a function of $t$.}
  \label{fig:mrho-t}
 \end{center}
\end{figure}

We extract pseudoscalar and vector meson masses $m_{PS}$ and $m_V$ 
at each $m_f$ by a single exponential fit with meson propagators.
Representative plots of effective masses are shown in
Figs.~\ref{fig:mpi-t} and \ref{fig:mrho-t}.
Fitting ranges chosen from inspection of such plots are 
$24\le t \le55$  and $14\le t \le30$ for pseudoscalar and vector 
meson masses, respectively.

\begin{figure}
 \begin{center}
\includegraphics[width=8cm]{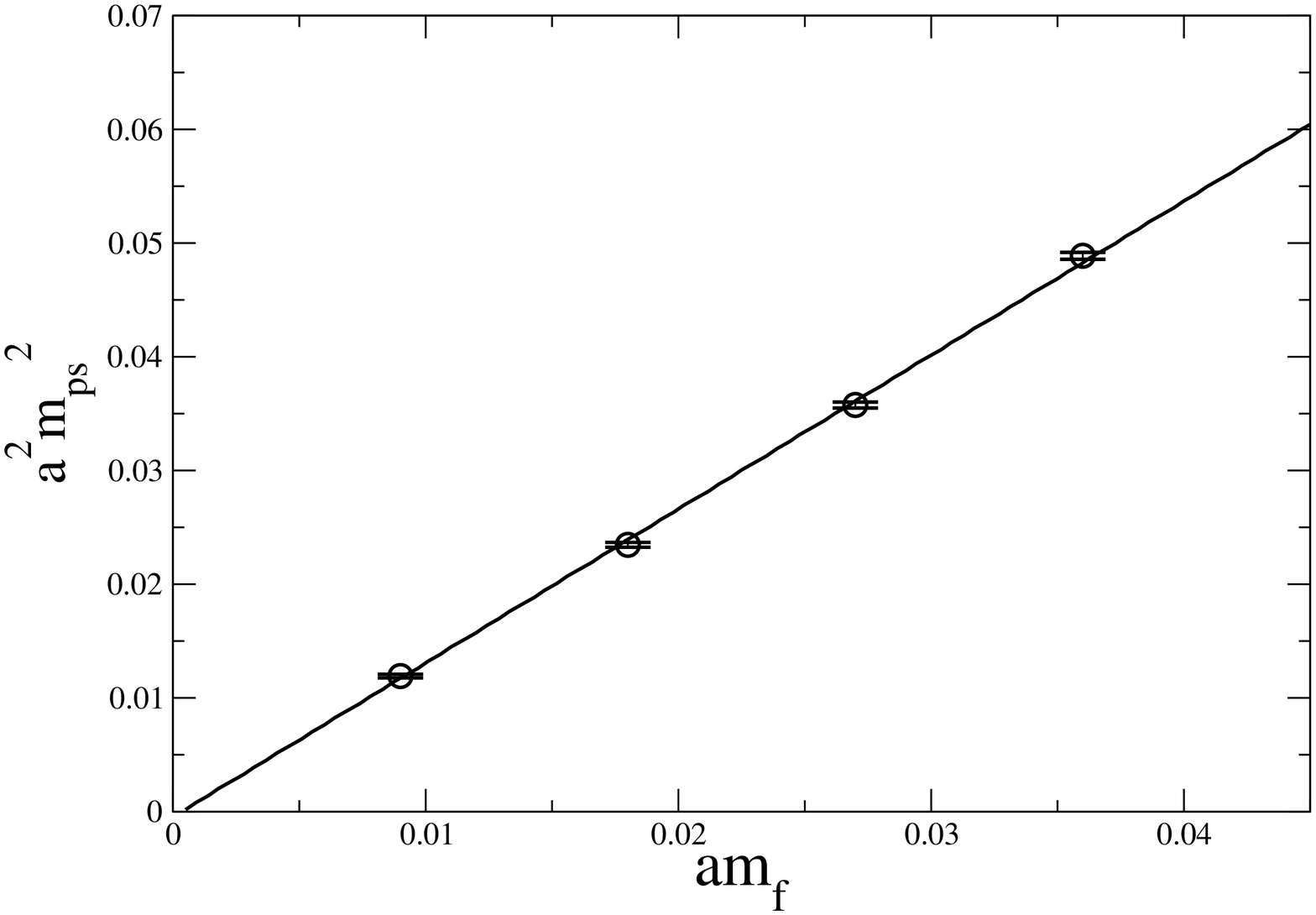}
\includegraphics[width=8cm]{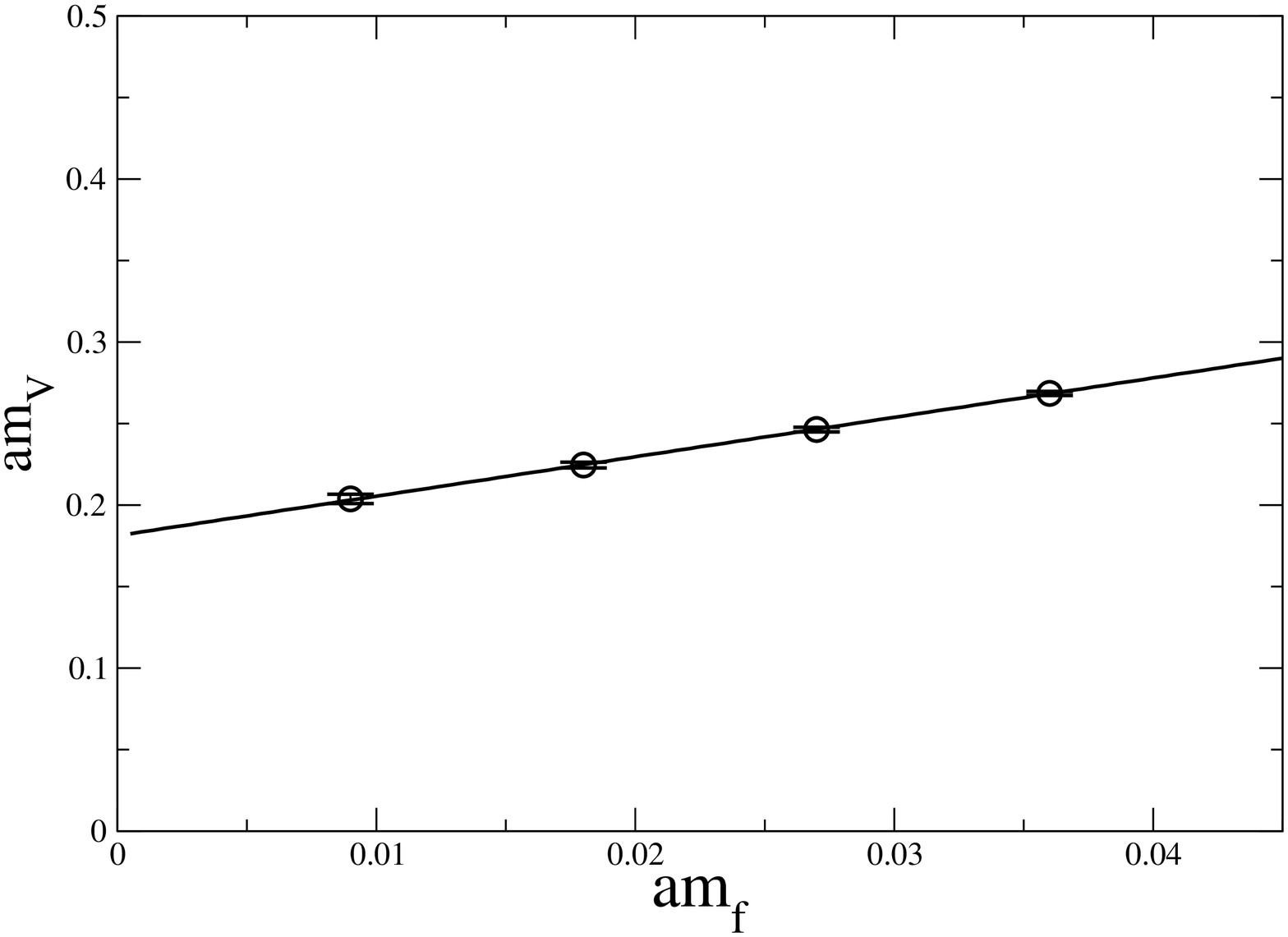}
  \caption{Pseudoscalar meson mass squared (left) and vector meson mass
  (right) as a function of bare quark mass $m_f a$.
  Lines show linear fits.}
  \label{fig:mpi2-mf}
 \end{center}
\end{figure}

For the chiral extrapolation we fit $m_{PS}^2$ and $m_V$ linearly in $m_f a$ 
as illustrated in Fig.~\ref{fig:mpi2-mf}.
Since pseudoscalar meson mass does not vanish at
$m_f=0$, we employ a fit of the form 
\begin{eqnarray}
m_{PS}^2a^2 &=& A_{PS}(m_fa+m_{\rm res}a),\label{eq:pion}\\
m_Va&=&A_V+B_Vm_fa \label{eq:rho}
\end{eqnarray}
and determine the parameters $A_{PS}, m_{\rm res}a$ for the pseudoscalar 
meson, and $A_V, B_V$ for the vector meson.  
The physical bare masses, $m_f^{ud}$ for the generated $u$ and $d$ quarks
and $m_f^{s}$ for the $s$ quark  
are determined by equations that
\begin{eqnarray}
\frac{\sqrt{ A_{PS}(m_f^{ud}a+m_{\rm res}a)}}{A_V+B_Vm_f^{ud}a}
&=&\frac{m_\pi}{m_\rho}=\frac{0.135}{0.77},\label{eq:udquark}
\\
\frac{\sqrt{ A_{PS}\left((m_f^{s}(K)a+m_f^{ud}a)/2+m_{\rm res}a\right)}}
{A_V+B_Vm_f^{ud}a}
&=&\frac{m_K}{m_\rho}=\frac{0.495}{0.77}, \label{eq:squark}
\\
\frac{A_V+B_Vm_f^s(\phi)a}{A_V+B_Vm_f^{ud}a}
&=&\frac{m_\phi}{m_\rho}=\frac{1.0194}{0.77}.
\label{eq:spquark}
\end{eqnarray}
For the $s$ quark, we extract two values of the $s$ quark mass,
$m_f^s(K)$ from the kaon mass input or $m_f^s(\phi)$ from the phi
meson mass input.
We then fix the lattice spacing $a$ by setting the vector meson mass 
at the physical quark mass point $m_f^{ud}$ to the experimental value 
$m_\rho=770$ MeV.
Numerical values of lattice spacing and quark masses
are listed in table \ref{tab:bare_bk}.

\subsection{Extraction of $B$ parameters}

In the course of our simulation we measure the kaon $B_K$
\eqn{eqn:BK}.
The $s$ and $d$ quark fields defining $B_K$ are the physical
fields given by (\ref{eq:physquark}), and the four-quark and bilinear
operators are taken to be local in the 4-dimensional space-time.  

\begin{figure}
 \begin{center}
\includegraphics[width=6.5cm]{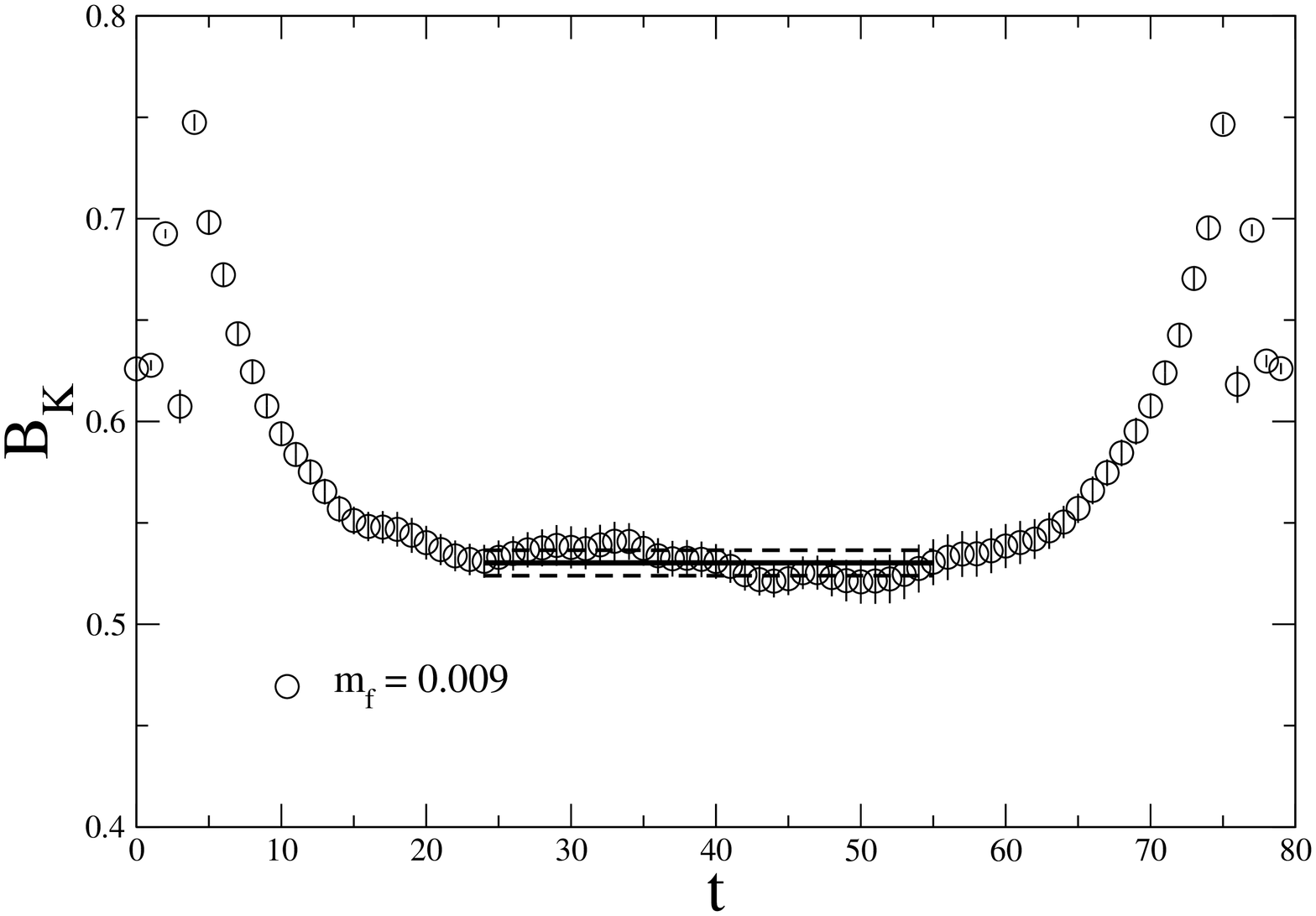}
\includegraphics[width=6.5cm]{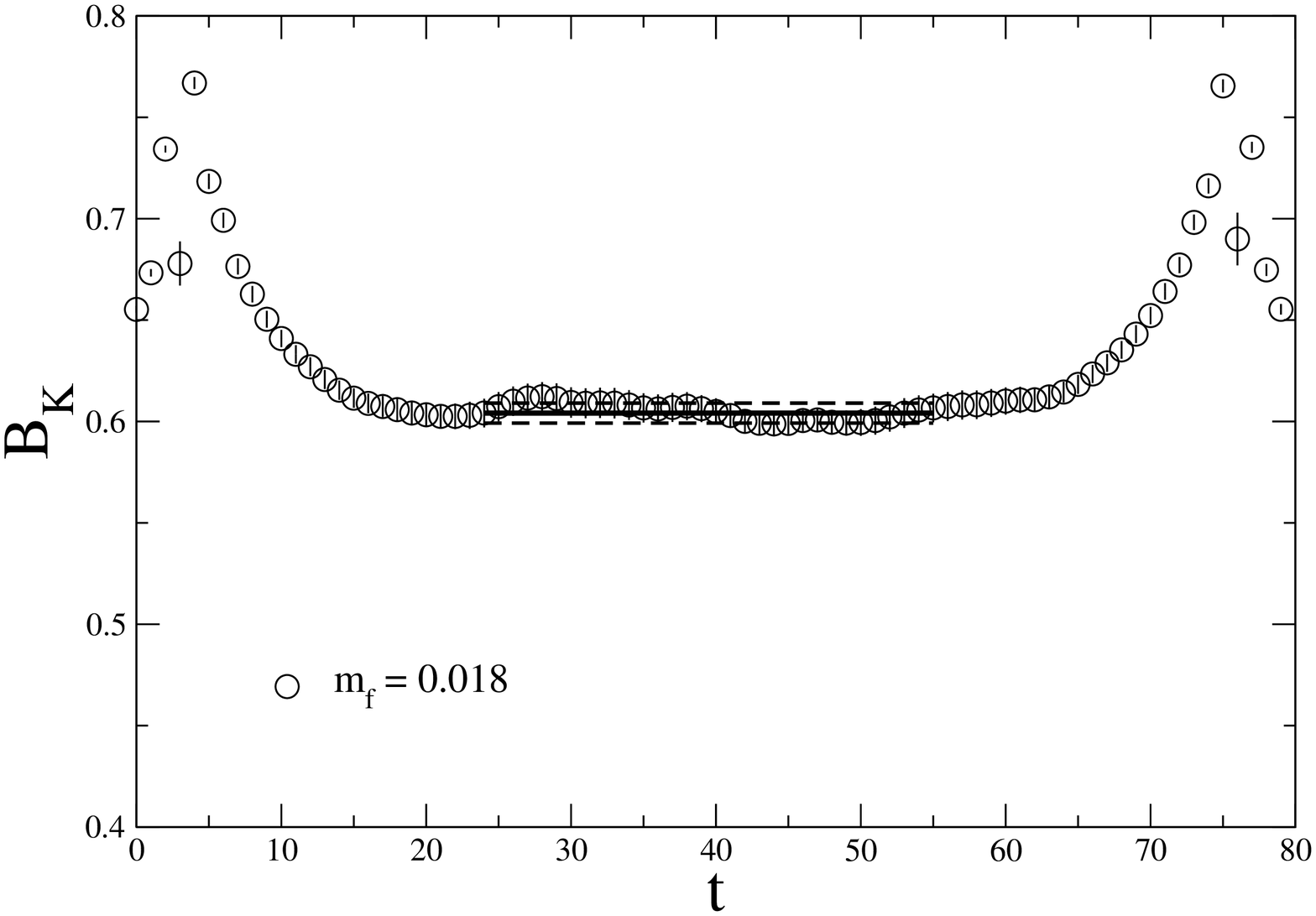}
\includegraphics[width=6.5cm]{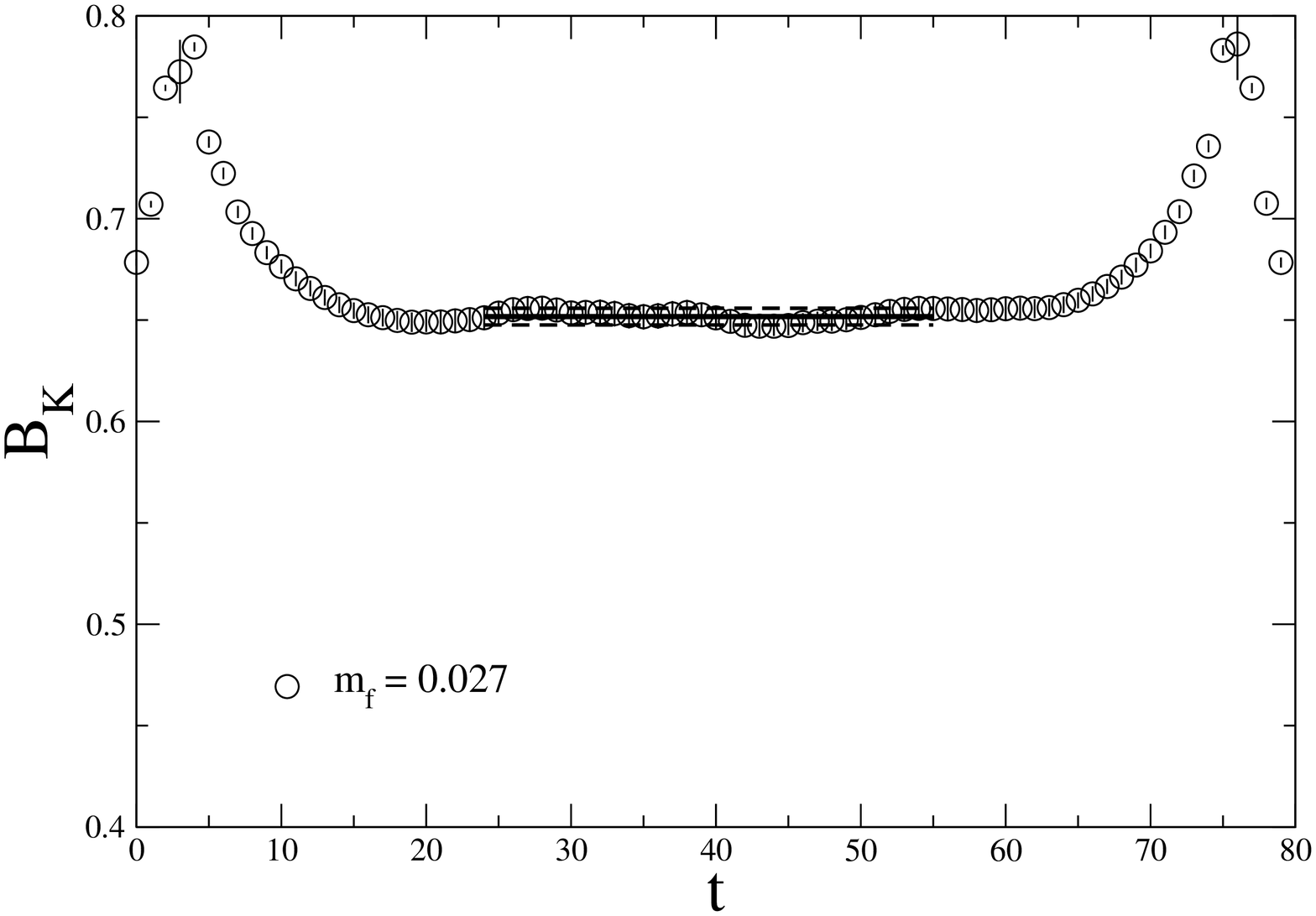}
\includegraphics[width=6.5cm]{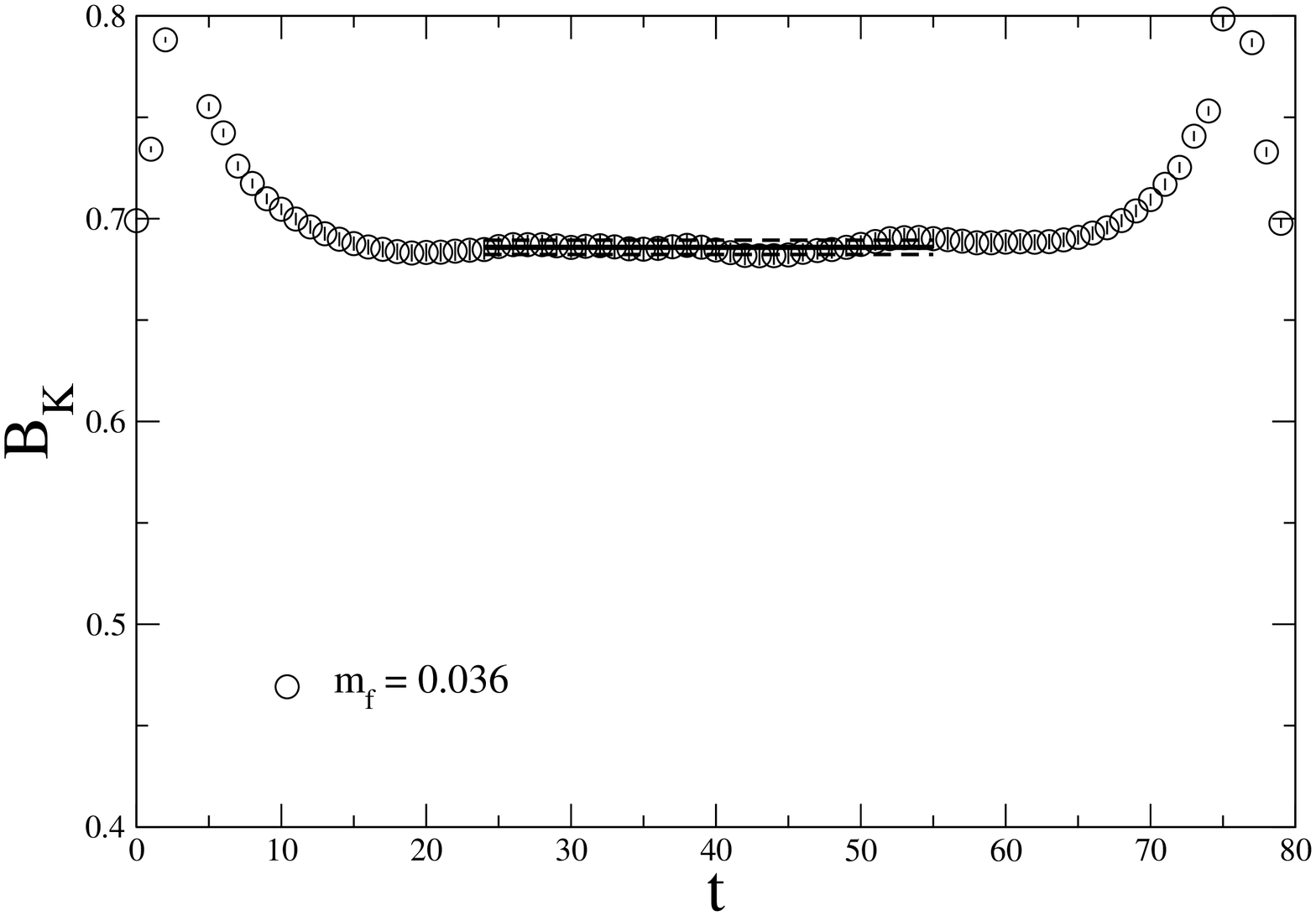}
  \caption{Ratios of the matrix element to the vacuum saturation
  \eqn{eqn:BK} as a function of $t$.}
  \label{fig:BK-t}
 \end{center}
\end{figure}

In Fig.~\ref{fig:BK-t} we show typical data for the ratio of kaon Green
functions for $B_K$ as a function of the temporal site $t$ of the weak
operator.
The values of these quantities at each $m_f$ are extracted by fitting
the plateau with a constant.
The fitting range, determined by the inspection of plots
for the ratio and those for the effective pseudo scalar meson mass, is
$24\le t \le 55$.

\begin{figure}
 \begin{center}
\includegraphics[width=8cm]{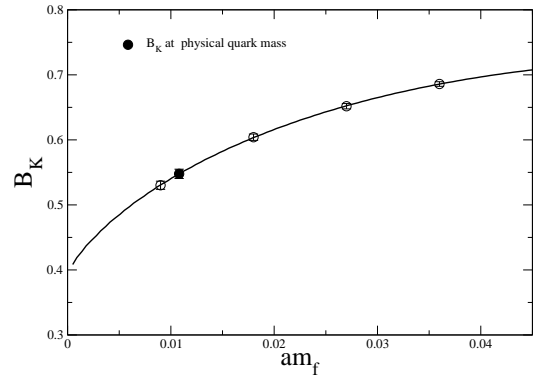}
  \caption{The bare $B_K$ as a function of $m_fa$.}
  \label{fig:BK-mf}
 \end{center}
\end{figure}

The bare value of $B_K$ is interpolated as a function of $m_fa$ using a 
formula suggested by chiral perturbation theory,
\begin{eqnarray}
B_K=B\left(1-3c\ m_fa\ \log(m_fa) + b\ m_fa\right).
\end{eqnarray}
This interpolation is illustrated in Fig.~\ref{fig:BK-mf}.
The physical value of $B_K$ is obtained at the point 
$m_f=(m_f^s(K)+m_f^{ud})/2$  estimated from the experimental value of 
$m_K/m_\rho$ (a solid circle in Fig.~\ref{fig:BK-mf}).
The result is given in table \ref{tab:bare_bk}.

\newpage
\clearpage
%
\begin{table}[htb]
 \begin{center}
  \begin{tabular}{|c|ccc|}
   \hline
   $\beta$ & $2.6$ & $2.9$ & $3.2$ \\
\hline
   $B_K^{(0)}$ & $0.5908(57)$ & $0.5655(69)$ & $0.5478(71)$ \\
\hline
$a^{-1}$ (GeV)& $1.807(37)$ & $2.807(55)$ & $4.186(65)$\\
\hline
${m}_{ud}$ (MeV)& $0.40(34)$ & $1.85(50)$ & $4.65(52)$\\
${m}_{ud}+{m}_{\rm res}$ (MeV) & $3.306(68)$ & $3.226(72)$ & $3.216(52)$\\
\hline
${m}_s(K)$ (MeV)& $83.8(19)$ & $83.2(20)$ & $85.8(14)$\\
${m}_s(K)+{m}_{\rm res}$ (MeV)& $86.7(18)$ & $84.6(19)$ & $84.3(14)$\\
${m}_s(\phi)$ & $114.1(93)$ (MeV)& $104.6(50)$ & $107.8(27)$\\
\hline
  \end{tabular}
  \caption{Numerical values of bare $B_K$ in DWQCD given by CP-PACS
  collaboration \cite{AliKhan:2001wr}.
  Values of lattice spacing and bare quark masses are also listed.
  Here $m_{ud}$ is the $u, d$ quark mass from the $\pi$ input, $m_s(K/\phi)$ is
  the strange quark mass from the $K$ $(\phi)$ input, and
  $-m_{\rm res}$ is a residual quark mass where the pion mass vanishes.
  Data at $\beta=3.2$ is new and not published in
  Ref.~\cite{AliKhan:2001wr}.}
  \label{tab:bare_bk}
 \end{center}
\end{table}

\begin{table}[htb]
\begin{center}
 \begin{tabular}{|c|c|c|c|c|c|c|c|c|}
  \hline
  $N_L$ & $6$ & $8$ & $10$ & $12$ & $14$ & $16$ & $18$ & $20$\\
  \hline
  $\beta$ & $2.4446$ & $2.6339$ & $2.7873$ & $2.9175$ & $3.0313$ &
  $3.1331$ & $3.2254$ & $3.3103$\\
  \hline
  num.~of conf.~for scheme $1\sim5$& $5165$ & $3632$ & $2000$ & $2188$ &
  $1000$ & $868$ & $778$ & $104$\\
  num.~of conf.~for scheme $6\sim9$& $1165$ & $1032$ & $1000$ & $670$ &
  $284$ & $312$ & $200$ & $104$\\
  \hline
 \end{tabular}
\caption{Values of $\beta$ which satisfies $aN_T=2L_{\rm max}$ and
 number of configurations for each lattice size.
 Data at $N_L=20$ is used only for the step scaling function.}
 \label{tab:beta-latsize}
\end{center}
\end{table}

\begin{table}[htb]
 \begin{center}
  \begin{tabular}{|c|lllllll|}
   \hline
   $N_L$   & $6$ & $8$ & $10$ & $12$ & $14$ & $16$ & $18$ \\
   \hline
  $\beta$ & $2.4446$ & $2.6339$ & $2.7873$ & $2.9175$ & $3.0313$ &
  $3.1331$ & $3.2254$\\
  \hline
$Z_{V}$ & $0.79580(23)$ & $0.85837(18)$ & $0.87048(15)$ & $0.90395(12)$ & $0.92286(30)$ & $0.94463(13)$ & $0.96184(12)$\\ 
\hline
$Z_{VA;1}^+$ & $0.69767(89)$ & $0.8778(14)$ & $0.9189(18)$ & $1.0179(21)$ & $1.0785(34)$ & $1.1388(37)$ & $1.1821(44)$\\ 
$Z_{VA;2}^+$ & $0.7196(13)$ & $0.9089(21)$ & $0.9679(28)$ & $1.0758(31)$ & $1.1521(53)$ & $1.2132(59)$ & $1.264(18)$\\ 
$Z_{VA;3}^+$ & $0.7398(11)$ & $0.9439(18)$ & $0.9956(23)$ & $1.1088(26)$ & $1.1804(44)$ & $1.2484(48)$ & $1.2994(58)$\\ 
$Z_{VA;4}^+$ & $0.6850(11)$ & $0.8528(17)$ & $0.9030(23)$ & $0.9982(24)$ & $1.0648(40)$ & $1.1196(45)$ & $1.1645(54)$\\ 
$Z_{VA;5}^+$ & $0.6819(11)$ & $0.8489(18)$ & $0.8985(23)$ & $0.9933(25)$ & $1.0595(42)$ & $1.1138(47)$ & $1.157(16)$\\ 
$Z_{VA;6}^+$ & $0.6382(18)$ & $0.7812(25)$ & $0.8224(26)$ & $0.8980(35)$ & $0.9592(53)$ & $1.0093(56)$ & $1.039(31)$\\ 
$Z_{VA;7}^+$ & $0.6817(16)$ & $0.8541(21)$ & $0.8930(23)$ & $0.9822(31)$ & $1.0450(49)$ & $1.1058(49)$ & $1.1409(66)$\\ 
$Z_{VA;8}^+$ & $0.6319(17)$ & $0.7702(23)$ & $0.8103(24)$ & $0.8838(32)$ & $0.9422(48)$ & $0.9913(50)$ & $1.0202(65)$\\ 
$Z_{VA;9}^+$ & $0.6292(18)$ & $0.7668(25)$ & $0.8061(26)$ & $0.8785(34)$ & $0.9395(51)$ & $0.9859(55)$ & $1.015(28)$\\ 
\hline
$Z_{VA;1}^-$ & $0.61284(68)$ & $0.7101(11)$ & $0.7024(13)$ & $0.7472(15)$ & $0.7587(25)$ & $0.7886(25)$ & $0.8061(30)$\\ 
$Z_{VA;2}^-$ & $0.6677(14)$ & $0.7791(22)$ & $0.7937(28)$ & $0.8516(28)$ & $0.8820(48)$ & $0.9074(53)$ & $0.934(13)$\\ 
$Z_{VA;3}^-$ & $0.69325(88)$ & $0.8290(15)$ & $0.8301(18)$ & $0.8945(20)$ & $0.9152(33)$ & $0.9532(35)$ & $0.9783(41)$\\ 
$Z_{VA;4}^-$ & $0.6082(11)$ & $0.6931(17)$ & $0.6974(21)$ & $0.7413(21)$ & $0.7630(35)$ & $0.7837(38)$ & $0.8019(42)$\\ 
$Z_{VA;5}^-$ & $0.6074(11)$ & $0.6910(17)$ & $0.6968(21)$ & $0.7400(20)$ & $0.7627(34)$ & $0.7823(38)$ & $0.802(10)$\\ 
$Z_{VA;6}^-$ & $0.5924(21)$ & $0.6679(28)$ & $0.6726(27)$ & $0.7135(36)$ & $0.7295(55)$ & $0.7537(56)$ & $0.769(23)$\\ 
$Z_{VA;7}^-$ & $0.6391(13)$ & $0.7506(20)$ & $0.7433(19)$ & $0.7958(27)$ & $0.8089(44)$ & $0.8418(44)$ & $0.8618(55)$\\ 
$Z_{VA;8}^-$ & $0.5614(19)$ & $0.6245(25)$ & $0.6244(24)$ & $0.6572(32)$ & $0.6698(48)$ & $0.6921(50)$ & $0.7018(55)$\\ 
$Z_{VA;9}^-$ & $0.5608(19)$ & $0.6233(25)$ & $0.6235(24)$ & $0.6570(32)$ & $0.6706(48)$ & $0.6909(50)$ & $0.703(19)$\\ 
  \hline
$Z_{B_K;1}$ & $1.1017(12)$ & $1.1914(19)$ & $1.2127(24)$ & $1.2457(25)$ & $1.2663(40)$ & $1.2763(42)$ & $1.2778(47)$\\ 
$Z_{B_K;2}$ & $1.1363(18)$ & $1.2336(28)$ & $1.2774(37)$ & $1.3166(38)$ & $1.3527(63)$ & $1.3596(66)$ & $1.366(19)$\\ 
$Z_{B_K;3}$ & $1.1682(15)$ & $1.2811(24)$ & $1.3139(30)$ & $1.3570(32)$ & $1.3860(52)$ & $1.3991(55)$ & $1.4045(63)$\\ 
$Z_{B_K;4}$ & $1.0816(15)$ & $1.1574(23)$ & $1.1918(29)$ & $1.2216(30)$ & $1.2502(48)$ & $1.2547(50)$ & $1.2587(58)$\\ 
$Z_{B_K;5}$ & $1.0768(15)$ & $1.1521(23)$ & $1.1857(30)$ & $1.2157(30)$ & $1.2441(50)$ & $1.2482(53)$ & $1.251(17)$\\ 
$Z_{B_K;6}$ & $1.0084(23)$ & $1.0605(32)$ & $1.0854(34)$ & $1.0990(42)$ & $1.1285(62)$ & $1.1313(62)$ & $1.123(33)$\\ 
$Z_{B_K;7}$ & $1.0772(20)$ & $1.1594(27)$ & $1.1787(30)$ & $1.2021(37)$ & $1.2294(57)$ & $1.2394(55)$ & $1.2329(72)$\\ 
$Z_{B_K;8}$ & $0.9984(22)$ & $1.0456(30)$ & $1.0695(31)$ & $1.0817(39)$ & $1.1086(56)$ & $1.1111(56)$ & $1.1025(71)$\\ 
$Z_{B_K;9}$ & $0.9942(23)$ & $1.0410(32)$ & $1.0640(33)$ & $1.0752(41)$ & $1.1054(60)$ & $1.1050(61)$ & $1.097(30)$\\ 
   \hline
$Z_{P}$ & $0.65512(64)$ & $0.66259(88)$ & $0.64990(99)$ & $0.6560(10)$ & $0.6580(16)$ & $0.6606(16)$ & $0.6627(18)$\\
$Z_m$ & $1.5264(15)$ & $1.5092(20)$ & $1.5387(23)$ & $1.5244(24)$ & $1.5198(37)$ & $1.5137(37)$ & $1.5091(42)$\\
   \hline
  \end{tabular}
  \caption{Numerical values of $Z_V(g_0)$,
  $Z_{VA+AV;s}^\pm(g_0,a\mu_{\rm min})$ and
  $Z_{B_K;s}(g_0,a\mu_{\rm min})$ at $2L_{\rm max}$.
  Values of $Z_P(g_0,a\mu_{\rm min})$ and $Z_m(g_0,a\mu_{\rm min})$ are
  also listed for the latter use.}
  \label{tab:zvaav.mumin}
 \end{center}
\end{table} 

\begin{table}[htb]
\begin{center}
 \begin{tabular}{|c|c|c|c|c||c|}
  \hline
  $N_L$ & $4$ & $6$ & $8$ & $10$ & $20$\\
  \hline
  $\beta$ & $2.6339$ & $2.9175$ & $3.1331$ & $3.3103$ & $3.3103$ \\
  \hline
$Z_{V}$ & $0.9905(21)$ & $0.87148(98)$ & $0.97930(77)$ & $0.97587(57)$ & $0.97898(26)$\\ 
\hline
$Z_{VA;1}^+$ & $1.0784(69)$ & $0.7937(48)$ & $1.0967(59)$ & $1.0853(70)$ & $1.258(11)$\\ 
$Z_{VA;2}^+$ & $1.0302(74)$ & $0.8061(63)$ & $1.0945(84)$ & $1.1054(91)$ & $1.349(16)$\\ 
$Z_{VA;3}^+$ & $1.1000(72)$ & $0.8215(56)$ & $1.1446(67)$ & $1.1384(83)$ & $1.376(14)$\\ 
$Z_{VA;4}^+$ & $1.0058(70)$ & $0.7807(56)$ & $1.0480(74)$ & $1.0555(76)$ & $1.246(12)$\\ 
$Z_{VA;5}^+$ & $1.0076(71)$ & $0.7811(57)$ & $1.0498(77)$ & $1.0555(76)$ & $1.240(12)$\\ 
$Z_{VA;6}^+$ & $1.0018(70)$ & $0.7703(54)$ & $1.0346(73)$ & $1.0366(73)$ & $1.1181(91)$\\ 
$Z_{VA;7}^+$ & $1.0797(69)$ & $0.7969(49)$ & $1.1024(59)$ & $1.0907(71)$ & $1.2145(89)$\\ 
$Z_{VA;8}^+$ & $0.9871(67)$ & $0.7574(51)$ & $1.0094(68)$ & $1.0113(66)$ & $1.0995(85)$\\ 
$Z_{VA;9}^+$ & $0.9890(68)$ & $0.7578(51)$ & $1.0112(72)$ & $1.0112(66)$ & $1.0942(88)$\\ 
\hline
$Z_{VA;1}^-$ & $1.0609(63)$ & $0.7389(30)$ & $0.9721(49)$ & $0.9349(57)$ & $0.8252(67)$\\ 
$Z_{VA;2}^-$ & $0.9831(82)$ & $0.7746(69)$ & $0.9775(93)$ & $0.9847(99)$ & $0.971(15)$\\ 
$Z_{VA;3}^-$ & $1.1135(69)$ & $0.7933(42)$ & $1.0644(60)$ & $1.0334(74)$ & $0.995(10)$\\ 
$Z_{VA;4}^-$ & $0.9423(74)$ & $0.7288(58)$ & $0.9043(79)$ & $0.9033(79)$ & $0.842(11)$\\ 
$Z_{VA;5}^-$ & $0.9435(74)$ & $0.7329(57)$ & $0.9088(79)$ & $0.9070(75)$ & $0.841(11)$\\ 
$Z_{VA;6}^-$ & $0.9559(77)$ & $0.7401(61)$ & $0.9239(83)$ & $0.9234(81)$ & $0.805(11)$\\ 
$Z_{VA;7}^-$ & $1.0929(66)$ & $0.7696(37)$ & $1.0252(54)$ & $0.9901(65)$ & $0.8777(72)$\\ 
$Z_{VA;8}^-$ & $0.9248(71)$ & $0.7070(53)$ & $0.8710(73)$ & $0.8654(70)$ & $0.7430(93)$\\ 
$Z_{VA;9}^-$ & $0.9260(71)$ & $0.7110(54)$ & $0.8753(74)$ & $0.8690(67)$ & $0.7425(93)$\\ 
\hline
$Z_{B_K;1}$ & $1.0992(51)$ & $1.0450(53)$ & $1.1436(61)$ & $1.1397(73)$ & $1.313(11)$\\ 
$Z_{B_K;2}$ & $1.0502(59)$ & $1.0615(74)$ & $1.1413(88)$ & $1.1608(96)$ & $1.407(17)$\\ 
$Z_{B_K;3}$ & $1.1213(55)$ & $1.0816(63)$ & $1.1935(71)$ & $1.1954(87)$ & $1.436(15)$\\ 
$Z_{B_K;4}$ & $1.0252(54)$ & $1.0280(64)$ & $1.0928(76)$ & $1.1084(80)$ & $1.300(13)$\\ 
$Z_{B_K;5}$ & $1.0271(55)$ & $1.0285(65)$ & $1.0947(80)$ & $1.1084(81)$ & $1.294(13)$\\ 
$Z_{B_K;6}$ & $1.0211(54)$ & $1.0142(62)$ & $1.0788(75)$ & $1.0885(76)$ & $1.1667(95)$\\ 
$Z_{B_K;7}$ & $1.1005(51)$ & $1.0493(54)$ & $1.1495(62)$ & $1.1453(74)$ & $1.2672(93)$\\ 
$Z_{B_K;8}$ & $1.0062(51)$ & $0.9972(57)$ & $1.0525(69)$ & $1.0619(69)$ & $1.1472(89)$\\ 
$Z_{B_K;9}$ & $1.0081(52)$ & $0.9977(58)$ & $1.0544(73)$ & $1.0619(70)$ & $1.1417(93)$\\ 
  \hline
$Z_P$ & $0.8718(36)$ & $0.7680(33)$ & $0.8160(36)$ & $0.8051(37)$ & $0.6712(53)$\\
$Z_m$ & $1.1471(48)$ & $1.3020(55)$ & $1.2254(54)$ & $1.2421(56)$ & $1.490(12)$\\
  \hline
 \end{tabular}
 \caption{Numerical values of $Z_V(g_0)$, 
 $Z_{VA+AV;s}^\pm(g_0,2a\mu_{\rm min})$, $Z_{B_K;s}(g_0,2a\mu_{\rm min})$,
 $Z_P(g_0,2a\mu_{\rm min})$ and $Z_m(g_0,2a\mu_{\rm min})$
 at $L_{\rm max}$ for SSF.
 A box size of $N_L=20$ lattice is $2L_{\rm max}$.}
  \label{tab:zvaav.ssf}
\end{center}
\end{table}

\begin{table}[htb]
\begin{center}
 \begin{tabular}{|c|c|c|c|c||c|c|c|c|}
  \hline
  $N_L$ & $4$ & $6$ & $8$ & $10$ & $8$ & $12$ & $16$ & $20$\\
  \hline
  $\beta$ & $2.6339$ & $2.9175$ & $3.1331$ & $3.3103$ & $2.6339$ & $2.9175$ & $3.1331$ & $3.3103$ \\
  \hline
$Z_{VA;1}^+$ & $0.7664(38)$ & $0.7669(49)$ & $0.7928(54)$ & $0.8157(50)$ & $0.8778(14)$ & $1.0179(21)$ & $1.1388(37)$ & $1.258(11)$\\
$Z_{VA;2}^+$ & $0.7602(44)$ & $0.7661(60)$ & $0.7977(71)$ & $0.8210(71)$ & $0.9089(21)$ & $1.0758(31)$ & $1.2132(59)$ & $1.349(16)$\\
$Z_{VA;3}^+$ & $0.7829(39)$ & $0.7937(55)$ & $0.8260(62)$ & $0.8629(96)$ & $0.9439(18)$ & $1.1088(26)$ & $1.2484(48)$ & $1.376(14)$\\
$Z_{VA;4}^+$ & $0.7440(42)$ & $0.7397(52)$ & $0.7655(60)$ & $0.801(14)$ & $0.8528(17)$ & $0.9982(24)$ & $1.1196(45)$ & $1.246(12)$\\
$Z_{VA;5}^+$ & $0.7445(42)$ & $0.7407(52)$ & $0.7670(63)$ & $0.802(14)$ & $0.8489(18)$ & $0.9933(25)$ & $1.1138(47)$ & $1.240(12)$\\
$Z_{VA;6}^+$ & $0.7379(41)$ & $0.7321(51)$ & $0.7542(58)$ & $0.7748(61)$ & $0.7812(25)$ & $0.8980(35)$ & $1.0093(56)$ & $1.1181(91)$\\
$Z_{VA;7}^+$ & $0.7675(38)$ & $0.7700(50)$ & $0.7957(53)$ & $0.8302(89)$ & $0.8541(21)$ & $0.9822(31)$ & $1.1058(49)$ & $1.2145(89)$\\
$Z_{VA;8}^+$ & $0.7294(40)$ & $0.7176(47)$ & $0.7374(52)$ & $0.771(13)$ & $0.7702(23)$ & $0.8838(32)$ & $0.9913(50)$ & $1.0995(85)$\\
$Z_{VA;9}^+$ & $0.7298(41)$ & $0.7186(48)$ & $0.7389(56)$ & $0.771(14)$ & $0.7668(25)$ & $0.8785(34)$ & $0.9859(55)$ & $1.0942(88)$\\
  \hline
$Z_{VA;1}^-$ & $0.8065(28)$ & $0.7611(37)$ & $0.7532(41)$ & $0.7437(39)$ & $0.7101(11)$ & $0.7472(15)$ & $0.7886(25)$ & $0.8252(67)$\\
$Z_{VA;2}^-$ & $0.8076(47)$ & $0.7693(60)$ & $0.7688(68)$ & $0.7573(81)$ & $0.7791(22)$ & $0.8516(28)$ & $0.9074(53)$ & $0.971(15)$\\
$Z_{VA;3}^-$ & $0.8465(34)$ & $0.8193(48)$ & $0.8221(52)$ & $0.8241(92)$ & $0.8290(15)$ & $0.8945(20)$ & $0.9532(35)$ & $0.995(10)$\\
$Z_{VA;4}^-$ & $0.7757(43)$ & $0.7211(49)$ & $0.7114(56)$ & $0.711(13)$ & $0.6931(17)$ & $0.7413(21)$ & $0.7837(38)$ & $0.842(11)$\\
$Z_{VA;5}^-$ & $0.7770(42)$ & $0.7247(47)$ & $0.7165(53)$ & $0.717(13)$ & $0.6910(17)$ & $0.7400(20)$ & $0.7823(38)$ & $0.841(11)$\\
$Z_{VA;6}^-$ & $0.7839(44)$ & $0.7352(51)$ & $0.7269(56)$ & $0.7147(70)$ & $0.6679(28)$ & $0.7135(36)$ & $0.7537(56)$ & $0.805(11)$\\
$Z_{VA;7}^-$ & $0.8299(32)$ & $0.7948(44)$ & $0.7920(45)$ & $0.7929(85)$ & $0.7506(20)$ & $0.7958(27)$ & $0.8418(44)$ & $0.8777(72)$\\
$Z_{VA;8}^-$ & $0.7605(41)$ & $0.6996(45)$ & $0.6854(50)$ & $0.684(12)$ & $0.6245(25)$ & $0.6572(32)$ & $0.6921(50)$ & $0.7430(93)$\\
$Z_{VA;9}^-$ & $0.7617(40)$ & $0.7031(43)$ & $0.6902(48)$ & $0.690(12)$ & $0.6233(25)$ & $0.6570(32)$ & $0.6909(50)$ & $0.7425(93)$\\
  \hline
$Z_{B_K;1}$ & $1.0096(33)$ & $1.0715(63)$ & $1.1041(69)$ & $1.1221(66)$ & $1.1914(19)$ & $1.2457(25)$ & $1.2763(42)$ & $1.313(11)$\\
$Z_{B_K;2}$ & $1.0014(41)$ & $1.0704(79)$ & $1.1110(94)$ & $1.1295(96)$ & $1.2336(28)$ & $1.3166(38)$ & $1.3596(66)$ & $1.407(17)$\\
$Z_{B_K;3}$ & $1.0314(36)$ & $1.1089(72)$ & $1.1504(81)$ & $1.187(13)$ & $1.2811(24)$ & $1.3570(32)$ & $1.3991(55)$ & $1.436(15)$\\
$Z_{B_K;4}$ & $0.9801(38)$ & $1.0334(67)$ & $1.0661(78)$ & $1.102(19)$ & $1.1574(23)$ & $1.2216(30)$ & $1.2547(50)$ & $1.300(13)$\\
$Z_{B_K;5}$ & $0.9807(39)$ & $1.0348(68)$ & $1.0683(83)$ & $1.103(19)$ & $1.1521(23)$ & $1.2157(30)$ & $1.2482(53)$ & $1.294(13)$\\
$Z_{B_K;6}$ & $0.9721(37)$ & $1.0228(65)$ & $1.0504(76)$ & $1.0659(82)$ & $1.0605(32)$ & $1.0990(42)$ & $1.1313(62)$ & $1.1667(95)$\\
$Z_{B_K;7}$ & $1.0112(33)$ & $1.0758(63)$ & $1.1082(68)$ & $1.142(12)$ & $1.1594(27)$ & $1.2021(37)$ & $1.2394(55)$ & $1.2672(93)$\\
$Z_{B_K;8}$ & $0.9608(36)$ & $1.0026(59)$ & $1.0270(68)$ & $1.060(18)$ & $1.0456(30)$ & $1.0817(39)$ & $1.1111(56)$ & $1.1472(89)$\\
$Z_{B_K;9}$ & $0.9615(37)$ & $1.0040(60)$ & $1.0291(73)$ & $1.061(18)$ & $1.0410(32)$ & $1.0752(41)$ & $1.1050(61)$ & $1.1417(93)$\\
  \hline
$Z_P$ & $0.8277(22)$ & $0.7670(30)$ & $0.7394(35)$ & $0.7240(31)$ & $0.66259(88)$ & $0.6560(10)$ & $0.6606(16)$ & $0.6712(53)$\\
$Z_m$ & $1.2082(32)$ & $1.3038(51)$ & $1.3524(65)$ & $1.3812(60)$ & $1.5092(20)$ & $1.5244(24)$ & $1.5137(37)$ & $1.490(12)$\\
  \hline
 \end{tabular}
 \caption{Numerical values of $Z_{VA+AV;s}^\pm(g_0,a\mu)$,
 $Z_{B_K;s}(g_0,a\mu)$, $Z_P(g_0,a\mu)$ and $Z_m(g_0,a\mu)$
 at $L_{\rm max}$ and $2L_{\rm max}$ at $M=1.4$ for SSF.}
  \label{tab:zvaav.ssf.M14}
\end{center}
\end{table}

\begin{table}[htb]
 \begin{center}
  \begin{tabular}{|c|lllllll|}
   \hline
   $N_L$   & $6$ & $8$ & $10$ & $12$ & $14$ & $16$ & $18$ \\
   \hline
  $\beta$ & $2.4446$ & $2.6339$ & $2.7873$ & $2.9175$ & $3.0313$ &
  $3.1331$ & $3.2254$\\
  \hline
$Z_{V}$ & $0.85609(24)$ & $0.88162(19)$ & $0.88822(16)$ & $0.91166(12)$ & $0.92852(30)$ & $0.94719(13)$ & $0.96359(12)$\\ 
   \hline
$Z_{VA;1}^+$ & $0.8285(11)$ & $0.9344(15)$ & $0.9645(19)$ & $1.0392(21)$ & $1.0951(34)$ & $1.1470(37)$ & $1.1880(44)$\\ 
$Z_{VA;2}^+$ & $0.8546(15)$ & $0.9675(23)$ & $1.0160(30)$ & $1.0983(32)$ & $1.1698(53)$ & $1.2218(59)$ & $1.270(18)$\\ 
$Z_{VA;3}^+$ & $0.8785(13)$ & $1.0047(19)$ & $1.0450(24)$ & $1.1320(27)$ & $1.1985(44)$ & $1.2573(49)$ & $1.3058(58)$\\ 
$Z_{VA;4}^+$ & $0.8134(13)$ & $0.9077(18)$ & $0.9479(24)$ & $1.0191(25)$ & $1.0811(41)$ & $1.1276(45)$ & $1.1703(54)$\\ 
$Z_{VA;5}^+$ & $0.8098(13)$ & $0.9036(19)$ & $0.9431(25)$ & $1.0141(26)$ & $1.0758(42)$ & $1.1218(47)$ & $1.163(16)$\\ 
$Z_{VA;6}^+$ & $0.7582(21)$ & $0.8317(26)$ & $0.8634(28)$ & $0.9169(35)$ & $0.9741(53)$ & $1.0166(56)$ & $1.044(31)$\\ 
$Z_{VA;7}^+$ & $0.8099(18)$ & $0.9093(23)$ & $0.9375(24)$ & $1.0029(31)$ & $1.0612(50)$ & $1.1138(50)$ & $1.1467(66)$\\ 
$Z_{VA;8}^+$ & $0.7507(20)$ & $0.8200(25)$ & $0.8507(26)$ & $0.9024(33)$ & $0.9568(48)$ & $0.9985(51)$ & $1.0254(66)$\\ 
$Z_{VA;9}^+$ & $0.7475(21)$ & $0.8164(26)$ & $0.8463(27)$ & $0.8970(35)$ & $0.9541(52)$ & $0.9930(55)$ & $1.020(28)$\\ 
   \hline
$Z_{VA;1}^-$ & $0.72775(80)$ & $0.7558(12)$ & $0.7373(14)$ & $0.7628(16)$ & $0.7704(25)$ & $0.7942(26)$ & $0.8101(30)$\\ 
$Z_{VA;2}^-$ & $0.7929(17)$ & $0.8293(23)$ & $0.8331(29)$ & $0.8694(29)$ & $0.8956(48)$ & $0.9139(53)$ & $0.938(13)$\\ 
$Z_{VA;3}^-$ & $0.8232(10)$ & $0.8825(16)$ & $0.8713(19)$ & $0.9132(20)$ & $0.9293(33)$ & $0.9600(36)$ & $0.9832(41)$\\ 
$Z_{VA;4}^-$ & $0.7223(13)$ & $0.7377(18)$ & $0.7320(22)$ & $0.7568(22)$ & $0.7748(35)$ & $0.7893(38)$ & $0.8058(43)$\\ 
$Z_{VA;5}^-$ & $0.7213(13)$ & $0.7355(18)$ & $0.7314(22)$ & $0.7555(21)$ & $0.7745(34)$ & $0.7879(38)$ & $0.806(10)$\\ 
$Z_{VA;6}^-$ & $0.7038(24)$ & $0.7111(30)$ & $0.7061(29)$ & $0.7285(36)$ & $0.7408(56)$ & $0.7592(56)$ & $0.773(23)$\\ 
$Z_{VA;7}^-$ & $0.7593(16)$ & $0.7991(21)$ & $0.7803(20)$ & $0.8126(28)$ & $0.8214(45)$ & $0.8479(44)$ & $0.8662(55)$\\ 
$Z_{VA;8}^-$ & $0.6669(22)$ & $0.6649(26)$ & $0.6555(25)$ & $0.6710(33)$ & $0.6802(49)$ & $0.6971(50)$ & $0.7054(56)$\\ 
$Z_{VA;9}^-$ & $0.6663(22)$ & $0.6636(27)$ & $0.6546(26)$ & $0.6708(33)$ & $0.6811(49)$ & $0.6959(50)$ & $0.706(19)$\\ 
  \hline
$Z_{B_K;1}$ & $1.1304(12)$ & $1.2021(19)$ & $1.2225(24)$ & $1.2504(25)$ & $1.2702(40)$ & $1.2784(42)$ & $1.2795(47)$\\ 
$Z_{B_K;2}$ & $1.1660(18)$ & $1.2447(28)$ & $1.2878(37)$ & $1.3215(38)$ & $1.3568(63)$ & $1.3619(66)$ & $1.368(19)$\\ 
$Z_{B_K;3}$ & $1.1987(15)$ & $1.2927(24)$ & $1.3246(31)$ & $1.3621(32)$ & $1.3902(52)$ & $1.4014(55)$ & $1.4064(63)$\\ 
$Z_{B_K;4}$ & $1.1098(15)$ & $1.1679(23)$ & $1.2015(30)$ & $1.2262(30)$ & $1.2540(48)$ & $1.2568(50)$ & $1.2604(59)$\\ 
$Z_{B_K;5}$ & $1.1049(16)$ & $1.1626(23)$ & $1.1954(31)$ & $1.2202(31)$ & $1.2479(50)$ & $1.2503(53)$ & $1.253(17)$\\ 
$Z_{B_K;6}$ & $1.0351(24)$ & $1.0703(32)$ & $1.0945(34)$ & $1.1032(42)$ & $1.1321(62)$ & $1.1333(62)$ & $1.124(33)$\\ 
$Z_{B_K;7}$ & $1.1056(20)$ & $1.1701(27)$ & $1.1885(30)$ & $1.2067(37)$ & $1.2334(58)$ & $1.2416(56)$ & $1.2347(72)$\\ 
$Z_{B_K;8}$ & $1.0248(22)$ & $1.0553(30)$ & $1.0784(31)$ & $1.0859(39)$ & $1.1121(56)$ & $1.1131(57)$ & $1.1041(71)$\\ 
$Z_{B_K;9}$ & $1.0204(24)$ & $1.0506(32)$ & $1.0729(33)$ & $1.0794(41)$ & $1.1089(60)$ & $1.1070(61)$ & $1.099(31)$\\ 
  \hline
$Z_{P}$ & $0.70476(69)$ & $0.68054(91)$ & $0.6631(10)$ & $0.6616(10)$ & $0.6620(16)$ & $0.6624(16)$ & $0.6639(18)$\\
$Z_m$ & $1.4189(14)$ & $1.4694(20)$ & $1.5080(23)$ & $1.5115(24)$ & $1.5106(37)$ & $1.5096(37)$ & $1.5063(42)$\\
   \hline
  \end{tabular}
  \caption{Numerical values of $Z_V(g_0)$, 
  $Z_{VA+AV;s}^\pm(g_0,a\mu_{\rm min})$ and
  $Z_{B_K;s}(g_0,a\mu_{\rm min})$ at $2L_{\rm max}$ with the tree level
  improved renormalization condition \eqn{eq:treeimp}.
  Values of $Z_P(g_0,a\mu_{\rm min})$ and $Z_m(g_0,a\mu_{\rm min})$ are
  also listed.}
  \label{tab:zvaav.imp}
 \end{center}
\end{table} 

\begin{table}[htb]
\begin{center}
 \begin{tabular}{|c|c|c|c|c||c|}
  \hline
  $N_L$ & $4$ & $6$ & $8$ & $10$ & $20$\\
  \hline
  $\beta$ & $2.6339$ & $2.9175$ & $3.1331$ & $3.3103$ & $3.3103$ \\
  \hline
$Z_{V}$ & $1.1071(24)$ & $0.9756(11)$ & $1.01320(80)$ & $1.00782(59)$ & $0.97970(26)$\\ 
\hline
$Z_{VA;1}^+$ & $1.3728(88)$ & $1.0330(63)$ & $1.1844(64)$ & $1.1716(76)$ & $1.261(11)$\\ 
$Z_{VA;2}^+$ & $1.3115(94)$ & $1.0493(83)$ & $1.1820(91)$ & $1.1933(98)$ & $1.352(16)$\\ 
$Z_{VA;3}^+$ & $1.4004(92)$ & $1.0693(72)$ & $1.2361(72)$ & $1.2289(90)$ & $1.380(14)$\\ 
$Z_{VA;4}^+$ & $1.2803(89)$ & $1.0162(73)$ & $1.1318(80)$ & $1.1395(82)$ & $1.249(12)$\\ 
$Z_{VA;5}^+$ & $1.2827(90)$ & $1.0167(74)$ & $1.1338(84)$ & $1.1394(82)$ & $1.243(12)$\\ 
$Z_{VA;6}^+$ & $1.2753(89)$ & $1.0026(71)$ & $1.1173(79)$ & $1.1190(79)$ & $1.1209(91)$\\ 
$Z_{VA;7}^+$ & $1.3744(88)$ & $1.0373(64)$ & $1.1906(64)$ & $1.1774(77)$ & $1.2175(89)$\\ 
$Z_{VA;8}^+$ & $1.2566(85)$ & $0.9858(66)$ & $1.0901(74)$ & $1.0917(71)$ & $1.1022(86)$\\ 
$Z_{VA;9}^+$ & $1.2590(87)$ & $0.9863(67)$ & $1.0920(77)$ & $1.0916(72)$ & $1.0968(89)$\\ 
\hline
$Z_{VA;1}^-$ & $1.3505(80)$ & $0.9618(40)$ & $1.0498(53)$ & $1.0092(62)$ & $0.8273(68)$\\ 
$Z_{VA;2}^-$ & $1.251(10)$ & $1.0083(90)$ & $1.056(10)$ & $1.063(11)$ & $0.973(15)$\\ 
$Z_{VA;3}^-$ & $1.4175(88)$ & $1.0325(55)$ & $1.1496(65)$ & $1.1156(79)$ & $0.997(10)$\\ 
$Z_{VA;4}^-$ & $1.1995(94)$ & $0.9486(75)$ & $0.9766(85)$ & $0.9751(85)$ & $0.844(11)$\\ 
$Z_{VA;5}^-$ & $1.2010(94)$ & $0.9539(75)$ & $0.9815(85)$ & $0.9791(81)$ & $0.843(11)$\\ 
$Z_{VA;6}^-$ & $1.2169(98)$ & $0.9634(79)$ & $0.9978(89)$ & $0.9968(87)$ & $0.807(11)$\\ 
$Z_{VA;7}^-$ & $1.3913(84)$ & $1.0017(48)$ & $1.1072(58)$ & $1.0688(70)$ & $0.8799(73)$\\ 
$Z_{VA;8}^-$ & $1.1773(90)$ & $0.9203(69)$ & $0.9407(79)$ & $0.9342(76)$ & $0.7448(94)$\\ 
$Z_{VA;9}^-$ & $1.1788(90)$ & $0.9254(70)$ & $0.9453(80)$ & $0.9381(72)$ & $0.7443(93)$\\ 
\hline
$Z_{B_K;1}$ & $1.1200(52)$ & $1.0854(55)$ & $1.1537(62)$ & $1.1535(74)$ & $1.314(11)$\\ 
$Z_{B_K;2}$ & $1.0700(60)$ & $1.1025(77)$ & $1.1514(89)$ & $1.1749(98)$ & $1.408(17)$\\ 
$Z_{B_K;3}$ & $1.1425(56)$ & $1.1234(65)$ & $1.2041(72)$ & $1.2099(88)$ & $1.437(15)$\\ 
$Z_{B_K;4}$ & $1.0446(55)$ & $1.0677(67)$ & $1.1025(77)$ & $1.1218(81)$ & $1.301(13)$\\ 
$Z_{B_K;5}$ & $1.0465(56)$ & $1.0682(68)$ & $1.1044(81)$ & $1.1218(82)$ & $1.295(13)$\\ 
$Z_{B_K;6}$ & $1.0404(55)$ & $1.0534(64)$ & $1.0884(76)$ & $1.1017(77)$ & $1.1678(95)$\\ 
$Z_{B_K;7}$ & $1.1213(52)$ & $1.0898(56)$ & $1.1597(62)$ & $1.1592(75)$ & $1.2685(93)$\\ 
$Z_{B_K;8}$ & $1.0252(52)$ & $1.0358(60)$ & $1.0619(70)$ & $1.0748(70)$ & $1.1483(89)$\\ 
$Z_{B_K;9}$ & $1.0271(53)$ & $1.0363(60)$ & $1.0637(74)$ & $1.0748(70)$ & $1.1427(93)$\\ 
  \hline
$Z_P$ & $0.9745(41)$ & $0.8598(36)$ & $0.8443(37)$ & $0.8315(38)$ & $0.6717(53)$\\
$Z_m$ & $1.0262(43)$ & $1.1631(49)$ & $1.1844(52)$ & $1.2027(55)$ & $1.489(12)$\\
  \hline
 \end{tabular}
 \caption{Numerical values of $Z_V(g_0)$,
 $Z_{VA+AV;s}^\pm(g_0,2a\mu_{\rm min})$, $Z_{B_K;s}(g_0,2a\mu_{\rm min})$,
 $Z_P(g_0,2a\mu_{\rm min})$ and $Z_m(g_0,2a\mu_{\rm min})$
 at $L_{\rm max}$ for SSF with the tree
 level improved renormalization condition \eqn{eq:treeimp}.
 A box size of $N_L=20$ lattice is $2L_{\rm max}$.}
  \label{tab:zvaav.ssf.imp}
\end{center}
\end{table}

\begin{table}[htb]
 \begin{center}
  \begin{tabular}{|c|c|c|c|}
   \hline
scheme & $\sigma_{VA+AV;s}^+$ & $\sigma_{VA+AV;s}^-$ & $\sigma_{B_K}$ \\
  \hline
1 & $1.136(22)$ & $0.841(16)$ & $1.116(23)$\\ 
2 & $1.221(33)$ & $0.988(32)$ & $1.197(34)$\\ 
3 & $1.164(29)$ & $0.889(22)$ & $1.143(30)$\\ 
4 & $1.182(30)$ & $0.944(28)$ & $1.158(31)$\\ 
5 & $1.180(30)$ & $0.936(27)$ & $1.154(32)$\\ 
6 & $1.120(24)$ & $0.905(26)$ & $1.099(25)$\\ 
7 & $1.096(22)$ & $0.835(18)$ & $1.076(23)$\\ 
8 & $1.113(24)$ & $0.887(25)$ & $1.091(25)$\\ 
9 & $1.111(25)$ & $0.880(25)$ & $1.087(26)$\\ 
  \hline
  \end{tabular}
 \caption{Value of the SSF in the continuum limit $\sigma_{VA+AV;s}^\pm$ and
 $\sigma_{B_K;s}$ at $L_{\rm max}$, obtained by a combined linear fit of
  data at $M=1.8$ with the tree level improved
  renormalization condition and at $M=1.4$.}
  \label{tab:ssf.cont}
 \end{center}
\end{table}

\begin{table}[htb]
 \begin{center}
  \begin{tabular}{|c|lllllll|}
   \hline
$\beta$ & $2.444602$ & $2.633865$ & $2.787275$ & $2.917468$ & $3.031335$ & $3.133065$ & $3.225406$ \\
  \hline
${\cal Z}_{B_K;1}$ & $1.256(22)$ & $1.336(23)$ & $1.358(23)$ & $1.389(24)$ & $1.411(25)$ & $1.420(25)$ & $1.422(25)$\\ 
${\cal Z}_{B_K;2}$ & $1.252(28)$ & $1.337(30)$ & $1.383(31)$ & $1.419(32)$ & $1.457(33)$ & $1.463(33)$ & $1.469(39)$\\ 
${\cal Z}_{B_K;3}$ & $1.208(23)$ & $1.303(25)$ & $1.335(25)$ & $1.373(26)$ & $1.401(27)$ & $1.413(27)$ & $1.418(27)$\\ 
${\cal Z}_{B_K;4}$ & $1.321(27)$ & $1.390(28)$ & $1.430(29)$ & $1.459(30)$ & $1.492(31)$ & $1.496(31)$ & $1.500(31)$\\ 
${\cal Z}_{B_K;5}$ & $1.294(25)$ & $1.361(27)$ & $1.400(28)$ & $1.429(28)$ & $1.461(29)$ & $1.464(29)$ & $1.467(35)$\\ 
${\cal Z}_{B_K;6}$ & $1.361(25)$ & $1.407(26)$ & $1.439(27)$ & $1.451(27)$ & $1.489(28)$ & $1.490(28)$ & $1.478(51)$\\ 
${\cal Z}_{B_K;7}$ & $1.273(21)$ & $1.347(22)$ & $1.368(23)$ & $1.389(23)$ & $1.420(24)$ & $1.429(24)$ & $1.421(25)$\\ 
${\cal Z}_{B_K;8}$ & $1.392(26)$ & $1.433(27)$ & $1.464(27)$ & $1.475(28)$ & $1.510(29)$ & $1.512(29)$ & $1.499(29)$\\ 
${\cal Z}_{B_K;9}$ & $1.365(20)$ & $1.406(20)$ & $1.435(21)$ & $1.444(21)$ & $1.484(23)$ & $1.481(23)$ & $1.470(46)$\\ 
  \hline
  \end{tabular}
  \caption{Numerical values of renormalization factors
  $\mathcal{Z}_{B_K;s}(g_0)$ for the RGI $B_K$.}
 \label{tab:RGIZBK.mumin}
 \end{center}
\end{table} 

\begin{table}[htb]
 \begin{center}
  \begin{tabular}{|c|ccc|c|c|}
   \hline
   $\beta$ & $2.6$ & $2.9$ & $3.2$ & continuum & $\chi^2/{\rm dof}$\\
\hline
$\mathcal{Z}_{B_K;1}(g_0)$ & $1.314(13)$ & $1.389(13)$ & $1.422(18)$&&\\
$\mathcal{Z}_{B_K;2}(g_0)$ & $1.321(17)$ & $1.419(18)$ & $1.470(26)$&&\\
$\mathcal{Z}_{B_K;3}(g_0)$ & $1.279(14)$ & $1.373(14)$ & $1.417(20)$&&\\
$\mathcal{Z}_{B_K;4}(g_0)$ & $1.378(16)$ & $1.459(16)$ & $1.502(22)$&&\\
$\mathcal{Z}_{B_K;5}(g_0)$ & $1.350(15)$ & $1.429(16)$ & $1.470(23)$&&\\
$\mathcal{Z}_{B_K;6}(g_0)$ & $1.400(15)$ & $1.458(15)$ & $1.493(28)$&&\\
$\mathcal{Z}_{B_K;7}(g_0)$ & $1.327(13)$ & $1.395(13)$ & $1.425(18)$&&\\
$\mathcal{Z}_{B_K;8}(g_0)$ & $1.428(15)$ & $1.482(15)$ & $1.508(21)$&&\\
$\mathcal{Z}_{B_K;9}(g_0)$ & $1.399(12)$ & $1.452(12)$ & $1.486(23)$&&\\
\hline
$\wh{B}_{K;1}$ & $0.777(11)$ & $0.786(12)$ & $0.779(14)$ & $0.7830(91)$ & $0.14$\\
$\wh{B}_{K;2}$ & $0.781(12)$ & $0.802(14)$ & $0.806(18)$ & $0.804(11)$ & $0.03$\\
$\wh{B}_{K;3}$ & $0.756(11)$ & $0.776(13)$ & $0.776(15)$ & $0.776(98)$ & $-$\\
$\wh{B}_{K;4}$ & $0.814(12)$ & $0.825(14)$ & $0.823(16)$ & $0.824(11)$ & $0.009$\\
$\wh{B}_{K;5}$ & $0.797(12)$ & $0.808(13)$ & $0.805(17)$ & $0.807(10)$ & $0.019$\\
$\wh{B}_{K;6}$ & $0.827(12)$ & $0.825(13)$ & $0.818(18)$ & $0.823(10)$ & $0.10$\\
$\wh{B}_{K;7}$ & $0.784(11)$ & $0.789(12)$ & $0.781(14)$ & $0.7856(91)$ & $0.19$\\
$\wh{B}_{K;8}$ & $0.844(12)$ & $0.838(13)$ & $0.826(16)$ & $0.833(10)$ & $0.34$\\
$\wh{B}_{K;9}$ & $0.827(11)$ & $0.821(12)$ & $0.814(16)$ & $0.8185(96)$ & $0.12$\\
\hline
  \end{tabular}
  \caption{Renormalization factors $\mathcal{Z}_{B_K;s}(g_0)$ and RGI
  $\wh{B}_K$ at three $\beta$'s with their continuum extrapolations for
  schemes $s=1,\cdots,9$.
  Values of $\chi^2/{\rm dof}$ are also listed.}
  \label{tab:RGIbk}
 \end{center}
\end{table}


\begin{table}[htb]
 \begin{center}
  \begin{tabular}{|c|ccc|c|c|}
   \hline
   $\beta$ & $2.6$ & $2.9$ & $3.2$ & continuum & $\chi^2/{\rm dof}$\\
\hline
$Z_{B_K;1}^{\msbar}(g_0,2 {\rm GeV})$ & $0.9494(96)$ & $1.003(10)$ & $1.027(13)$&&\\
$Z_{B_K;2}^{\msbar}(g_0,2 {\rm GeV})$ & $0.954(12)$ & $1.025(13)$ & $1.062(19)$&&\\
$Z_{B_K;3}^{\msbar}(g_0,2 {\rm GeV})$ & $0.924(10)$ & $0.991(11)$ & $1.023(15)$&&\\
$Z_{B_K;4}^{\msbar}(g_0,2 {\rm GeV})$ & $0.995(12)$ & $1.054(12)$ & $1.085(16)$&&\\
$Z_{B_K;5}^{\msbar}(g_0,2 {\rm GeV})$ & $0.975(11)$ & $1.032(12)$ & $1.062(17)$&&\\
$Z_{B_K;6}^{\msbar}(g_0,2 {\rm GeV})$ & $1.011(11)$ & $1.053(11)$ & $1.078(20)$&&\\
$Z_{B_K;7}^{\msbar}(g_0,2 {\rm GeV})$ & $0.9582(94)$ & $1.0080(98)$ & $1.029(13)$&&\\
$Z_{B_K;8}^{\msbar}(g_0,2 {\rm GeV})$ & $1.032(11)$ & $1.070(12)$ & $1.089(16)$&&\\
$Z_{B_K;9}^{\msbar}(g_0,2 {\rm GeV})$ & $1.0108(90)$ & $1.0490(91)$ & $1.073(17)$&&\\
   \hline
$B_{K;1}^{\msbar}({\rm NDR},2{\rm GeV})$ & $0.5609(78)$ & $0.5675(89)$ & $0.563(10)$ & $0.5655(66)$ & $0.11$\\
$B_{K;2}^{\msbar}({\rm NDR},2{\rm GeV})$ & $0.5638(92)$ & $0.580(10)$ & $0.582(13)$ & $0.5807(79)$ & $0.015$\\
$B_{K;3}^{\msbar}({\rm NDR},2{\rm GeV})$ & $0.5458(80)$ & $0.5607(92)$ & $0.561(11)$ & $0.5608(71)$ & $0.0004$\\
$B_{K;4}^{\msbar}({\rm NDR},2{\rm GeV})$ & $0.5880(90)$ & $0.596(10)$ & $0.594(12)$ & $0.5952(77)$ & $0.016$\\
$B_{K;5}^{\msbar}({\rm NDR},2{\rm GeV})$ & $0.5759(87)$ & $0.5837(97)$ & $0.582(12)$ & $0.5830(75)$ & $0.012$\\
$B_{K;6}^{\msbar}({\rm NDR},2{\rm GeV})$ & $0.5974(88)$ & $0.5956(97)$ & $0.591(14)$ & $0.5941(80)$ & $0.073$\\
$B_{K;7}^{\msbar}({\rm NDR},2{\rm GeV})$ & $0.5661(78)$ & $0.5700(89)$ & $0.564(10)$ & $0.5673(66)$ & $0.20$\\
$B_{K;8}^{\msbar}({\rm NDR},2{\rm GeV})$ & $0.6095(89)$ & $0.6053(99)$ & $0.597(12)$ & $0.6019(76)$ & $0.28$\\
$B_{K;9}^{\msbar}({\rm NDR},2{\rm GeV})$ & $0.5972(78)$ & $0.5932(89)$ & $0.588(12)$ & $0.5914(71)$ & $0.12$\\
   \hline
  \end{tabular}
  \caption{Renormalization factors $Z_{B_K;s}^{\msbar}(g_0,2 {\rm GeV})$
  and renormalized ${B}_K$ in $\msbar$ scheme with NDR at three
  $\beta$'s with their continuum extrapolations for schemes $s=1,\cdots,9$.
  Values of $\chi^2/{\rm dof}$ are also listed.}
  \label{tab:MSbk}
 \end{center}
\end{table}

\begin{table}[htb]
 \begin{center}
   \begin{tabular}{|clllllll|}
  \hline
  \hline
      $L$   & $6$ & $8$ & $10$ & $12$ & $14$ & $16$ & $18$ \\
  \hline
   $\beta$ & $2.444602$ & $2.633865$ & $2.787275$ & $2.917468$ & $3.031335$ & $3.133065$ & $3.225406$ \\
  \hline
$Z_{VV+AA;1}^+$&0.7240(31)&0.8799(37)&0.9230(46)&1.0158(94)&1.077(10)&1.138(10)&1.176(13) \\
$Z_{VV+AA;2}^+$&0.7460(43)&0.9086(55)&0.9720(72)&1.072(14)&1.156(15)&1.213(17)&1.236(18) \\ 
$Z_{VV+AA;3}^+$&0.7669(37)&0.9475(47)&1.0012(58)&1.103(12)&1.178(12)&1.251(13)&1.284(16) \\
$Z_{VV+AA;4}^+$&0.7102(37)&0.8527(45)&0.9054(58)&0.997(11)&1.069(11)&1.118(13)&1.142(14) \\
$Z_{VV+AA;5}^+$&0.7071(38)&0.8473(45)&0.9007(60)&0.992(12)&1.061(12)&1.109(13)&1.138(14) \\
$Z_{VV+AA;6}^+$&0.6639(33)&0.7827(37)&0.8224(48)&0.9061(96)&0.9581(89)&1.001(11)&1.029(11) \\
$Z_{VV+AA;7}^+$&0.7095(29)&0.8578(34)&0.8957(41)&0.9865(84)&1.0390(75)&1.1008(91)&1.137(11) \\ 
$Z_{VV+AA;8}^+$&0.6570(32)&0.7720(35)&0.8099(45)&0.8912(88)&0.9431(80)&0.9838(97)&1.0103(99) \\ 
$Z_{VV+AA;9}^+$&0.6542(33)&0.7671(37)&0.8058(47)&0.8866(94)&0.9361(89)&0.976(10)&1.008(11) \\
\hline
$Z_{VV+AA;1}^-$ &0.6328(24)&0.7153(30)&0.7056(34)&0.7480(64)&0.7614(73)&0.7928(86)&0.8108(94) \\ 
$Z_{VV+AA;2}^-$ &0.6903(47)&0.7766(58)&0.7986(75)&0.848(14)&0.912(14)&0.909(15)&0.907(14) \\ 
$Z_{VV+AA;3}^-$ &0.7156(31)&0.8347(39)&0.8368(47)&0.8842(91)&0.9246(92)&0.964(11)&0.975(13) \\ 
$Z_{VV+AA;4}^-$ &0.6299(38)&0.6916(45)&0.7004(56)&0.741(11)&0.7914(96)&0.787(11)&0.7836(99) \\ 
$Z_{VV+AA;5}^-$ &0.6278(38)&0.6884(43)&0.6990(55)&0.743(11)&0.7829(97)&0.781(11)&0.7856(99) \\ 
$Z_{VV+AA;6}^-$ &0.6142(37)&0.6690(43)&0.6757(53)&0.716(11)&0.7559(92)&0.750(10)&0.7550(96) \\ 
$Z_{VA+AV;7}^-$ &0.6621(26)&0.7557(32)&0.7486(37)&0.7904(70)&0.8157(75)&0.8480(91)&0.863(10) \\ 
$Z_{VV+AA;8}^-$ &0.5828(34)&0.6261(37)&0.6266(46)&0.6626(94)&0.6982(76)&0.6928(94)&0.6935(85) \\ 
$Z_{VV+AA;9}^-$ &0.5808(34)&0.6233(38)&0.6253(46)&0.6645(98)&0.6906(81)&0.6873(95)&0.6952(88) \\ 
    \hline
    \hline
   \end{tabular}
  \caption{Numerical values of renormalization factors
  $Z_{VV+AA;s}^+(g_0,a\mu_{\rm min})$ and
  $Z_{VV+AA;s}^-(g_0,a\mu_{\rm min})$ for the parity even operator with
  the chirally rotated scheme.
  A number of configurations is $100$ on each lattice size.}
\label{tab:zvvaa1.mumin}
     \end{center}
   \end{table}

\begin{table}[htb]
 \begin{center}
  \begin{tabular}{|c|lllllll|}
   \hline
$\beta$ & $2.444602$ & $2.633865$ & $2.787275$ & $2.917468$ & $3.031335$ & $3.133065$ & $3.225406$ \\
  \hline
${\cal Z}_m$ & $1.642(17)$ & $1.700(18)$ & $1.745(18)$ & $1.749(18)$ & $1.748(19)$ & $1.747(19)$ & $1.743(19)$\\
  \hline
  \end{tabular}
  \caption{Numerical values of the renormalization factor
  $\mathcal{Z}_{m}(g_0)$ for the RGI quark mass.}
 \label{tab:RGIZm}
 \end{center}
\end{table} 

\begin{table}[htb]
 \begin{center}
  \begin{tabular}{|c|ccc|c|c|}
   \hline
   $\beta$ & $2.6$ & $2.9$ & $3.2$ & continuum & $\chi^2/{\rm dof}$\\
\hline
${\cal Z}_m$ & $1.694(10)$ & $1.749(10)$ & $1.742(13)$ & &\\
\hline
$\hat{m}_{ud}$ (MeV) & $0.68(58)$ & $3.23(87)$ & $8.11(90)$ & $12.2(15)$ & $2.0$\\
$\hat{m}_s(K)$ (MeV) & $141.9(33)$ & $145.5(37)$ & $149.4(27)$ & $154.8(52)$ & $0.036$\\
\hline
$\hat{m}_{ud}+\hat{m}_{\rm res}$ (MeV)& $5.60(12)$ & $5.64(13)$ & $5.60(10)$ & $5.613(66)$ & $0.033$\\
$\hat{m}_s(K)+\hat{m}_{\rm res}$ (MeV)& $146.9(31)$ & $147.9(34)$ & $146.9(26)$ & $147.1(17)$ & $0.033$\\
$\hat{m}_s(\phi)$ (MeV) & $193(16)$ & $183.0(88)$ & $187.8(49)$ & $187.1(41)$ & $0.20$\\
\hline
  \end{tabular}
  \caption{The renormalization factor $\mathcal{Z}_{m}(g_0)$ and RGI
  light quark masses at three $\beta$'s with their continuum
  linear extrapolations.
  Values of $\chi^2/{\rm dof}$ are also listed.}
  \label{tab:RGImass}
 \end{center}
\end{table}

\begin{table}[htb]
 \begin{center}
  \begin{tabular}{|c|ccc|c|c|}
   \hline
   $\beta$ & $2.6$ & $2.9$ & $3.2$ & continuum & $\chi^2/{\rm dof}$\\
\hline
$Z_m^{\msbar}$ (MeV)& $1.2153(72)$ & $1.2545(74)$ & $1.2497(96)$ & &\\
\hline
${m}_{ud}^{\msbar}$ (MeV) & $0.49(42)$ & $2.31(63)$ & $5.81(65)$ & $8.8(10)$ & $2.0$\\
${m}_s^{\msbar}(K)$ (MeV) & $101.8(23)$ & $104.4(26)$ & $107.2(19)$ & $111.0(37)$ & $0.036$\\
\hline
$\left({m}_{ud}+{m}_{\rm res}\right)^{\msbar}$ (MeV)& $4.018(86)$ & $4.047(94)$ & $4.020(72)$ & $4.026(48)$ & $0.033$\\
$\left({m}_s(K)+{m}_{\rm res}\right)^{\msbar}$ (MeV)& $105.3(23)$ & $106.1(25)$ & $105.4(19)$ & $105.6(12)$ & $0.033$\\
${m}_s^{\msbar}(\phi)$ (MeV)& $139(11)$ & $131.2(63)$ & $134.7(35)$ & $134.2(30)$ & $0.20$\\
\hline
  \end{tabular}
  \caption{The renormalization factor ${Z}_m^{\msbar}(g_0)$ and renormalized
  light quark masses in $\msbar$ scheme at three $\beta$'s with their
  continuum linear extrapolations.
  Values of $\chi^2/{\rm dof}$ are also listed.}
  \label{tab:MSmass}
 \end{center}
\end{table}

\clearpage

\begin{figure}
  \begin{center}
  \begin{tabular}{ll}
  \begin{minipage}{65mm}
    \begin{center}
      \scalebox{0.45}{\includegraphics{fig/zvvaa/06z8m.eps}}
    \end{center}
  \end{minipage}&
\hspace{0.35cm}
  \begin{minipage}{65mm}
    \begin{center}
      \scalebox{0.45}{\includegraphics{fig/zvvaa/08z8m.eps}}
    \end{center}
  \end{minipage}\\[3.cm]
  \begin{minipage}{65mm}
    \begin{center}
      \scalebox{0.45}{\includegraphics{fig/zvvaa/10z8m.eps}}
    \end{center}
  \end{minipage}&
\hspace{0.35cm}
  \begin{minipage}{65mm}
    \begin{center}
      \scalebox{0.45}{\includegraphics{fig/zvvaa/12z8m.eps}}
    \end{center}
  \end{minipage}\\[3.cm]
  \begin{minipage}{65mm}
    \begin{center}
      \scalebox{0.45}{\includegraphics{fig/zvvaa/14z8m.eps}}
    \end{center}
  \end{minipage}&
\hspace{0.35cm}
  \begin{minipage}{65mm}
    \begin{center}
      \scalebox{0.45}{\includegraphics{fig/zvvaa/16z8m.eps}}
    \end{center}
  \end{minipage}\\[3.cm]
  \begin{minipage}{65mm}
    \begin{center}
      \scalebox{0.45}{\includegraphics{fig/zvvaa/18z8m.eps}}
    \end{center}
  \end{minipage}&
   \end{tabular}
\end{center}
 \caption{A comparison between two renormalization factors
 $Z_{VA+AV;1}^-(g_0,a\mu_{\rm min})$(open circle) and
 $Z_{VV+AA;1}^-(g_0,a\mu_{\rm min})$(open up triangle) as
 a function of $x_0$ for the scheme $8$ at various lattice sizes.
}
     \label{fig:z8mzvvaa}
\end{figure}

\clearpage

\end{document}